\pdfoutput=1
\documentclass[fleqn,10pt]{wlscirep}
\usepackage[utf8]{inputenc}
\usepackage[T1]{fontenc} 
\usepackage{bm}
\usepackage{longtable}
\usepackage[english]{babel}
\usepackage{glossaries}
\usepackage{graphicx}
\usepackage{xcolor}
\usepackage{amsmath}
\newcommand{\angstrom}{\text{\normalfont\AA}}
\usepackage{amssymb}
\usepackage{multirow}
\usepackage{booktabs}
\usepackage{hyperref}
\usepackage{etoolbox}
\usepackage{textcomp}
\usepackage{xfrac}
\usepackage{caption}
\usepackage{subcaption}
\usepackage{chngcntr}
\usepackage[version=4]{mhchem}
\usepackage{listofitems}
\usepackage{xr} 

\definecolor{DARKMAGENTA}{HTML}{AF2F40}
\definecolor{DARKGRAY}{HTML}{555555}
\hypersetup{
    colorlinks=true,
    linkcolor=DARKMAGENTA,
    citecolor=DARKMAGENTA,
    urlcolor=DARKMAGENTA,
}

\usepackage{listings}
\newacronym{dft}{DFT}{density-functional theory}
\newacronym{eos}{EOS}{equation of state}
\newacronym{api}{API}{application programming interface}
\newacronym{gui}{GUI}{graphical user interface}
\newacronym{bcc}{BCC}{body-centered cubic}
\newacronym{fcc}{FCC}{face-centered cubic}
\newacronym{sc}{SC}{simple cubic}
\newacronym{hcp}{HCP}{hexagonal close-packed}
\newacronym{ae}{AE}{all-electron}

\newcommand\qe{\textsc{Quantum ESPRESSO}}
\newcommand\siesta{\textsc{Siesta}}
\newcommand\sirius{SIRIUS}
\newcommand\cptwok{CP2K}
\newcommand\fleur{FLEUR}
\newcommand\abinit{\textsc{Abinit}}
\newcommand\vasp{VASP}
\newcommand\bigdft{BigDFT}
\newcommand\castep{CASTEP}
\newcommand\wientwok{WIEN2k}
\newcommand\gpaw{GPAW}

\newcommand\qelong{\texttt{Quantum ESPRESSO@PW|SSSP-prec-v1.3}}
\newcommand\siestalong{\texttt{SIESTA@AtOrOptDiamond|PseudoDojo-v0.4}}
\newcommand\siriuslong{\texttt{SIRIUS/CP2K@PW|SSSP-prec-v1.2}}
\newcommand\cptwoklong{\texttt{CP2K/Quickstep@TZV2P|GTH}}
\newcommand\fleurlong{\texttt{FLEUR@LAPW+LO}}
\newcommand\abinitlong{\texttt{ABINIT@PW|PseudoDojo-v0.5}}
\newcommand\vasplong{\texttt{VASP@PW|GW-PAW54*}}
\newcommand\bigdftlong{\texttt{BigDFT@DW|HGH-K(Valence)}}
\newcommand\casteplong{\texttt{CASTEP@PW|C19MK2}}
\newcommand\wientwoklong{\texttt{WIEN2k@(L)APW+lo+LO}}
\newcommand\gpawlong{\texttt{GPAW@PW|PAW-v0.9.20000}}

\makeatletter
\newcommand*{\addFileDependency}[1]{
\typeout{(#1)}
\@addtofilelist{#1}
\IfFileExists{#1}{}{\typeout{No file #1.}}
}\makeatother

\newcommand*{\myexternaldocument}[1]{%
\externaldocument{#1}%
\addFileDependency{#1.tex}%
\addFileDependency{#1.aux}%
}

\newif\iftwofiles\twofilesfalse

\iftwofiles
\usepackage{xr-hyper}
\myexternaldocument{supplementary/supplementary-content}
\fi

\title{How to verify the precision of density-functional-theory implementations via reproducible and universal workflows}

\author[1]{Emanuele Bosoni}

\author[2]{Louis Beal}

\author[3]{Marnik Bercx}

\author[4]{Peter Blaha}

\author[5]{Stefan Bl\"ugel}

\author[5,6]{Jens Br\"oder}

\author[7,8,9]{Martin Callsen}

\author[7,8]{Stefaan Cottenier}

\author[2]{Augustin Degomme}

\author[1]{Vladimir Dikan}

\author[3]{Kristjan Eimre}

\author[10,11]{Espen Flage-Larsen}

\author[12]{Marco Fornari}

\author[1]{Alberto Garcia}

\author[2]{Luigi Genovese}

\author[13]{Matteo Giantomassi}

\author[3,14]{Sebastiaan P. Huber}

\author[5]{Henning Janssen}

\author[15]{Georg Kastlunger}

\author[16]{Matthias Krack}

\author[17,18]{Georg Kresse}

\author[19,20]{Thomas D. K\"uhne}

\author[8,21]{Kurt Lejaeghere}

\author[4]{Georg K. H. Madsen}

\author[17,18]{Martijn Marsman}

\author[3,16]{Nicola Marzari}

\author[5]{Gregor Michalicek}

\author[22]{Hossein Mirhosseini}

\author[23]{Tiziano M. A. M\"uller}

\author[13]{Guido Petretto}

\author[24,25]{Chris J. Pickard}

\author[13]{Samuel Ponc\'e}

\author[13]{Gian-Marco Rignanese}

\author[26]{Oleg Rubel}

\author[4,8]{Thomas Ruh}

\author[7,8,27]{Michael Sluydts}

\author[7,28]{Danny E.P.\ Vanpoucke}

\author[15]{Sudarshan Vijay}

\author[17,18]{Michael Wolloch}

\author[5]{Daniel Wortmann}

\author[29]{Aliaksandr V. Yakutovich}

\author[3,16]{Jusong Yu}

\author[3]{Austin Zadoks}

\author[30,31]{Bonan Zhu}

\author[3,16,*]{Giovanni Pizzi}

\affil[1]{Institut de Ci\`encia de Materials de Barcelona, ICMAB-CSIC, Campus UAB, 08193 Bellaterra, Spain} 
\affil[2]{Univ. Grenoble-Alpes, CEA, IRIG-MEM-L\_Sim, 38000 Grenoble, France} 
\affil[3]{Theory and Simulation of Materials (THEOS) and National Centre for Computational Design and Discovery of Novel Materials (MARVEL), \'Ecole Polytechnique F\'ed\'erale de Lausanne (EPFL), CH-1015 Lausanne, Switzerland} 
\affil[4]{Institute for Materials  Chemistry, Technical University of Vienna, Getreidemarkt 9/165-TC, A-1060 Vienna, Austria} 
\affil[5]{Peter Gr\"unberg Institut and Institute for Advanced Simulation, Forschungszentrum J\"ulich and JARA, D-52425 J\"ulich, Germany} 
\affil[6]{Institute for Advanced Simulation, Materials Data Science and Informatics (IAS-9), Forschungszentrum J\"ulich, D-52425 J\"ulich, Germany} 
\affil[7]{Department of Electromechanical, Systems and Metal Engineering, Ghent University, Belgium} 
\affil[8]{Center for Molecular Modeling (CMM), Ghent University, Belgium} 
\affil[9]{Institute of Atomic and Molecular Sciences, Academia Sinica, Taipei 10617, Taiwan} 
\affil[10]{Norwegian EuroHPC Competence Center, Sigma2 AS,
Norway}
\affil[11]{SINTEF Industry, Materials Physics, Oslo, Norway} 
\affil[12]{Department of Physics and Science of Advanced Materials Program, Central Michigan University, Mount Pleasant, Michigan 48859, USA} 
\affil[13]{Institut de la Mati\`ere Condens\'ee et des Nanosciences (IMCN), Universit\'e catholique de Louvain, Chemin des \'Etoiles~8, Louvain-la-Neuve 1348, Belgium} 
\affil[14]{National Centre of Competence in Research (NCCR) Catalysis, École Polytechnique Fédérale de Lausanne (EPFL), CH-1015, Lausanne, Switzerland} 
\affil[15]{Center for Catalysis Theory (Cattheory), Department of Physics, Technical University of Denmark (DTU), 2800 Kongens Lyngby, Denmark} 
\affil[16]{Laboratory for Materials Simulations (LMS), Paul Scherrer Institut (PSI), CH-5232 Villigen PSI, Switzerland} 
\affil[17]{University of Vienna, Faculty of Physics and Center for Computational Materials Science, Kolingasse 14-16, A-1090 Vienna, Austria} 
\affil[18]{VASP Software GmbH, Sensengasse 8, A-1090 Vienna, Austria} 
\affil[19]{Center for Advanced Systems Understanding (CASUS) and Helmholtz-Zentrum Dresden-Rossendorf, D-02826 Görlitz, Germany} 
\affil[20]{Paderborn Center for Parallel Computing (PC2) and Center for Sustainable Systems Design, University of Paderborn, D-33098 Paderborn, Germany} 
\affil[21]{OCAS NV/ArcelorMittal Global R\&D Gent, Pres.\ J.\ F.\ Kennedylaan 3, Zelzate B-9060, Belgium} 
\affil[22]{Dynamics of Condensed Matter, Chair of Theoretical Chemistry, University of Paderborn, D-33098 Paderborn, Germany} 
\affil[23]{HPE HPC EMEA Research Lab, CH-4051 Basel, Switzerland} 
\affil[24]{Department of Materials Science \& Metallurgy, University of Cambridge, 27 Charles Babbage Road, Cambridge CB3 0FS, United Kingdom} 
\affil[25]{Advanced Institute for Materials Research, Tohoku University 2-1-1 Katahira, Aoba, Sendai, 980-8577, Japan} 
\affil[26]{Department of Materials Science and Engineering, McMaster University, 1280 Main Street West, Hamilton, Ontario L8S 4L8, Canada} 
\affil[27]{ePotentia, Frans van Dijckstraat 59, 2100 Deurne Antwerpen, Belgium} 
\affil[28]{Institute for Materials Research (IMO-IMOMEC), UHasselt - Hasselt University, Belgium} 
\affil[29]{Swiss Federal Laboratories for Materials Science and Technology (Empa),  nanotech@surfaces laboratory, CH-8600 D\"ubendorf, Switzerland} 
\affil[30]{Department of Chemistry, University College London, 20 Gordon St, Bloomsbury, London WC1H 0AJ, United Kingdom} 
\affil[31]{The Faraday Institution, Didcot OX11 0RA, United Kingdom} 
\affil[*]{e-mail: giovanni.pizzi@psi.ch}

\begin{abstract}
In the past decades many density-functional theory methods and codes adopting periodic boundary conditions have been developed and are now extensively used in condensed matter physics and materials science research. 
Only in 2016, however, their precision (i.e., to which extent properties computed with different codes agree among each other) was systematically assessed on elemental crystals: a first crucial step to evaluate the reliability of such computations. 
We discuss here general recommendations for verification studies aiming at further testing precision and transferability of density-functional-theory computational approaches and codes. 
We illustrate such recommendations using a greatly expanded protocol covering the whole periodic table from Z=1 to 96 and characterizing 10 prototypical cubic compounds for each element: 4 unaries and 6 oxides, spanning a wide range of coordination numbers and oxidation states. The primary outcome is a reference dataset of 960 equations of state cross-checked between two all-electron codes, then used to verify and improve nine pseudopotential-based approaches. Such effort is facilitated by deploying AiiDA common workflows that perform automatic input parameter selection, provide identical input/output interfaces across codes, and ensure full reproducibility. Finally, we discuss the extent to which the current results for total energies can be reused for different goals (e.g., obtaining formation energies).
\end{abstract}
\begin{document}

\flushbottom
\maketitle

\keywords{DFT, verification, pseudopotentials, automation, equation of state}

\thispagestyle{empty}

\noindent \textbf{Key points:} 
\begin{itemize}
\item Verification efforts are critical to assess the reliability of density-functional theory (DFT) simulations and provide results with properly quantified uncertainties.
\item Developing standard computation protocols to perform verification studies and publishing curated and FAIR reference datasets can significantly facilitate their use to improve codes and computational approaches.
\item The use of fully automated workflows with common interfaces between codes can guarantee uniformity, transferability, and reproducibility of results.
\item A careful description of the numerical and methodological details needed to compare with the reference datasets is essential; we discuss and illustrate this point with a dataset of 960 all-electron equations of state.
\item Reference datasets should always include an explanation of the target property for which they were generated, and a discussion of their limits of applicability.
\item Further extensions of DFT verification efforts are needed to cover more functionals, more computational approaches, and the treatment of magnetic and relativistic (spin-orbit) effects. They should also aim at concurrently delivering optimized protocols that, not only target ultimate precision, but also optimize the computational cost for a target accuracy.
\end{itemize}

\noindent \textbf{Website summary:} Verification efforts of DFT calculations are of crucial importance to evaluate the reliability of simulation results.  We discuss general recommendations for performing such studies and illustrate them with an all-electron reference dataset of 960 equations of state covering the whole periodic table (hydrogen to curium). The importance of verification for the improvement of pseudopotential codes is also demonstrated. 

\vspace{2cm}

The fast improvement of hardware, methods, and tools for \gls{dft} calculations in periodic boundary conditions has greatly advanced the field  of condensed matter physics and computational materials science, paving the way for an effective use of the ``materials design process'' that accelerates the discovery, development and deployment of new materials thanks to the aid of simulations\cite{Alberi_2019,Marzari2021}. Efficient software infrastructures\cite{Pizzi:2016, Huber:2020, ONG2013314, CPE:CPE3505, atomate, ASE1, ASE2, CURTAROLO2012218, pyiron-paper, HTTK-Armiento2020, Gonze:2020}  facilitate, nowadays, large high-throughput calculations of a panoply of material properties which are often made available to the public in large repositories\cite{mat-project, CMR, OQMD-Kirklin2015, TCOD-Merkys2017, NREL-MatDB, Talirz:2020, NOMAD-Draxl2018, CURTAROLO2012227}.
Most datasets aspire to be findable, accessible, interoperable, and reusable (FAIR)\cite{Wilkinson:2016} in order to accelerate materials discovery, possibly with the aid of machine learning.
They are queryable with {\it ad hoc} \glspl{api} or, for many of them, via a single common \gls{api} thanks to the recent efforts of the OPTIMADE\cite{Andersen_2021} consortium.
However, full integration of different data is often limited by considerations related to uncertainty quantification\cite{ieee-8055462,Wang2021,Carbogno2022,Alberi_2019, PONCE2014341}. In this work, we discuss recommendations on how to quantify to which extent properties (total energies and derived quantities) obtained by different \gls{dft} codes agree among each other.

In principle, \gls{dft} applies the fundamental laws of quantum physics to predict properties of a material, with no other inputs than the chemical composition and the crystal structure.
In reality, the electronic-structure calculations involve a variety of choices to solve the equations prescribed by \gls{dft} and introduce several levels of approximation.
Those choices, reflected in the resulting data, range from the specific flavor of \gls{dft} (e.g., the approach used for the exchange-correlation functional) to the discretization assumptions (e.g., the basis set), to the specific computational parameters needed by the codes.
Some approaches are more reliable, and therefore often slower, while others make more substantial approximations in order to gain computational speed and enable the study of systems with more atoms.
Furthermore, even when formally the same choices have been made in different codes, these may provide slightly different results due to the details of their implementations.
The importance of verifying the precision of codes has been long recognized~\cite{popleNobel}.
Despite this, when considering \gls{dft} codes adopting periodic boundary conditions, a first systematic assessment of their precision was performed in 2016, where the consistency of 40 computational approaches was assessed by calculating the \gls{eos} (i.e., the energy-versus-volume curve) for a test set of 71 elemental crystals\cite{Lejaeghere:2016,deltasite}.
This so-called ``$\Delta$-project'' led to the conclusion that the mainstream codes were in very good agreement with each other, which was not the case a decade before.
Despite being already a large project by itself, the ``$\Delta$-project'' was only the first step towards a careful verification of \gls{dft} calculations, which requires a much larger diversity of structural and chemical variables, as also discussed in the outlook of Ref.~\citenum{Lejaeghere:2016}.

In this Expert Recommendation, we list a set of guiding principles to perform new verification studies of \gls{dft} calculations (see Box 1), as well as a recommendation (see Box 2) for users of \gls{dft} codes, encouraging them to refer to quantitative sources on the reliability and precision of the codes and computational approaches used in their publications.
In order to illustrate these recommendations, 
we create a curated reference set of highly converged results for the \gls{eos} of 960 crystals, using two independent state-of-the-art \gls{ae} \gls{dft} codes (\fleur{}\cite{fleurCode,fleurSource} and \wientwok\cite{WIEN2k,WIEN20}). 
These 960 crystals cover all elements and a wide variety of structural and chemical environments in the form of four unary compounds and six oxides. The resulting data are shared on the Materials Cloud\cite{Talirz:2020} according to the FAIR\cite{Wilkinson:2016} principles.
A key feature of our work is that the thousands of computations performed are implemented within a reproducible and automatic infrastructure. Specifically, the launching and management of all the \gls{dft} calculations is carried out using AiiDA\cite{Pizzi:2016,Huber:2020,Uhrin:2021}. The choice of code-specific inputs and numerical parameters (called ``protocols'' in the following) are implemented in the publicly available {\tt aiida-common-workflows} (ACWF) package\cite{Huber2021,acwf} together with a number of error handlers to recover automatically from typical failure modes of each code. This setup enables to easily generate new datasets and to extend the current work for the verification of other computational approaches (see also Box 4). 

 As we discuss later, some choices of numerical parameters (such as the smearing type and size, or the k-point integration mesh) must be performed consistently in order to make correct use of the dataset. The suggestions regarding how to use our reference dataset are summarized in Box 3. One of our recommendations for verification efforts is to develop metrics to quantify discrepancies between codes that depend on physically measurable quantities. We implement this recommendation by defining two new metrics (in addition to the $\Delta$ metric introduced in Ref.~\citenum{Lejaeghere:2014}) to facilitate quantitative comparison of \gls{eos} results for pairs of codes or computational approaches, and we discuss their benefits. 
 
Using these metrics, we then compare the \gls{eos} results of our reference dataset to the results obtained by a number of pseudopotential codes.
The latter are designed to enhance computational efficiency by considering explicitly only ``valence'' electrons, which contribute to bonding\cite{GPP,RMARTIN1, COHENLOUIE}.
 The codes considered here are:
\abinit{}\cite{Gonze:2016,Romero:2020,Gonze:2020},
\bigdft\cite{Ratcliff2020},
\castep{}\cite{Clark:2005},
\cptwok{}\cite{cp2k,Kuehne:2020},
\gpaw{}\cite{GPAW1,GPAW2},
\qe{}\cite{Giannozzi:2009,Giannozzi:2017},
\siesta{}\cite{Soler:2002,Garcia:2020}, the \sirius{}\cite{sirius} library (via its \cptwok{} interface)\ and
\vasp{}\cite{Kresse:1996,Kresse:1999}.
The numerical basis sets implemented in these codes include plane waves, Gaussians combined with plane waves, Daubechies wavelets, and atomic orbitals.
For this reason, we do not label our results simply with the code name, but with a short string also indicating a few additional relevant parameters to better specify the details of the computational approach.
We stress that the aim of this study is not to provide a ranking or to evaluate the quality of different codes, but to illustrate with a few examples the value of curated datasets generated following our recommendations. In particular, we illustrate its use to improve existing pseudopotentials and to assess the consistency of results of several computational approaches to compute the \gls{eos} within \gls{dft}.

Finally, in our Outlook, we discuss a set of recommendations (summarized in Box 4) on future extensions of verification efforts.
On the one hand, we suggest to cover more exchange--correlation functionals, computational approaches, and treatment of magnetic and relativistic (spin-orbit) effects.
On the other hand, we highlight how future studies should  not only target ultimate precision, but also aim at delivering protocols that optimize the computational cost for a target accuracy.
We stress that, in this Expert Recommendation, we limit all discussions to verification efforts: i.e., investigating code precision, that is, how codes reproduce the ideal theoretical results given by \gls{dft} (e.g., with a given choice of exchange--correlation functional).
We do not discuss validation, i.e., accuracy with respect to the experimental results. While this is also an highly relevant topic (and we briefly mention it in Box 2), it is beyond the scope of this Expert Recommendation.

\begin{figure}[t]
\noindent\fbox{%
    \parbox{\textwidth}{
{\bfseries Expert Recommendation Box 1, Summary of recommendations to perform verification studies of DFT calculations} 
\par\noindent\rule{\linewidth}{0.4pt}
\begin{itemize}
    \item Quantitatively estimate the precision of DFT computational approaches and implementations with respect to exact numerical results.
    Provide adequate details of the verification protocols to ensure reproducibility of the results and a correct reuse in data-driven research, e.g., clarifying their range of applicability and specifying which parameters need to be fixed --- independent of the approach --- to ensure comparable results.
    \item Develop fully automated workflows to guarantee uniformity and transferability of parameters between computational approaches.
    This includes the definition and use of ``standard protocols'', i.e., automated selection of numerical parameters --- often specific to each computational approach --- that can ensure numerically precise results.
    \item Publish curated reference datasets from systematic verification studies.
    Facilitate their use to improve other codes by making the datasets FAIR: findable and accessible on open repositories, interoperable by using standard formats and clear annotations, and reusable by specifying all parameters needed to reproduce the results.
    See Box 3 for an example.
    \item Organize the reference data in appropriate subsets by recognizing the diversity of focus and the non-uniform capabilities of available computational approaches (e.g., if some systems require additional effort to be supported by all codes). 
\end{itemize}
}}
\end{figure}

\begin{figure}[t]
\noindent\fbox{%
    \parbox{\textwidth}{
{\bfseries Expert Recommendation Box 2, Summary of recommendations for users of DFT codes} 
\par\noindent\rule{\linewidth}{0.4pt}
\begin{itemize}
\item When publishing research that makes use of DFT codes, refer as much as possible to quantitative sources that document the precision of the numerical implementation (all-electron vs. pseudopotential, basis-set type and size, \ldots).
\item Equally important is a validation statement that refers to the accuracy of the chosen exchange--correlation functional to correctly and accurately address the physics at hand. Note that, however, this is beyond the scope of the current Expert Recommendation focusing on the precision of numerical implementations.
\item Always cite the exact pseudopotentials that are used in published simulations, including the exchange--correlation functional, the library from which they were obtained and the exact library version, together with all the essential numerical parameters of the calculations (e.g., k-point integration mesh and smearing, basis set type and size or plane-wave cutoffs, \ldots). Lack of this information results in essentially non-reproducible simulations.
\end{itemize}
}}
\end{figure}

\section*{AE reference dataset for EOS parameters}\label{aec}

In this section we discuss our reference dataset of \gls{eos} calculations, that we use to illustrate, with a practical example, how to implement the recommendations of Box 1.
The results are obtained with the \gls{ae} codes \fleur{} and \wientwok{}, using the PBE\cite{Perdew:1996} exchange-correlation functional. The two codes use the linearized augmented plane waves plus local orbitals method, but differ in details of the basis set and some computational setup parameters.  

\subsection*{Crystal-structures dataset}
We compute the \gls{eos} on a dataset of 960 cubic crystal structures.
In order to provide a chemically comprehensive dataset, we consider all elements in the periodic table from $Z=1$ (hydrogen) to $Z=96$ (curium).
Furthermore, we systematically scan structural diversity and investigate the transferability to more complex chemical environments by examining, for each element, 4 mono-elemental cubic crystals (``unaries dataset'') and six cubic oxides (``oxides dataset'').

Specifically, the unaries dataset considers all elements in the \gls{fcc}, \gls{bcc}, \gls{sc} and diamond crystal structure, thus covering a wide range of coordination numbers (12, 8, 6, and 4, respectively); a total of 384 systems. More information about the crystal structures are presented in the Supplementary Information (SI) Table~\ref{unaries-table}.
The oxides dataset is composed of six cubic oxides for each of the 96 elements X, with chemical formula X$_2$O, XO, X$_2$O$_3$, XO$_2$, X$_2$O$_5$ and XO$_3$, thus totaling $576$ additional structures, whose crystal structures are detailed in SI Table~\ref{oxides-table}. The oxide stoichiometries are chosen such that the formal oxidation state of the element considered varies from +1 to +6.
We note that the actual oxidation state is typically different (see the discussion with a Hirshfeld-I\cite{VanpouckeDannyEP:2013aJComputChem, VanpouckeDannyEP:2013bJComputChem, BultinckHI2007} analysis in SI Sec.~\ref{SI:sec-hirshfeld}), but shows a good correlation with calculated charges being on average about half the formal oxidation state. The X-O distance varies rather systematically over these 6 oxides (see SI Fig.~\ref{sifig:first-neigh-dist}), typically with the smallest distance for XO$_3$ and the largest for X$_2$O. This indicates that XO$_3$ could be a proxy for systems with very short bond lengths, such as in high-pressure studies. 
The two datasets of unaries and oxides (jointly called ``full dataset''; a total of 960 systems) complement each other in covering chemical and structural variety for each element.

In addition to the criteria above, all structures have been chosen to be cubic and such that forces on all atoms are zero by symmetry. Therefore, the only free parameter is the unit cell volume $V$ or, equivalently, the lattice parameter. As a consequence, the \gls{eos} results can be compared with any code able to compute total energies, with no requirement on the capability of computing forces and stresses.

It is important to note that most structures are not stable in nature (in particular under our constraint of cubic spacegroup symmetry).
Still, they can be used to assess that all codes reproduce the same \gls{dft} result, with the advantage of providing a consistent set across the whole periodic table.

\subsection*{Computation of EOS parameters and comparison between AE codes}

The \gls{eos} has been traditionally used to determine the computational parameters and study convergence of \gls{dft} calculations. By fitting the \gls{dft} energy vs.\@ cell volume to an \gls{eos}, it is possible to extract the theoretical predictions of the equilibrium volume $V_0$, the bulk modulus $B_0$, and its derivative with respect to the pressure, $B_1$. The Birch--Murnaghan \gls{eos}\cite{Birch1947}
\begin{equation}
E(V) = E_0 + \frac{9V_0B_0}{16}
\left\{
\left[\left(\frac{V_0}{V}\right)^\frac{2}{3}-1\right]^3B_1+ \left[\left(\frac{V_0}{V}\right)^\frac{2}{3}-1\right]^2
\left[6-4\left(\frac{V_0}{V}\right)^\frac{2}{3}\right]\right\}
\label{b-m}
\end{equation}
was used in the $\Delta$-project\cite{Lejaeghere:2016,deltasite} and we follow the same approach by performing a fit of $E(V)$ of Eq.~\eqref{b-m} using calculations of the total energy corresponding to 7 equidistant constant volumes between 94\% and 106\% of a reference central volume $\tilde V_0$ (for each Structure). We emphasize that the results are quite sensitive to the precise choice of volume range, reference central volume, and even of fitting algorithm, as we discuss in SI Sec.~\ref{SI:stability_eos}.
In this work, the reference central volumes $\tilde V_0$ for each of the 960 crystals have been chosen after an iterative process of performing more and more accurate simulations with the two AE codes considered here, until the difference between the reference central volume and the equilibrium volume of the EOS fit was smaller than the 2\% volume spacing between total-energy calculations. These central reference volumes are tabulated in SI Sec.~\ref{SIsec:structures} and the corresponding crystal-structure files are available in the data entry associated to this Recommendation\cite{MCA-ACWF}. These volumes have no physical significance, but for precise comparison between computational approaches, each of them should use the same reference volumes.

\begin{figure*}[tb]
\centering
 \includegraphics[width=\linewidth] {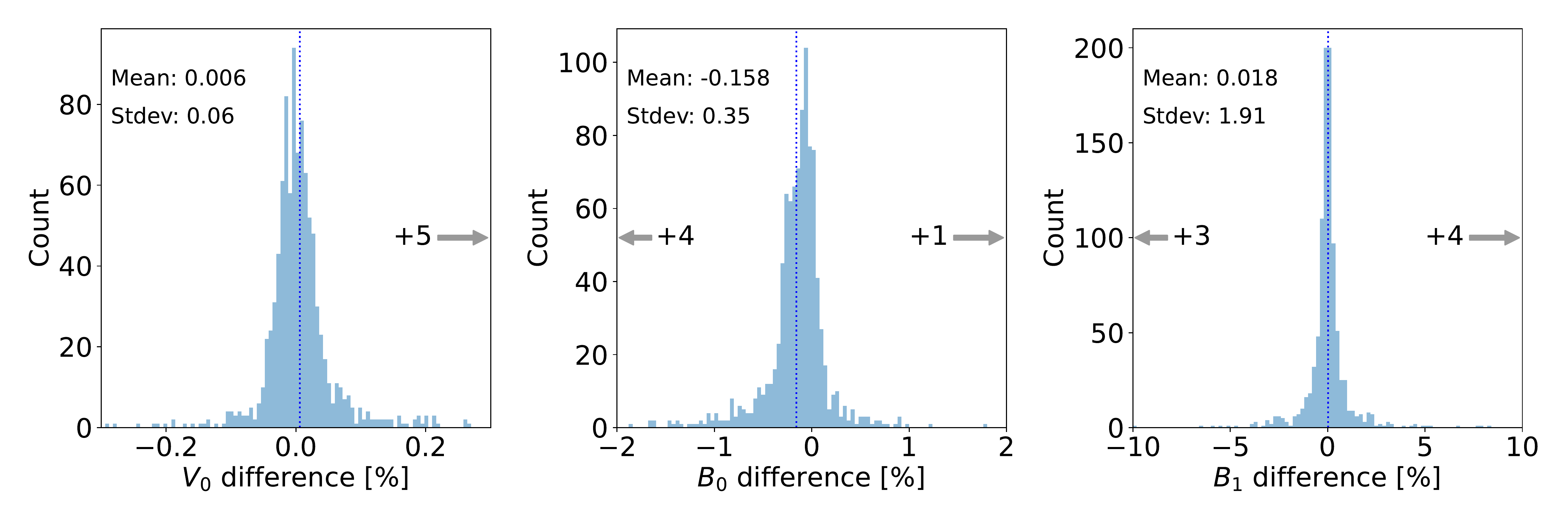}
 \caption{\textbf{Histograms of the percentage difference between the results of the two all-electron codes (\fleur{} and \wientwok{}) with respect to their average for the three parameters of the \gls{eos}: $V_0$, $B_0$ and $B_1$, for the full dataset of unaries and oxides.}
 Positive values indicate larger values for \wientwok{} with respect to \fleur{}.
 Mean and standard deviation (stdev) of the distributions are reported on the top left of each panel.
 The number near the arrows indicate the number of outliers outside of the $x$-axis range.
 The relative difference on $V_0$ is below 0.1\% for 93\% of all structures in our dataset; the relative difference on $B_0$ is below 1\% for 97\% of the structures; and the relative difference on $B_1$ is below 2\% for 92\% of the structures.
 The 5 outliers for  $V_0$  are NeO$_3$ (0.302\%), RbO$_3$ (0.343\%), Cs$_2$O$_5$ (0.323\%), Fr$_2$O$_5$ (0.645\%) and  Ra$_2$O$_5$ (0.333\%),
 corresponding to lattice parameters for \fleur{}/\wientwok{} of 4.320/4.324\AA, 4.783/4.789\AA, 6.247/6.254\AA, 6.120/6.133\AA{} and 6.238/6.244\AA, respectively.
 \label{fig:ae-histograms}}
\end{figure*}

The results obtained with the \gls{ae} codes \fleur{} and \wientwok{} constitute our reference data.
Figure \ref{fig:ae-histograms} shows the distributions of the percentage difference between \fleur{} and \wientwok{} for $V_0$, $B_0$ and $B_1$ with respect to their average, for instance the $V_0$ difference (in \%) is given by:
\begin{equation}
    100\cdot \frac{V_0^{\text{\wientwok}} - V_0^{\text{\fleur}}}{(V_0^{\text{\wientwok}} + V_0^{\text{\fleur}})/2}.
\end{equation}
Although the histograms do not carry material-specific information, they clearly highlight the agreement between the two \gls{ae} codes. The relative difference on the equilibrium volume is below 0.3\% for all the materials except for 5 oxides (see SI Sec.~\ref{SI:results-ae} and the raw data in Ref.~\citenum{MCA-ACWF} for the full dataset).
The discrepancies for $B_0$ and $B_1$ are larger; this is not surprising, because they originate from higher derivatives of the \gls{eos} curves (see also discussion in SI Sec.~\ref{SI:stability_eos}).
We emphasize that these values, obtained after careful convergence of all numerical parameters related to the basis-set choices in the two codes, are of extremely high precision, with a spread that can even be an order of magnitude smaller than the typical discrepancies that we observe between pseudopotential codes (see discussion later).

The complete list of numerical parameters used for the \gls{ae} calculations is presented in SI Sec.~\ref{SI:parameters-ae}.
We highlight here that the exact choice of the electronic-state smearing and of the k-point integration mesh, as well as the specific quantity considered as the energy $E(V)$ (internal energy, or free energy including the entropic smearing contribution as we do here), are of crucial importance for a reliable comparison among codes and must be fully consistent; therefore, we discuss those in detail in section ``Using the All-Electron Reference Dataset'' and in Box 3.

\subsection*{Average AE dataset: the reference for further studies}
In addition to the data for each of the two codes and in order to provide a single comparison reference, we also provide a ``reference average all-electron dataset'' obtained by averaging the values of $V_0$, $B_0$, and $B_1$ for each of the 960 systems in the full dataset.
The corresponding values are in SI Sec.~\ref{SI:results-ae} and
published according to the FAIR principles in Ref.~\citenum{MCA-ACWF}.
Considering the very good agreement between the two codes, this average dataset constitutes an excellent reference, and we use this average to compare with the pseudopotential codes in Section ``Comparison with Pseudopotential-based Computational Approaches''.
In addition, if error bars are desired, the spread between the results of the two \gls{ae} codes can be used as an estimate of our dataset precision.

\section*{Metrics for EOS comparison}
In Refs.~\citenum{Lejaeghere:2016,deltasite}, the ``$\Delta$'' metric was used to compare the \gls{eos} computed with two different \gls{dft} computational approaches $a$ and $b$.
There, $\Delta = \Delta(a,b)$ was defined as:
\begin{equation}
\Delta(a,b) = \sqrt{\frac{1}{V_M - V_m} \int_{V_m}^{V_M} [E_a(V) - E_b(V)]^2 ~ dV},\label{eq:delta}
\end{equation}
where $E_a(V)$ and $E_b(V)$ are the Birch--Murnaghan fits of the data points obtained from approaches $a$ and $b$ respectively, the two \gls{eos} have been lined up with respect to their minimum energy, and as discussed earlier the integral spans a $\pm 6\%$ volume range centered at a central volume $\tilde V_0$ (with $\tilde V_0$ values tabulated in SI Sec.~\ref{SIsec:structures}), i.e., $V_m = 0.94 \tilde V_0$ and $V_M = 1.06 \tilde V_0$.

The use of a single metric to compare two \gls{eos} curves simplifies the data analysis, since it can be used instead of the difference of the Birch--Murnaghan parameters $V_0$, $B_0$, and $B_1$, as we did in Fig.~\ref{fig:ae-histograms}.
However the value of $\Delta(a,b)$, that has the units of energy, has the shortcoming of being too sensitive to the value of the bulk modulus of the material: visually similar discrepancies between two curves result in larger $\Delta$ values for materials with larger $B_0$.
This was already recognized in Ref.~\citenum{Jollet:2014}, where a modified metric $\Delta_1$ was suggested, renormalized to a reference value of $V_0$ and $B_0$.
In addition, the $E_a(V)$ and $E_b(V)$ quantities in Eq.~\eqref{eq:delta} are typically renormalized by the number of atoms in the unit cell, to provide a ``$\Delta$/atom'' metric, independent of the choice of the simulation cell size.
Since we expand our analysis to two-component oxides, generalizations might be required (e.g., by normalizing instead per formula unit).

We propose and recommend here two new metrics that we label $\varepsilon$ and $\nu$, and we discuss their pros and cons.
We first define the following shorthand notation for the integral average of a quantity $f(V)$ over the volume range $\left[V_m, \, V_M \right]$:
\begin{equation}\label{eq:average-volume-integral}
    \langle f \rangle = \frac{1}{V_{M}-V_{m}}\int_{V_{m}}^{V_{M}} f(V) ~ dV.
\end{equation}
Using this notation, we can simply write $\Delta(a,b) = \sqrt{\langle [E_{a}(V) - E_{b}(V)]^2 \rangle}$.
The first metric $\varepsilon(a,b)$ that we define is a renormalized dimensionless version of $\Delta$:
\begin{equation}\label{eq:epsilin_integral}
\varepsilon(a,b) = \sqrt{
    \frac{  \langle[E_{a}(V) - E_{b}(V)]^2 \rangle}
    {\sqrt{\langle [E_{a}(V) - \langle E_{a} \rangle]^2 \rangle  \langle [ E_{b}(V) - \langle E_{b} \rangle]^2 \rangle}} }.
\end{equation}
This metric, similarly to the $\Delta_1$ of Ref.~\citenum{Jollet:2014} or the subsequently defined $\Delta_{rel}$ available in the DeltaCodesDFT package\cite{deltasite}, is insensitive to the magnitude of the bulk modulus (see SI Sec.~\ref{SI:epsilon_indipendent_vol}).
In addition, it is independent of the use of a ``per-formula-unit'' or ``per-atom'' definition of the \gls{eos} (see SI Sec.~\ref{SI:epsilon_indipendent_vol}). Therefore, $\varepsilon(a,b)$ provides a uniform metric across the variety of structural and chemical environments under investigation, given the requirement that it must be calculated with the same relative volume range for every material. As the list of central reference volumes has been fixed (see SI Sec.~\ref{SIsec:structures}), and as we use the same $\pm 6\%$ volume range as in Ref.~\citenum{Lejaeghere:2016,deltasite}, the 960 intervals $\left[V_m, \, V_M \right]$ are unambiguously defined.  
We highlight, in passing, that the discrete form of Eq.~\eqref{eq:epsilin_integral}, i.e.:
\begin{equation}\label{eq:epsilin_discrete}
\varepsilon(a,b) = \sqrt{
    \frac{  \sum_i [E_{a}(V_i) - E_{b}(V_i)]^2}
    {\sqrt{ \sum_i [E_{a}(V_i) - \langle E_{a} \rangle]^2 ~ \sum_i [ E_{b}(V_i) - \langle E_{b} \rangle]^2}} }
\end{equation}
where the index $i$ runs over the explicit calculations of $E(V)$ from \gls{dft}, provides a reasonably good approximation to the value of Eq.~\eqref{eq:epsilin_integral} as long as the minima of the $E_{a}(V_i)$ and $E_{b}(V_i)$ data points are aligned on the energy scale, with the advantage that it can be used to directly compare raw \gls{dft} total energy data without requiring an \gls{eos} fitting. Nevertheless, we stress that in the rest of this work we use the expression of Eq.~\eqref{eq:epsilin_integral} and not its discrete version of Eq.~\eqref{eq:epsilin_discrete}.
Eq.~\eqref{eq:epsilin_discrete} is grounded in the definition of the coefficient of determination (or $R^2$) in statistics\cite{R2test} as a fraction of variance unexplained. We can interpret the value of $1-\varepsilon^2$ as the coefficient of determination (i.e. $1-\varepsilon^2 \approx R^2$) in a situation when one \gls{eos} $E_{a}(V)$ is treated as a fit for the other \gls{eos} $E_{b}(V)$. More precisely, since we want to define a symmetric metric $\varepsilon(a,b)=\varepsilon(b,a)$, our $\varepsilon^2$ is the $1-R^2$ value that one would obtain treating $E_{a}(V)$ as a fit for $E_{b}(V)$ and vice versa with the geometric mean of both data variances.
We note that the interpretation $R^2 \approx 1-\varepsilon^2$ holds in very good approximation when the value of $\varepsilon$ is much smaller than one (i.e., for very similar $E(V)$ curves).  

We discuss the sensitivity of $\varepsilon$ to perturbations of the Birch--Murnaghan parameters in SI Sec.~\ref{sisec:perturbations-eos}.
The main outcome is that $\varepsilon$ is mostly sensitive to the variations of $V_0$, and much less of $B_0$ and $B_1$.
For some applications, however, some of the parameters are more relevant than others (e.g., if one is mostly interested in accurate bulk moduli).
For these cases, we recommend (see also Box 4) to define metrics of discrepancy that depend directly on physically measurable quantities.
Since an \gls{eos} is very well described by the three parameters $V_0$, $B_0$ and $B_1$, we suggest a second metric $\nu$ that directly captures the relative difference of these three parameters between two computational approaches $a$ and $b$, using appropriate weights $w_{V_0}$, $w_{B_0}$, $w_{B_1}$:
\begin{equation}
     \nu_{w_{V_0},w_{B_0},w_{B_1}}(a,b) = 100
     \sqrt{	\sum_{Y=V_0,B_0,B_1} \left[ w_{Y} \cdot \frac{Y_{a}-Y_{b}}{(Y_{a}+Y_{b})/2} \right] ^2
     },\label{eq:nu-definition}
\end{equation}
where, for instance, $(V_0)_a$ indicates the value of $V_0$ obtained by fitting the data of approach $a$, and so on. The (arbitrary) prefactor 100 is chosen to obtain values with similar order of magnitude to those of $\varepsilon$. Furthermore, it also helps in interpreting the value of $\nu$ as an estimate of percentage errors (rather than relative errors) on the fit parameters.
We highlight that $\nu$ depends on the weights ($w_{Y}$), that in turn could be chosen to satisfy particular applications.
Here, since we aim to be application-agnostic, we choose weights based only on the sensitivity of our fitting procedure to random numerical noise applied to the energy values of the \gls{eos} data points.
The detailed procedure to determine the weights is described in SI Sec.~\ref{SI:stability_eos}; we just report here the final choice $w_{V_0} = 1$, $w_{B_0} = 1/20$ and $w_{B_1} = 1/400$.
Intuitively, the reduced weights are consistent with the expected increase of numerical uncertainty propagated in the fit when estimating higher-order derivatives of the \gls{eos}.
Similarly to $\varepsilon$, also these weights depend on the volume range of the datapoints used for the \gls{eos} fit, as well as the specific choice of the fitting algorithm (see details in SI Sec.~\ref{SI:stability_eos}).
In the rest of the manuscript, we refer to $\nu$ assuming this specific choice of weights, i.e., $\nu \equiv \nu_{1,\sfrac{1}{20},\sfrac{1}{400}}$.
In SI Sec.~\ref{sisec:perturbations-eos} we also discuss an intuitive interpretation of the $\nu$ metric: it is the percentage error on the equilibrium volume between the two approaches $a$ and $b$, when $B_0$ and $B_1$ are the same in the two approaches; otherwise, when $B_0$ and $B_1$ differ, it corresponds to an equivalent percentage error on $V_0$ that would result in quantitatively similar changes to the \gls{eos} curve, within the $\pm 6\%$ volume range considered here.

\begin{figure*}[tb]
\begin{center}
    \resizebox{.9\textwidth}{!}{%
		\includegraphics[height=5cm]{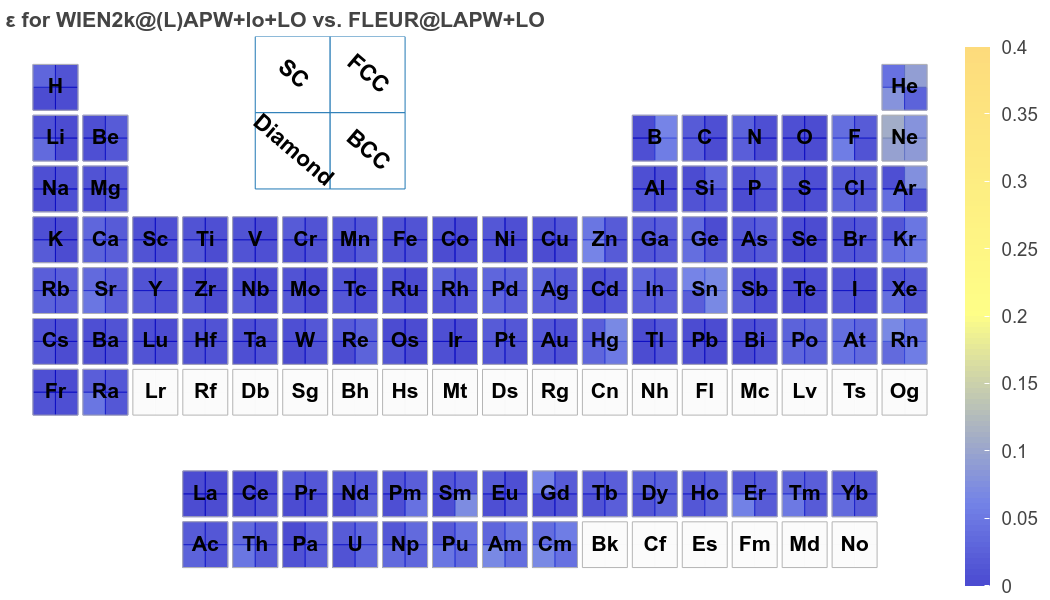}%
		\includegraphics[height=5cm]{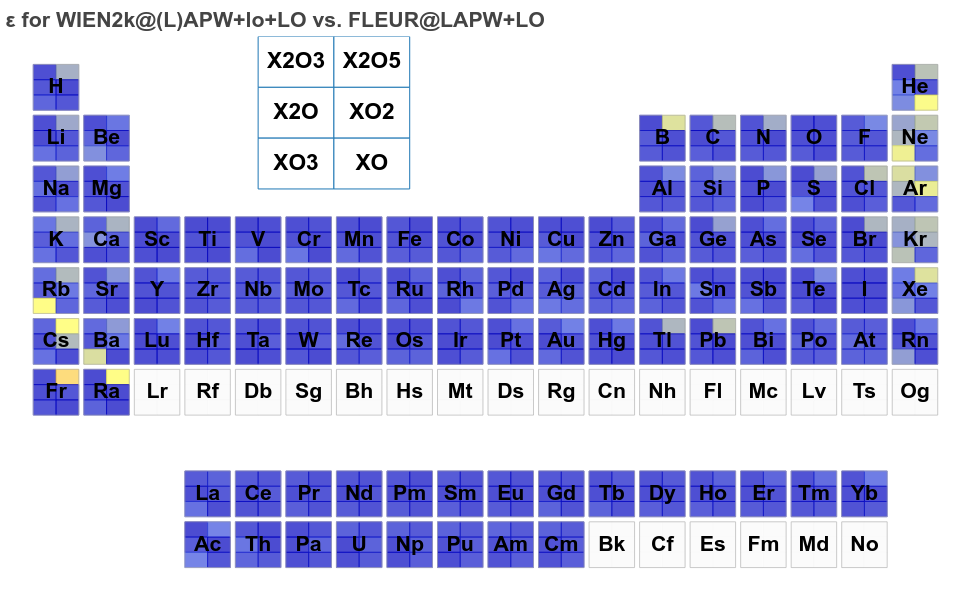}%
	}
	\resizebox{.9\textwidth}{!}{%
        \includegraphics[height=5cm]{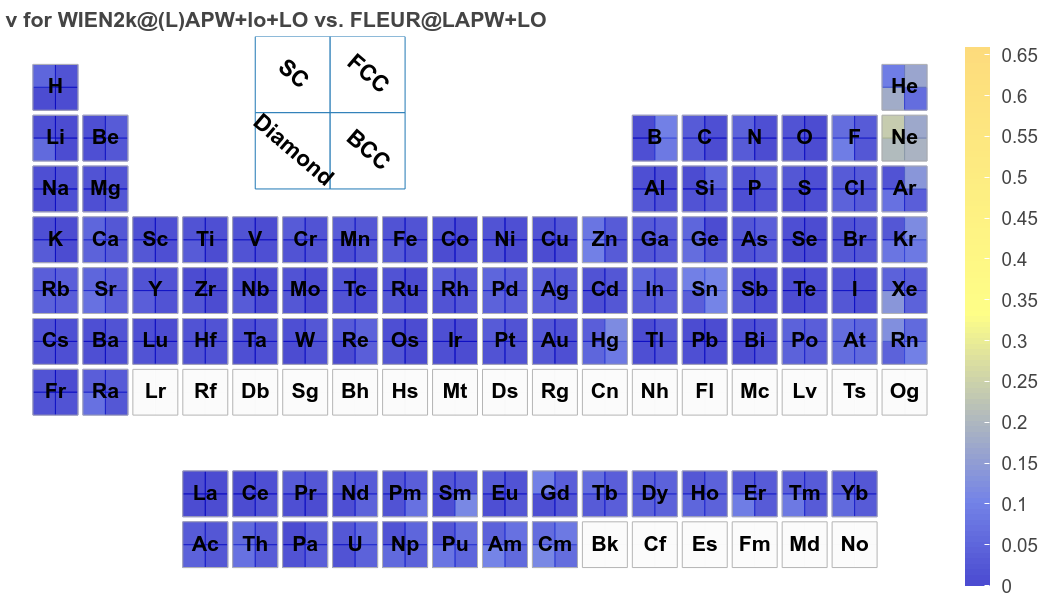}%
        \includegraphics[height=5cm]{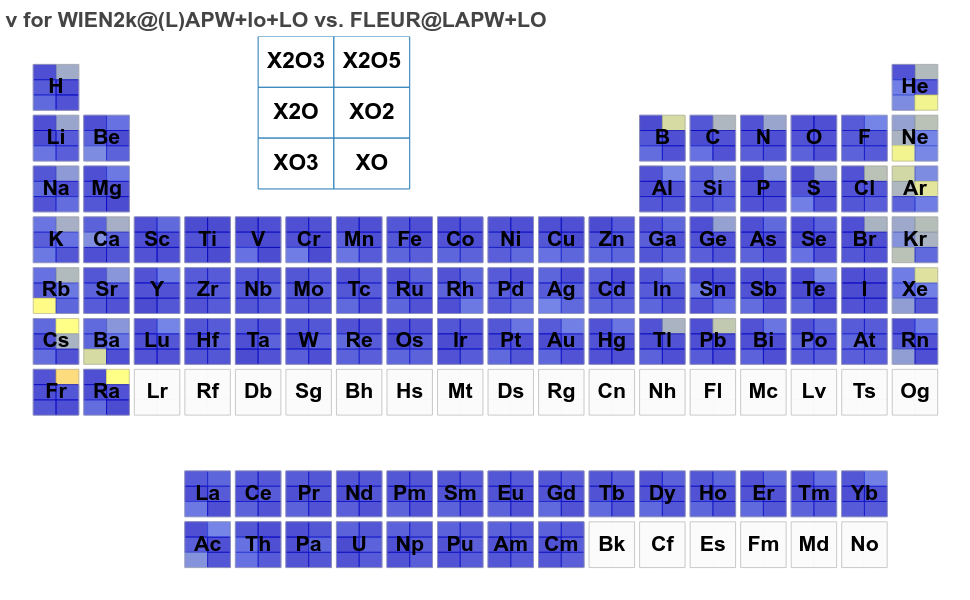}%
    }
\end{center}
 \caption{\textbf{Discrepancy between the \gls{ae} results (obtained with \wientwok{} and \fleur{}) in our reference dataset, measured either with the $\varepsilon$ metric (top panels) or the $\nu$  metric (bottom panels), for all 96 elements considered.}
 Each square for a given element is subdivided in 4 (6) in the left (right) panel, each referring to one of the unary (oxide) structures, as indicated in the legend presented in each panel. The color scale is the same for each pair (unaries and oxides) of periodic tables for the same metric ($\varepsilon$ or $\nu$). All structures are within our threshold for good agreement except Cs$_2$O$_5$ ($\varepsilon=0.20$, $\nu = 0.33$), Fr$_2$O$_5$ ($\varepsilon=0.40$, $\nu = 0.66$), Ra$_2$O$_5$ ($\varepsilon=0.21$, $\nu=0.33$) and RbO$_3$ ($\varepsilon=0.21$, $\nu=0.37$).
 \label{fig:ae-periodic-tables}}
\end{figure*}

The  metrics $\varepsilon$ and $\nu$ allow to compare a pair of codes for each material in the dataset.
Fig.~\ref{fig:ae-periodic-tables} reports the results for the pair (\fleur{}, \wientwok{}) across the entire set of structures under investigation in the form of periodic tables, enabling a quick identification of the most problematic elements in each set.
For instance, as one might expect the agreement is generally worse for noble gases: weakly bonded systems with a very small bulk modulus and thus more susceptible to numerical errors due to the choice of the basis set and of other computational parameters.

We emphasize that with our choice of the volume range and weights for $\nu$, the two metrics provide very consistent information, highlighting the importance of properly defining metrics based on physically measurable quantities, and on careful analysis of the error propagation of the fitting procedure, as we recommend in Box 4 (see also SI Sec.~\ref{sisec:perturbations-eos}, where we discuss quantitatively the effect of perturbations on the EOS parameters to the values of $\varepsilon$, $\nu$ and $\Delta$). 
Indeed, although $\varepsilon$ and $\nu$ are constructed according to quite different principles, they turn out to contain nearly identical information (in SI Sec.~\ref{sisec:correlation-metrics} we show that they are to a good extent linearly correlated for $\nu \lesssim 1$, with $\nu \approx 1.65 \varepsilon$). This has the consequence that periodic tables for $\varepsilon$ or $\nu$ will be almost identical if the range of the color scale is taken according to this linear correlation (as it is the case, e.g., in Fig.~\ref{fig:ae-periodic-tables} and is discussed in more detail in SI Sec.~\ref{sisec:periodic-tables-per-code}).

Finally, we identify (and report in Box 3, see also discussion in SI Sec.~\ref{sisec:perturbations-eos}) indicative thresholds on $\epsilon$ and $\nu$ to represent an excellent agreement between two \gls{eos} curves if $\varepsilon \lesssim 0.06$ or $\nu \lesssim 0.1$, or a good agreement (noticeable, but still relatively small) if $\varepsilon \lesssim 0.2$ or $\nu \lesssim 0.33$. 
As discussed earlier, we can interpret the two thresholds $\varepsilon=0.06$ ($\varepsilon=0.2$) approximately as a determination coefficient $R^2\approx 1-0.06^2=0.9964$ ($R^2\approx 1-0.2^2=0.96$) if one \gls{eos} is treated as a fit for the other. 
The data from the two \gls{ae} codes shows an overall excellent agreement: only four systems out of 960 have one or both metrics outside of the ``good-agreement'' range (Cs$_2$O$_5$, Fr$_2$O$_5$, Ra$_2$O$_5$ and RbO$_3$) when comparing the two \gls{ae} codes of our reference dataset, and 883 out of 960 systems have an excellent agreement for both $\varepsilon$ and $\nu$ according to the thresholds discussed above.

\section*{Using the AE reference dataset}

In this section we detail several aspects that must be carefully considered when using the reference dataset presented in the previous section. We then show how the dataset has been used to evaluate the precision of several computational approaches based on pseudopotentials and to improve a number of pseudopotential libraries.

\subsection*{Recommendations on how to use the dataset}\label{sec:rec-dataset}

\begin{figure}[tb]
    \centering
    \includegraphics[width=0.78\linewidth] {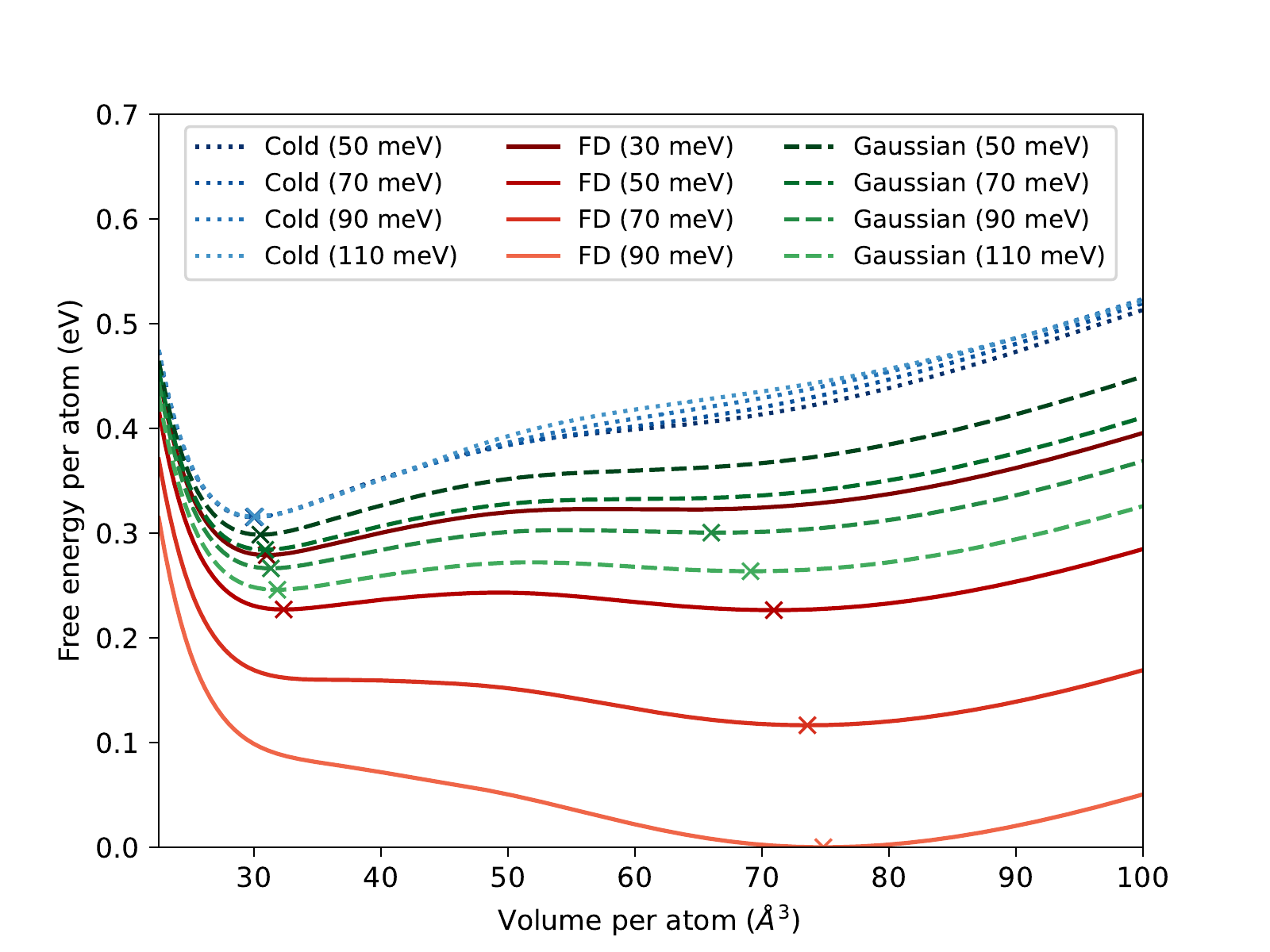}
\caption{\label{fig:various-smearings}\textbf{Effect of different choices of smearing on the EOS of erbium in the diamond structure computed with the \qe{} code.}
Er in the diamond structure is one of the systems in which the effect of smearing is most pronounced. 
In the legend, ``Cold'', ``FD'' and ``Gaussian'' indicate, respectively, cold smearing\cite{MV-cold-smearing}, Fermi--Dirac smearing and Gaussian smearing.
Two alternative minimum-energy solutions can exist for the FD and Gaussian smearings at very different volumes (as indicated by the crosses, when local minima exist in the curves); which one is selected depends on the choice of smearing and broadening.
Note also the much reduced sensitivity of the cold smearing to the broadening temperature, and how FD and Gaussian smearings are essentially equivalent after a renormalisation of the FD broadening by a factor $\approx 2.565$, as discussed in Ref.~\citenum{dossantos2022}. }
\end{figure}

When comparing our reference dataset with results from other codes, either for verification purposes or as a reference to improve basis sets and pseudopotentials, it is essential to use the same approximations (such as the exchange--correlation functional or the treatment of spin) and numerical choices (smearing and k-point integration mesh), as these parameters significantly affect the \gls{eos} results.
Therefore, we discuss here (and summarize in Box 3) specific recommendations on which parameters should not be changed when generating new data to compare with.

All calculations are performed in periodic boundary conditions using the PBE\cite{Perdew:1996} exchange-correlation functional, without including spin-polarization effects (non-magnetic calculations) and within a scalar-relativistic approximation (no spin-orbit coupling) for the orbitals treated as valence states.
The reciprocal-space integration is performed with a Monkhorst--Pack uniform k-point grid including the $\Gamma$ point, chosen as the smallest integration mesh guaranteeing a linear spacing of at most 0.06~\AA{}$^{-1}$ in each of the three reciprocal-space directions  for the smallest volume, and the same set of k-points (in scaled units) for all other volumes.
A Fermi--Dirac smearing of electronic states with a broadening of 0.0045 Ry ($\approx 61.2$ meV) is used in all cases, requiring the high-density k-points sampling mentioned above.
In addition, the quantity $E(V)$ that is fitted with the Birch--Murnaghan expression of Eq.~\eqref{b-m} is not the internal energy, but the free energy that includes the entropic contribution $-TS$ introduced by the smearing (where $T$ is the effective temperature given by the smearing broadening).

We stress that, in general, two codes using a different smearing distribution are expected to return comparable results only in the limit of an infinite number of k-points and an infinitesimal smearing.
However, for the purpose of verification, we do not need to reach this computationally expensive limit, provided that the same parameters among codes are chosen.
As a consequence, our results should not be considered a prediction of the zero-temperature (i.e., no smearing) limit.
We still highlight, however, that our choice of the k-point density results in a very dense and almost converged integration mesh (for fixed broadening): e.g., all values of $V_0$ computed with \wientwok{} change by less than 0.07\% when comparing with a denser k-point integration mesh with linear spacing of 0.045~{\AA}$^{-1}$, except in two cases (RbO$_3$: 3.7\% change, and HeO, 0.16\% change).
More details are reported in SI Sec.~\ref{sisec:kpt-convergence}.

To emphasize the sensitivity of the \gls{eos} to the choice of smearing, we show in Fig.~\ref{fig:various-smearings} one of the most pathological cases of our dataset, erbium in the diamond crystal structure.
In this case, the \gls{eos} does not have a simple shape but displays instead, for the case of Fermi--Dirac and Gaussian smearing, two minima at very different volumes.
Which one is favored in energy depends on the type of smearing and the value of the broadening.
This behavior can be explained by the presence of a set of narrow $f$ bands close to the Fermi level, shown in SI Sec.~\ref{sisec:Er-dia-bands}, whose filling strongly depends on the smearing.
If we are after an improved erbium pseudopotential, trying to optimize it with a different smearing (and thus possibly for a different minimum) will result in an incorrect pseudopotential.
A similar reasoning holds for the choice of using the free energy instead of the internal energy for the \gls{eos} (see SI Sec.~\ref{sisec:TS-vs-no-TS}).
We also highlight that we adopted a scalar-relativistic treatment of valence electrons for our dataset.
In most pseudopotential codes, this can be applied only for the valence electrons that are treated explicitly, while the treatment for the core electrons is implicitly included in the pseudopotential used. Even for \gls{ae} codes, electrons are typically partitioned into a core (treated fully relativistically) and a valence set (treated scalar relativistically). We highlight that the two \gls{ae} codes used in this work do not adopt the same core/valence assignment for all crystals (see SI Sec.~\ref{SI:parameters-ae}), yet they agree very well, illustrating that the core/valence assignment might not lead to ambiguities in the calculated results, provided all other numerical parameters are chosen consistently.

Finally, we note that many additional code-specific parameters exist, such as the type and size of the basis set or the pseudopotential family.
These choices are implemented in our automated common workflows\cite{Huber2021} and can be selected using a new protocol defined for this work and named {\tt verification-PBE-v1}.
Specific details for each code are reported in SI Sec.~\ref{SI:parameters-ae} for the \gls{ae} codes, and in SI Sec.~\ref{SI:sec_codes_param} for all the other codes.

Before showing an example of the comparison of our reference \gls{ae} dataset with nine computational approaches based on pseudopotentials, we discuss an additional recommendation.
Our dataset was generated to provide reference \gls{eos} for each of the 960 structures.
One might be tempted to reuse our dataset for different purposes.
For instance, since the values of the minimum energy of the \gls{eos} curves are also available from the fits, one could imagine using them to compare total energies of various oxides of the same element X, estimating their relative stability and the corresponding formation energies.
However, while this approach often results in sensible values, some notable cases lead to significantly off results, even by 1~eV/atom (see results in SI Sec.~\ref{sisec:formation-energies}).
The reason is that we designed our workflows and protocols for the \gls{eos}, in order to guarantee that simulation parameters are chosen consistently for all volumes of a given material, but this is not necessarily true among different materials.
As an example, since oxides of the same element might have very different interatomic spacings, the choices of atomic radii (and the corresponding  core/valence separation) for the \gls{ae} codes might be different in different systems, which precludes direct comparison between total energies.
From a more general point of view, one needs to be aware of the context in which data was produced, and consider implications and limitations when using them for different applications, as we discuss in the Outlook section.

\begin{figure}[t]
\noindent\fbox{%
    \parbox{\textwidth}{
{\bfseries Expert Recommendation Box 3, Summary of details to properly compare with the reference dataset presented in this work} 
\par\noindent\rule{\linewidth}{0.4pt}
\begin{itemize}
    \item Use the PBE exchange--correlation functional, do not include spin-polarization effects, and consider a scalar-relativistic treatment of (valence) electrons.
    \item Use Fermi--Dirac smearing with a value of 0.0045 Ry.
    While this choice does not lead to zero-smearing results (which would require extrapolation and extremely dense k-point integration meshes), using the same values ensures that results are comparable.
    In extreme cases, using a different smearing may affect significantly the equilibrium volume and the overall shape of the EOS.
    \item Compute the equations of state (EOS) using as the proper variational functional the free energy $E - TS$, where $E$ is the internal energy and $-TS$ is the smearing-energy entropic contribution.
    Other choices, such as the internal energy $E$, or the extrapolated energy for zero smearing (e.g., the expression $E-TS/2$, valid for Fermi--Dirac or Gaussian smearings\cite{Gillan1989,dossantos2022}) can result in significant changes of the EOS, including large variations of the minimum-energy volume.
    \item Use the same protocol to fit the \gls{eos} curves: 7 equally spaced points in a volume range of $\pm 6\%$ around the specified central volume. With these choices, values $\varepsilon \lesssim 0.06$ or $\nu \lesssim 0.1$ can be considered to indicate an excellent agreement, and $\varepsilon \lesssim 0.2$ or $\nu \lesssim 0.33$ a good agreement (with a noticeable, but still relatively small discrepancy between them). A different volume range will affect these thresholds and require a different choice of weights for $\nu$ to capture differences that are not purely statistical in nature. In addition, a different volume range will affect the k-point integration mesh (see next point).
    \item Use the exact same choice of the k-point integration mesh: regular grid including the $\Gamma$ point and the smallest mesh guaranteeing a spacing between points of at most 0.06 \AA$^{-1}$ along each of the three reciprocal-space directions for the smallest volume, and the same set of k-points (in scaled units) for all other volumes.
    This is typically converged for most systems and ensures that results can be compared even in the rare case of an unconverged k-point integration mesh. 
    \item Do not transfer the choices performed for this reference dataset to a different context, since it might lead to incorrect conclusions.
    For instance, extracting formation energies from our reference dataset can provide inaccurate results, since the parameters used in our simulations are guaranteed to be consistent only for different volumes of the same material, but not necessarily among different materials.
    \end{itemize}
}}
\end{figure}

\begin{figure*}[tbp]
    \centering
    \includegraphics[width=\textwidth] {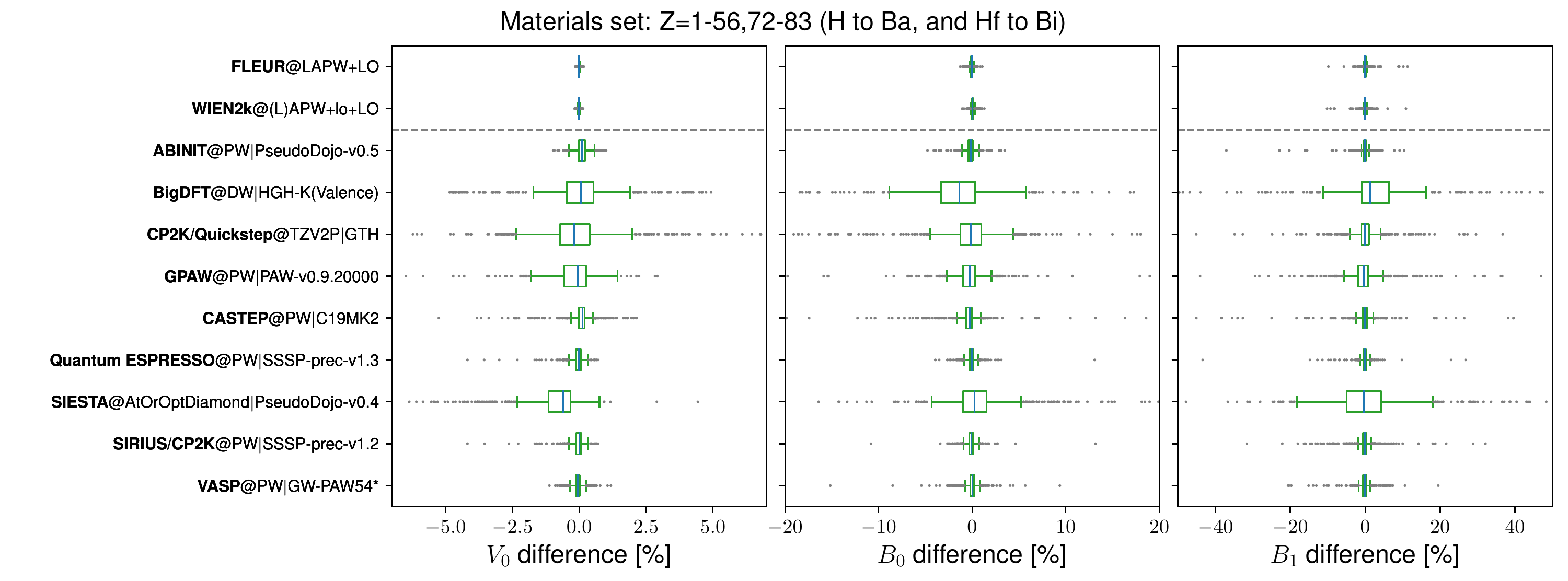}
     \includegraphics[width=\textwidth] {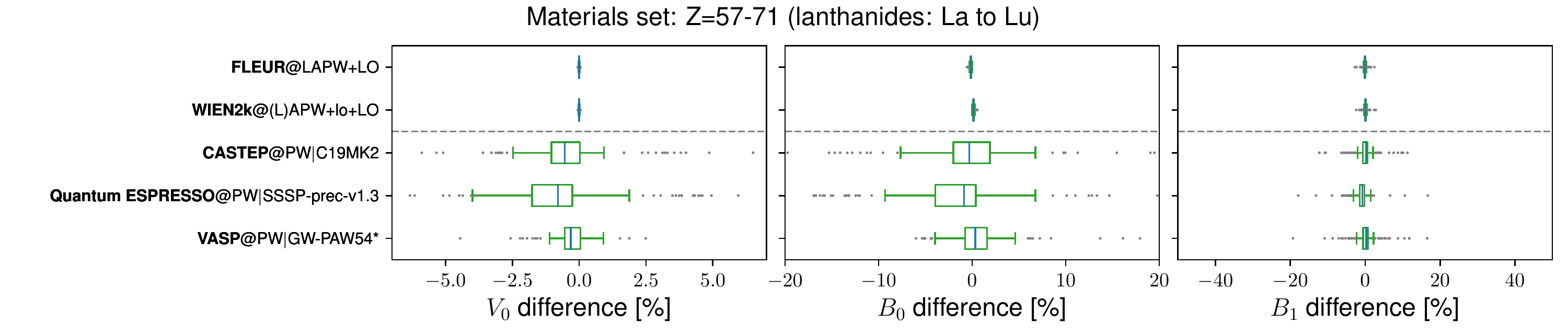}
     \includegraphics[width=\textwidth] {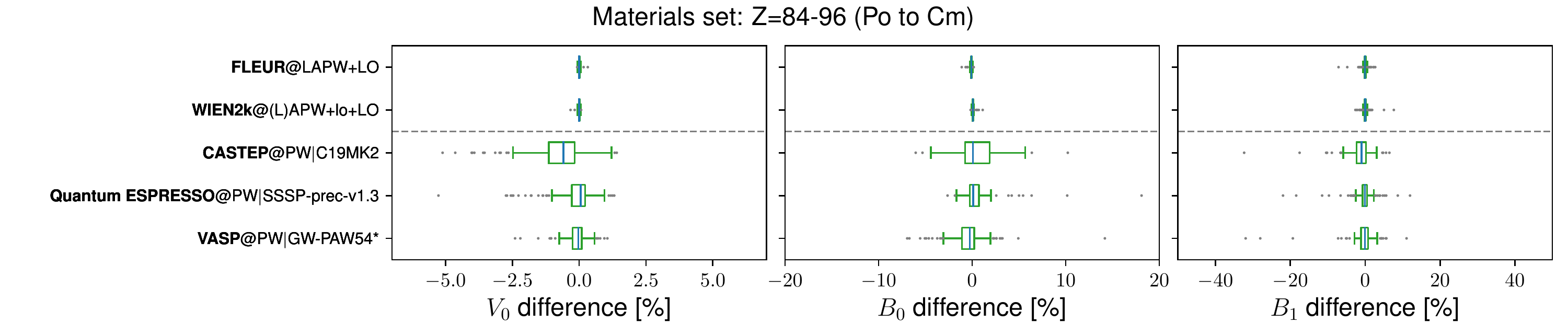}
     \caption{\textbf{Box-and-whisker plots comparing the $V_0$, $B_0$ and $B_1$ discrepancy of each computational approach involved in this work with respect to the average all-electron reference dataset.} The two \gls{ae} codes are also reported at the top of the plot above the dashed line, for comparison.
     In the box-and-whisker plots, the central blue line represents the median and the box extends between the first quartile $Q1$ and the third quartile $Q3$.
     The ``whiskers'' extend between $Q1 - 1.5\cdot$IQR  and $Q3 + 1.5\cdot$IQR (where IQR is the inter-quartile range $Q3 - Q1$). Outliers are represented as grey points.
     Note that some of the outliers are outside of the visible axis range in order to facilitate comparison between codes on the same axis range.
     Each row corresponds to a different subset of materials, where only the computational approaches that could compute those materials are included (since not all approaches include pseudopotentials for rare-earths).
     Specifically, the top row includes all materials from H to Bi excluding the lanthanide elements from La to Lu (68 elements in total).
     For this set, 295 crystals are missing for \bigdftlong, Na (FCC) is missing for \cptwoklong, Hg (FCC) and RbO$_3$ are missing for \siestalong, and all 10 crystals containing Tc are missing for \gpawlong.
     The central row reports the results for lanthanides only (from La to Lu), and the bottom row for all materials from Po to Cm (i.e., heavy elements, including actinides up to Cm). 
     \label{fig:all-codes-box-whisker}}
    \end{figure*}
    
\subsection*{Comparison with pseudopotential-based computational approaches}\label{ppc}
Using the recommendations of the previous section and of Box 3, we now compare our reference dataset with the results obtained with nine computational approaches based on pseudopotentials.
As discussed earlier, each approach is not only defined by the choice of the code, but also by the pseudopotentials used (and, where applicable, by the type of basis set).
Therefore, we summarize here briefly the meaning of the labels used for every computational approach.
The two \gls{ae} codes, \fleur{} and \wientwok, are labeled with their code name, followed by an indication of the basis set they use: \fleurlong{} and \wientwoklong{}, respectively (see SI Sec.~\ref{SI:parameters-ae} for more details). All other labels also include, at the end, the name of the pseudopotential library that was used. In particular: 
\abinitlong{} indicates the \abinit{} code, adopting a plane-wave (PW) basis set, using norm-conserving pseudopotentials from the PseudoDojo standard library version 0.5\cite{Setten:2018, PseudoDojoSite}; \bigdftlong{} indicates a (partial) set of structures with valence-only Hartwigsen--Goedecker--Hutter pseudopotentials\cite{Hartwigsen1998} calculated with the \bigdft{} code, adopting a basis set of Daubechies wavelets (DW), \casteplong{} indicates the \castep{} code using on-the-fly generated core-corrected ultrasoft pseudopotentials from the \texttt{C19} library with updated settings for the $f$ block elements, \cptwoklong{} indicates the \cptwok{} Quickstep code using Goedecker--Teter--Hutter pseudopotentials\cite{goedecker1996separable,krack2005pseudopotentials} and a molecularly optimized TZV2P-type basis set\cite{vandevondele2007gaussian}, \gpawlong{} indicates the \gpaw{}\cite{GPAW1,GPAW2} code used in its plane-wave mode using GPAW's PAW pseudopotentials included in the setup release 0.9.20000\cite{gpaw-setups-link}, \qelong{} indicates the \qe{} code using the Standard Solid-State Pseudopotentials (SSSP) library (PBE precision version 1.3)\cite{Prandini:2018,SSSP_1_3}, \siestalong{} indicates the \siesta{} code using norm-conserving pseudopotentials from the PseudoDojo standard library version 0.4 in psml format\cite{Setten:2018,PseudoDojoSite,Garcia:2018} and localized basis sets in which the orbitals for each element are taken from a partial optimization, considering just the unary Diamond structure for that element (therefore no optimization for the chemical environment of each material has been performed), \siriuslong{} indicates the \sirius{} library code (run via its interface to \cptwok{}) using the SSSP pseudopotential library (PBE precision version 1.2)\cite{Prandini:2018}
and \vasplong{} indicates the \vasp{} code (v6.3) using the PAW GW PBE pseudopotentials released in the dataset \texttt{potpaw$\_$PBE.54}, except for the lanthanides (see Section ``Pseudopotentials Improvement'').
The exact versions of the codes and libraries, together with the other code-specific choices implemented in the {\tt verification-PBE-v1} protocol, are detailed in SI Sec.~\ref{SI:sec_codes_param}.
The choices of computational approaches (for each code) listed above have been selected by the workflow developers of each code, trying to identify converged parameters and limiting to choices commonly available to users (or in some cases improving upon them, as we discuss later in the section ``Pseudopotentials improvement'').

The results are presented in Fig.~\ref{fig:all-codes-box-whisker} in the form of box-and-whisker plots for the percentage error of $V_0$, $B_0$ and $B_1$, with respect to the reference average \gls{ae} dataset.
Applying one of our general recommendations of Box 1, we partition our results in three groups (considering separately rare earths and/or heavy elements), in order to highlight the non-uniform capabilities of the various computational approaches.
Indeed, the narrow bands originating from the localized $f$ electrons are very
challenging to be described accurately with plain \gls{dft}\cite{Topsakal:2014}.
Therefore, often pseudopotentials for these elements are not available and thus several approaches cannot produce data for rare earths.
Even when available, those pseudopotentials might be less tested and thus deliver a lower precision.
By separating the results, we also enable a fairer comparison of approaches for the common set of elements (from H to Bi excluding the lanthanide elements from La to Lu).

Our results show that different numerical approaches have different precision; in general, the spread of the parameters of pseudopotential approaches are significantly larger than those between our two \gls{ae} codes.
In addition, the results indicate that additional work is required to obtain a high precision with approaches employing localized basis sets (that, on the other hand, are typically faster and scale better with system size) with respect to those using a plane-wave basis set (in the present case, the approaches listed above using the \abinit{}, \castep{}, \gpaw{}, \qe{}, \vasp{} and \sirius{}/CP2K codes).
Indeed, while a plane-wave basis set can be tuned with a single numerical parameter (the energy cutoff), systematically improving localized basis sets (and the associated pseudopotentials) requires dedicated efforts, that we recommend in Box 4. Verification projects such as this will facilitate these efforts by providing appropriate benchmarks. Another example of verification is presented in SI Sec.~\ref{sisec:plane-wave}, where we discuss the agreement of different codes adopting the same computational approach, in particular with the same plane-wave basis set and the same pseudopotential library. In this case, the results show an agreement that is similar in precision to the one between the two \gls{ae} codes.

Periodic tables (similar to Fig.~\ref{fig:ae-periodic-tables}) for each code are provided in SI Sec.~\ref{sisec:periodic-tables-per-code}, allowing for a closer inspection of the results resolved per chemical element and crystal-structure type. 
These tables also show that using a larger crystal-structure set (960 systems here) with respect to the set of 71 of Ref.~\citenum{Lejaeghere:2016} helps in highlighting possible shortcomings of pseudopotentials, as we discuss in more detail in SI Sec.~\ref{sisec:71-vs-960}.
The results of each code are also available in Ref.~\citenum{MCA-ACWF}, and can be visually displayed and compared directly online on the Materials Cloud\cite{Talirz:2020} at \url{https://acwf-verification.materialscloud.org}.

\subsection*{Pseudopotentials improvement}\label{pseudo-improvement}
Curated datasets such as the one presented in this Recommendation can drive efforts to improve pseudopotentials, ultimately delivering more precise computational approaches.
To illustrate this, we briefly summarize examples of pseudopotential enhancements that we performed to improve the comparison with our \gls{ae} results (more technical details are discussed in SI Sec.~\ref{sisec:additional-pseudos}) and used in the generation of the data of Fig.~\ref{fig:all-codes-box-whisker}.

The results for \abinitlong{} for elements around the 4$f$ block (from Te to Ba, and from Tl to Rn) were not giving ideal agreement using available pseudopotentials from PseudoDojo (version 0.4).
In almost all cases, we found that the accuracy of the pseudopotentials is significantly improved
by including a projector for the unbound $f$ state, at the expense of an increase of the computational cost when applying the non-local part of the Hamiltonian $V_{nl}$ (this can, however, be mitigated by the use of Legendre polynomials).
Without this projector, the local part of the pseudopotential cannot reproduce the all-electron scattering properties
of the $f$ angular momentum (see SI Sec.~\ref{sisec:abinit-pseudo-improvement}). This
led to the creation of a new PseudoDojo table (version 0.5), used here.

For \casteplong, starting from the on-the-fly pseudopotential generation settings for the built-in \texttt{C19} library, pseudopotentials for the lanthanide and actinide elements were improved by systematically changing the core radii, adding additional projectors, and adding fractional occupations of states that are empty in the reference atomic  configurations.
While making these changes did result in improvements, we note that no iterative optimization has been carried out to fit to the \gls{ae} results.

For \qelong, the pseudopotentials of SSSP PBE Precision version 1.1.2\cite{SSSP_1_1_2} have been updated for elements Na, Cu, Cs, Cd, Ba, As, Te, I, Hg, Ne, Ar, Kr, Xe, Rn; these new pseudopotentials have been released in the new SSSP PBE Precision 1.2\cite{SSSP_1_2}.
The new pseudopotentials have been selected by re-verifying the precision of pseudopotentials from various external libraries against the \gls{ae} reference dataset discussed in this Recommendation, and replacing those displaying significant discrepancies with pseudopotentials from other libraries that displayed a better agreement (lower $\varepsilon$ and $\nu$). Moreover, in SSSP PBE Precision version 1.3\cite{SSSP_1_3} (used here) new pseudopotentials have been included for actinides (Th-Lr) from Ref.~\citenum{sachs2021dft}, as well as for Ac, At, Ra, and Fr from PSlibrary\cite{dal2014pseudopotentials}.

For \vasplong, the latest available PAW potential set (version 5.4) was improved by reducing by about 20\% the core radii for lanthanides (other than La, Ce, and Lu).
Furthermore, placing two electrons in the 6$s$ shell, half an electron in the 5$d$ shell and the rest in the $f$ shell led to the most balanced description.
For Tm, Er, and Yb, three $f$ projectors were required to accurately describe the $f$ scattering properties.
The optimization was continued until very accurate scattering properties were obtained and agreement with very small core potentials was excellent, in turn resulting in a significant improvement of the agreement with the \gls{ae} reference dataset.

\begin{figure}[t]
\noindent\fbox{%
    \parbox{\textwidth}{
{\bfseries Expert Recommendation Box 4, Summary of recommendations to extend the verification effort presented in this work} 
\par\noindent\rule{\linewidth}{0.4pt}
\begin{itemize}
    \item Extend the current study to more computational approaches (codes, basis or pseudopotential sets, etc.), adopting the same reference crystal-structure set presented here.
    \item Extend the current study to more properties (forces, phonons, Kohn--Sham band structures, formation energies, $\ldots$). 
    Choose properties (and materials) that maximize the number of codes that can compute them (e.g., here for the \gls{eos}, only single-point DFT simulations are required; forces and stresses were not used).
    \item     Investigate the generality of optimal protocols and develop new ones for each property being computed, generalizing how to select a consistent set of parameters for multiple runs. For instance, for the \gls{eos} it is important to use the same k-point integration mesh at all volumes, but for a formation energy one wants a mesh that has reached a threshold accuracy for each component. In addition, use the same core/valence assignment, core radii and any other approach-specific precaution needed to compare total energies of different crystals.
    \item Create additional curated sets needed to generate improved pseudopotentials. E.g., extend to fully relativistic simulations and consider other exchange--correlation functionals in addition to PBE, such as the local-density approximation (LDA) and PBEsol, but also a selection of hybrids and meta-generalized-gradient approximations (meta-GGAs) for instance. Provide also curated sets needed to generate core-hole pseudopotentials for the simulation of core-level spectroscopies.
    \item Beside targeting improved pseudopotentials, develop dedicated efforts to optimize localized basis sets when these cannot be systematically improved by just tuning one or a few numerical parameters.
    \item Develop new protocols aiming at ``good-enough data'': i.e., not only targeting ultimate numerical precision (needed for verification), but also optimizing the computational cost for a target accuracy.
    \item Disseminate these protocols to the broad simulation community to optimize energy and CPU time and to expand the computational feasibility of DFT computational approaches in high-throughput studies or for expensive post-DFT methods (e.g., many-body perturbation theory).
    \item For new verification protocols, define metrics (such as $\epsilon$ and $\nu$ discussed here for the \gls{eos}) that depend on physically measurable quantities. Using such metrics, that condense in a single quantity the precision of computational approaches on a property of interest, one can easily define precision thresholds, compare approaches quantitatively, and evaluate the uniformity of results in a dataset. If fitting procedures are needed, assess the robustness of the chosen algorithms and estimate the uncertainty on the fitted parameters, using the results to define error bars.
\end{itemize}
}}
\end{figure}

\section*{Outlook}
This work constitutes a next step in a grand scheme of actions aiming at controlling the numerical aspects of electronic structure calculations, where the diversity of computational approaches and codes provides an opportunity for pairwise verification.
Compared to earlier work\cite{Lejaeghere:2016,deltasite}, we define here more discriminative metrics (crystals where two approaches would agree according to $\Delta$ might agree less according to $\varepsilon$ or $\nu$, see SI Fig.~\ref{fig:eos-sensitivity-3}a,b) and consider many more crystals, leading to more stringent testing. 
While major conclusions based on previous work remain valid (see SI Sec.~\ref{sisec:71-vs-960} and SI Sec.~\ref{sisec:comparing-delta}), the dataset presented here---together with the clear set of recommendations on how to reuse the data---provides a more refined and valuable reference for verification, uncertainty quantification and pseudopotential optimization.

Additionally, by formulating recommendations on how to perform further validation studies, and by providing and sharing universal common workflow interfaces (based on the AiiDA workflow infrastructure) to reproduce our calculations and perform new ones, we facilitate the community in taking new steps towards a better control of the uncertainty quantification in electronic structure calculations. 
There are several directions in which those steps could be taken. First, we recommend the creation of similar datasets for other commonly used exchange--correlation functionals (such as LDA and PBEsol, but possibly also a selection of hybrid and meta-GGA functionals), as well as for fully relativistic simulations.
For these studies, we recommend to use the same initial set of crystal structures discussed here, possibly only adapting the central point of the volume interval $[V_m, V_M]$ if the equilibrium volume $V_0$ for the functional does not lie anymore roughly in the middle of the interval.
Indeed, the set is fairly complete and systematic, and using the same structures facilitates the comparison between different computational approaches and approximations.
In addition, we recommend to test and verify codes also for magnetic materials.

Once such datasets are available, efforts to further improve pseudopotentials (and basis sets) should be initiated or continued, with the aim of making the results easily available to the broad simulation community.
One useful outcome could be, for instance, the generation of new reliable fully relativistic pseudopotential datasets for rare-earths (especially of the norm-conserving type, often required by many codes computing advanced materials properties).
Another relevant example, involving also the generation of additional bespoke \gls{ae} reference datasets, is the generation of pseudopotentials with a hole in the core, needed to predict the outcome of X-ray photoelectron spectroscopy (XPS) or X-ray absorption spectroscopy (XAS) experiments.

As a note, we highlight here that some of our structures are unrealistic.
When generating a new pseudopotential for a given chemical element, one might want to accept a compromise and not reproduce precisely the \gls{eos} of all 10 unaries and oxides, in order to obtain a computationally cheaper pseudopotential (e.g., with less projectors, more electrons in the core, or requiring a smaller energy cutoff), as long as the results are precise enough for the intended applications.

Other properties beyond the \gls{eos} are relevant to characterize materials and might benefit from tailored verification efforts; these include, e.g., formation energies, electronic band structures and phonon frequencies.
As we already highlighted, the simulation protocols might be significantly different for each property.
We therefore recommend that these protocols are well designed, documented and discussed, together with their limit of applicability.
In particular, especially if limiting to a scalar-relativistic approach as we did here, we recommend to further investigate the relevance of the choice of which electrons are included in the core or in the valence, as this can be of higher relevance than for the \gls{eos} (e.g., for formation energies, see also SI Sec.~\ref{sisec:formation-energies}).
Moreover, new metrics should be designed to quantitatively compare results, ideally directly dependent on physically measurable quantities.
Error propagation through any fitting procedure or data analysis should be carefully assessed, as we did in SI Sec.~\ref{SI:stability_eos}, to be able to define appropriate error bars.

Finally, we emphasize that while the goal here was ultimate precision in order to provide a reference dataset and obtain the best agreement possible between computational approaches, in real simulations one needs to optimize also the computational cost for a target accuracy, to obtain ``good-enough data'' for their scientific purpose.
This is especially true for high-throughput runs or when the DFT simulations are the first step of more expensive post-DFT methods.
We thus encourage to develop protocols to automatically define or select optimally converged parameters that at the same time minimize energy and CPU time, and then disseminate these to the whole community, so that they become easily accessible and usable by a broad range of users.

\section*{Acknowledgments}
This work is supported in part by the European Union's Horizon 2020 research and innovation programme under grant agreement No.~824143 (European MaX Centre of Excellence ``Materials design at the Exascale'') and by NCCR MARVEL, a National Centre of Competence in Research, funded by the Swiss National Science Foundation (grant number 205602). For the purpose of Open Access, a CC BY public copyright licence is applied to any Author Accepted Manuscript (AAM) version arising from this submission.

We acknowledge Flaviano Jos\'e dos Santos for useful discussions on the analysis of the smearing types and k-point convergence, and Xavier Gonze, Marc Torrent, and François Jollet for useful discussions on PAW pseudopotentials.

M.F. and N.M. acknowledge the contribution of Sadasivan Shankar in early discussions about the use of 6 prototype oxides as general platform to explore the transferability of pseudopotentials.

Work at ICMAB (E.B., A.G., V.D.) is supported by the Severo Ochoa Centers of Excellence Program (MCIN CEX2019-000917-S), by grant PGC2018-096955-B-C44 of MCIN/AEI/10.13039/501100011033, “ERDF A way of making Europe”, and by GenCat 2017SGR1506.
We also thank the Barcelona Supercomputer Center for computational resources.
V.D. acknowledges support from DOC-FAM, European Union's Horizon 2020 research and innovation programme under the Marie Sklodowska-Curie grant agreement No 754397.

O.R. acknowledges travel support from WIEN2k (Technical University of Vienna).

The J\"ulich team (S.B., J.B., H.J., G.M., D.W) acknowledges support by the Joint Lab Virtual Materials Design (JL-VMD) of the Forschungszentrum J\"ulich, the Helmholtz Platform for Research Software Engineering - Preparatory Study (HIRSE\_PS), and we gratefully acknowledge the computing time granted through JARA on the supercomputers JURECA\cite{thornig2021jureca} at Forschungszentrum J\"ulich and CLAIX at RWTH Aachen University.

H.M and T.D.K (University of Paderborn) gratefully acknowledge the Gauss Centre for Supercomputing e.V. (www.gauss-centre.eu) for funding this project by providing computing time on the GCS Supercomputer JUWELS at J\"ulich Supercomputing Centre (JSC).

S.P. and G.-M.R. (Universit\'e catholique de Louvain) acknowledge support from the F.R.S.-FNRS.
Computational resources have been provided by the PRACE-21 resources MareNostrum at BSC-CNS and by the Consortium des \'Equipements de Calcul Intensif (C\'ECI), funded by the Fonds de la Recherche Scientifique de Belgique (F.R.S.-FNRS) under Grant No. 2.5020.11 and by the Walloon Region as well as computational resources awarded on the Belgian share of the EuroHPC LUMI supercomputer. 

G.Ka. and S.V. received funding from the VILLUM Centre for the Science of Sustainable Fuels and Chemicals (9455) from VILLUM FONDEN. Computational resources were provided by the Niflheim supercomputing cluster at the Technical University of Denmark (DTU).
They also thank Jens Jørgen Mortensen and Ask H. Larsen for the valuable discussions on optimizing the workflow for the GPAW code.

S.C.\ acknowledges financial support from
OCAS NV by an OCAS-endowed chair at Ghent University.
The computational resources and services used at Ghent University were provided the VSC (Flemish Supercomputer Center), funded by the Research Foundation Flanders (FWO) and the Flemish Government – department EWI.

M.W. gratefully acknowledges computational resources provided by the Vienna Scientific Cluster (VSC).
This research was funded in part by the Austrian Science Fund (FWF) [P 32711].

E.F.L. would like to acknowledge resources provided by Sigma2 - the National Infrastructure for High Performance Computing and Data Storage in Norway and support from the Norwegian Research Infrastructure Services (NRIS).

B.Z. is grateful to the UK Materials and Molecular Modelling Hub for computational resources, which is partially funded by EPSRC (EP/P020194/1 and EP/T022213/1) and acknowledge the use of the UCL Myriad and Kathleen High Performance Computing Facility (Myriad@UCL, Kathleen@UCL), and associated support services, in the completion of this work.

N.M., G.Pi., and A.G. acknowledge support from the European Union's Horizon 2020 research and innovation programme under grant agreement No. 957189 (BIG-MAP), also part of the BATTERY 2030+ initiative under grant agreement No. 957213.

G.P., J.Y. and G.-M.R. acknowledge support by the Swiss National Science Foundation (SNSF) and by the Fonds de la Recherche Scientifique de Belgique (F.R.S.-FNRS) through the ``FISH4DIET'' Project (SNSF grant 200021E\_206190 and F.R.S.-FNRS grant T.0179.22).

G.P. acknowledges support by the Open Research Data Program of the ETH Board, under the Establish project ``PREMISE''.

J.Y. acknowledges support from the European Union's Horizon 2020 research and innovation programme under grant agreement No. 760173 (MARKETPLACE).

\section*{Author contributions}

S.C. and K.L. developed an initial early design of this research topic, and analyzed together with M.C. the first exploratory datasets.
M.S. was responsible for job, queue and data management in the first exploratory phase.
E.B. and G.Pi. contributed the idea to use the AiiDA and \texttt{aiida-common-workflow} infrastructure to carry on the thousands of \gls{dft} simulations required by the project. 
E.B. and G.Pi. coordinated the whole project.

M.F. and N.M. contributed the idea of using six oxides to test pseudopotentials across different coordinations and chemistries.
P.B., G.M., and O.R. performed the iterative refinement of the input parameters for the \gls{ae} calculations, that ultimately resulted in the generation of the central volumes of our dataset.
O.R. proposed the $\varepsilon$ metric. N.M. proposed the $\nu$ metric. S.P. raised the issue of the smearing selection, that ultimately led to the decision of a fixed k-point integration mesh and smearing broadening.
K.E. contributed to the data analysis and the conversion of the data into a dynamic website.
D.E.P.V. contributed the Hirshfeld-I calculated charges and their analysis. 
E.B., M.W. and G.Pi. performed the analysis of the error propagation in the fit and the estimation of the parameters of the $\nu$ metric.
E.B., S.C., O.R., and G.Pi. analyzed in detail the dependency and sensitivity of the metrics $\Delta$, $\varepsilon$ and $\nu$.

A.Z. and S.P. developed the \abinit{} implementation of the common workflow which relies on the \texttt{aiida-abinit} plugin developed and maintained by A.Z., G.Pe. and S.P.
M.G. created new pseudopotentials used for \abinit{} and improved the parameter profile. A.Z. and S.P. performed the \abinit{} calculations, including verification tests.
The work on \abinit{} was supervised by G.M.R. and S.P.

L.B., A.D. and L.G. contributed to the \bigdft{}-related parts of the work.
A.D. developed the \bigdft{} implementation of the common workflows.
L.B. and A.D. generated the \bigdft{} results under the supervision of L.G.

B.Z. developed the \castep{} implementation of the common workflow which relies on the \texttt{aiida-castep} plugin which is also maintained by B.Z., and performed all  \castep{} simulations.
C.J.P. created new on-the-fly generated pseudopotentials using the verification tests performed by B.Z.

M.K., T.D.K., H.M., T.M.A.M. and A.V.Y. contributed to all \cptwok{}-related parts of this work.
A.V.Y. implemented the workflows and performed preliminary calculations.
The workflows rely on the \texttt{aiida-cp2k} plugin developed by A.V.Y., T.M.A.M., and others.
M.K. performed preliminary calculations, created new pseudopotentials and contributed to the design of the protocol, T.D.K. contributed to the \cptwok{} setup, discussion of the results and supervised the calculations, H.M. conducted all AiiDA calculations and analyzed the results, T.M.A.M. provided implementations of \cptwok{} input and output parsers.

S.B., J.B., H.J., G.M. and D.W. contributed the FLEUR-related parts of this work\footnote{Corresponding contact person for the FLEUR contributions: Gregor Michalicek, g.michalicek@fz-juelich.de}.
G.M. hereby developed the parameter profile and performed the calculations. J.B. and H.J. adapted and extended the AiiDA-FLEUR plugin and the related parts of the AiiDA common-workflows package.
S.B. and D.W. contributed to the analysis and discussion of the FLEUR results.

G.Ka. and S.V. contributed to the GPAW-related parts of the work.
S.V. developed the GPAW implementation of the common workflows, which relies on the \texttt{aiida-ase} plugin and ran the calculations.
G.Ka. and S.V. analyzed the GPAW calculations.

M.B., S.P.H., N.M., J.Y. and G.Pi. contributed to all \qe{}-related parts of this work. M.B. and S.P.H. developed the \qe{} implementation of the common workflow which relies on the \texttt{aiida-quantumespresso} plugin developed and maintained by M.B., S.P.H., G.Pi. and others.
J.Y. generated and tested new pseudopotentials used for \qe{}.
The work on \qe{} was supervised by G.Pi. and N.M.

H.M. and T.D.K. contributed to all \sirius{}/CP2K-related parts of this work.
The workflows rely on the \texttt{aiida-cp2k} plugin developed by A.V.Y., T.M.A.M., and others. H.M. conducted all AiiDA calculations and analyzed the results. T.D.K. contributed to the \sirius{}/CP2K setup, discussion of the results and supervised the calculations. 

E.B., V.D. and A.G. contributed to the \siesta{}-related parts of the work.
E.B. developed the \siesta{} implementation of the common workflows, that relies on the \texttt{aiida-siesta} plugin developed by E.B., A.G., V.D. and others.
E.B. generated the \siesta{} results in collaboration with A.G.

M.W., M.M. and E.F.L. performed the execution and analysis of the \vasp-related workflows used to generate the data for this work.
G.Kr. generated updated potentials for the lanthanides.
E.F.L. maintains the \vasp{} implementation of the common workflows project and the \texttt{aiida-vasp} plugin (developed by a community of contributors, see full contributor list in the plugin documentation) which is used to execute the \vasp{} calculations.

P.B., G.K.H.M., O.R. and T.R. contributed the WIEN2k-related parts of this work.
T.R. performed preliminary calculations, P.B. created the setup of the WIEN2k calculations and supervised and analyzed the results, G.K.H.M. contributed the conversion of AiiDA structures to a WIEN2k struct file, and O.R. developed the \texttt{aiida-wien2k} plugin and performed all AiiDA-WIEN2k calculations.

E.B., M.F. and G.Pi. wrote the first version of the manuscript, and all authors contributed to the editing and revision of the manuscript.

\section*{Competing interests}
G.Pe. and G.-M.R. are shareholders and Directors of Matgenix SRL.
G.Kr. is shareholder of the VASP Software GmbH, and M.W. and M.M. are part-time employees of the VASP Software GmbH.
C.J.P. is an author of the CASTEP code and receives income from its commercial sales.

\section*{Publisher’s note}
Springer Nature remains neutral with regard to jurisdictional claims in published maps and institutional affiliations.

\section*{Code availability}
\label{sec:codeavailability}
The source code of the common workflows is released under the MIT open-source license and is made available on GitHub (\href{https://github.com/aiidateam/aiida-common-workflows}{https://github.com/aiidateam/aiida-common-workflows}).
It is also distributed as an installable package through the Python Package Index (\href{https://pypi.org/project/aiida-common-workflows}{https://pypi.org/project/aiida-common-workflows}).
The source code of the scripts to generate the plots is released under the MIT open-source license and is made available on GitHub (\href{https://github.com/aiidateam/acwf-verification-scripts}{https://github.com/aiidateam/acwf-verification-scripts}). All codes to generate the figures of this paper are available in the data entry of Ref.~\citenum{MCA-ACWF}.

\section*{Data availability}
\label{sec:dataavailability}
The data and the scripts used to create all the images in this work are available on the Materials Cloud Archive\cite{MCA-ACWF}.
Note that the data includes the entire AiiDA provenance graph of each workflow execution presented in the main text (including therefore all input files and output files of all simulations, as well as their logical relationship, in AiiDA format), as well as the curated data that is extracted from that database in order to produce the images.

\iftwofiles
\relax
\else
\clearpage

\newcommand{\setmergedsupplementary}{\newif \ifsupplementaryonly \supplementaryonlyfalse}
\setmergedsupplementary
\providecommand{\setmergedsupplementary}{\newif \ifsupplementaryonly \supplementaryonlytrue}
\setmergedsupplementary

\ifsupplementaryonly
\pdfoutput=1
\documentclass[fleqn,10pt]{wlscirep}

\myexternaldocument{../acwf-verification-main}

\begin{document}
\fi

\renewcommand\thesection{S\arabic{section}}
\renewcommand\thefigure{S\arabic{section}.\arabic{figure}}
\renewcommand\thetable{S\arabic{section}.S\arabic{table}}
\counterwithin{figure}{section}
\counterwithin{table}{section}
\renewcommand\theequation{S\arabic{equation}}
\setcounter{equation}{0}
	
\graphicspath{{./}{./supplementary/}}
\makeatletter
\providecommand*{\input@path}{}
\g@addto@macro\input@path{{./}{./supplementary/}}
\makeatother

\begin{center}
{\LARGE \bfseries Supplementary information for ``How to verify the precision of density-functional-theory implementations via reproducible and universal workflows''}

\vspace{1ex}

Emanuele Bosoni, Louis Beal, Marnik Bercx, Peter Blaha, Stefan Bl\"ugel, Jens Br\"oder, Martin Callsen, Stefaan Cottenier, Augustin Degomme, Vladimir Dikan, Kristjan Eimre, Espen Flage-Larsen, Marco Fornari, Alberto Garcia, Luigi Genovese, Matteo Giantomassi, Sebastiaan P. Huber, Henning Janssen, Georg Kastlunger, Matthias Krack, Georg Kresse, Thomas D. K\"uhne, Kurt Lejaeghere, Georg K. H. Madsen, Martijn Marsman, Nicola Marzari, Gregor Michalicek, Hossein Mirhosseini, Tiziano M. A. M\"uller, Guido Petretto, Chris J. Pickard, Samuel Ponc\'e, Gian-Marco Rignanese, Oleg Rubel, Thomas Ruh, Michael Sluydts, Danny E. P. Vanpoucke, Sudarshan Vijay, Michael Wolloch, Daniel Wortmann, Aliaksandr V. Yakutovich, Jusong Yu, Austin Zadoks, Bonan Zhu, Giovanni Pizzi
\end{center}

\section{Structures under investigation}\label{SIsec:structures}
This section reports some details on the crystal structures used in the verification study. As already explained in the main text, we consider two subsets: the ``unaries dataset'' and the ``oxides dataset''.

\begin{figure}[tbp]
    \centering
    \begin{subfigure}[b]{0.3\textwidth}
    \centering\includegraphics[width=0.7\linewidth]{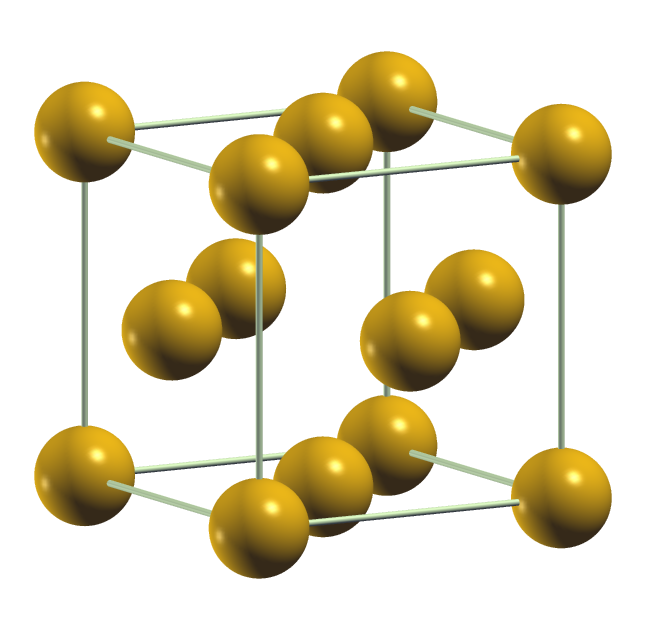} 
    \caption{\gls{fcc} crystal (conventional cell).}
    \end{subfigure}
    \begin{subfigure}[b]{0.3\textwidth}
        \centering\includegraphics[width=0.7\linewidth]{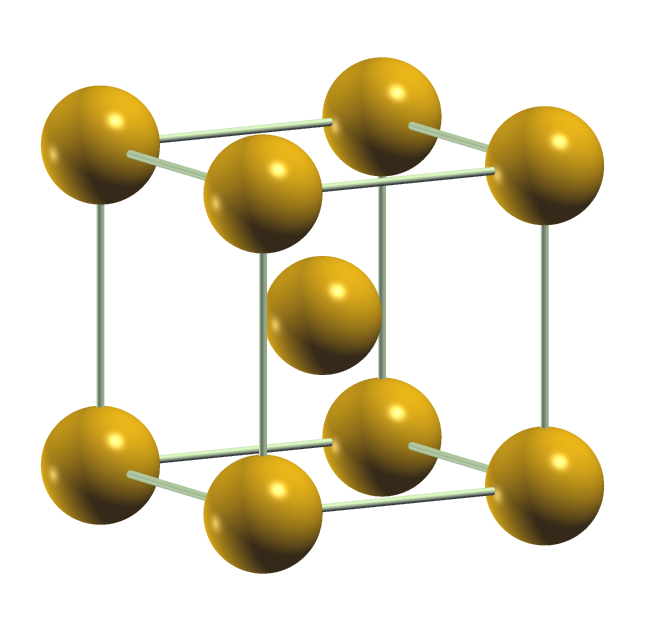} 
        \caption{\gls{bcc} crystal (conventional cell).}
    \end{subfigure}

    \begin{subfigure}[b]{0.3\textwidth}
        \centering\includegraphics[width=0.7\linewidth]{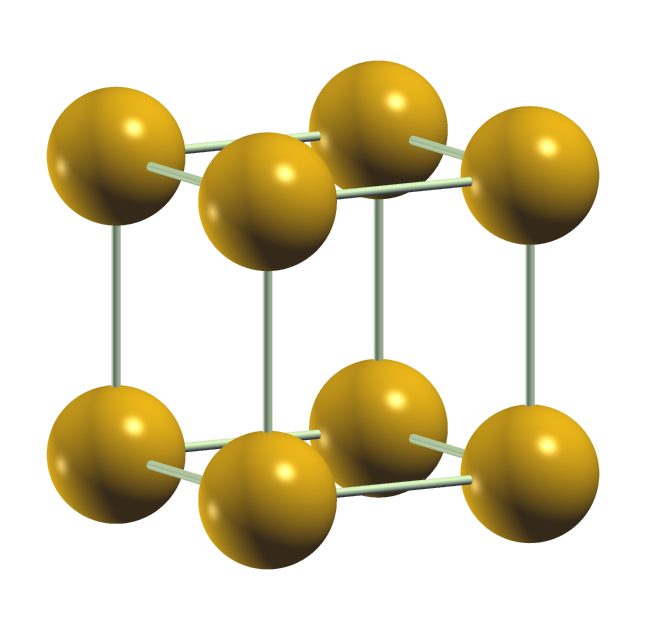} 
        \caption{\gls{sc} crystal (conventional cell).}
        \end{subfigure}
        \begin{subfigure}[b]{0.3\textwidth}
            \centering\includegraphics[width=0.7\linewidth]{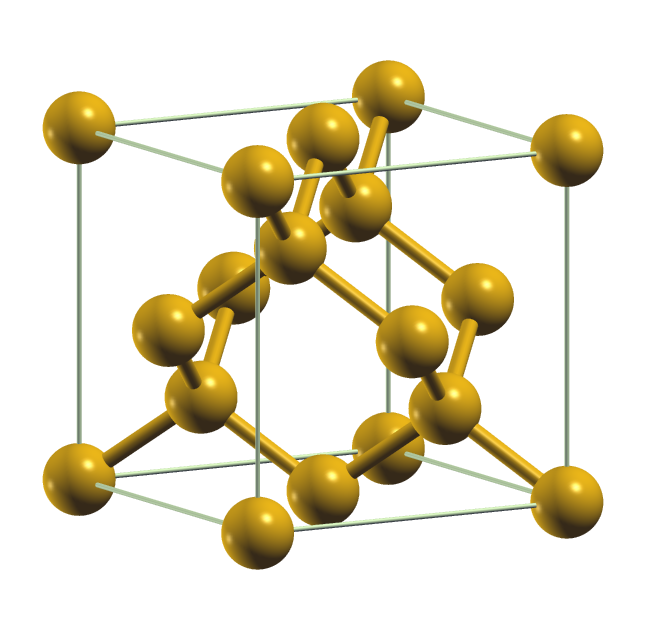} 
            \caption{Diamond crystal (conventional cell).}
        \end{subfigure}
    \caption{Conventional cells of the 4 unary prototypes used in this work. Images generated using XCrysDen\cite{Kokalj1999}.\label{sifig:unaries-conventional-cells}}
\end{figure}

The ``unaries dataset'' consists of 4 monoelemental cubic crystals for every element from Z=1 (hydrogen) to Z=96 (curium), in the well-known structures face-centered cubic, body-centered cubic, simple cubic and in the diamond structure. The details of each of the 4 monoelemental cubic crystals are described in Table \ref{unaries-table}, together with the indication of a prototype belonging to each category. A visualization of the crystal structures is reported in SI Fig.~\ref{sifig:unaries-conventional-cells}.

\begin{table*}[h!]
    \caption{\label{unaries-table}Description of the four unary crystals under investigation, with a prototypical crystalline example and the corresponding ID from the ICSD database\cite{ICSD}. The quantity $l$ is the length of the primitive-cell lattice vectors, $a$ the cubic conventional-cell side, and $d_{nn}$ is the nearest-neighbor distance.}
\begin{center}
\begin{tabular}{ccccccc}
 & Prototype  & \multirow{2}{*}{Space group} & Wyckoff site    & Coordination & \multirow{2}{*}{$l$} & \multirow{2}{*}{$d_{nn}$} \\
 & (ICSD number)  &         & (site symmetry) &    number        & & \\ 

 \hline\hline

 \multirow{2}{*}{\gls{fcc}} & Al & Fm$\bar3$m & 4a & \multirow{2}{*}{12} & \multirow{2}{*}{$a/\sqrt{2}$} & \multirow{2}{*}{$a/\sqrt{2}$}\\ 
          & (43423) & (225) & (m$\bar 3$m)  &  &  \\ \hline 

\multirow{2}{*}{\gls{bcc}} & V & Im$\bar 3$m & 2a & \multirow{2}{*}{8} & \multirow{2}{*}{$\sqrt{3} a /2$} & \multirow{2}{*}{$\sqrt{3} a /2$}\\ 
& (43420) & (229) & (m$\bar3$m)\\ \hline

\multirow{2}{*}{\gls{sc}} & $\alpha-$Po & Pm$\bar3$m  &  1a & \multirow{2}{*}{6} & \multirow{2}{*}{$a$} & \multirow{2}{*}{$a$} \\ 
 & (43211) & (221) & (m$\bar3$m) \\ \hline

\multirow{2}{*}{Diamond} & C (diamond)  & Fd$\bar3$m & 8a & \multirow{2}{*}{4} & \multirow{2}{*}{$a/\sqrt{2}$} & \multirow{2}{*}{$\sqrt{3} a /4$} \\
& (28857) & (227) & ($\bar4$3m) & 
\end{tabular}
\end{center}
\end{table*}

\begin{figure}[tbp]
    \centering
    \begin{subfigure}[b]{0.3\textwidth}
    \includegraphics[width=\linewidth]{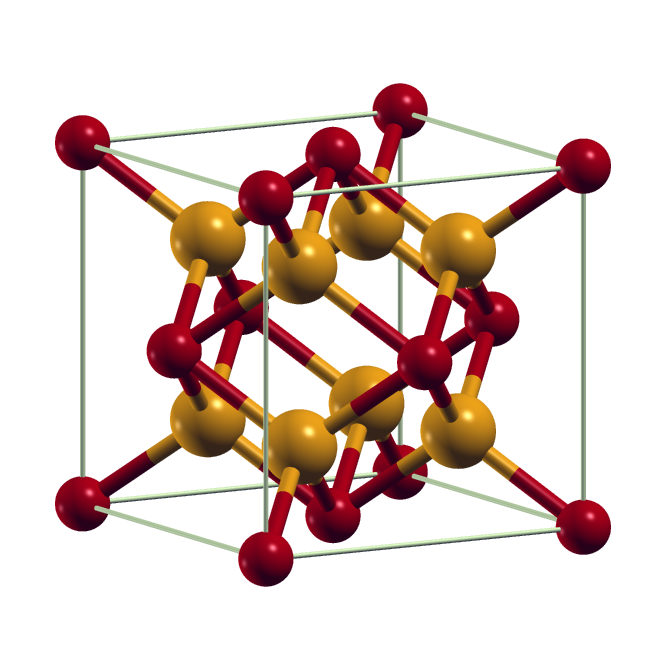} 
    \caption{X$_2$O crystal (conventional cell).}
    \end{subfigure}
    \begin{subfigure}[b]{0.3\textwidth}
        \includegraphics[width=\linewidth]{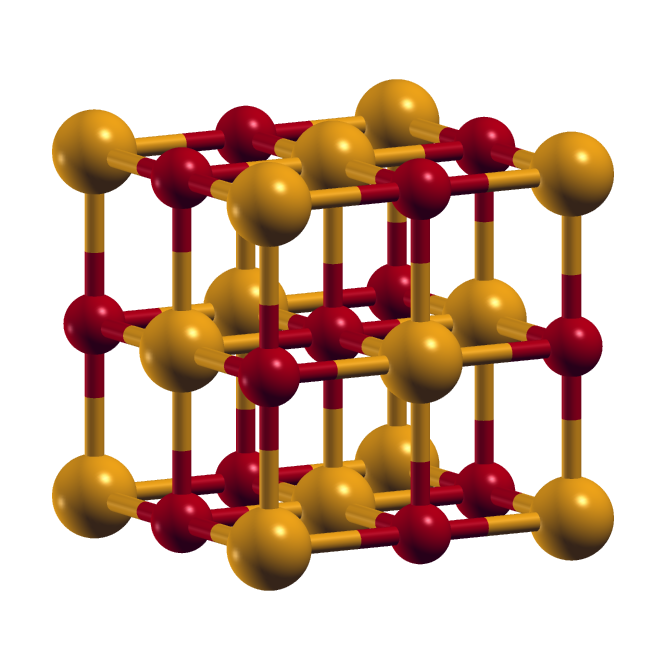} 
        \caption{XO crystal (conventional cell).}
    \end{subfigure}
    \begin{subfigure}[b]{0.3\textwidth}
        \includegraphics[width=\linewidth]{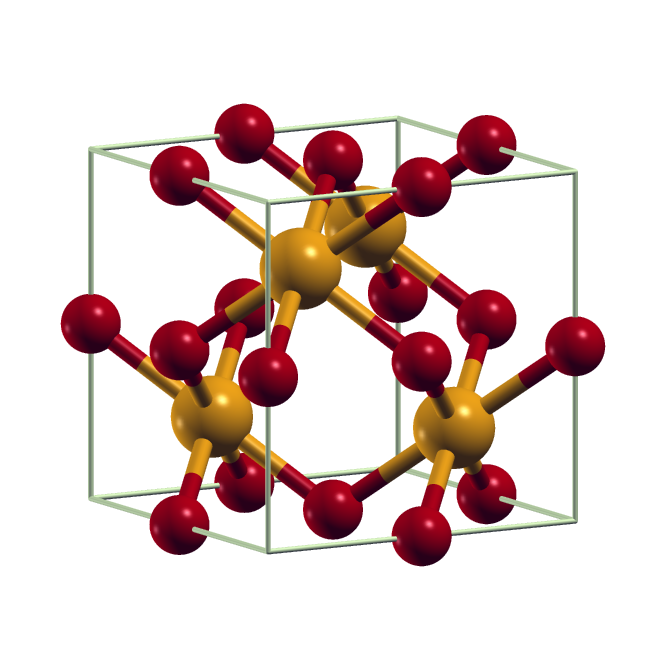} 
        \caption{X$_2$O$_3$ crystal (conventional cell).}
    \end{subfigure}

    \begin{subfigure}[b]{0.3\textwidth}
        \includegraphics[width=\linewidth]{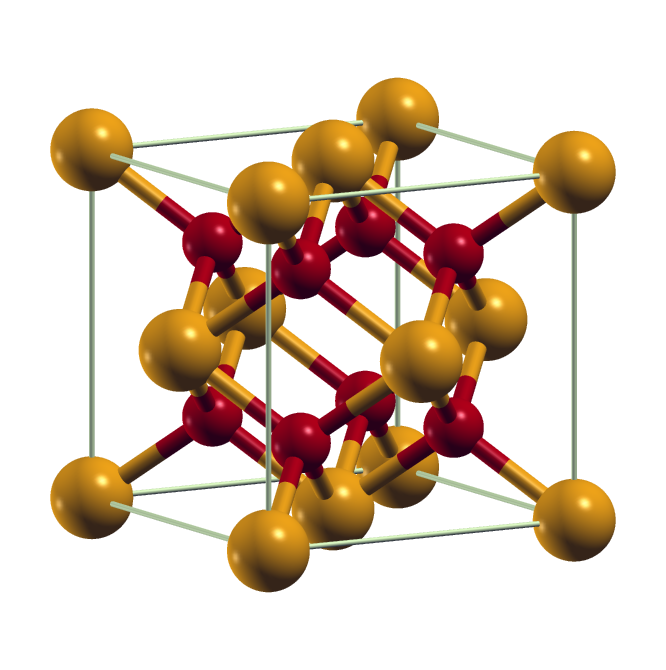} 
        \caption{XO$_2$ crystal (conventional cell).}
        \end{subfigure}
        \begin{subfigure}[b]{0.3\textwidth}
            \includegraphics[width=\linewidth]{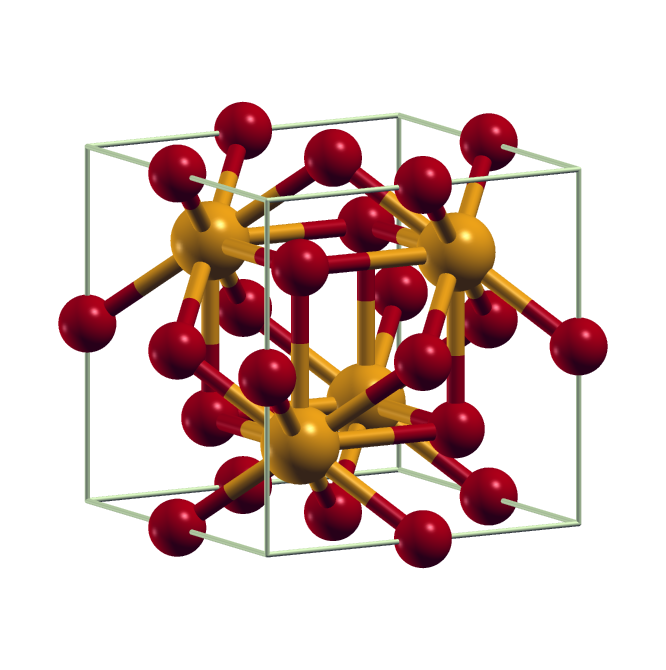} 
            \caption{X$_2$O$_5$ crystal (conventional cell).}
        \end{subfigure}
        \begin{subfigure}[b]{0.3\textwidth}
            \includegraphics[width=\linewidth]{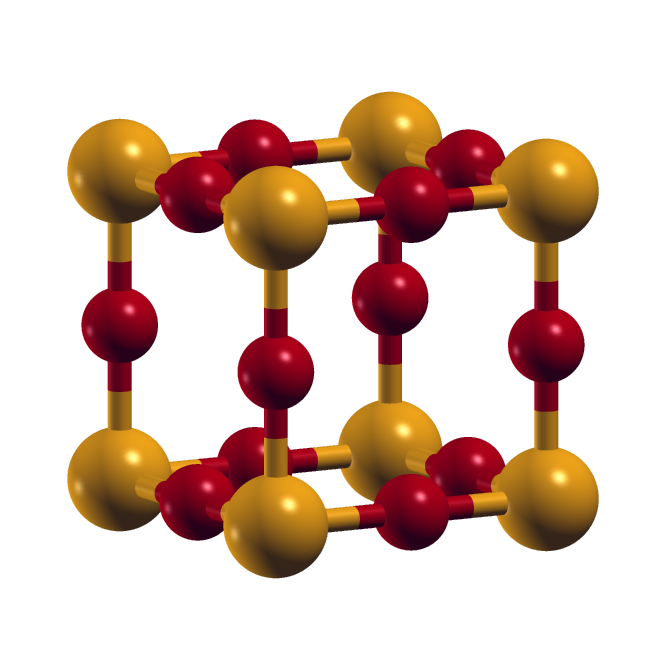} 
            \caption{XO$_3$ crystal (conventional cell).}
        \end{subfigure}
    \caption{Conventional cells of the 6 oxide prototypes used in this work. Oxygen atoms are represented as red atoms, while X atoms as orange atoms. Images generated using XCrysDen\cite{Kokalj1999}.\label{sifig:oxides-conventional-cells}}
\end{figure}

The ``oxides dataset'' is composed by six cubic oxides with chemical formula  X$_2$O, XO, X$_2$O$_3$, XO$_2$, X$_2$O$_5$ and XO$_3$, where X goes also in this case from hydrogen to curium. The details of each of these structures are reported in Table \ref{oxides-table}, that also includes the formal oxidation number that is expected for X in the structure (we stress that the actual oxidation state is different from the formal charge, see SI Sec.~\ref{SI:sec-hirshfeld}). A visualization of the crystal structures is reported in SI Fig.~\ref{sifig:oxides-conventional-cells}.

\begin{table*}[h!]
    \caption{\label{oxides-table} Description of the crystal structure of the six cubic oxides, with a prototypical crystalline example and the corresponding ID from the ICSD database\cite{ICSD}. The quantity $l$ is the length of the primitive-cell lattice vectors, $a$ the cubic conventional-cell side, and $d_{nn}$ is the distance of the X atom to its nearest-neighbor (oxygen) atom.}
\begin{center}
\begin{tabular}{cccccccc}
& Formal & Prototype  & Space group & Wyckoff site & Coordination &  $l$ & $ d_{nn}$ \\ 

& oxidation of X & (ICSD number) & & site symmetry of X & of X  & &\\ 

\hline\hline

X$_2$O & +1 & Na$_2$O & Fm$\overline{3}$m  & X=8c, O=4a & 4 & $a/\sqrt2$ & $a \sqrt3/4$\\ 
& & (60435) &(225) & -43m  & & & \\ \hline

XO & +2 & NaCl & Fm$\overline{3}$m  & X=4a, O=4b & 6  & $a/\sqrt2$ & $a/2$ \\ 
& & (18189) & (225) & m-3m &  & & \\ \hline

X$_2$O$_3$ & +3 & Ag$_2$O$_3$ & Pn$\overline{3}$m  & X=4b, O=6d & 6 & $a$ & $a \sqrt3/4$\\
& &(15999) & (224)  & -3m & & &\\ \hline

XO$_2$ & +4 & ZrO$_2$  & Fm$\overline{3}$m  & X=4a, O=8c & 8 & $a/\sqrt2$ & $a \sqrt3/4$\\ 
& & (105553) &(225) & m-3m  &   & & \\ \hline

X$_2$O$_5$ & +5 & --- & Pn$\overline{3}$m  & X=4b, O=4c+6d & 6 & $a$ & $a \sqrt3/4$ \\ 
& &  & (224) & -3m  & & & \\ \hline

XO$_3$ & +6 & ReO$_3$  & Pm$\overline{3}$m  & X=1a, O=3d & 6  & $a$ & $a/2$  \\ 
& & (647352) & (221) & m-3m &   & & \\ \hline

\end{tabular}
\end{center}
\end{table*}

For every material, the primitive cell is provided as input of the verification study (except when X is oxygen; in this case, the same cell is used as for all other oxides where X is different from oxygen, even if in the case of oxygen some of these cells might not be the smallest primitive cell, due to the increased symmetry). However, some codes might prefer to perform the actual calculation of the \gls{eos} on the cubic conventional cell, or use the actual primitive cell in the case of X = oxygen. The actual number of atoms in the simulation is reported inside the JSON files with the results of this verification study, available in Ref.~\citenum{MCA-ACWF}.

In the following, we will often refer to quantities (energy, volume, etc.) per formula unit. To avoid ambiguity, we explicitly list here the number of atoms in the formula unit for each of our 10 prototypes: FCC (1), BCC(1), SC(1), Diamond (2), X$_2$O (3), XO (2), X$_2$O$_3$ (5), XO$_2$ (3), X$_2$O$_5$ (7), XO$_3$ (4). We highlight that these numbers also correspond to the number of atoms in the primitive cell, except for X$_2$O$_3$ and X$_2$O$_5$ that have 10 and 14 atoms in the primitive cell, respectively.

\begin{figure}[h!]
\centering
\begin{subfigure}[b]{\textwidth}
    \centering
    \includegraphics[width=1\columnwidth] {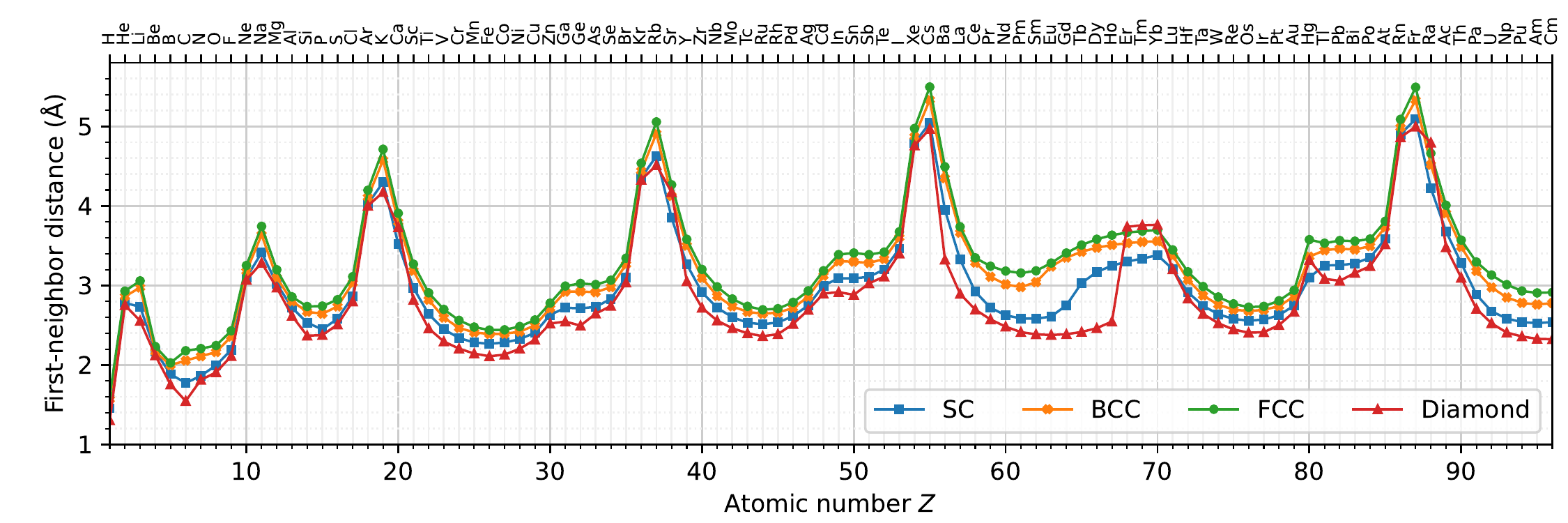}
    \caption{\label{sifig:first-neigh-dist-unaries}First-neighbor distance for the unaries dataset.}
\end{subfigure}

\begin{subfigure}[b]{\textwidth}
    \centering
    \includegraphics[width=\columnwidth] {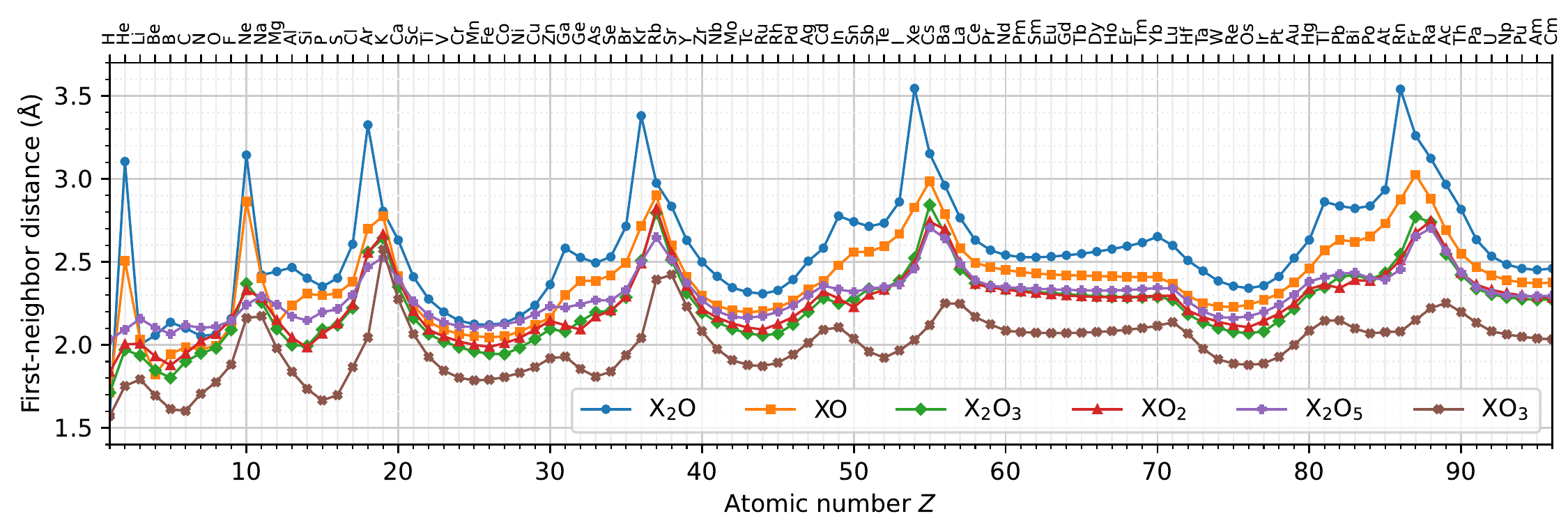}
    \caption{First-neighbor distance of the X atom for the oxides dataset (the first neighbor is, in all cases considered here, an oxygen atom).}
\end{subfigure}
\caption{First-neighbor distance of the X atom to its closest neighbor for all 960 systems in our dataset.\label{sifig:first-neigh-dist}}
\end{figure}

Finally, we report the central volumes for all 960 structures used for the calculation of the \gls{eos} data in SI Table~\ref{table-starting-volume}. In order to compare results, the same central volumes (and the same volume range of $\pm 6\%$, with 7 points) should be used when generating additional datasets.
To visualize the data, we report in SI Fig.~\ref{sifig:first-neigh-dist} the distance of the X atom from its closest (oxygen) neighbor, across the whole periodic table and for the 10 prototypes.

{\footnotesize
\begin{center}
\begin{longtable}{c|cccc|cccccc}
\caption{\label{table-starting-volume}  Table with the central volumes used for the calculation of the \gls{eos} datapoints. Volumes are expressed in \AA$^3$ per formula unit (see definition of the formula unit in SI Sec.~\ref{SIsec:structures}). The reference structures having these central volumes are available in Ref.~\citenum{MCA-ACWF}.} \\
  & FCC & BCC & SC & Diamond & X$_2$O & X$_2$O$_5$ & XO$_2$ & X$_2$O$_3$ & XO & XO$_3$ \\
\hline
\endfirsthead
\caption{\label{table-starting-volume}  (continued)  Table with the central volumes used for the calculation of the \gls{eos} datapoints. Volumes are expressed in \AA$^3$ per formula unit (see definition of the formula unit in SI Sec.~\ref{SIsec:structures}). The reference structures having these central volumes are available in Ref.~\citenum{MCA-ACWF}.} \\
  & FCC & BCC & SC & Diamond & X$_2$O & X$_2$O$_5$ & XO$_2$ & X$_2$O$_3$ & XO & XO$_3$ \\
\hline
\endhead
H & 2.96383 & 2.96392 & 3.08364 & 6.84867 & 11.96463 & 51.67201 & 19.10477 & 30.91989 & 10.02535 & 31.07589 \\
He & 17.83621 & 18.12465 & 21.38414 & 64.32268 & 92.12727 & 56.39662 & 24.77499 & 47.12856 & 31.51877 & 43.04663 \\
Li & 20.21287 & 20.26593 & 20.40472 & 51.36159 & 24.72203 & 61.84374 & 24.92385 & 44.77362 & 16.82806 & 46.06591 \\
Be & 7.87403 & 7.81517 & 10.26455 & 29.37901 & 26.83653 & 57.32171 & 22.18348 & 38.79261 & 12.12488 & 39.00478 \\
B & 5.89415 & 6.14152 & 6.6991 & 16.62498 & 30.02954 & 54.37317 & 20.37072 & 35.98325 & 14.71073 & 33.62805 \\
C & 7.31505 & 6.69648 & 5.60717 & 11.39533 & 28.55083 & 58.60995 & 22.75464 & 42.26604 & 15.70062 & 32.91818 \\
N & 7.60577 & 7.23569 & 6.48497 & 18.35414 & 26.67062 & 57.22806 & 25.50994 & 45.6704 & 15.3358 & 39.68942 \\
O & 8.00192 & 7.7972 & 7.94802 & 21.36412 & 27.11225 & 57.93205 & 27.11223 & 47.87245 & 15.90618 & 44.79083 \\
F & 10.14406 & 10.08192 & 10.52097 & 29.004 & 30.96209 & 60.78463 & 30.83431 & 56.31851 & 18.85462 & 53.30451 \\
Ne & 24.26591 & 24.70382 & 29.68755 & 89.0995 & 95.7889 & 69.51806 & 38.90827 & 81.94501 & 47.11781 & 80.77935 \\
Na & 37.10691 & 36.99607 & 39.7588 & 108.85725 & 43.71875 & 74.35023 & 36.67184 & 70.84884 & 27.7619 & 82.17703 \\
Mg & 23.11539 & 22.93116 & 27.59134 & 80.79618 & 44.83726 & 69.448 & 30.59043 & 56.77165 & 19.2497 & 62.13962 \\
Al & 16.48998 & 16.92508 & 20.17082 & 55.21453 & 46.20985 & 63.07075 & 26.30127 & 49.23828 & 22.4582 & 49.74088 \\
Si & 14.4803 & 14.66715 & 16.23082 & 40.92143 & 42.69026 & 61.00716 & 24.05335 & 48.82219 & 24.59467 & 41.777 \\
P & 14.58744 & 14.29218 & 14.60552 & 41.27791 & 40.01409 & 65.27141 & 27.20059 & 56.43141 & 24.30168 & 37.00066 \\
S & 15.88301 & 15.73711 & 17.19755 & 48.58823 & 42.6917 & 67.07762 & 29.73936 & 58.62747 & 24.68481 & 39.1327 \\
Cl & 21.29569 & 21.46346 & 23.45819 & 67.51736 & 54.55214 & 75.16606 & 34.69697 & 67.62641 & 26.95723 & 52.00364 \\
Ar & 52.33201 & 53.50562 & 65.51497 & 198.12738 & 113.32949 & 92.69732 & 51.32183 & 103.20479 & 39.41786 & 68.32473 \\
K & 73.99534 & 73.80511 & 79.47128 & 224.24589 & 68.00111 & 99.39122 & 58.36793 & 113.79592 & 42.74207 & 136.60539 \\
Ca & 42.20189 & 42.15587 & 43.70466 & 160.093 & 56.09468 & 84.42406 & 42.31153 & 79.93599 & 28.18987 & 94.27512 \\
Sc & 24.6858 & 24.88426 & 26.11875 & 68.84236 & 43.1186 & 71.37218 & 32.98765 & 62.21668 & 22.3317 & 70.50445 \\
Ti & 17.39633 & 17.26807 & 18.40137 & 45.88764 & 36.3163 & 63.72954 & 28.1791 & 54.26031 & 19.61645 & 57.35797 \\
V & 13.9076 & 13.46008 & 14.68791 & 37.2808 & 32.71378 & 59.96306 & 26.57435 & 50.7807 & 18.31774 & 50.25823 \\
Cr & 11.89373 & 11.55544 & 12.80026 & 33.09544 & 30.44097 & 58.12757 & 25.42189 & 48.27373 & 17.6635 & 46.89194 \\
Mn & 10.75345 & 10.78666 & 11.9053 & 30.35903 & 29.56957 & 57.65839 & 24.64946 & 46.45968 & 17.2926 & 45.62581 \\
Fe & 10.26671 & 10.50643 & 11.65681 & 28.93599 & 29.39392 & 58.00272 & 24.1901 & 45.43045 & 17.13572 & 45.9389 \\
Co & 10.31329 & 10.54766 & 11.90016 & 29.77725 & 29.83543 & 58.94975 & 25.03188 & 45.30703 & 17.27042 & 47.13241 \\
Ni & 10.83846 & 10.90046 & 12.56525 & 33.01991 & 31.63922 & 60.71118 & 26.20001 & 47.99648 & 17.99342 & 49.20696 \\
Cu & 11.96066 & 12.00521 & 13.9456 & 38.3573 & 34.5396 & 64.24158 & 28.23319 & 51.90164 & 19.10594 & 52.00277 \\
Zn & 15.15266 & 15.35236 & 18.21695 & 49.37022 & 40.67075 & 68.51634 & 30.3891 & 56.63194 & 20.27908 & 56.55698 \\
Ga & 18.89945 & 19.1961 & 20.12767 & 50.86204 & 53.04849 & 67.83879 & 29.26061 & 55.49587 & 24.35231 & 57.48398 \\
Ge & 19.61105 & 19.26408 & 19.92661 & 47.84474 & 49.67078 & 69.75991 & 28.19282 & 60.60137 & 27.12181 & 51.09293 \\
As & 19.25156 & 19.06952 & 20.35448 & 57.08944 & 47.81491 & 72.02994 & 31.50589 & 65.30072 & 27.1941 & 47.3416 \\
Se & 20.38999 & 20.33796 & 22.67444 & 63.49704 & 49.91744 & 71.9333 & 33.07453 & 66.14337 & 28.35245 & 49.87193 \\
Br & 26.41028 & 26.78091 & 29.81278 & 86.15579 & 61.62718 & 77.98233 & 37.01169 & 73.66672 & 31.10869 & 58.2754 \\
Kr & 66.18624 & 67.66229 & 82.81744 & 250.49559 & 118.94061 & 96.19024 & 47.53319 & 97.16926 & 40.08181 & 67.9248 \\
Rb & 91.38789 & 91.27765 & 99.14298 & 283.10731 & 81.07467 & 114.34806 & 69.0703 & 134.5233 & 48.78267 & 109.64128 \\
Sr & 54.91091 & 54.05117 & 57.38684 & 224.08214 & 70.15512 & 97.78005 & 51.29072 & 97.64817 & 35.05586 & 113.81785 \\
Y & 32.47792 & 33.03014 & 34.81815 & 87.6147 & 56.02664 & 82.48242 & 40.19378 & 76.25924 & 28.0575 & 88.92292 \\
Zr & 23.22672 & 22.85337 & 24.67007 & 61.95106 & 48.0859 & 71.98006 & 33.4764 & 65.13987 & 24.28322 & 72.31405 \\
Nb & 18.76368 & 18.12949 & 20.16049 & 51.64695 & 43.28804 & 65.77641 & 31.24816 & 60.07849 & 22.37503 & 61.62135 \\
Mo & 16.04515 & 15.79339 & 17.60535 & 46.04579 & 39.71891 & 62.89519 & 29.7259 & 56.76111 & 21.51546 & 55.61563 \\
Tc & 14.50906 & 14.62353 & 16.24577 & 42.51631 & 38.36272 & 62.29306 & 28.72035 & 54.60343 & 21.21284 & 53.12124 \\
Ru & 13.84099 & 14.24038 & 15.84808 & 40.6141 & 37.88283 & 63.27681 & 28.19283 & 53.64005 & 21.40643 & 52.5239 \\
Rh & 14.05529 & 14.47873 & 16.32474 & 41.91926 & 38.86199 & 65.4046 & 29.59307 & 54.41932 & 22.00732 & 54.2774 \\
Pd & 15.31609 & 15.44184 & 17.88203 & 49.04068 & 42.21056 & 68.95044 & 31.39346 & 59.01692 & 23.3105 & 58.57564 \\
Ag & 17.83932 & 18.00008 & 20.82095 & 60.08343 & 48.38454 & 74.95147 & 34.38486 & 65.63846 & 25.51029 & 65.19006 \\
Cd & 22.85103 & 23.39168 & 26.91335 & 74.58927 & 53.08317 & 80.62228 & 38.5362 & 73.25114 & 27.09207 & 73.17106 \\
In & 27.48501 & 27.76645 & 29.54359 & 76.27387 & 65.89698 & 78.42019 & 36.47078 & 70.18826 & 30.50678 & 74.75339 \\
Sn & 27.92759 & 27.62156 & 29.43402 & 73.68474 & 63.45403 & 76.77808 & 34.0064 & 72.49854 & 33.52195 & 67.66217 \\
Sb & 27.49335 & 27.16815 & 29.94869 & 85.55535 & 61.56831 & 78.86722 & 37.56726 & 78.6904 & 33.62451 & 60.24063 \\
Te & 28.31403 & 28.53875 & 32.78185 & 92.82855 & 62.89378 & 79.75651 & 39.06131 & 78.71778 & 34.89162 & 56.91725 \\
I & 35.12009 & 35.98158 & 41.54866 & 121.14185 & 72.20499 & 81.49821 & 41.44808 & 83.86273 & 38.00947 & 60.84655 \\
Xe & 87.15115 & 89.27395 & 109.89372 & 332.24175 & 137.22109 & 91.77459 & 47.37046 & 98.94124 & 45.25347 & 66.83409 \\
Cs & 117.71338 & 116.59594 & 128.22933 & 377.80616 & 96.47952 & 122.11939 & 63.56002 & 141.69282 & 53.27331 & 76.18877 \\
Ba & 64.22484 & 63.32039 & 61.52071 & 113.27682 & 79.80533 & 113.54449 & 60.31047 & 116.85012 & 43.27865 & 91.66471 \\
La & 36.95535 & 37.81167 & 37.01799 & 74.70356 & 65.16399 & 94.98527 & 47.88165 & 91.26986 & 34.42877 & 90.59307 \\
Ce & 26.53359 & 27.2707 & 24.80597 & 60.39358 & 56.09746 & 84.04452 & 40.7946 & 82.12996 & 31.03708 & 81.66735 \\
Pr & 24.09713 & 23.11993 & 20.26124 & 52.47588 & 52.30443 & 80.56703 & 39.81497 & 79.99604 & 30.07713 & 76.65652 \\
Nd & 22.76384 & 20.98371 & 18.16723 & 47.14628 & 50.5552 & 79.71193 & 39.09564 & 78.63461 & 29.46021 & 72.64543 \\
Pm & 22.24361 & 20.2416 & 17.37482 & 43.35196 & 49.83014 & 79.14646 & 38.53757 & 77.59916 & 29.02086 & 71.71784 \\
Sm & 22.8249 & 21.62656 & 17.16717 & 41.84622 & 49.72505 & 78.72225 & 38.07924 & 76.77967 & 28.71252 & 71.28495 \\
Eu & 24.97468 & 26.1259 & 17.68798 & 41.41038 & 49.99783 & 78.37384 & 37.68823 & 76.15461 & 28.5026 & 71.08958 \\
Gd & 27.96256 & 28.92878 & 20.74434 & 41.94188 & 50.44511 & 78.06943 & 37.3575 & 75.49738 & 28.36053 & 71.10647 \\
Tb & 30.53338 & 30.8832 & 27.62122 & 43.44373 & 51.03094 & 77.80866 & 37.10079 & 74.83874 & 28.25829 & 71.34317 \\
Dy & 32.47158 & 32.24109 & 31.62626 & 46.07452 & 51.787 & 77.66792 & 36.94349 & 74.27473 & 28.174 & 71.78804 \\
Ho & 33.88587 & 33.24726 & 34.02416 & 50.89551 & 52.71344 & 77.74286 & 36.8788 & 73.8801 & 28.09541 & 72.37003 \\
Er & 34.81162 & 33.9149 & 35.72581 & 160.65962 & 53.80493 & 78.00809 & 36.90264 & 73.68319 & 28.02352 & 73.17841 \\
Tm & 35.32142 & 34.35289 & 36.91694 & 163.2943 & 55.1443 & 78.45266 & 37.02748 & 73.68387 & 27.97309 & 74.23839 \\
Yb & 35.68954 & 34.45601 & 38.29539 & 164.06073 & 57.46052 & 79.24563 & 37.29979 & 73.91316 & 27.99156 & 75.71634 \\
Lu & 28.96169 & 29.57967 & 32.89043 & 101.18911 & 54.06305 & 78.94317 & 37.2627 & 72.22017 & 26.55736 & 78.07351 \\
Hf & 22.56668 & 22.30091 & 24.73374 & 70.20567 & 48.64474 & 71.33364 & 33.12187 & 64.18594 & 24.23282 & 70.82176 \\
Ta & 18.83578 & 18.29148 & 20.70598 & 56.84159 & 45.0229 & 65.63914 & 31.39846 & 59.96546 & 22.85916 & 61.7502 \\
W & 16.45344 & 16.14682 & 18.44138 & 49.57763 & 41.78609 & 62.77976 & 30.14777 & 57.13456 & 22.23519 & 56.04598 \\
Re & 15.0181 & 15.10498 & 17.14394 & 45.16071 & 40.19475 & 62.23732 & 29.3041 & 55.34484 & 22.11252 & 53.81481 \\
Os & 14.34475 & 14.78799 & 16.73456 & 42.93 & 39.57612 & 63.20103 & 28.80205 & 54.71102 & 22.54512 & 53.11224 \\
Ir & 14.51798 & 15.07236 & 17.01051 & 43.22586 & 40.36403 & 65.48808 & 30.42919 & 55.53976 & 23.42604 & 53.88197 \\
Pt & 15.65559 & 15.8485 & 18.10254 & 48.25544 & 43.22013 & 69.26345 & 32.29065 & 60.62094 & 24.64978 & 57.336 \\
Au & 17.96337 & 18.01979 & 20.75903 & 58.53091 & 49.44229 & 75.17109 & 35.00364 & 67.13364 & 26.87643 & 63.90256 \\
Hg & 32.36324 & 29.07647 & 30.07776 & 113.00826 & 56.10435 & 83.26519 & 39.30435 & 76.32712 & 29.80446 & 72.55805 \\
Tl & 31.19774 & 31.4643 & 34.37303 & 90.43011 & 72.24321 & 86.01923 & 40.72926 & 79.68706 & 33.96315 & 79.0134 \\
Pb & 32.13111 & 31.99849 & 34.45976 & 88.02959 & 70.3268 & 88.26763 & 39.60503 & 86.34487 & 36.51455 & 79.29789 \\
Bi & 31.77084 & 31.66197 & 35.18324 & 97.08873 & 69.26622 & 89.00017 & 42.09222 & 87.47546 & 36.02854 & 74.10038 \\
Po & 32.54441 & 32.88941 & 37.58851 & 104.9496 & 70.35802 & 85.35526 & 41.79777 & 84.65548 & 37.34068 & 70.91823 \\
At & 39.02559 & 39.94292 & 46.15152 & 133.91052 & 77.8056 & 84.55797 & 43.85025 & 88.77385 & 40.78895 & 71.68554 \\
Rn & 93.1132 & 95.57693 & 117.94773 & 355.33994 & 136.67664 & 91.11763 & 48.68438 & 101.57056 & 47.53838 & 72.06566 \\
Fr & 117.20593 & 116.47957 & 132.25863 & 384.39398 & 106.69693 & 114.96558 & 58.91507 & 131.0799 & 55.38287 & 79.53102 \\
Ra & 71.59113 & 70.96816 & 75.36362 & 339.34601 & 93.80196 & 121.53488 & 63.85 & 126.55639 & 47.84186 & 87.62984 \\
Ac & 45.55131 & 45.9684 & 50.16201 & 129.36355 & 80.39637 & 103.88226 & 52.95356 & 101.65853 & 38.99538 & 91.45268 \\
Th & 32.20115 & 32.67168 & 35.25242 & 92.51762 & 68.79224 & 89.10514 & 44.32081 & 87.47888 & 33.14217 & 84.96607 \\
Pa & 25.30196 & 24.73132 & 24.06085 & 61.37914 & 56.29059 & 79.83198 & 40.64908 & 78.8075 & 30.08799 & 77.67439 \\
U & 21.70953 & 20.2118 & 19.2531 & 49.63674 & 50.133 & 76.26324 & 39.02558 & 75.40504 & 28.35782 & 72.23741 \\
Np & 19.28905 & 17.78811 & 17.25306 & 42.95581 & 47.24644 & 74.80568 & 38.06636 & 73.6267 & 27.33128 & 70.29134 \\
Pu & 17.80178 & 16.59418 & 16.41493 & 40.17108 & 45.85351 & 74.2317 & 37.40308 & 72.67609 & 26.8416 & 68.83836 \\
Am & 17.36314 & 16.20862 & 16.14903 & 38.63027 & 45.41542 & 74.27276 & 36.96336 & 72.29916 & 26.66371 & 67.84802 \\
Cm & 17.48759 & 16.44718 & 16.39497 & 38.19162 & 45.8595 & 74.58951 & 36.65683 & 72.29245 & 26.79178 & 67.29113 \\
\end{longtable}

\end{center}
}

\clearpage

\section{Hirshfeld-I charges}\label{SI:sec-hirshfeld}
Six different oxide crystals are imposed, in order to force the element $X$ into 6 different formal oxidation states. The hope is that this will bring each element into 6 chemically sufficiently different environments. In this section, we analyze whether this expectation has been realized. This is done by monitoring the Hirshfeld-I charges throughout this oxide set, as a proxy for the chemical environment.

\subsection{Hirshfeld-I atoms-in-molecules methodology}
Atomic charges have been calculated within the context of an atoms-in-molecules (AIM) approach. The basic goal of these approaches is to divide the electrons, or more specifically the electron density, of a multi-atom system into subunits associated with chemical atoms. One can either start from the calculated wavefunctions (e.g., Mulliken charges\cite{Mulliken_a, Mulliken_b} ) or from the electron density distribution (EDD) (e.g., Hirshfeld\cite{Hirshfeld1977} or Bader\cite{Bader1991} charges). In this work, the iterative Hirshfeld approach (HI), which is an improvement of the Hirshfeld approach, is used. This approach alleviates the dependence on the chosen initial references of the original Hirshfeld method\cite{VanpouckeDannyEP:2013aJComputChem, VanpouckeDannyEP:2013bJComputChem, BultinckHI2007}. In Hirshfeld (and other stockholder) methods, the EDD in each point in space is divided over all the nearby AIM, in contrast to, for example, the Bader method\cite{Bader1991}, which assigns the entire electron density of a given point in space to a single AIM. This gives rise to smooth AIM which overlap in real-space. The Hirshfeld weights for an atom A are defined as:
\begin{equation}
w_A^H (r)=\frac{\rho_A^{AIM}(r)}{\rho_{mol}(r)},
\end{equation}
where $\rho_A^{AIM} (r)$ and  $\rho_{mol}(r)$ are the EDDs for the AIM and the molecule respectively. This however creates a circular reference, as the AIM EDD is calculated using the Hirshfeld weights. As a solution, Hirshfeld suggested the use of a reference EDD defined as the spherical average of the EDD of the free atom in a chosen reference state\cite{Hirshfeld1977}. To make sure weights at every point in space remain normalized to unity, the molecular density is replaced by the sum of the atomic reference EDD, giving rise to a so-called \textit{promolecular} EDD. Within this setup, one has to chose suitable atomic reference states, and it was found that for different atomic reference states, different atomic charges were obtained. This issue was resolved by Bultinck \textit{et al.}\cite{BultinckHI2007}, who proposed an extension of the scheme by iterative modification of the reference state. Starting with, for example, neutral reference EDDs, $\rho_A^0 (r)$, the $w_A^H (r)$ are calculated. From these the AIM EDDs are calculated as:
\begin{equation}
\rho_A^{AIM} (r)=\frac{\rho_A^0 (r)}{\rho_{promol}^0 (r)} \rho_{mol} (r).  
\end{equation}
With these AIM EDDs the atomic charge, $x$, is calculated through integration over the entire system. In the following step, the reference EDD is constructed as the linear interpolation of EDDs with atomic charge $I<x<I+1$, with $I$ the integer value of the ionic charge lower than $x$. Using these reference EDDs, $\rho_A^x (r)$, new weights, AIM EDDs, and atomic charges are calculated. This scheme is then iterated to convergence of the atomic charges.\\
For periodic systems, the problem of the infinite size of the system is resolved by only considering the atoms of the unit cell for calculation of the charges and periodic copies which are ``\textit{nearby}"\cite{VanpouckeDannyEP:2013aJComputChem}.
The Hirshfeld weights are calculated for all grid points (of an atom centered Becke grid\cite{Becke1988}) associated with the atoms of the unit cell, and all other grid points which are located in the same spatial region. Atoms contributing to the weights are thus the unit cell atoms, as well as periodic copies within a limited range\cite{VanpouckeDannyEP:2013aJComputChem, VanpouckeDannyEP:2013bJComputChem}.
Furthermore, it was found that the EDDs of the valence electrons (i.e., all electrons not included in the frozen core) can be used without loss of quality compared to all-electron EDDs, while using a much coarser grid.

\subsection{Computational settings}
In this work, HI-charges are calculated using the previous implementation for periodic systems,\cite{VanpouckeDannyEP:2013aJComputChem, VanpouckeDannyEP:2013bJComputChem} as found in the HIVE package\cite{HIVE}.
The calculations for generating the EDDs are performed using the VASP package. Reference atomic densities are calculated using a small unit cell of $20\times 20\times 20$ \AA$^3$ for the cations, while a large $40\times 40\times 40$ \AA$^3$ cell is used for the tail correction of the anions. The plane wave kinetic energy cut off is set to $1000$ eV. The EDDs of the oxide and unary systems are obtained from static calculations using a $33\times 33\times 33$ $\Gamma$-centered k-point integration mesh and a kinetic energy cut off of $1000$ eV, using the PBE functional as defined in \ref{VASP_SI}.
The atomic charges of the systems are calculated using the HI partitioning scheme with a charge convergence criterion of $1.0\times 10^{-4}$ electron. Charges are integrated on a logarithmic radial grid with atom-centered spherical shells of 1202 Lebedev--Laikov grid points\cite{Becke1988, LebedevLaikov}.

\subsection{Discussion of Hirshfeld-I atomic charges and their relation to formal oxidation states}
By imposing the topology of the oxides, every element X should exist in a predetermined formal oxidation state covering all integer values from $+1$ to $+6$. This formal oxidation state, however does not correspond one-to-one to the local configuration of the charge density around element X that describes how element X binds to the surrounding oxygens. This is particularly relevant for those oxides that are ‘\textit{exotic}’ (e.g., hydrogen in \ce{HO3} has a formal oxidation state of $+6$, whereas even in a complete ionic picture hydrogen can only donate a single electron). In order to survey the actual chemical environment of every element X, we therefore calculated the Hirshfeld-I charges for X in all oxides and unaries (not shown), near the equilibrium volume\cite{VanpouckeDannyEP:2013aJComputChem, VanpouckeDannyEP:2013bJComputChem, BultinckHI2007}.
As any other AIM scheme to define charges, Hirshfeld-I charges have their limitations. A common limitation all AIM schemes have to deal with is the fact that atomic charge is not a quantum mechanical observable. As such, there exists no absolute true value to find. The choice of the specific AIM scheme to calculate charges is therefore guided by the wish to satisfy other requirements. Attractive features of the Hirshfeld-I charges in this context are: They are (1) basis-set independent\cite{BultinckHIbasisset2007}, (2) very robust, meaning charges are not structure dependent if the chemical environment remains the same, while very sensitive to changes in the chemical environment or oxidation state\cite{WolffisJJVanpouckeDEP:MicroMesoMater2019, BeukenVanpoucke:AngewChemIntEd2015, Verstraelen2012}, (3) and large, though always smaller than the formal charge. 

\begin{figure*}[t!]
    \centering
    \begin{subfigure}[t]{0.5\textwidth}
        \centering
        \includegraphics[width=\columnwidth]{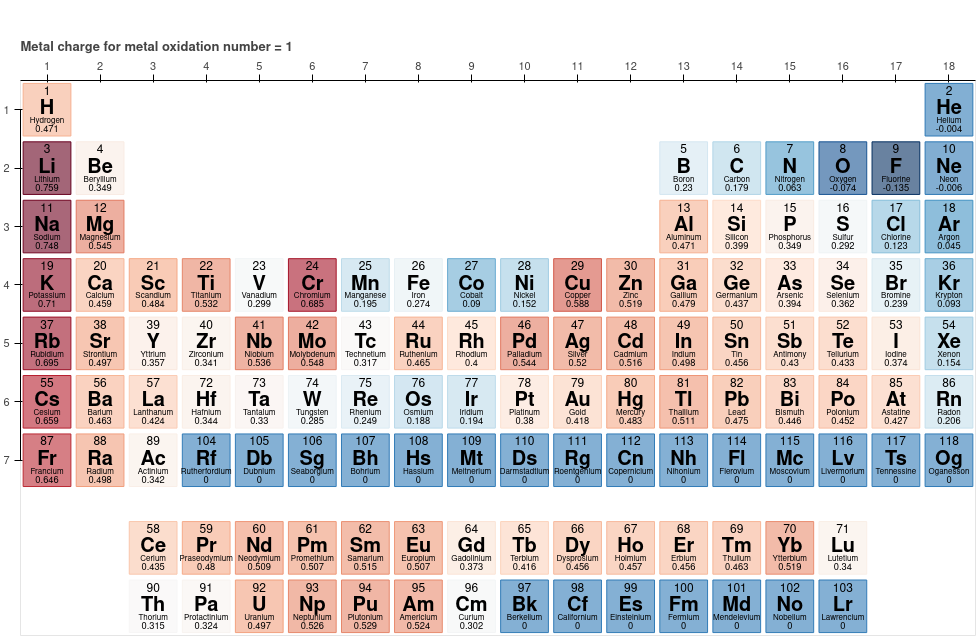}
        \caption{ }
    \end{subfigure}%
    ~ 
    \begin{subfigure}[t]{0.5\textwidth}
        \centering
        \includegraphics[width=\columnwidth]{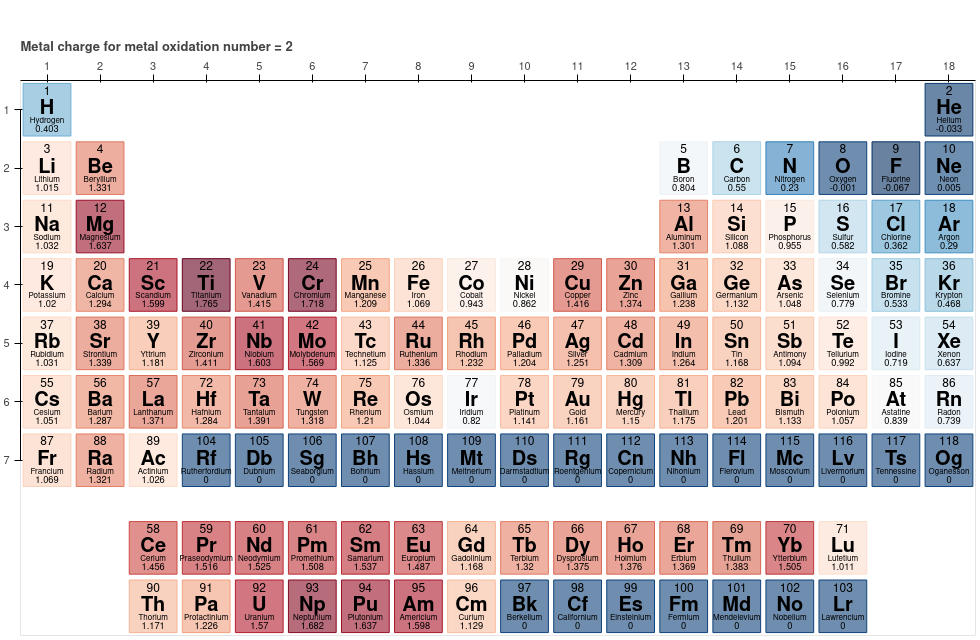}
        \caption{ }
    \end{subfigure}
    \begin{subfigure}[t]{0.5\textwidth}
        \centering
        \includegraphics[width=\columnwidth]{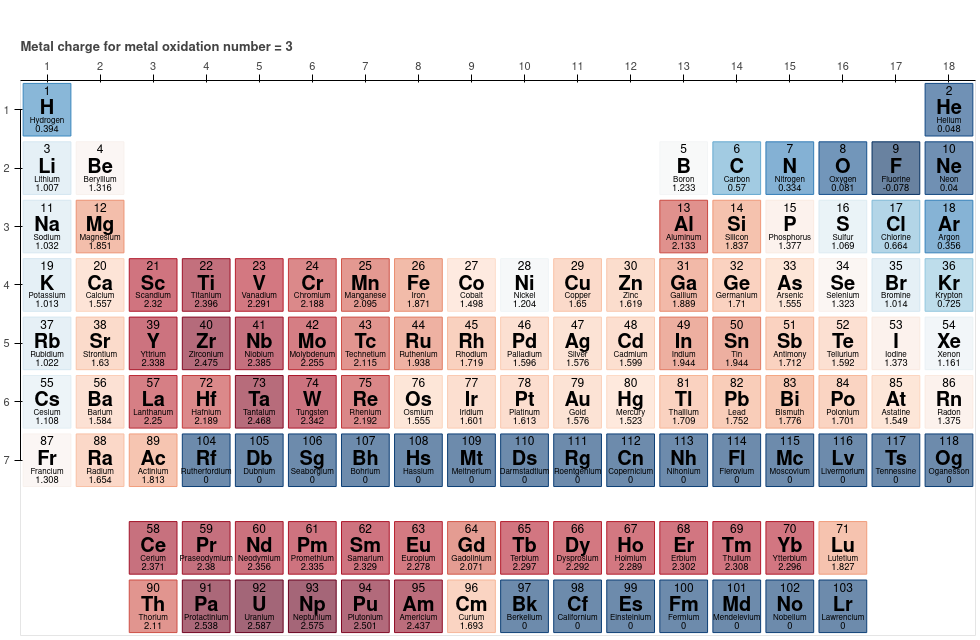}
        \caption{ }
    \end{subfigure}%
    ~ 
    \begin{subfigure}[t]{0.5\textwidth}
        \centering
        \includegraphics[width=\columnwidth]{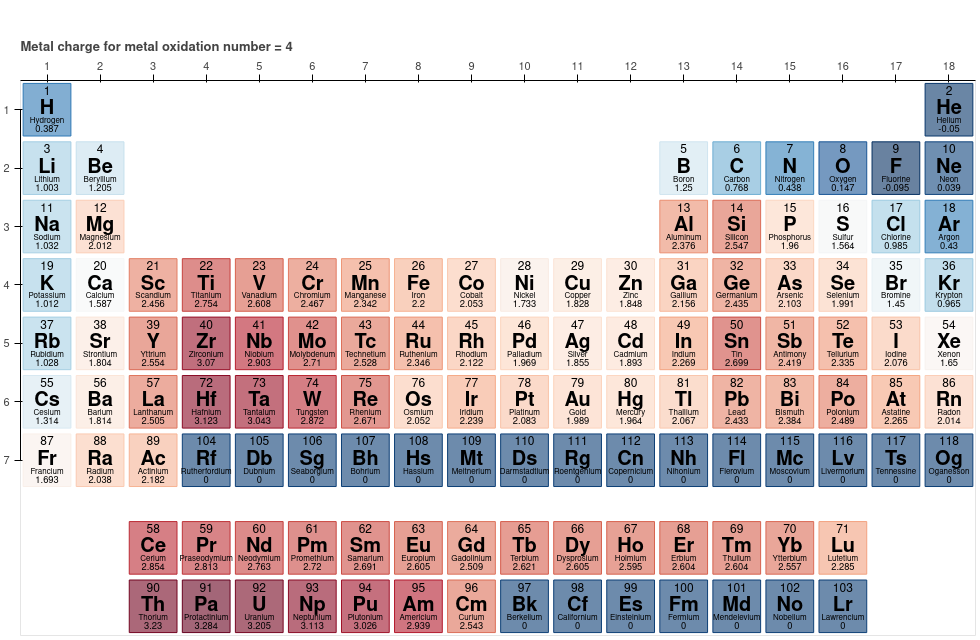}
        \caption{ }
    \end{subfigure}
\begin{subfigure}[t]{0.5\textwidth}
        \centering
        \includegraphics[width=\columnwidth]{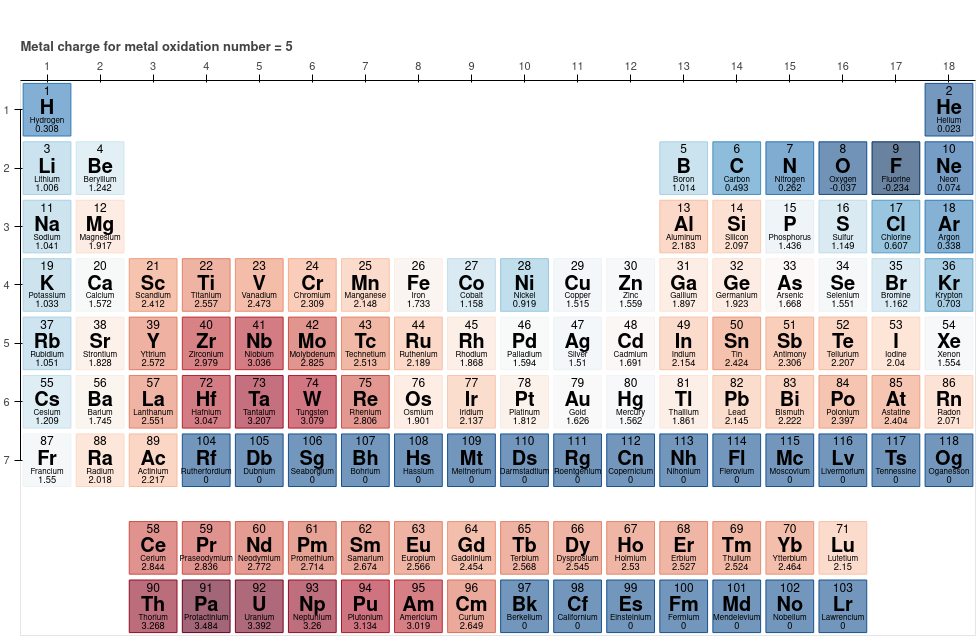}
        \caption{ }
    \end{subfigure}%
    ~ 
    \begin{subfigure}[t]{0.5\textwidth}
        \centering
        \includegraphics[width=\columnwidth]{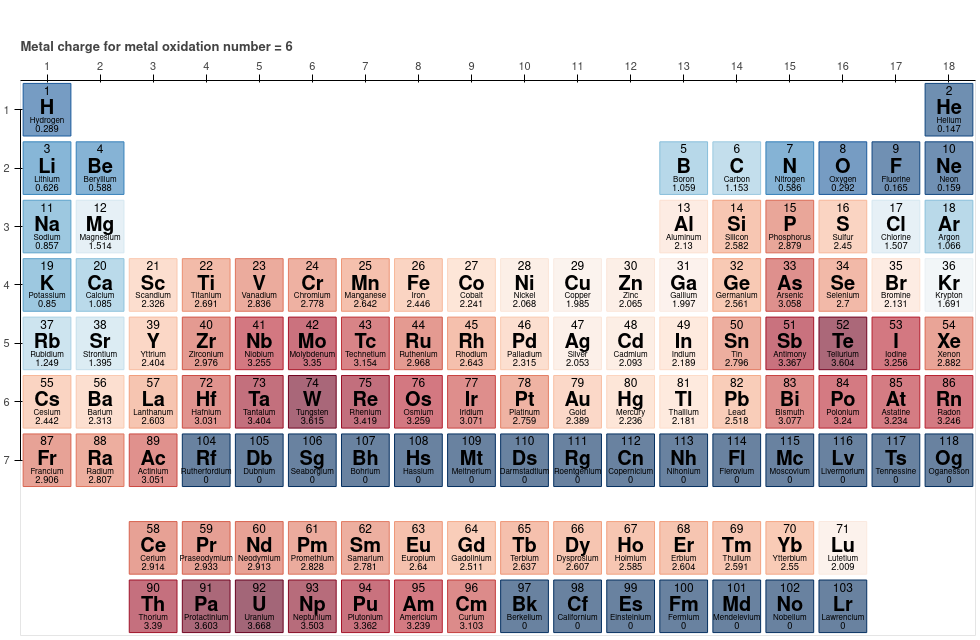}
        \caption{ }
    \end{subfigure}
    
    \caption{The average Hirshfeld-I charge on the metal atom for the oxides dataset according to the formal charge. Formal charge $+1$ (a), $+2$ (b), $+3$ (c), $+4$ (d), $+5$ (e), and $+6$ (f). Brownish colors indicate elements that have a Hirshfeld-I charge that is in line with their formal charge. \label{sifig:HI_PTables}}
\end{figure*}

The results in SI Fig.~\ref{sifig:HI_PTables} show the variation of the oxygen and metal charges over the periodic table as function of the formal oxidation state. These pictures reflect several trends that are intuitively expected: for X$_2$O (formal oxidation state +1), the elements with a HI charge closest to 1 are the alkali elements, while for XO (formal oxidation state +2) the elements with a HI charge closest to 2 are the earth-alkaline elements. On the other hand, the HI charges are clearly limited and are often about one half of the formal oxidation state. For instance, for XO$_3$ (formal oxidation state +6), the HI charges are often in the range 2.5-3.5.

\begin{figure*}[t!]
    \centering
    \begin{subfigure}[t]{0.5\textwidth}
        \centering
        \includegraphics[width=\columnwidth]{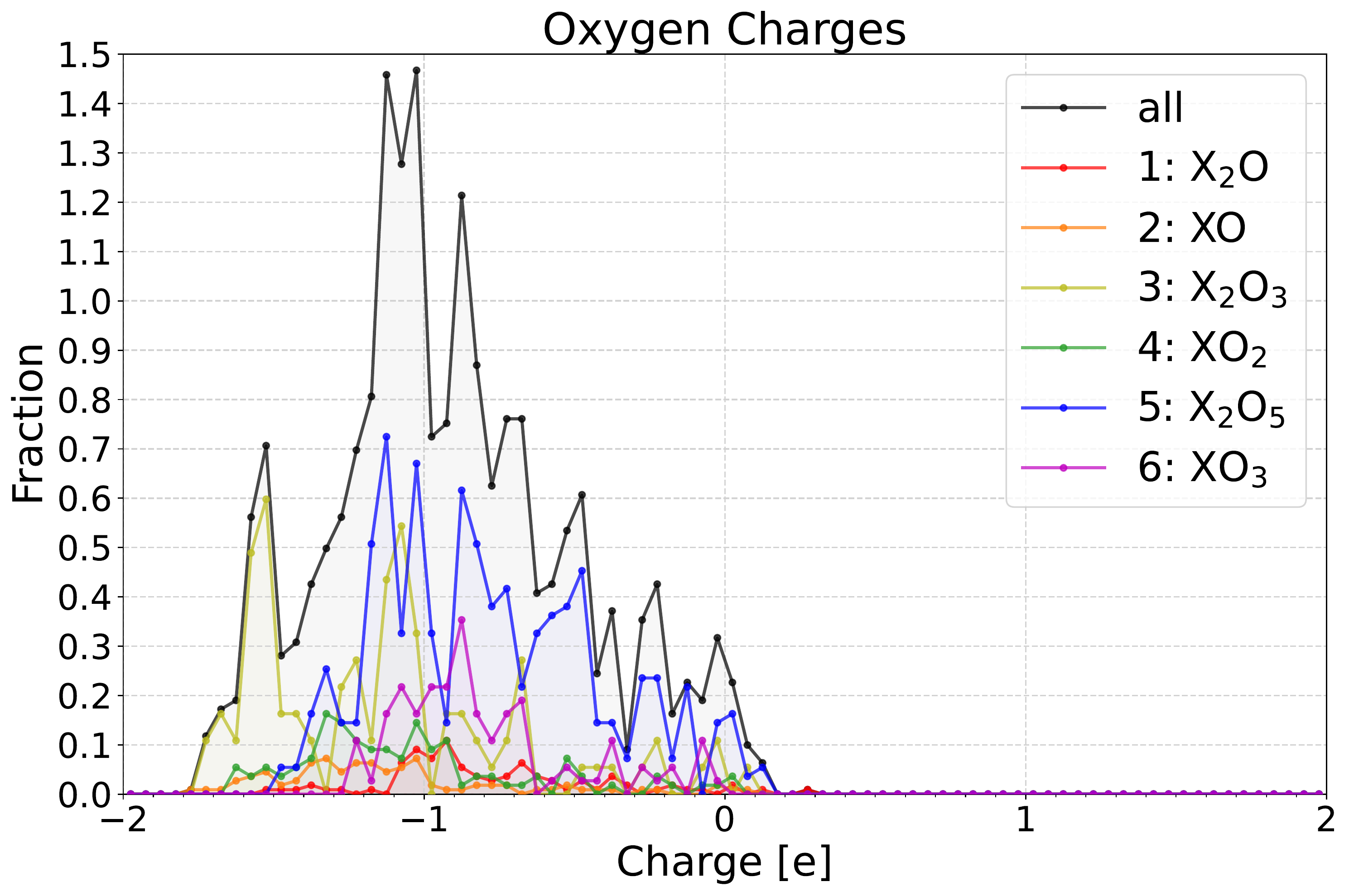}
        \caption{ }
    \end{subfigure}%
    ~ 
    \begin{subfigure}[t]{0.5\textwidth}
        \centering
        \includegraphics[width=\columnwidth]{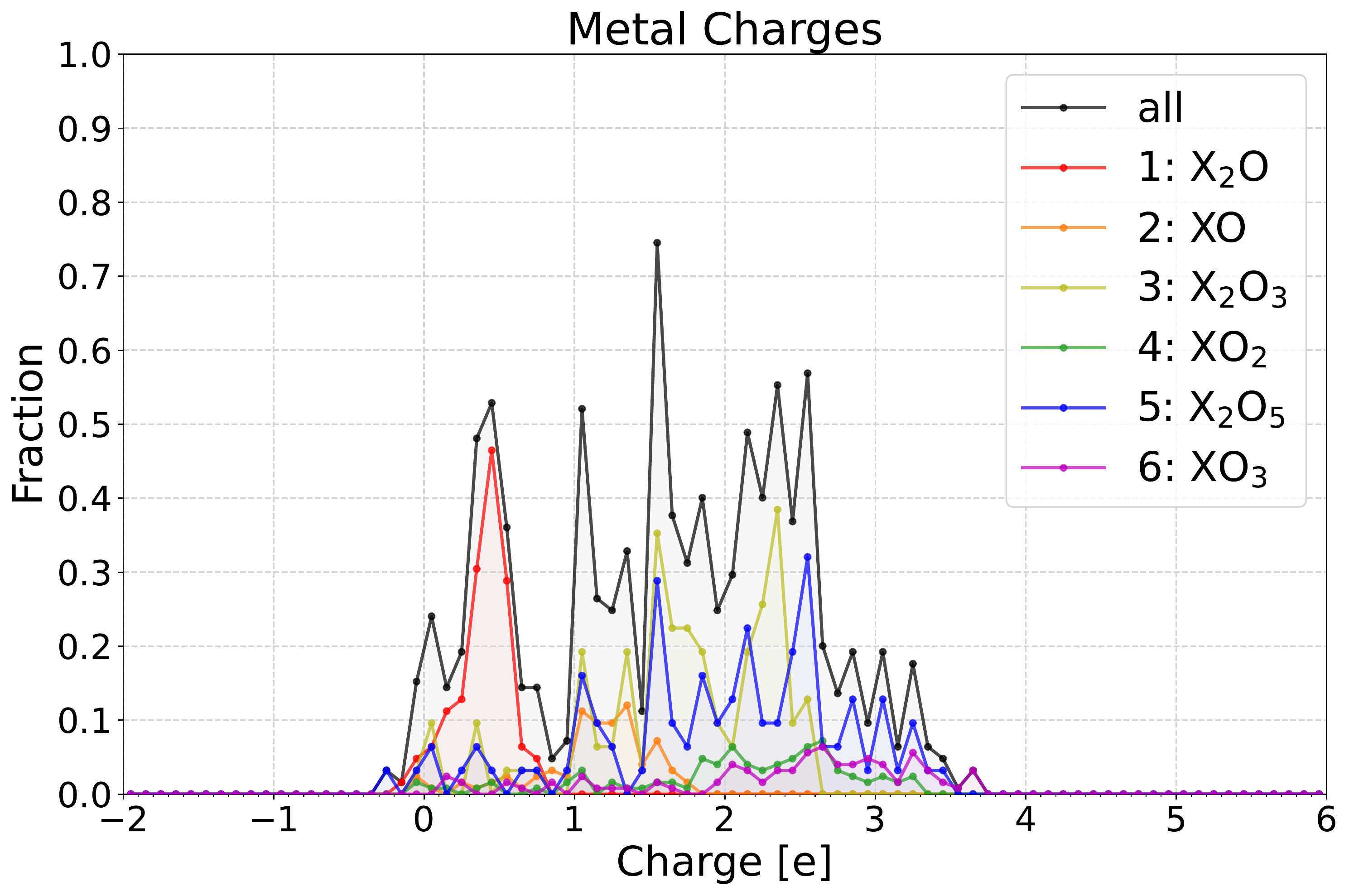}
        \caption{ }
    \end{subfigure}
    \caption{(a) Distribution of oxygen charges according to the Hirshfeld-I method for the oxides dataset. (b) Distribution of metal charges according to the Hirshfeld-I method for the oxides dataset.\label{sifig:HI_MOcharge}}
\end{figure*}

\begin{figure*}[t!]
    \centering
    \includegraphics[width=0.8\textwidth]{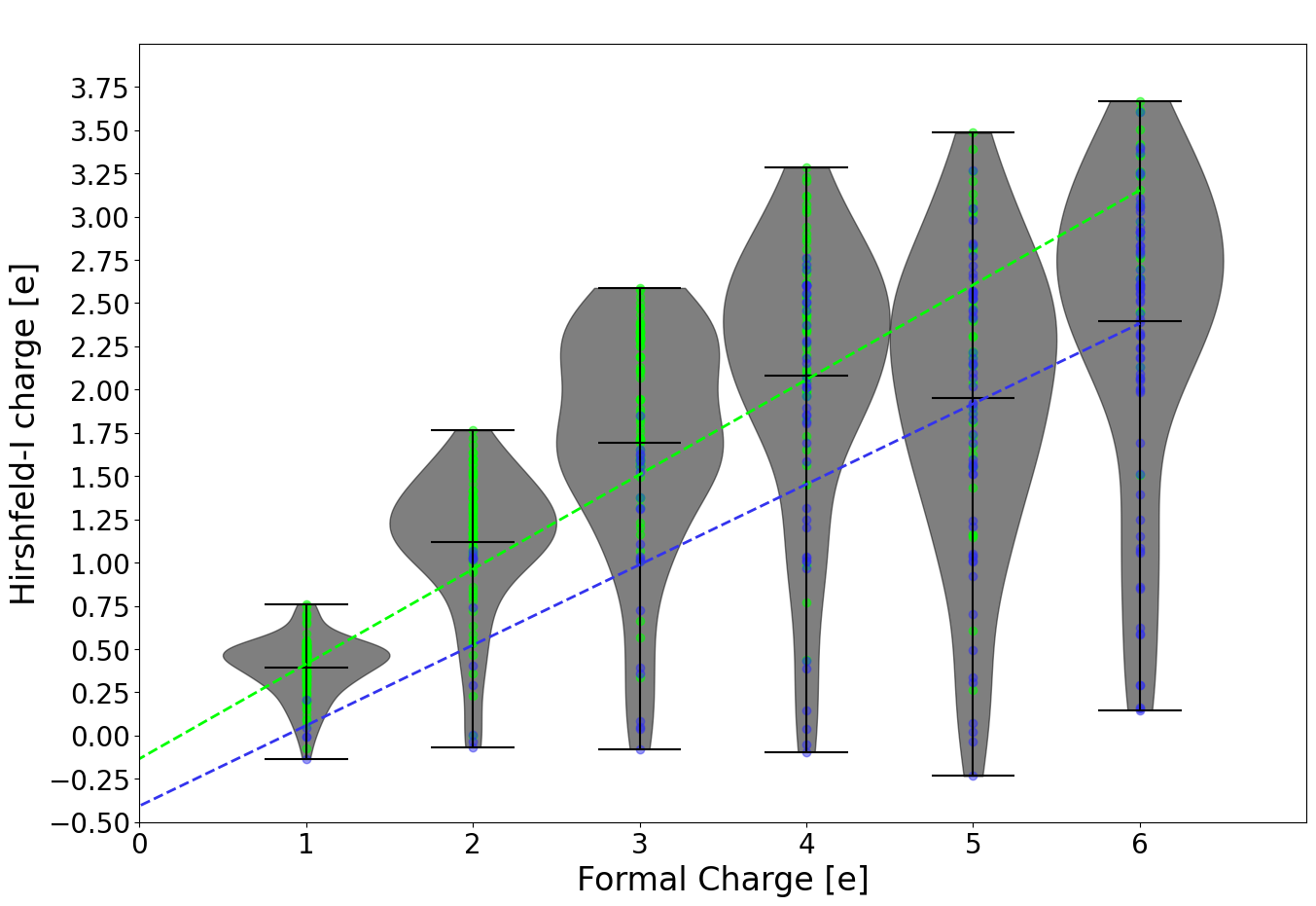}
    \caption{Violin plot of the HI charges of X in the oxides, as function of the formal oxidation state. Green dots and green linear fit: all oxides for which the formal oxidation state of X in this oxide is less than or equal to the maximal common formal oxidation state of X (as listed in Ref.~\citenum{HirBook}). Blue dots and blue linear fit: all oxides for which the formal oxidation state of X in this oxide is larger than the maximal common formal oxidation state of X (as listed  in Ref.~\citenum{HirBook}). The green subset has therefore all oxides for which the formal oxidation state of X is `reasonable', the blue subset represents oxides for which the formal oxidation state of X is `exotic'. \label{sifig:HI_violin}}
\end{figure*}

Looking at the entire distribution of the atomic charges over the entire oxide dataset, (see SI Fig.~\ref{sifig:HI_MOcharge}) shows that the average metal charge gradually shifts to higher charges with increasing formal charge. In case of the oxygen charges, note that the oxygen charge decreases in size with increasing formal charge of the metal, which is a consequence of the fact that the formal charge is never actually fully transferred. Visualizing the results per formal oxidation state of the oxide system (see SI Fig.~\ref{sifig:HI_violin}) shows an increasing trend, as expected. More interestingly, if the materials are split in two subsets -- those with a `reasonable' formal oxidation state for X and those with an `exotic' formal oxidation stated for X (see caption of SI Fig.~\ref{sifig:HI_violin}) -- it becomes clear that for the `reasonable' subset the calculated Hirshfeld-I charge is on average about half of the formal charge. In the case of the other subset, a lower value is found. Taken together, the different features of SI Fig.~\ref{sifig:HI_violin} consistently express that even though the nominal formal charges are not obtained, the six different crystal structures for the oxides give rise to systematically different chemical environments. This was exactly the purpose of imposing these six different oxide crystal structures.

\clearpage

\section{Determination of the weights of the metric $\nu$ based on the error propagation on the Birch--Murnaghan fit parameters\label{SI:stability_eos}}

In this Section, we motivate the choice of weights $w_{V_0} = 1$, $w_{B_0} = 1/20$ and $w_{B_1} = 1/400$ discussed in the text for the $\nu$ metric.

When computing the \gls{eos} curves $E(V)$, the results from any simulation are affected by numerical noise, originating from many different sources (finiteness of the k-point integration mesh, basis set discretization, thresholds to stop the self-consistent convergence cycle, \ldots).
When these points are fitted to a Birch--Murnaghan equation of state, the error propagates to the resulting fit parameters. This has been investigated in detail in Ref.~\citenum{volume-error}, and several of the observations mentioned underneath are in line with the conclusions reached there.
Intuitively, one can already expect that the numerical error will be larger for those parameters that are associated with higher-order derivatives. For instance, $V_0$ is the minimum of the \gls{eos} curve (i.e., the zero of the first derivative) is expected to be affected by a smaller error with respect to $B_0$ that is related to the curvature of the \gls{eos} curve close to the minimum (thus, to its second-order derivative).
A first observation is that the error on all parameters will increase for increasing input noise on the energy datapoints. However, our goal in this section is not to quantify the error on each of these properties independently, but rather to understand if the error on pairs of fit parameters is related.
In particular we will show that errors on $B_0$ ($B_1$) are typically 20 (400) times larger than those on $V_0$; by arbitrarily setting $w_{V_0} = 1$ (a change to this would result only in a global multiplicative factor), this will justify our final choice of weights.

\begin{figure}
    \centering
    \includegraphics[width=0.4\textwidth]{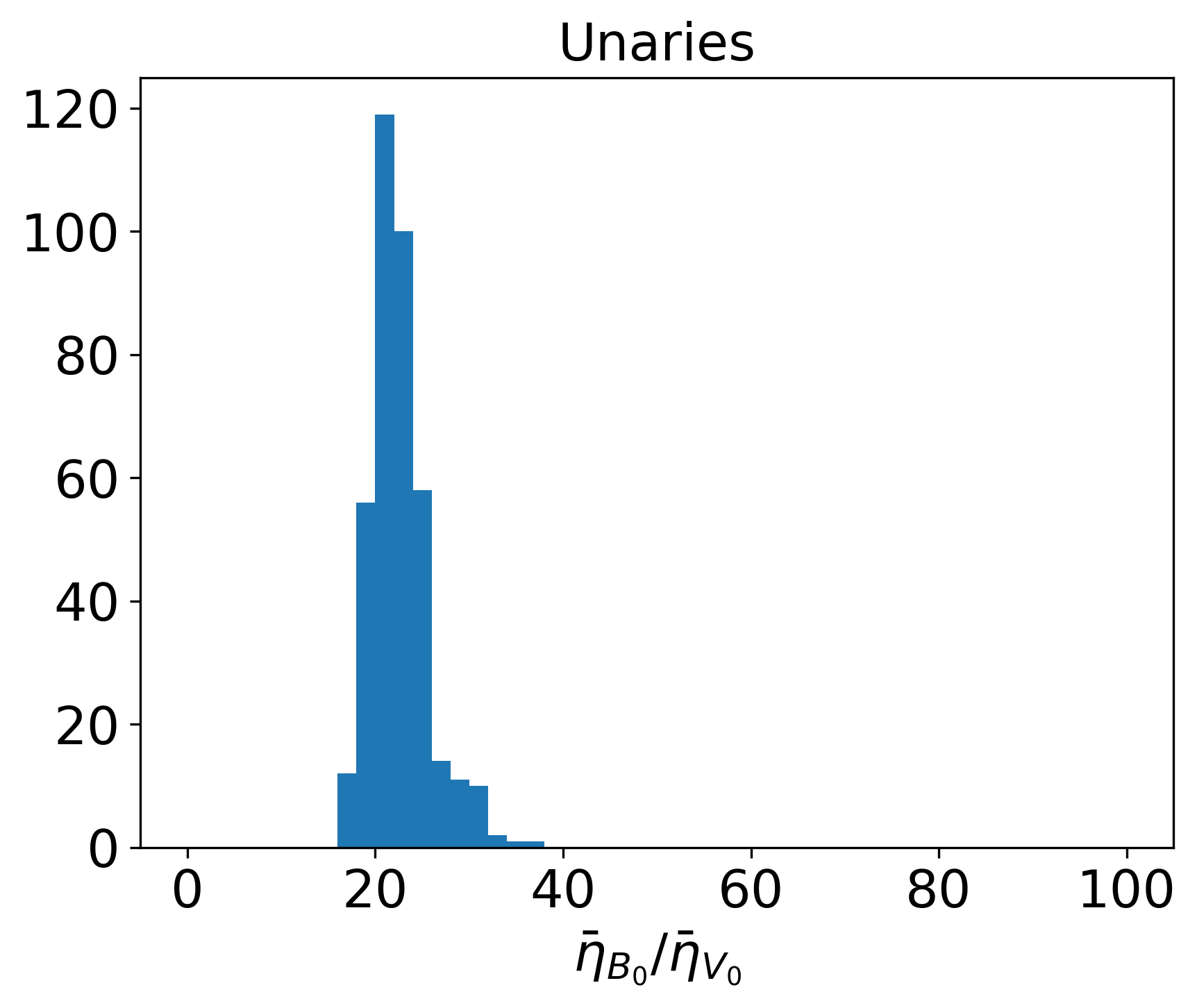}
    \includegraphics[width=0.4\textwidth]{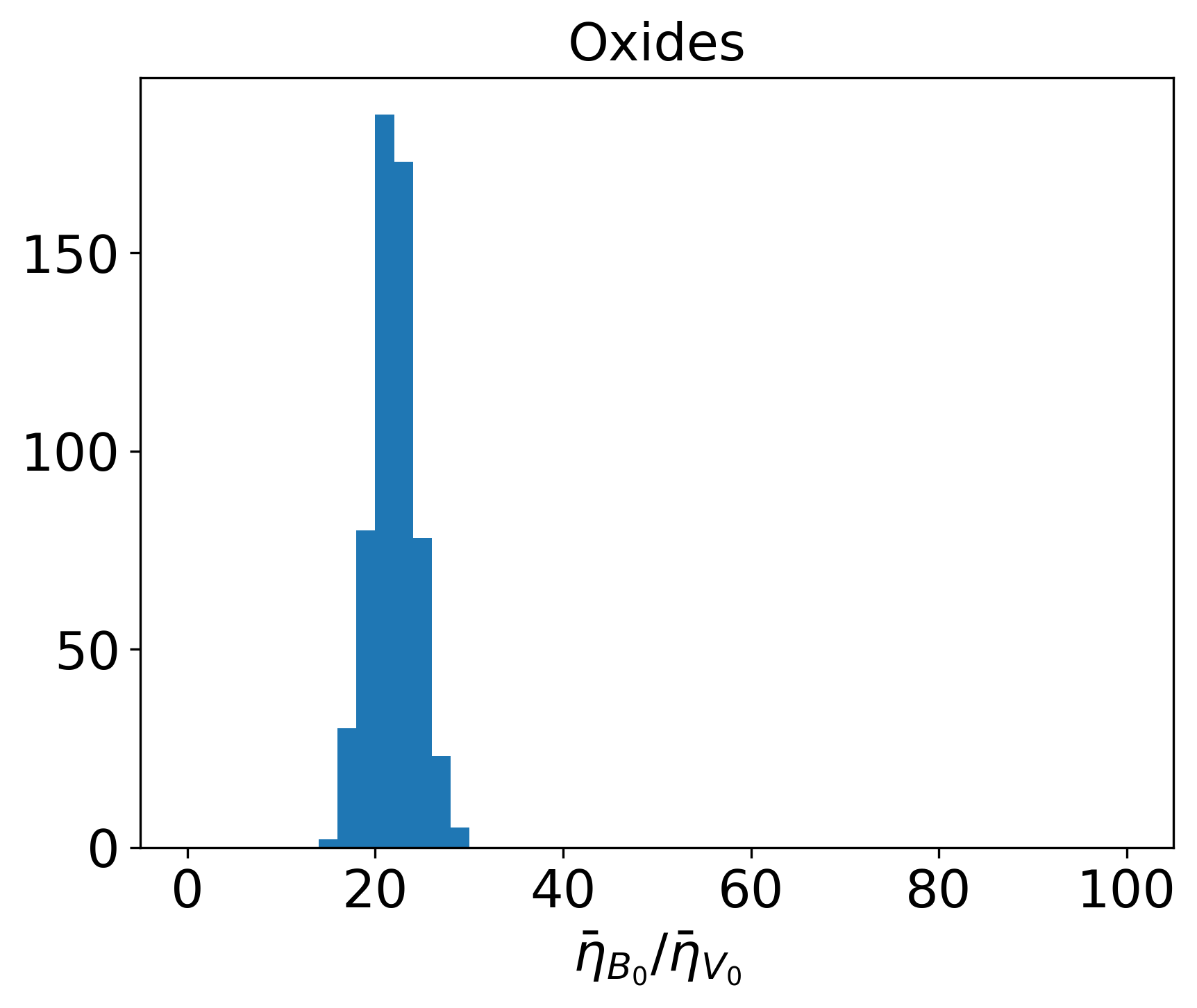}

    \includegraphics[width=0.4\textwidth]{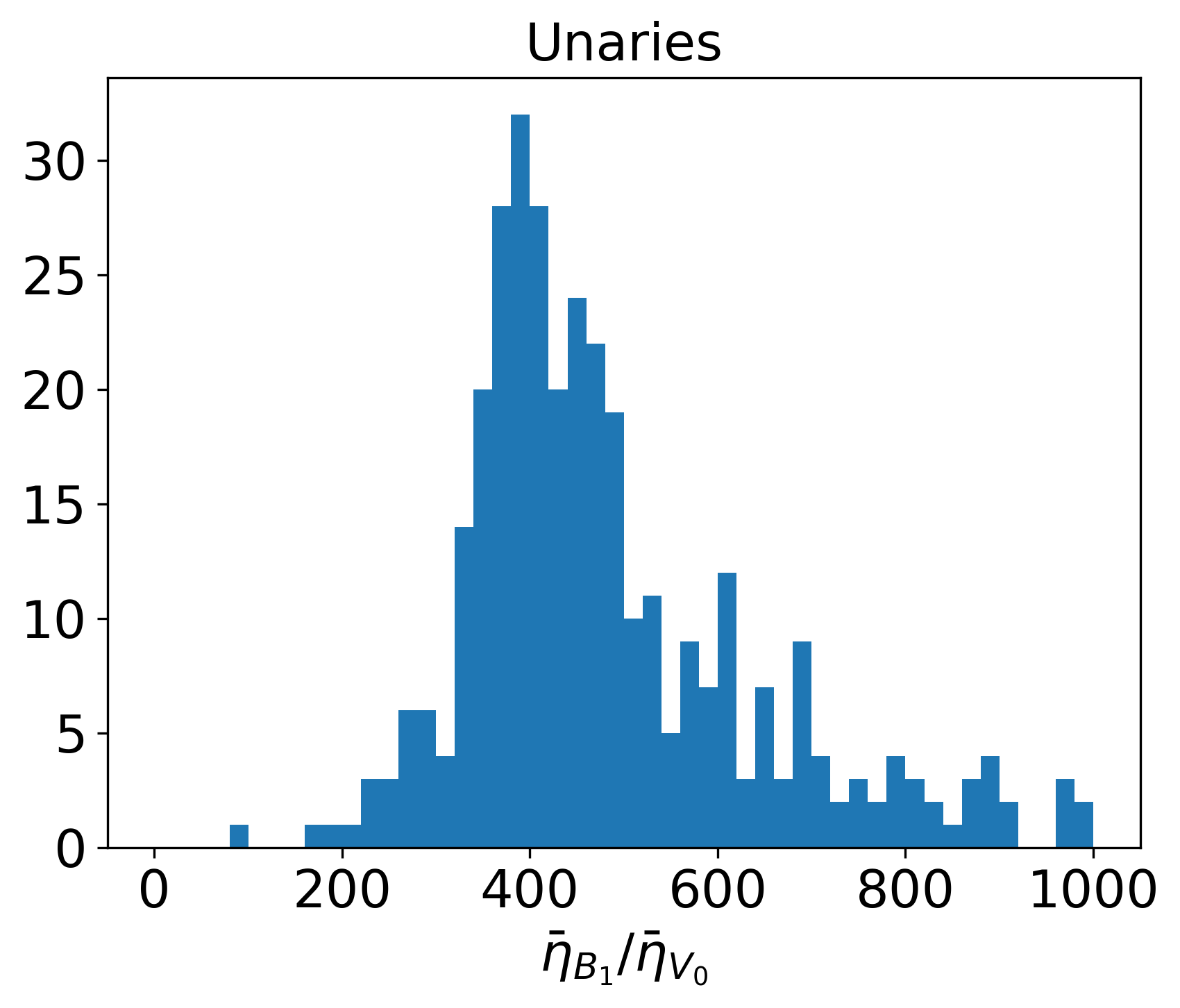}
    \includegraphics[width=0.4\textwidth]{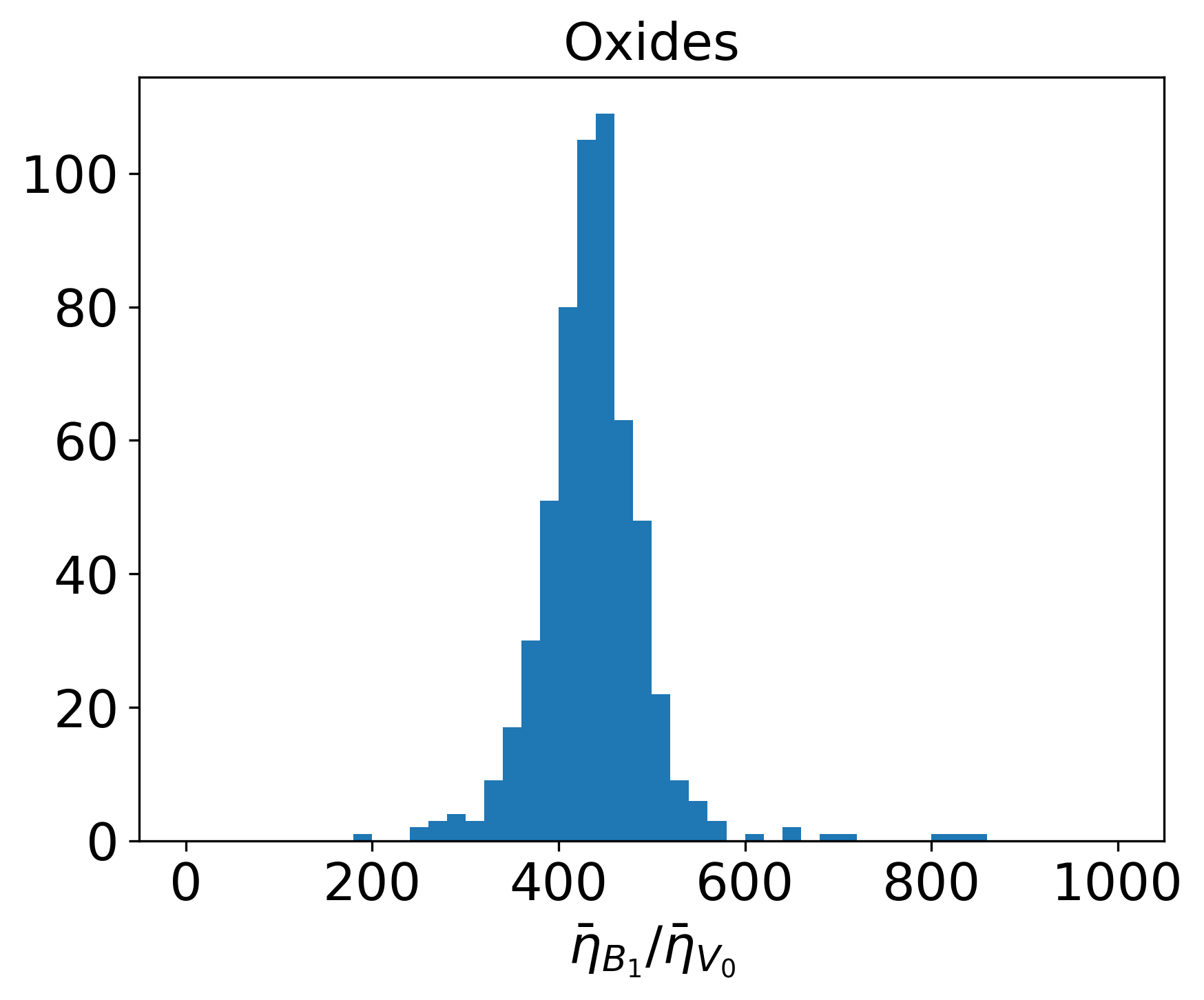}
    \caption{Histograms of the typical error ratios of $B_0$ vs. $V_0$ (${\bar \eta_{B_0}}/{\bar \eta_{V_0}}$, top row) and of $B_1$ vs. $V_0$ (${\bar \eta_{B_1}}/{\bar \eta_{V_0}}$, bottom row) for the unaries set (left column) and the oxides set (right column). The simulations were run for $N_s=100$ random samples and a standard deviation on the error of the energy on the datapoints of $n_\sigma = 10^{-5}$ eV. The histograms indicate that the error of $B_0$ ($B_1$) is approximately 20 (400) times larger than the error on $V_0$, justifying our choice of weights for the metric $\nu$.
    \label{fig:histograms-nu-weights}}
\end{figure}

We extract these relative weights using the following approach. We start from our reference \gls{ae} dataset and consider, for each of the 960 materials, the fitted parameters $V^{ref}_0$, $B^{ref}_0$ and $B^{ref}_1$.
Rather than using the datapoints from the \gls{ae} simulations, however, we generate a new ``perfect'' dataset (i.e., not affected by any numerical noise) by creating, for every curve, 7 fictitious points lying exactly on the Birch--Murnaghan curve, with the same volume spacing as discussed in the main text (spacing of 2\% in volume between 94\% and 106\% of the tabulated central volume). This removes from our analysis any existing numerical noise of the \gls{ae} simulations that is due to the numerical approximations in the two specific codes, rather than originating from the fitting procedure.
We then select a reference average numerical error $n_\sigma$ for the energy value of each point, and randomly displace each energy by a random value following a normal distribution with zero mean and standard deviation $n_\sigma$. 
We fit these noisy datapoints with the Birch--Murnaghan curve, thus obtaining fitted values of $V_0$, $B_0$ and $B_1$, that will be different from the initial reference ones $V^{ref}_0$, $B^{ref}_0$ and $B^{ref}_1$. We therefore compute the relative errors on each of them with respect to their average (similar to what is done for $\varepsilon$ and $\nu$): e.g., for the minimum volume $\eta_{V_0} = (V_0 - V^{ref}_0)/((V_0 + V^{ref}_0)/2)$, for the bulk modulus
$\eta_{B_0} = (B_0 - B^{ref}_0)/((B_0 + B^{ref}_0)/2)$
and for the derivative of the bulk modulus
$\eta_{B_1} = (B_1 - B^{ref}_1)/((B_1 + B^{ref}_1)/2)$.
We repeat the procedure for $N_s$ random samples, and finally compute the average of the absolute value of the three relative errors $\eta_{V_0}$, $\eta_{B_0}$ and $\eta_{B_1}$ on the $N_s$ samples (for each material):
\begin{equation}
\bar \eta_{V_0} = \sum_{i=1}^{N_s} \frac{|\eta_{V_0}(i)|}{N_s}
\end{equation}
and similarly for $\bar \eta_{B_0}$ and $\bar \eta_{B_1}$, where $i$ denotes each of the individual independent random noise samples.
The values $\bar \eta_{V_0}$, $\bar \eta_{B_0}$ and $\bar \eta_{B_1}$ quantify the typical average errors on the three fit parameters for a numerical noise of magnitude $n_s$.
By producing histograms of the three quantities over the whole dataset of 960 structures, we obtain a peaked distribution that represents the range of values typical of our materials dataset ($V_0\approx 3-400$ \AA$^3$, $B_0\approx 0.001-2.7$ eV/\AA$^3$ and $B_1\approx 0.4-12$).
The position of the peak, as we expected, depends on the noise magnitude $n_\sigma$. As we discussed, however, we do not consider the histograms of these three quantities but we produce, instead, histograms for the two ``relative'' quantities
\begin{equation}
\frac{\bar \eta_{B_0}}{\bar \eta_{V_0}}\qquad\text{and}\qquad\frac{\bar \eta_{B_1}}{\bar \eta_{V_0}}.
\end{equation}
The positions of the peaks of these histograms will represent the quantities we wish to determine: the typical ratio of numerical error on pairs of fit parameters.
The results of our simulations can be summarized as follows:
\begin{itemize}
    \item $N_s=100$ samples are already enough to converge the statistics and the histograms for our goal of identifying the peaks of the histograms;
    \item the positions of the peaks of $\bar \eta_{V_0}$, $\bar \eta_{B_0}$ and $\bar \eta_{B_1}$ are roughly proportional to the input noise $n_\sigma$; however, the position of the peaks of $\frac{\bar \eta_{B_0}}{\bar \eta_{V_0}}$ and $\frac{\bar \eta_{B_1}}{\bar \eta_{V_0}}$ are, to a good approximation, independent of $n_\sigma$ for the noises that we considered (in the range $10^{-4}-10^{-6}$ eV) (see also Ref.~\citenum{volume-error} for similar conclusions on the $\Delta$ metric);
    \item for our choice of volume range (94\%--106\%), the two histograms (see SI Fig.~\ref{fig:histograms-nu-weights}) display clear peaks at positions that can be rounded to 20 and 400, respectively. The peak positions are consistent when considering independently unaries and oxides (even if the spread of the peaks is different in the two cases). Hence, we choose the weights as $w_{V_0} = 1$, $w_{B_0} = 1/20$ and $w_{B_1} = 1/400$.
    \item The peak positions are insensitive to the number of datapoints, as long as the total volume range is not modified. Instead, they change significantly if the volume range is changed. For instance, using a volume range of 90\%--110\% would result in values closer to 15 and 200 for the two ratios, respectively. This can also be intuitively explained: $B_1$, for instance, is related to the non-parabolicity of the Birch--Murnaghan curve away from its minimum. If we consider a very small volume range, the curve will be very close to parabolic, and we therefore expect a large error on $B_1$ since the fit has very little information on the non-parabolic behavior. For larger volume ranges, the curve starts to deviate significantly from a parabola, thus providing more information to the fitting algorithm on the actual value of $B_1$, in turn resulting into a smaller relative error on $B_1$ vs. $B_0$ or $V_0$.
    \item The stability of the fit, especially on $B_1$, is significantly affected by the choice of fitting algorithm.
    For instance, we realized that if one uses the \texttt{optimize.curve\_fit} subroutine of SciPy (\url{https://www.scipy.org}), which is \emph{not} the algorithm used in this work, the choice of the fitting starting point is very important, and we also observe that iterating the procedure a few times (using the results of the previous step as starting points for the next fit) improves the stability. Instead, the function used in this work (that is the same also used in Ref.~\citenum{Lejaeghere:2016,deltasite}) is a non-iterative fitting algorithm that proves to be much more robust. 
\end{itemize}

\clearpage

\section{Reference all-electron results for $V_0$, $B_0$ and $B_1$}\label{SI:results-ae}

This section reports the complete reference dataset of the \gls{eos} parameters obtained with the two all-electron codes \fleur{} and \wientwok, and the absolute value of their percentage difference (that we indicate with $\eta$). Moreover, it reports the averaged parameters among them,
that constitutes our reference average dataset presented in this manuscript. 
Data is divided in 10 tables, one for each crystal structure (4 unaries and 6 oxides).
The agreement for $V_0$ is within 0.3\% for all materials except Cs$_2$O$_5$ (0.323\%),
Fr$_2$O$_5$ (0.645\%), Ra$_2$O$_5$ (0.333\%), NeO$_3$ (0.302\%) and RbO$_3$ (0.343\%). Not surprisingly, these are 5 crystals with very small bulk moduli B$_0$: it has been shown in Ref.~\citenum{volume-error} that the error in the volume scales inversely with the value of the bulk modulus.

Parameters are expressed per formula unit (see also SI Sec.~\ref{SIsec:structures}). Note that, for X$_2$O$_3$ and for X$_2$O$_5$, the primitive cell has twice the number of atoms (10 and 14, respectively) than the number of atoms in the formula unit (5 and 7, respectively). This is reflected in a factor of 0.5 in the volumes reported in this table with respect to the volume of the unit cells in the input files available in Ref.~\citenum{MCA-ACWF}. 

{\footnotesize
\centering
\begin{center}
\begin{longtable}{c|ccc|ccc|ccc|ccc}
\caption{\label{sitab:table-ae-X/FCC}Table with all calculated \gls{eos} parameters for the FCC structures obtained with \fleur{} and \wientwok.} \\ 
& \multicolumn{3}{c|}{ \fleur } & \multicolumn{3}{c|}{ \wientwok } & \multicolumn{3}{c|}{Abs. percentage difference [\%]} & \multicolumn{3}{c}{ Average set }\\ 
& $V_0$ [\AA$^3$] & $B_0$ [eV/\AA$^3$]  & $B_1$ & $V_0$ [\AA$^3$] & $B_0$ [eV/\AA$^3$] & $B_1$ & $|\eta(V_0)|$ & $|\eta(B_0)|$ & $|\eta(B_1)|$ & $V_0$ [\AA$^3$] & $B_0$ [eV/\AA$^3$] & $B_1$ \\ 
\hline
\endfirsthead
\caption{(continued) Table with all calculated \gls{eos} parameters for the FCC structures obtained with \fleur{} and \wientwok.} \\ 
& \multicolumn{3}{c|}{ \fleur } & \multicolumn{3}{c|}{ \wientwok } & \multicolumn{3}{c|}{Abs. percentage difference [\%]} & \multicolumn{3}{c}{ Average set }\\ 
& $V_0$ [\AA$^3$] & $B_0$ [eV/\AA$^3$]  & $B_1$ & $V_0$ [\AA$^3$] & $B_0$ [eV/\AA$^3$] & $B_1$ & $|\eta(V_0)|$ & $|\eta(B_0)|$ & $|\eta(B_1)|$ & $V_0$ [\AA$^3$] & $B_0$ [eV/\AA$^3$] & $B_1$ \\ 
\hline
\endhead
H & 2.9651 & 0.6768  & 3.2596 & 2.9646 & 0.6764 & 3.2598& 0.017 & 0.048 &  0.007 & 2.9648 & 0.6766 & 3.2597\\
He & 17.7867 & 0.0053  & 6.208 & 17.7585 & 0.0054 & 6.6423& 0.159 & 0.404 &  6.759 & 17.7726 & 0.0053 & 6.4251\\
Li & 20.2246 & 0.0862  & 3.3304 & 20.2243 & 0.0862 & 3.3298& 0.001 & 0.001 &  0.019 & 20.2245 & 0.0862 & 3.3301\\
Be & 7.8705 & 0.7437  & 3.4867 & 7.8728 & 0.7437 & 3.471& 0.029 & 0.006 &  0.449 & 7.8716 & 0.7437 & 3.4789\\
B & 5.8886 & 1.6978  & 3.7481 & 5.8945 & 1.6993 & 3.7554& 0.099 & 0.088 &  0.196 & 5.8916 & 1.6985 & 3.7517\\
C & 7.3221 & 0.9364  & 3.5885 & 7.3212 & 0.9358 & 3.5906& 0.012 & 0.07 &  0.057 & 7.3216 & 0.9361 & 3.5896\\
N & 7.6017 & 1.1174  & 4.1748 & 7.6009 & 1.1161 & 4.1633& 0.011 & 0.119 &  0.276 & 7.6013 & 1.1167 & 4.169\\
O & 7.9988 & 0.8755  & 4.7639 & 7.9987 & 0.8745 & 4.7427& 0.002 & 0.115 &  0.446 & 7.9988 & 0.875 & 4.7533\\
\end{longtable}
\end{center}
\begin{center}
\begin{longtable}{c|ccc|ccc|ccc|ccc}
\caption{\label{sitab:table-ae-X/BCC}Table with all calculated \gls{eos} parameters for the BCC structures obtained with \fleur{} and \wientwok.} \\ 
& \multicolumn{3}{c|}{ \fleur } & \multicolumn{3}{c|}{ \wientwok } & \multicolumn{3}{c|}{Abs. percentage difference [\%]} & \multicolumn{3}{c}{ Average set }\\ 
& $V_0$ [\AA$^3$] & $B_0$ [eV/\AA$^3$]  & $B_1$ & $V_0$ [\AA$^3$] & $B_0$ [eV/\AA$^3$] & $B_1$ & $|\eta(V_0)|$ & $|\eta(B_0)|$ & $|\eta(B_1)|$ & $V_0$ [\AA$^3$] & $B_0$ [eV/\AA$^3$] & $B_1$ \\ 
\hline
\endfirsthead
\caption{(continued) Table with all calculated \gls{eos} parameters for the BCC structures obtained with \fleur{} and \wientwok.} \\ 
& \multicolumn{3}{c|}{ \fleur } & \multicolumn{3}{c|}{ \wientwok } & \multicolumn{3}{c|}{Abs. percentage difference [\%]} & \multicolumn{3}{c}{ Average set }\\ 
& $V_0$ [\AA$^3$] & $B_0$ [eV/\AA$^3$]  & $B_1$ & $V_0$ [\AA$^3$] & $B_0$ [eV/\AA$^3$] & $B_1$ & $|\eta(V_0)|$ & $|\eta(B_0)|$ & $|\eta(B_1)|$ & $V_0$ [\AA$^3$] & $B_0$ [eV/\AA$^3$] & $B_1$ \\ 
\hline
\endhead
H & 2.9668 & 0.6748  & 3.2642 & 2.9667 & 0.6747 & 3.2615& 0.004 & 0.023 &  0.083 & 2.9668 & 0.6748 & 3.2629\\
He & 18.0345 & 0.0051  & 6.3686 & 18.0262 & 0.0051 & 6.4241& 0.046 & 0.903 &  0.868 & 18.0304 & 0.0051 & 6.3964\\
Li & 20.2674 & 0.0867  & 3.3459 & 20.2675 & 0.0867 & 3.3469& 0.0 & 0.009 &  0.029 & 20.2675 & 0.0867 & 3.3464\\
Be & 7.8148 & 0.7736  & 3.3532 & 7.8167 & 0.7736 & 3.3517& 0.024 & 0.001 &  0.044 & 7.8158 & 0.7736 & 3.3525\\
B & 6.1368 & 1.4325  & 4.8875 & 6.1419 & 1.4311 & 4.9017& 0.083 & 0.094 &  0.29 & 6.1394 & 1.4318 & 4.8946\\
C & 6.6858 & 1.1343  & 4.407 & 6.6856 & 1.1346 & 4.4061& 0.002 & 0.032 &  0.022 & 6.6857 & 1.1345 & 4.4066\\
N & 7.2349 & 1.1666  & 4.0231 & 7.2346 & 1.1658 & 4.0177& 0.004 & 0.068 &  0.133 & 7.2347 & 1.1662 & 4.0204\\
O & 7.7864 & 0.9447  & 4.714 & 7.7862 & 0.9437 & 4.7018& 0.002 & 0.106 &  0.258 & 7.7863 & 0.9442 & 4.7079\\
\end{longtable}
\end{center}
\begin{center}
\begin{longtable}{c|ccc|ccc|ccc|ccc}
\caption{\label{sitab:table-ae-X/SC}Table with all calculated \gls{eos} parameters for the SC structures obtained with \fleur{} and \wientwok.} \\ 
& \multicolumn{3}{c|}{ \fleur } & \multicolumn{3}{c|}{ \wientwok } & \multicolumn{3}{c|}{Abs. percentage difference [\%]} & \multicolumn{3}{c}{ Average set }\\ 
& $V_0$ [\AA$^3$] & $B_0$ [eV/\AA$^3$]  & $B_1$ & $V_0$ [\AA$^3$] & $B_0$ [eV/\AA$^3$] & $B_1$ & $|\eta(V_0)|$ & $|\eta(B_0)|$ & $|\eta(B_1)|$ & $V_0$ [\AA$^3$] & $B_0$ [eV/\AA$^3$] & $B_1$ \\ 
\hline
\endfirsthead
\caption{(continued) Table with all calculated \gls{eos} parameters for the SC structures obtained with \fleur{} and \wientwok.} \\ 
& \multicolumn{3}{c|}{ \fleur } & \multicolumn{3}{c|}{ \wientwok } & \multicolumn{3}{c|}{Abs. percentage difference [\%]} & \multicolumn{3}{c}{ Average set }\\ 
& $V_0$ [\AA$^3$] & $B_0$ [eV/\AA$^3$]  & $B_1$ & $V_0$ [\AA$^3$] & $B_0$ [eV/\AA$^3$] & $B_1$ & $|\eta(V_0)|$ & $|\eta(B_0)|$ & $|\eta(B_1)|$ & $V_0$ [\AA$^3$] & $B_0$ [eV/\AA$^3$] & $B_1$ \\ 
\hline
\endhead
H & 3.0874 & 0.6395  & 3.2502 & 3.0859 & 0.6375 & 3.2341& 0.047 & 0.312 &  0.497 & 3.0867 & 0.6385 & 3.2421\\
He & 21.4945 & 0.0037  & 6.3014 & 21.476 & 0.0037 & 6.8211& 0.086 & 0.081 &  7.921 & 21.4853 & 0.0037 & 6.5613\\
Li & 20.4075 & 0.0764  & 3.3063 & 20.4101 & 0.0765 & 3.312& 0.013 & 0.064 &  0.173 & 20.4088 & 0.0765 & 3.3091\\
Be & 10.2668 & 0.4705  & 3.3596 & 10.2677 & 0.4705 & 3.3576& 0.009 & 0.007 &  0.057 & 10.2672 & 0.4705 & 3.3586\\
B & 6.7003 & 1.4616  & 3.6331 & 6.7006 & 1.4619 & 3.6397& 0.004 & 0.022 &  0.181 & 6.7005 & 1.4618 & 3.6364\\
C & 5.583 & 2.0602  & 4.3081 & 5.5814 & 2.0606 & 4.3233& 0.028 & 0.021 &  0.353 & 5.5822 & 2.0604 & 4.3157\\
N & 6.481 & 1.2978  & 3.4544 & 6.4789 & 1.289 & 3.4364& 0.032 & 0.678 &  0.522 & 6.4799 & 1.2934 & 3.4454\\
O & 7.9532 & 1.0734  & 4.5927 & 7.9538 & 1.0704 & 4.5533& 0.008 & 0.287 &  0.861 & 7.9535 & 1.0719 & 4.573\\
\end{longtable}
\end{center}
\begin{center}
\begin{longtable}{c|ccc|ccc|ccc|ccc}
\caption{\label{sitab:table-ae-X/Diamond}Table with all calculated \gls{eos} parameters for the Diamond structures obtained with \fleur{} and \wientwok.} \\ 
& \multicolumn{3}{c|}{ \fleur } & \multicolumn{3}{c|}{ \wientwok } & \multicolumn{3}{c|}{Abs. percentage difference [\%]} & \multicolumn{3}{c}{ Average set }\\ 
& $V_0$ [\AA$^3$] & $B_0$ [eV/\AA$^3$]  & $B_1$ & $V_0$ [\AA$^3$] & $B_0$ [eV/\AA$^3$] & $B_1$ & $|\eta(V_0)|$ & $|\eta(B_0)|$ & $|\eta(B_1)|$ & $V_0$ [\AA$^3$] & $B_0$ [eV/\AA$^3$] & $B_1$ \\ 
\hline
\endfirsthead
\caption{(continued) Table with all calculated \gls{eos} parameters for the Diamond structures obtained with \fleur{} and \wientwok.} \\ 
& \multicolumn{3}{c|}{ \fleur } & \multicolumn{3}{c|}{ \wientwok } & \multicolumn{3}{c|}{Abs. percentage difference [\%]} & \multicolumn{3}{c}{ Average set }\\ 
& $V_0$ [\AA$^3$] & $B_0$ [eV/\AA$^3$]  & $B_1$ & $V_0$ [\AA$^3$] & $B_0$ [eV/\AA$^3$] & $B_1$ & $|\eta(V_0)|$ & $|\eta(B_0)|$ & $|\eta(B_1)|$ & $V_0$ [\AA$^3$] & $B_0$ [eV/\AA$^3$] & $B_1$ \\ 
\hline
\endhead
H & 6.8313 & 0.5712  & 3.187 & 6.8309 & 0.5707 & 3.1744& 0.005 & 0.083 &  0.395 & 6.8311 & 0.5709 & 3.1807\\
He & 64.261 & 0.0017  & 6.4088 & 64.1688 & 0.0017 & 6.9241& 0.144 & 2.148 &  7.73 & 64.2149 & 0.0017 & 6.6665\\
Li & 51.373 & 0.0336  & 2.9432 & 51.4004 & 0.0337 & 2.9633& 0.053 & 0.235 &  0.681 & 51.3867 & 0.0337 & 2.9533\\
Be & 29.3729 & 0.2736  & 3.442 & 29.3751 & 0.2736 & 3.4413& 0.008 & 0.001 &  0.019 & 29.374 & 0.2736 & 3.4417\\
B & 16.6272 & 1.1051  & 3.5246 & 16.6255 & 1.1056 & 3.5309& 0.01 & 0.039 &  0.178 & 16.6263 & 1.1054 & 3.5277\\
C & 11.3933 & 2.7062  & 3.7134 & 11.3897 & 2.7041 & 3.7087& 0.031 & 0.078 &  0.126 & 11.3915 & 2.7052 & 3.7111\\
N & 18.3562 & 0.9779  & 3.3817 & 18.3504 & 0.9761 & 3.3695& 0.032 & 0.192 &  0.36 & 18.3533 & 0.977 & 3.3756\\
O & 21.3621 & 0.7423  & 4.4205 & 21.3637 & 0.7391 & 4.4212& 0.007 & 0.429 &  0.017 & 21.3629 & 0.7407 & 4.4209\\
\end{longtable}
\end{center}
\begin{center}
\begin{longtable}{c|ccc|ccc|ccc|ccc}
\caption{\label{sitab:table-ae-X2O}Table with all calculated \gls{eos} parameters for the X$_2$O structures obtained with \fleur{} and \wientwok.} \\ 
& \multicolumn{3}{c|}{ \fleur } & \multicolumn{3}{c|}{ \wientwok } & \multicolumn{3}{c|}{Abs. percentage difference [\%]} & \multicolumn{3}{c}{ Average set }\\ 
& $V_0$ [\AA$^3$] & $B_0$ [eV/\AA$^3$]  & $B_1$ & $V_0$ [\AA$^3$] & $B_0$ [eV/\AA$^3$] & $B_1$ & $|\eta(V_0)|$ & $|\eta(B_0)|$ & $|\eta(B_1)|$ & $V_0$ [\AA$^3$] & $B_0$ [eV/\AA$^3$] & $B_1$ \\ 
\hline
\endfirsthead
\caption{(continued) Table with all calculated \gls{eos} parameters for the X$_2$O structures obtained with \fleur{} and \wientwok.} \\ 
& \multicolumn{3}{c|}{ \fleur } & \multicolumn{3}{c|}{ \wientwok } & \multicolumn{3}{c|}{Abs. percentage difference [\%]} & \multicolumn{3}{c}{ Average set }\\ 
& $V_0$ [\AA$^3$] & $B_0$ [eV/\AA$^3$]  & $B_1$ & $V_0$ [\AA$^3$] & $B_0$ [eV/\AA$^3$] & $B_1$ & $|\eta(V_0)|$ & $|\eta(B_0)|$ & $|\eta(B_1)|$ & $V_0$ [\AA$^3$] & $B_0$ [eV/\AA$^3$] & $B_1$ \\ 
\hline
\endhead
H & 11.9643 & 1.2194  & 3.8782 & 11.9627 & 1.2122 & 3.8508& 0.013 & 0.593 &  0.708 & 11.9635 & 1.2158 & 3.8645\\
He & 92.2182 & 0.003  & 6.5355 & 92.1387 & 0.003 & 6.363& 0.086 & 0.848 &  2.674 & 92.1785 & 0.003 & 6.4493\\
Li & 24.7216 & 0.4951  & 3.8922 & 24.7224 & 0.4948 & 3.8849& 0.003 & 0.051 &  0.189 & 24.722 & 0.4949 & 3.8885\\
Be & 26.8278 & 0.672  & 3.2268 & 26.8396 & 0.6725 & 3.2323& 0.044 & 0.071 &  0.172 & 26.8337 & 0.6722 & 3.2296\\
B & 30.0246 & 0.4338  & 1.9828 & 30.0324 & 0.4342 & 1.9891& 0.026 & 0.083 &  0.318 & 30.0285 & 0.434 & 1.986\\
C & 28.5544 & 0.6939  & 3.606 & 28.5543 & 0.6934 & 3.6067& 0.0 & 0.082 &  0.02 & 28.5543 & 0.6937 & 3.6064\\
N & 26.6671 & 0.8843  & 4.2476 & 26.6692 & 0.8829 & 4.2274& 0.008 & 0.158 &  0.477 & 26.6682 & 0.8836 & 4.2375\\
O & 27.1117 & 0.8128  & 4.643 & 27.1151 & 0.8112 & 4.6361& 0.012 & 0.195 &  0.149 & 27.1134 & 0.812 & 4.6396\\
\end{longtable}
\end{center}
\begin{center}
\begin{longtable}{c|ccc|ccc|ccc|ccc}
\caption{\label{sitab:table-ae-XO}Table with all calculated \gls{eos} parameters for the XO structures obtained with \fleur{} and \wientwok.} \\ 
& \multicolumn{3}{c|}{ \fleur } & \multicolumn{3}{c|}{ \wientwok } & \multicolumn{3}{c|}{Abs. percentage difference [\%]} & \multicolumn{3}{c}{ Average set }\\ 
& $V_0$ [\AA$^3$] & $B_0$ [eV/\AA$^3$]  & $B_1$ & $V_0$ [\AA$^3$] & $B_0$ [eV/\AA$^3$] & $B_1$ & $|\eta(V_0)|$ & $|\eta(B_0)|$ & $|\eta(B_1)|$ & $V_0$ [\AA$^3$] & $B_0$ [eV/\AA$^3$] & $B_1$ \\ 
\hline
\endfirsthead
\caption{(continued) Table with all calculated \gls{eos} parameters for the XO structures obtained with \fleur{} and \wientwok.} \\ 
& \multicolumn{3}{c|}{ \fleur } & \multicolumn{3}{c|}{ \wientwok } & \multicolumn{3}{c|}{Abs. percentage difference [\%]} & \multicolumn{3}{c}{ Average set }\\ 
& $V_0$ [\AA$^3$] & $B_0$ [eV/\AA$^3$]  & $B_1$ & $V_0$ [\AA$^3$] & $B_0$ [eV/\AA$^3$] & $B_1$ & $|\eta(V_0)|$ & $|\eta(B_0)|$ & $|\eta(B_1)|$ & $V_0$ [\AA$^3$] & $B_0$ [eV/\AA$^3$] & $B_1$ \\ 
\hline
\endhead
H & 10.0288 & 1.0197  & 4.1229 & 10.0269 & 1.0187 & 4.1254& 0.019 & 0.096 &  0.06 & 10.0278 & 1.0192 & 4.1242\\
He & 31.5746 & 0.0241  & 6.0688 & 31.4823 & 0.0236 & 5.979& 0.293 & 2.034 &  1.491 & 31.5284 & 0.0239 & 6.0239\\
Li & 16.8226 & 0.4482  & 4.1939 & 16.8291 & 0.4481 & 4.1855& 0.039 & 0.014 &  0.201 & 16.8258 & 0.4481 & 4.1897\\
Be & 12.1214 & 1.5087  & 3.7923 & 12.1269 & 1.5059 & 3.7899& 0.045 & 0.188 &  0.064 & 12.1241 & 1.5073 & 3.7911\\
B & 14.7095 & 1.1486  & 3.5065 & 14.7115 & 1.1488 & 3.5147& 0.014 & 0.021 &  0.235 & 14.7105 & 1.1487 & 3.5106\\
C & 15.7031 & 1.0655  & 3.8408 & 15.7004 & 1.0641 & 3.8476& 0.017 & 0.13 &  0.179 & 15.7018 & 1.0648 & 3.8442\\
N & 15.3394 & 1.1522  & 4.1425 & 15.3367 & 1.1491 & 4.1256& 0.018 & 0.269 &  0.409 & 15.3381 & 1.1507 & 4.134\\
O & 15.9065 & 1.0734  & 4.5758 & 15.9076 & 1.0704 & 4.5541& 0.007 & 0.282 &  0.475 & 15.9071 & 1.0719 & 4.5649\\
\end{longtable}
\end{center}
\begin{center}
\begin{longtable}{c|ccc|ccc|ccc|ccc}
\caption{\label{sitab:table-ae-X2O3}Table with all calculated \gls{eos} parameters for the X$_2$O$_3$ structures obtained with \fleur{} and \wientwok. Note that for X$_2$O$_3$ the primitive cell includes two formula units, therefore the volume of the primitive cells are twice those reported in this table.} \\ 
& \multicolumn{3}{c|}{ \fleur } & \multicolumn{3}{c|}{ \wientwok } & \multicolumn{3}{c|}{Abs. percentage difference [\%]} & \multicolumn{3}{c}{ Average set }\\ 
& $V_0$ [\AA$^3$] & $B_0$ [eV/\AA$^3$]  & $B_1$ & $V_0$ [\AA$^3$] & $B_0$ [eV/\AA$^3$] & $B_1$ & $|\eta(V_0)|$ & $|\eta(B_0)|$ & $|\eta(B_1)|$ & $V_0$ [\AA$^3$] & $B_0$ [eV/\AA$^3$] & $B_1$ \\ 
\hline
\endfirsthead
\caption{(continued) Table with all calculated \gls{eos} parameters for the X$_2$O$_3$ structures obtained with \fleur{} and \wientwok. Note that for X$_2$O$_3$ the primitive cell includes two formula units, therefore the volume of the primitive cells are twice those reported in this table.} \\ 
& \multicolumn{3}{c|}{ \fleur } & \multicolumn{3}{c|}{ \wientwok } & \multicolumn{3}{c|}{Abs. percentage difference [\%]} & \multicolumn{3}{c}{ Average set }\\ 
& $V_0$ [\AA$^3$] & $B_0$ [eV/\AA$^3$]  & $B_1$ & $V_0$ [\AA$^3$] & $B_0$ [eV/\AA$^3$] & $B_1$ & $|\eta(V_0)|$ & $|\eta(B_0)|$ & $|\eta(B_1)|$ & $V_0$ [\AA$^3$] & $B_0$ [eV/\AA$^3$] & $B_1$ \\ 
\hline
\endhead
H & 30.9245 & 0.9742  & 4.1995 & 30.9217 & 0.9724 & 4.1695& 0.009 & 0.177 &  0.718 & 30.9231 & 0.9733 & 4.1845\\
He & 47.1431 & 0.2595  & 5.2575 & 47.1461 & 0.2587 & 5.2709& 0.006 & 0.3 &  0.254 & 47.1446 & 0.2591 & 5.2642\\
Li & 44.7336 & 0.5089  & 4.3172 & 44.7367 & 0.5078 & 4.3366& 0.007 & 0.206 &  0.449 & 44.7351 & 0.5084 & 4.3269\\
Be & 38.7978 & 1.0917  & 3.8856 & 38.8015 & 1.0919 & 3.8876& 0.01 & 0.016 &  0.052 & 38.7996 & 1.0918 & 3.8866\\
B & 35.986 & 1.4594  & 4.1701 & 35.9842 & 1.4596 & 4.1679& 0.005 & 0.015 &  0.051 & 35.9851 & 1.4595 & 4.169\\
C & 42.2741 & 0.9343  & 4.2434 & 42.264 & 0.9326 & 4.2485& 0.024 & 0.181 &  0.121 & 42.269 & 0.9335 & 4.246\\
N & 45.6783 & 0.9213  & 4.1362 & 45.6711 & 0.919 & 4.1271& 0.016 & 0.244 &  0.22 & 45.6747 & 0.9202 & 4.1316\\
O & 47.8736 & 0.8285  & 4.5295 & 47.8768 & 0.8264 & 4.5414& 0.007 & 0.248 &  0.262 & 47.8752 & 0.8275 & 4.5355\\
\end{longtable}
\end{center}
\begin{center}
\begin{longtable}{c|ccc|ccc|ccc|ccc}
\caption{\label{sitab:table-ae-XO2}Table with all calculated \gls{eos} parameters for the XO$_2$ structures obtained with \fleur{} and \wientwok.} \\ 
& \multicolumn{3}{c|}{ \fleur } & \multicolumn{3}{c|}{ \wientwok } & \multicolumn{3}{c|}{Abs. percentage difference [\%]} & \multicolumn{3}{c}{ Average set }\\ 
& $V_0$ [\AA$^3$] & $B_0$ [eV/\AA$^3$]  & $B_1$ & $V_0$ [\AA$^3$] & $B_0$ [eV/\AA$^3$] & $B_1$ & $|\eta(V_0)|$ & $|\eta(B_0)|$ & $|\eta(B_1)|$ & $V_0$ [\AA$^3$] & $B_0$ [eV/\AA$^3$] & $B_1$ \\ 
\hline
\endfirsthead
\caption{(continued) Table with all calculated \gls{eos} parameters for the XO$_2$ structures obtained with \fleur{} and \wientwok.} \\ 
& \multicolumn{3}{c|}{ \fleur } & \multicolumn{3}{c|}{ \wientwok } & \multicolumn{3}{c|}{Abs. percentage difference [\%]} & \multicolumn{3}{c}{ Average set }\\ 
& $V_0$ [\AA$^3$] & $B_0$ [eV/\AA$^3$]  & $B_1$ & $V_0$ [\AA$^3$] & $B_0$ [eV/\AA$^3$] & $B_1$ & $|\eta(V_0)|$ & $|\eta(B_0)|$ & $|\eta(B_1)|$ & $V_0$ [\AA$^3$] & $B_0$ [eV/\AA$^3$] & $B_1$ \\ 
\hline
\endhead
H & 19.1041 & 0.9886  & 4.4002 & 19.1051 & 0.9874 & 4.3784& 0.005 & 0.112 &  0.496 & 19.1046 & 0.988 & 4.3893\\
He & 24.7738 & 0.4151  & 4.9973 & 24.7801 & 0.414 & 5.0015& 0.025 & 0.261 &  0.084 & 24.777 & 0.4146 & 4.9994\\
Li & 24.9191 & 0.5833  & 4.5324 & 24.9255 & 0.582 & 4.5315& 0.026 & 0.218 &  0.02 & 24.9223 & 0.5827 & 4.5319\\
Be & 22.1853 & 1.1333  & 4.0661 & 22.1855 & 1.1321 & 4.081& 0.001 & 0.102 &  0.365 & 22.1854 & 1.1327 & 4.0735\\
B & 20.3748 & 1.6109  & 4.1373 & 20.3723 & 1.6108 & 4.2092& 0.012 & 0.006 &  1.722 & 20.3735 & 1.6109 & 4.1732\\
C & 22.7585 & 1.047  & 4.6709 & 22.7543 & 1.0455 & 4.6565& 0.019 & 0.15 &  0.308 & 22.7564 & 1.0462 & 4.6637\\
N & 25.5102 & 1.0007  & 4.2193 & 25.5107 & 0.9983 & 4.2054& 0.002 & 0.237 &  0.329 & 25.5104 & 0.9995 & 4.2123\\
O & 27.1117 & 0.8128  & 4.643 & 27.1151 & 0.8112 & 4.6361& 0.012 & 0.195 &  0.149 & 27.1134 & 0.812 & 4.6396\\
\end{longtable}
\end{center}
\begin{center}
\begin{longtable}{c|ccc|ccc|ccc|ccc}
\caption{\label{sitab:table-ae-X2O5}Table with all calculated \gls{eos} parameters for the X$_2$O$_5$ structures obtained with \fleur{} and \wientwok. Note that for X$_2$O$_5$ the primitive cell includes two formula units, therefore the volume of the primitive cells are twice those reported in this table.} \\ 
& \multicolumn{3}{c|}{ \fleur } & \multicolumn{3}{c|}{ \wientwok } & \multicolumn{3}{c|}{Abs. percentage difference [\%]} & \multicolumn{3}{c}{ Average set }\\ 
& $V_0$ [\AA$^3$] & $B_0$ [eV/\AA$^3$]  & $B_1$ & $V_0$ [\AA$^3$] & $B_0$ [eV/\AA$^3$] & $B_1$ & $|\eta(V_0)|$ & $|\eta(B_0)|$ & $|\eta(B_1)|$ & $V_0$ [\AA$^3$] & $B_0$ [eV/\AA$^3$] & $B_1$ \\ 
\hline
\endfirsthead
\caption{(continued) Table with all calculated \gls{eos} parameters for the X$_2$O$_5$ structures obtained with \fleur{} and \wientwok. Note that for X$_2$O$_5$ the primitive cell includes two formula units, therefore the volume of the primitive cells are twice those reported in this table.} \\ 
& \multicolumn{3}{c|}{ \fleur } & \multicolumn{3}{c|}{ \wientwok } & \multicolumn{3}{c|}{Abs. percentage difference [\%]} & \multicolumn{3}{c}{ Average set }\\ 
& $V_0$ [\AA$^3$] & $B_0$ [eV/\AA$^3$]  & $B_1$ & $V_0$ [\AA$^3$] & $B_0$ [eV/\AA$^3$] & $B_1$ & $|\eta(V_0)|$ & $|\eta(B_0)|$ & $|\eta(B_1)|$ & $V_0$ [\AA$^3$] & $B_0$ [eV/\AA$^3$] & $B_1$ \\ 
\hline
\endhead
H & 51.6294 & 0.8182  & 4.3453 & 51.7182 & 0.8239 & 4.3852& 0.172 & 0.69 &  0.915 & 51.6738 & 0.821 & 4.3653\\
He & 56.3417 & 0.6362  & 4.7394 & 56.4553 & 0.6425 & 4.7791& 0.201 & 0.994 &  0.836 & 56.3985 & 0.6393 & 4.7593\\
Li & 61.7966 & 0.6087  & 4.5281 & 61.8956 & 0.6113 & 4.5471& 0.16 & 0.43 &  0.418 & 61.8461 & 0.61 & 4.5376\\
Be & 57.3047 & 0.9393  & 4.1197 & 57.3444 & 0.9407 & 4.1334& 0.069 & 0.148 &  0.332 & 57.3245 & 0.94 & 4.1265\\
B & 54.302 & 1.0186  & 4.46 & 54.4447 & 1.0179 & 4.4191& 0.262 & 0.063 &  0.922 & 54.3733 & 1.0182 & 4.4396\\
C & 58.5562 & 0.848  & 4.2161 & 58.6738 & 0.8484 & 4.234& 0.201 & 0.045 &  0.422 & 58.615 & 0.8482 & 4.2251\\
N & 57.1867 & 0.9674  & 4.4911 & 57.2811 & 0.9708 & 4.5012& 0.165 & 0.355 &  0.225 & 57.2339 & 0.9691 & 4.4962\\
O & 57.9311 & 0.9  & 4.6713 & 57.9351 & 0.8985 & 4.6548& 0.007 & 0.168 &  0.354 & 57.9331 & 0.8993 & 4.6631\\
\end{longtable}
\end{center}
\begin{center}
\begin{longtable}{c|ccc|ccc|ccc|ccc}
\caption{\label{sitab:table-ae-XO3}Table with all calculated \gls{eos} parameters for the XO$_3$ structures obtained with \fleur{} and \wientwok.} \\ 
& \multicolumn{3}{c|}{ \fleur } & \multicolumn{3}{c|}{ \wientwok } & \multicolumn{3}{c|}{Abs. percentage difference [\%]} & \multicolumn{3}{c}{ Average set }\\ 
& $V_0$ [\AA$^3$] & $B_0$ [eV/\AA$^3$]  & $B_1$ & $V_0$ [\AA$^3$] & $B_0$ [eV/\AA$^3$] & $B_1$ & $|\eta(V_0)|$ & $|\eta(B_0)|$ & $|\eta(B_1)|$ & $V_0$ [\AA$^3$] & $B_0$ [eV/\AA$^3$] & $B_1$ \\ 
\hline
\endfirsthead
\caption{(continued) Table with all calculated \gls{eos} parameters for the XO$_3$ structures obtained with \fleur{} and \wientwok.} \\ 
& \multicolumn{3}{c|}{ \fleur } & \multicolumn{3}{c|}{ \wientwok } & \multicolumn{3}{c|}{Abs. percentage difference [\%]} & \multicolumn{3}{c}{ Average set }\\ 
& $V_0$ [\AA$^3$] & $B_0$ [eV/\AA$^3$]  & $B_1$ & $V_0$ [\AA$^3$] & $B_0$ [eV/\AA$^3$] & $B_1$ & $|\eta(V_0)|$ & $|\eta(B_0)|$ & $|\eta(B_1)|$ & $V_0$ [\AA$^3$] & $B_0$ [eV/\AA$^3$] & $B_1$ \\ 
\hline
\endhead
H & 31.0974 & 0.7662  & 4.3422 & 31.0812 & 0.7635 & 4.477& 0.052 & 0.348 &  3.056 & 31.0893 & 0.7648 & 4.4096\\
He & 43.0362 & 0.2461  & 5.777 & 43.0829 & 0.2483 & 5.6076& 0.108 & 0.909 &  2.977 & 43.0595 & 0.2472 & 5.6923\\
Li & 46.0735 & 0.3228  & 4.8151 & 46.1048 & 0.3246 & 4.6981& 0.068 & 0.559 &  2.462 & 46.0892 & 0.3237 & 4.7566\\
Be & 38.9692 & 0.7063  & 4.2122 & 39.015 & 0.7037 & 4.1621& 0.117 & 0.368 &  1.197 & 38.9921 & 0.705 & 4.1872\\
B & 33.6256 & 1.2128  & 4.1217 & 33.6321 & 1.1989 & 4.1092& 0.019 & 1.149 &  0.304 & 33.6288 & 1.2059 & 4.1154\\
C & 32.9199 & 1.0243  & 4.6901 & 32.9186 & 1.0098 & 4.5753& 0.004 & 1.429 &  2.478 & 32.9192 & 1.017 & 4.6327\\
N & 39.7156 & 0.6921  & 4.0531 & 39.7033 & 0.6898 & 4.1545& 0.031 & 0.333 &  2.469 & 39.7094 & 0.691 & 4.1038\\
O & 44.7719 & 0.5504  & 4.6516 & 44.7872 & 0.5481 & 4.529& 0.034 & 0.427 &  2.671 & 44.7796 & 0.5493 & 4.5903\\
\end{longtable}
\end{center}

}

\clearpage

\section{Simulation parameters for \fleur{} and \wientwok}\label{SI:parameters-ae}

The \gls{ae} codes used for producing the reference results in this work implement the (linearized) augmented-plane-wave + local orbitals ((L)APW+LO) method. This method is named after a family of different, but related basis sets, each having different characteristics and demands on cutoff parameters. It is based on a partitioning of the unit cell into muffin-tin (MT) spheres at the atom positions and an interstitial region in between the spheres. The MT sphere radii hereby have to be adapted to each structure to avoid overlapping spheres, but are typically chosen to be nearly touching, because large spheres reduce demands on the LAPW basis set cutoff parameter. The exact choice of the radii typically differs between (L)APW+LO codes.

The choice of the core/valence electron separation is related to the MT sphere radii. To avoid instabilities in the calculations and to obtain precise results, the extent of the core-electron states beyond the MT sphere boundary has to be limited. The \gls{ae} codes employed here share the same core/valence separation for many structures, but differ in this choice for a considerable amount of other structures. Taking into account the overall excellent agreement between the results from the two codes it can be deduced that there is no significant dependence of the results on the differing description of physics for core and valence electrons, as long as the core/valence separation is within a reasonable range.

The valence electrons are represented by the (L)APW+LO basis chosen by the respective code. The different possible choices of basis sets nevertheless share a common set of parameters with which they are specified. These are cutoff parameters for the basis set size and the angular momentum expansion in the MT spheres, as well as energy parameters defining the linearization centers for each atom. Additionally, the local orbital setup has to be defined for each MT sphere. The two \gls{ae} codes employed in this work make use of different kinds of (L)APW+LO basis sets and their choices for the setup aspects discussed here differ strongly. A sketch on the parameter setup recipes is provided in the following two subsections. A detailed list of the setup parameter choices for each structure is available in the supplementary data, as referenced in the respective sections below.

\subsection{FLEUR}
The FLEUR code\cite{fleurCode,fleurSource} is an open-source implementation of the all-electron full-potential linearized augmented-plane-wave (FLAPW) method\cite{PhysRevB.12.3060,PhysRevB.24.864}. The calculations with this code make use of a conventional LAPW basis in combination with local orbitals (LOs) of different types to describe semicore states\cite{PhysRevB.43.6388} and to eliminate the linearization error for the valence states\cite{Michalicek20132670,PhysRevB.74.045104,PhysRevB.83.045105}. The employed setup profile defines global parameters, identical for all calculations and element-specific parameters. Further parameters are automatically adapted to each investigated structure.

In the context of this profile, global parameters are the reciprocal LAPW basis set cutoff parameter $K_\text{max} = 5.0~a_0^{-1}$ and the plane-wave cutoff parameters $G_\text{max} = G_\text{max,XC} = 25.0~a_0^{-1}$, where $a_0$ is the Bohr radius. The latter cutoff parameter covers the expansions of the plane-wave part of the density and the potential, the interstitial-region indicator function, and the exchange-correlation contribution to the potential. The Fermi--Dirac smearing and the k-point density are also fixed to the common choices of this work. Element-specific setup aspects include the core-valence separation of the electron states and the LO setup. The radii of the MT spheres centered on atom $\alpha$, $R_\text{MT}^{\alpha}$, are adapted to the smallest unit cell within an equation-of-states (EOS) workflow. For this, element-specific initial MT sphere radii are expanded to cover up to about $92\%$ of the distance between the atoms, with a limit of a maximal MT radius of $2.66~a_0$. The procedure also implies an adaption of the angular momentum cutoffs in the spheres to $l_\text{max}^\alpha \approx K_\text{max} R_\text{MT}^{\alpha}$. A detailed description of the different parameters is available in the FLEUR user guide\cite{fleurCode,fleurSource}.

The parameter profile is the result of an iterative refinement process aiming at precision and stability on the basis of the structures investigated in this work, i.e., structures with a wide range of neighboring atom distances. However, it is not designed to provide an absolute convergence of the total energy, which may still be affected by structure-adapted numerical parameters like the MT radii. The determination of these parameters for each structure on the basis of the smallest unit cell solves this issue for the EOS workflow. Comparing the total-energy results from the EOS workflows for different structures may also yield reasonable numbers, but these do not reflect the precision capabilities of the code when used in an adequate way to perform such a comparison, see also discussion in SI Sec.~\ref{sisec:formation-energies}.

With the exception of the k-point integration-mesh generation, the used parameter profile is implemented in the openly available releases of FLEUR starting with the MaX-R6.1 release. The results presented here have been obtained using the development version as of 2022/03/31. The profile is employed by invoking the FLEUR input generator with the command line option "\texttt{-profile oxides\_validation}". Beyond using this profile, the AiiDA common workflows package\cite{Huber2021} protocol "\texttt{verification-pbe-v1}" in combination with AiiDA-FLEUR\cite{Broeder:2019,AiiDA.FLEUR.1.3} also sets the k-point integration mesh.

The resulting parametrization for each structure is discussed in the \texttt{all-electron-setups} folder of the supplementary data available in Ref.~\citenum{MCA-ACWF}, in particular in the files
\begin{itemize}
    \item \texttt{setup-oxides-verification-PBE-v1-fleur.json}
    \item \texttt{setup-unaries-verification-PBE-v1-fleur.json} 
\end{itemize}
that are documented in the file \texttt{all-electron-data.md}.

In an ongoing effort the data from this work is also used to define further FLEUR parameter profiles for different precision levels and computational effort. This is done by reducing in a controlled way cutoff parameters and changing other aspects of the setup and relating the corresponding results to those presented here.

\subsection{WIEN2k}

The WIEN2k calculations\cite{WIEN2k, WIEN20} employ the (linearized) augmented plane wave + local orbitals  ((L)APW+lo) method\cite{Madsen01} with additional local orbitals\cite{PhysRevB.43.6388} (LOs) for states with an energy 1 Ry below the Fermi energy (semicore states)  and high-derivative LOs (HDLOs)\cite{Karsai17} for all ``chemical'' $l$ values (except when there is already a semicore LO). The APW+lo basis set is used for all ``chemical'' $l$ values ($s$, $p$, $d$ or $f$ - depending on the atom), while LAPW is used for higher angular momentum up to $l_\text{max}$~=~10. 

The WIEN2k calculations have been initialized (see the WIEN2k users guide\cite{WIEN2k}) for the smallest volume of each case using:\vspace{6pt}

{\footnotesize\texttt{init\_lapw -b -prec 3 -nokshift -fermits 0.0045 -red 3 -numk -1 0.0317506}\vspace{6pt}}

For subsequent volumes we use:\vspace{6pt}

{\footnotesize\texttt{init\_lapw -b -prec 3 -nokshift -fermits 0.0045 -red Element:RMT -numk 0 kx ky kz -fft ix iy iz}\vspace{6pt}}

\noindent where \texttt{Element:RMT}, \texttt{kx ky kz}, and \texttt{ix iy iz} are inserted from the output of the first volume to ensure identical parameters.

This high-precision setup limits the maximal atomic sphere radius $R_\text{MT}^{\alpha}$ to $2.35~a_0$, but otherwise sets the sphere sizes automatically depending on nearest neighbor distances and type of atom (largest for $f$ elements, intermediate for $d$ elements and smallest for $sp$ elements). Also the plane wave cutoff $R_\text{MT, min}^{\alpha} K_\text{max}$ is set automatically depending on the type of atom and the smallest atomic sphere radius $R_\text{MT, min}^{\alpha}$ and varies from 7.08 (H$_2$O$_5$) to 11. All states with an energy above $-6$~Ry or with a charge density of more than 0.01~$e^-$ outside the atomic sphere are considered as valence states and treated scalar-relativistically, while lower energy states are considered as core and solved numerically with a radial symmetric Dirac equation.  Note that with this choice the definition of core states for an element may change depending on its $R_\text{MT}^{\alpha}$ and this makes the calculation of formation energies in certain cases unrealistic as discussed also in SI Sec.~\ref{sisec:formation-energies}. An SCF cycle was considered converged when both the change in total energy was less than $10^{-6}$~Ry and the change in the electron charge density within $R_\text{MT}^{\alpha}$ was less than $10^{-6}~e$.

The charge density and potential inside spheres is expanded into lattice harmonics up to $L_\text{max}$~=~6 and for the non-spherical Hamiltonian matrix elements the angular momentum of the wave functions is restricted to $l^{ns}_\text{max}$~=~8. In the interstitial region the density/potential is expanded into a Fourier series with cutoff parameter $G_\text{max}~=~25~a_0^{-1}$ (except for alkali metals, noble gases, and Hg, where $G_\text{max}~=~40~a_0^{-1}$).

The resulting parametrization for each structure is discussed in the \texttt{all-electron-setups} folder of the supplementary data available in Ref.~\citenum{MCA-ACWF}, in particular in the files
\begin{itemize}
    \item \texttt{setup-oxides-verification-PBE-v1-wien2k.json}
    \item \texttt{setup-unaries-verification-PBE-v1-wien2k.json} 
\end{itemize}
that are documented in the file \texttt{all-electron-data.md}.

\clearpage

\section{Dependence of the metrics on the size of the simulation cell and on bond stiffness}\label{SI:epsilon_indipendent_vol}
The equation of state is often expressed in terms of the absolute energy $E$ and volume $V$ of the simulation cell, but other times quantities "per-formula-unit" or "per-atom" are considered. We want here to demonstrate that the new metrics $\varepsilon$ and $\nu$, introduced here, are intrinsic quantities, i.e., they are independent of the simulation cell size (while the original $\Delta$ metric is extensive).

\textbf{Dependence on the number of atoms in the simulation cell}: We first show that the original metric $\Delta$ is an extensive quantity that depends linearly on the number of atoms (or, equivalently, on the volume) of the simulation cell.

Let us consider a supercell where the number of atoms is increased by a factor of $C$. The volume and the total energy (assuming $E_0 = 0$) will scale accordingly as
\begin{equation}
    E' = CE, \quad V' = CV.
\end{equation}
The integrand in Eq.~\eqref{eq:delta} for the $\Delta$ metric will scale as
\begin{equation}
    [E'_a(V') - E'_b(V')]^2~dV' = C^3 [E_a(V) - E_b(V)]^2~dV
\end{equation}
and the denominator as
\begin{equation}
    V'_M - V'_m = C (V_M - V_m).
\end{equation}
The effect of increasing the number of atoms on the $\Delta$ metric is
\begin{align} 
    \Delta'(a,b) & = 
    \sqrt{\frac{1}{V'_M - V'_m} \int_{V'_m}^{V'_M} [E'_a(V') - E'_b(V')]^2 ~ dV'} =
    \sqrt{\frac{C^3}{C(V_M - V_m)} \int_{V_m}^{V_M} [E_a(V) - E_b(V)]^2 ~ dV}\\
   \notag & = C \sqrt{\frac{1}{V_M - V_m} \int_{V_m}^{V_M} [E_a(V) - E_b(V)]^2 ~ dV} = C \Delta(a,b).
\end{align}
Thus, $\Delta$ scales \textit{linearly} with the number of atoms in the simulation cell. This shortcoming is typically addressed by computing the $\Delta$ metric renormalized per atom.

For the $\varepsilon$ metric we proceed in the same manner to demonstrate, instead, its independence from the simulation cell size.
\begin{align}
    \varepsilon'(a,b) & =
    \sqrt{
        \frac{
            \int_{V'_m}^{V'_M} [E'_a(V') - E'_b(V')]^2 ~ dV'
        }{
            \sqrt{
                \left(
                    \int_{V'_m}^{V'_M} [E'_a(V') - \langle E'_a \rangle]^2 ~ dV'
                \right)
                ~
                \left(
                    \int_{V'_m}^{V'_M} [E'_b(V') - \langle E'_b \rangle]^2 ~ dV'
                \right)
            }
        }
    }
    \\
   \notag & =
    \sqrt{
        \frac{
            C^3 \int_{V_m}^{V_M} [E_a(V) - E_b(V)]^2 ~ dV
        }{
            \sqrt{
                \left(
                    C^3 \int_{V_m}^{V_M} [E_a(V) - \langle E_a \rangle]^2 ~ dV
                \right)
                ~
                \left(
                    C^3 \int_{V_m}^{V_M} [E_b(V) - \langle E_b \rangle]^2 ~ dV
                \right)
            }
        }
    }
    \\
   \notag & =
    \sqrt{
        \frac{
            \int_{V_m}^{V_M} [E_a(V) - E_b(V)]^2 ~ dV
        }{
            \sqrt{
                \left(
                    \int_{V_m}^{V_M} [E_a(V) - \langle E_a \rangle]^2 ~ dV
                \right)
                ~
                \left(
                    \int_{V_m}^{V_M} [E_b(V) - \langle E_b \rangle]^2 ~ dV
                \right)
            }
        }
    } =\varepsilon(a,b).
\end{align}
Since $C$ cancels out, we have proven that $\varepsilon$ is independent of the number of atoms in a simulation cell considered.

Analogously, it is also easy to see that $\nu$ is also an intrinsic quantity, independent of the number of atoms in the simulation cell: indeed, $\nu$ is defined as a function of the relative errors of the parameters $V_0$, $B_0$ and $B_1$, that are all intrinsic quantities.

We stress that this means that, while for $\Delta$ is recommended to report it normalized (e.g., per atom, or per formula unit), $\varepsilon$ and $\nu$ should \emph{not} be normalized.

\textbf{Sensitivity to the value of the bulk modulus}:  Let us compare the metrics obtained comparing results for two different materials. We assume that the first material has bulk modulus $B_0$ and the second is identical except for its bulk modulus, that is scaled by factor of $C$, i.e., $B_0' = C B_0$. We assume that there is no other difference between the two materials (same $V_0$ and $B_1$) and we are considering the same simulation volume (or number of atoms in the simulation cell). The total energy of the second material will then scale as
\begin{equation}
    E' = CE
\end{equation}
according to Eq.~\eqref{b-m} and assuming $E_0=0$ (minimum energy of both materials have been shifted to zero). The integrand in Eq.~\eqref{eq:delta} for the $\Delta$ metric scales as
\begin{equation}
    [E'_a(V) - E'_b(V)]^2~dV = C^2 [E_a(V) - E_b(V)]^2~dV.
\end{equation}
Following similar steps as for the number of atoms above, we arrive to the conclusion that $\Delta$ scales also \textit{linearly} with the bulk modulus, while the $\varepsilon$ metric is insensitive to the it. 
Similarly, also $\nu$ is insensitive to the reference value of the bulk modulus of the material, since it only depends on the relative change of the bulk modulus $B_0$ between the two systems, that does not depend on the factor $C$.

We stress that this fact does not mean that $\varepsilon$ or $\nu$ will not capture a difference between bulk moduli when comparing two computational approaches $a$ and $b$. 
It means, instead, that two datasets with a similar discrepancy in the bulk moduli (say 2\%) will result in the same $\varepsilon$ or $\nu$ irrespective of overall stiffness of their chemical bonds (i.e., their bulk modulus). We highlight that this shortcoming of the $\Delta$ metric was already recognized in the literature and addressed by defining a modified metric $\Delta_1$ in Ref.~\citenum{Jollet:2014}.

\clearpage

\section{Sensitivity of $\Delta$, $\varepsilon$ and $\nu$ to perturbations of the \gls{eos} parameters and choice of thresholds for excellent and good agreement\label{sisec:perturbations-eos}}

Any of the three metrics $\Delta$, $\varepsilon$ or $\nu$ expresses the difference between two \gls{eos} curves by a single number. It is not \emph{a~priori} obvious, however, when those numbers can be considered small or large, and which features of the \gls{eos} have the largest impact on the value. In this section, we address these points.

\begin{figure*}[h!]
    \centering
     \includegraphics[width=\linewidth] {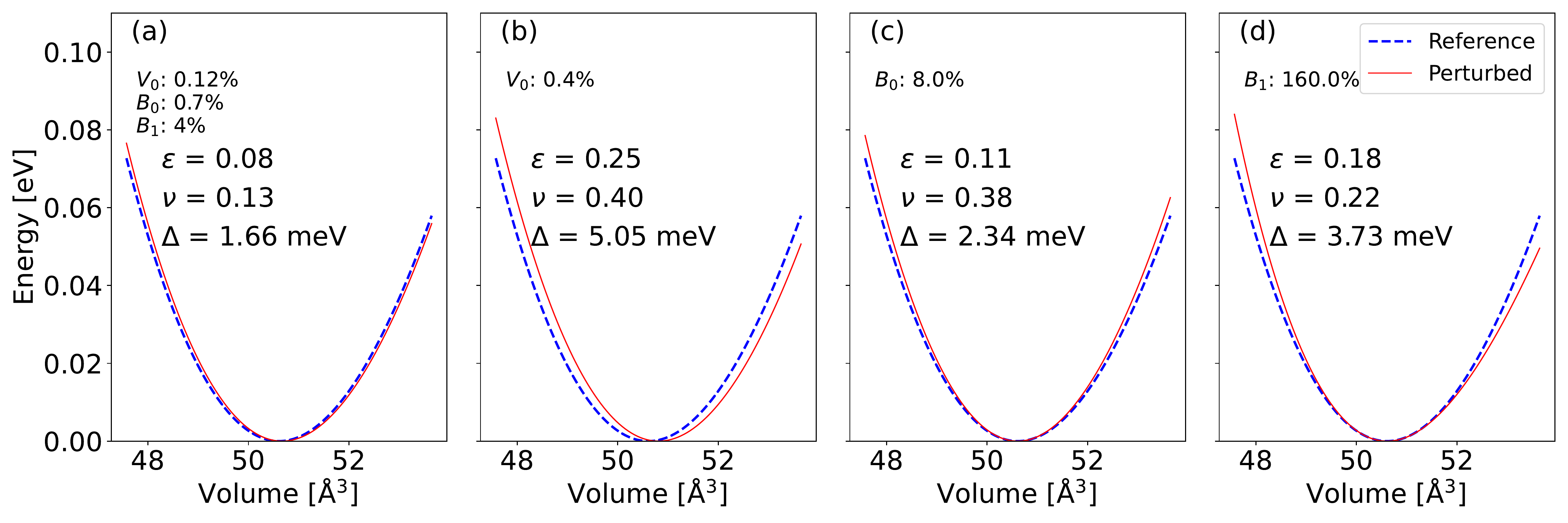}
     \caption{\gls{eos} for a hypothetical material  with $V_0=50.61\ \angstrom^{3}$ per formula unit, $B_0 = 0.71\ \text{eV}/\angstrom^3$, and $B_1 = 4.67$ (``reference'' curve, dashed blue), compared with the \gls{eos} where a perturbation has been applied to some of the \gls{eos}-defining parameters (``perturbed'' curve, solid red).
    Each panel reports the resulting value for the three metrics $\varepsilon$, $\nu$ and $\Delta$ obtained comparing the two curves. 
    The perturbed parameters and the magnitude of the perturbation are also indicated in each panel. In particular, in panel (a) a perturbation of 0.12\% is applied to $V_0$, of 0.7\% to $B_0$ and of 4\% to $B_1$. 
    Panels (b), (c) and (d) present the cases when a perturbation is applied independently to $V_0$ (0.4\%), $B_0$ (8\%) or $B_1$ (160\%), respectively, while the other parameters are kept unchanged.
    \label{fig:eos-sensitivity}}
\end{figure*}

In SI Fig.~\ref{fig:eos-sensitivity} we compare the \gls{eos} of a hypothetical material with the \gls{eos} obtained after four different perturbations of the material (see caption for details). The values for $\Delta$, $\varepsilon$ and $\nu$ that express the difference between the original and perturbed \gls{eos} are listed for each case. Analyzing these results allows to associate typical orders of magnitude to each of the metrics, and quantify their variation with respect to changes in $V_0$, $B_0$, and $B_1$.

The hypothetical reference material that is used in SI Fig.~\ref{fig:eos-sensitivity} has $V_0=50.61$ \angstrom$^3$ per formula unit, $B_0 = 0.71$ eV/\angstrom$^3$, and $B_1 = 4.67$. These values are obtained as averages over the entire crystals set and thus represent a hypothetical ``average'' \gls{eos}.

In panel (a), a perturbation is applied to all three parameters, namely 0.12\% to $V_0$, of 0.7\% to $B_0$ and of 4\% to $B_1$.
These values are twice the standard deviations of the discrepancies between the two \gls{ae} codes in our reference dataset, see Fig.~\ref{fig:ae-histograms} in the main text.
Since the two EOS curves are almost undistinguishable, this result highlights the high level of agreement between our two \gls{ae} codes. Based on this observation, we define a qualitative range of $\varepsilon \lesssim 0.06$ or $\nu \lesssim 0.1$ for which we consider two codes display excellent agreement. The threshold of $\varepsilon=0.06$ (approximately) corresponds to a determination coefficient $R^2 \approx 1 - \varepsilon^2 = 0.9964$ when one EOS curve is considered as a fit to the other.

\begin{figure*}[h!]
    \centering
     \includegraphics[width=\linewidth] {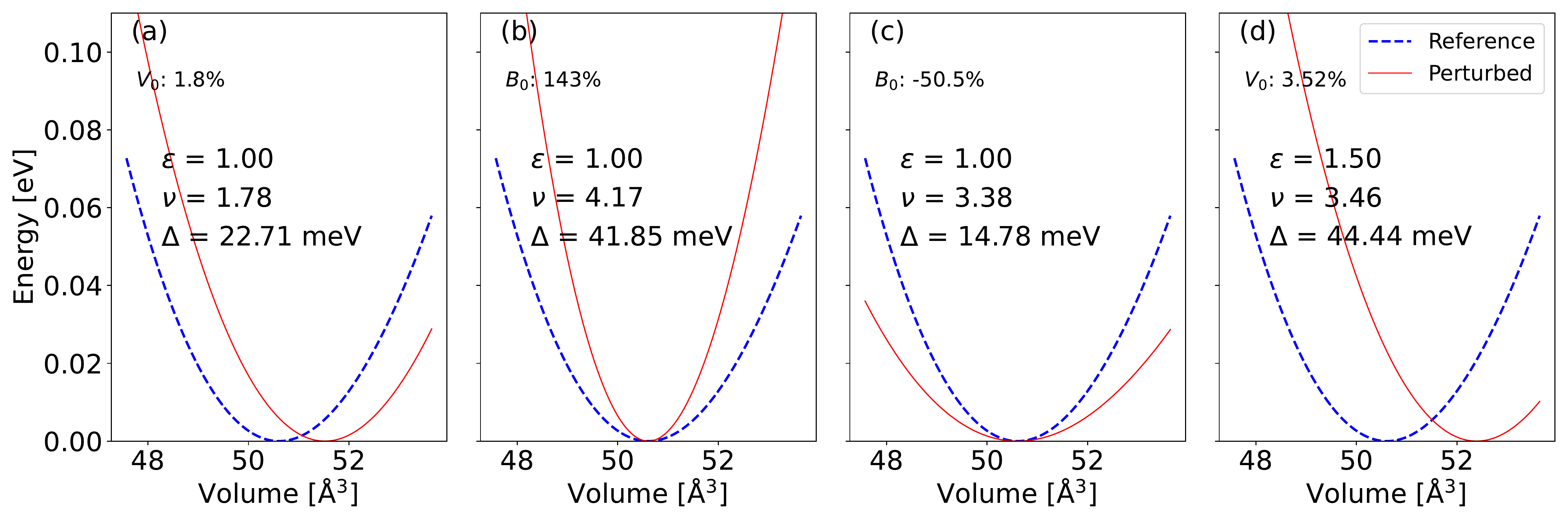}
     \caption{\gls{eos} for a hypothetical material  with $V_0=50.61\ \angstrom^{3}$ per formula unit, $B_0 = 0.71\ \text{eV}/\angstrom^3$, and $B_1 = 4.67$ (``reference'' curve, dashed blue), compared with the \gls{eos} where a perturbation has been applied to some of the \gls{eos}-defining parameters (``perturbed'' curve, solid red).
    With respect to SI Fig.~\ref{fig:eos-sensitivity}, we highlight here typical values for the metrics for large changes of the parameters that make the \gls{eos} curves clearly different.
    Each panel reports the resulting value for the three metrics $\varepsilon$, $\nu$ and $\Delta$ obtained comparing the two curves. 
    The perturbed parameters and the magnitude of the perturbation are also indicated in each panel. In particular, in panel (a) a perturbation is applied to $V_0$ to obtain a value of $\varepsilon=1$; panels (b) and (c) apply a (positive and negative, respectively) perturbation to $B_0$ resulting in $\varepsilon=1$, and finally in panel (d) an even larger perturbation is applied to $V_0$ to obtain a value of $\varepsilon=1.5$. 
    \label{fig:eos-sensitivity-large}}
\end{figure*}

In panels (b), (c) and (d), instead, a larger perturbation is applied to only one of the three parameters $V_0$, $B_0$ or $B_1$, respectively. The perturbation to $V_0$ is chosen as $0.4\%$, and the magnitude of the perturbations for $B_0$ and $B_1$ is scaled following the inverse ratios 1/20/400 of the weights for the $\nu$ metric (see SI Sec.~\ref{SI:stability_eos}).
This results in visually similar discrepancies between the two curves. This is an expected result, and is another way to interpret the results discussed in SI Sec.~\ref{SI:stability_eos} for the weights of $\nu$.
Indeed, those weights were obtained by inferring the error propagated on the fitted parameters from a given amount of random noise on the datasets; the inverse weights can be conversely interpreted, intuitively, as the relative magnitude of the perturbation to each of the parameters required to induce similar changes to the \gls{eos} curve.
These panels help us make a number of observations:
\begin{itemize}
    \item All metrics ($\Delta$ and $\epsilon$ intrinsically, and $\nu$ by explicit definition of the weights) give a much stronger importance to changes of the equilibrium volume $V_0$ than to changes of the other parameters. This is a positive feature of the metrics, as the \gls{eos} shape is mostly sensitive to $V_0$ as well. The metric $\nu$ has the additional advantage, as already discussed, that weights can be tuned to give more importance to other parameters, if an application requires it.
    \item With our definitions of $\epsilon$ and $\nu$, the two metric often return similar values for a given pair of \gls{eos} curves, with $\varepsilon$ typically slightly smaller (in the SI Sec.~\ref{sisec:correlation-metrics}, we actually identify an approximate proportionality ratio between the two, valid for small values of the metrics).
    \item The values of the metrics on these last three panels allow us to define threshold values for noticeable, but still relatively small, changes between two EOS curves: $\varepsilon \lesssim 0.2$ or $\nu \lesssim 0.33$. Therefore, we take the ranges of $0.06 < \varepsilon \lesssim 0.20$ or $0.10 < \nu \lesssim 0.33$ as the signature for good (but not excellent) agreement between two codes. The upper end of this range $\varepsilon=0.20$ (approximately) corresponds to a determination coefficient $R^2 \approx 1 - \varepsilon^2 = 0.96$ when one EOS curve is considered as a fit to the other.
    
    \item We can assign an intuitive meaning to the metric $\nu$. If two \gls{eos} only differ in the equilibrium volume, its numerical value corresponds to the percentage error on the equilibrium volume between the two curves. If also $B_0$ and $B_1$ change, then $\nu$ will also take into account the discrepancies on these two parameters, rescaled so that similar contributions to $\nu$ result to similar quantitative changes to the \gls{eos} curve (in the volume range of interest, $\pm 6\%$ in this work).
    \item We note that in panels (b), (c) and (d) the value of $\nu$ is not exactly 0.4 as one might naively expect, because the perturbation that we apply refers to the reference curve, but the $\nu$ metric is defined in a symmetric way, with percentage differences with respect to the average of the two curves.
    \item SI Fig.~\ref{fig:eos-sensitivity-large}, finally, illustrates the clear disagreement between \gls{eos} curves when $\varepsilon \ge 1.0$ or $\nu \ge 1.65$ (these values are used as upper limit for the colorbars of the figures in SI Sec.~\ref{sisec:periodic-tables-per-code}). 
    As a note, $\varepsilon = 1$ is an estimator for the situation where the coefficient of determination $R^2 \approx 1 - \varepsilon^2$ starts to be negative (even if the approximation $R^2 \approx 1 - \varepsilon^2$ does not hold exactly anymore for such large values of $\varepsilon$). We highlight that a negative $R^2$ value indicates that a horizontal line at the average value of the data provides a better fit than the actual fit function.
    It is clear from SI Fig.~\ref{fig:eos-sensitivity-large} that in such cases there is no agreement at all between the results of two codes. Therefore, when $\varepsilon > 1.0$ or $\nu > 1.65$, two codes are said to be clearly different in SI Sec.~\ref{sisec:periodic-tables-per-code}.
\end{itemize}    

\clearpage

\section{Mutual correlation between the metrics\label{sisec:correlation-metrics}}

\begin{figure*}[hb!]
    \centering
     \includegraphics[width=0.4\linewidth]{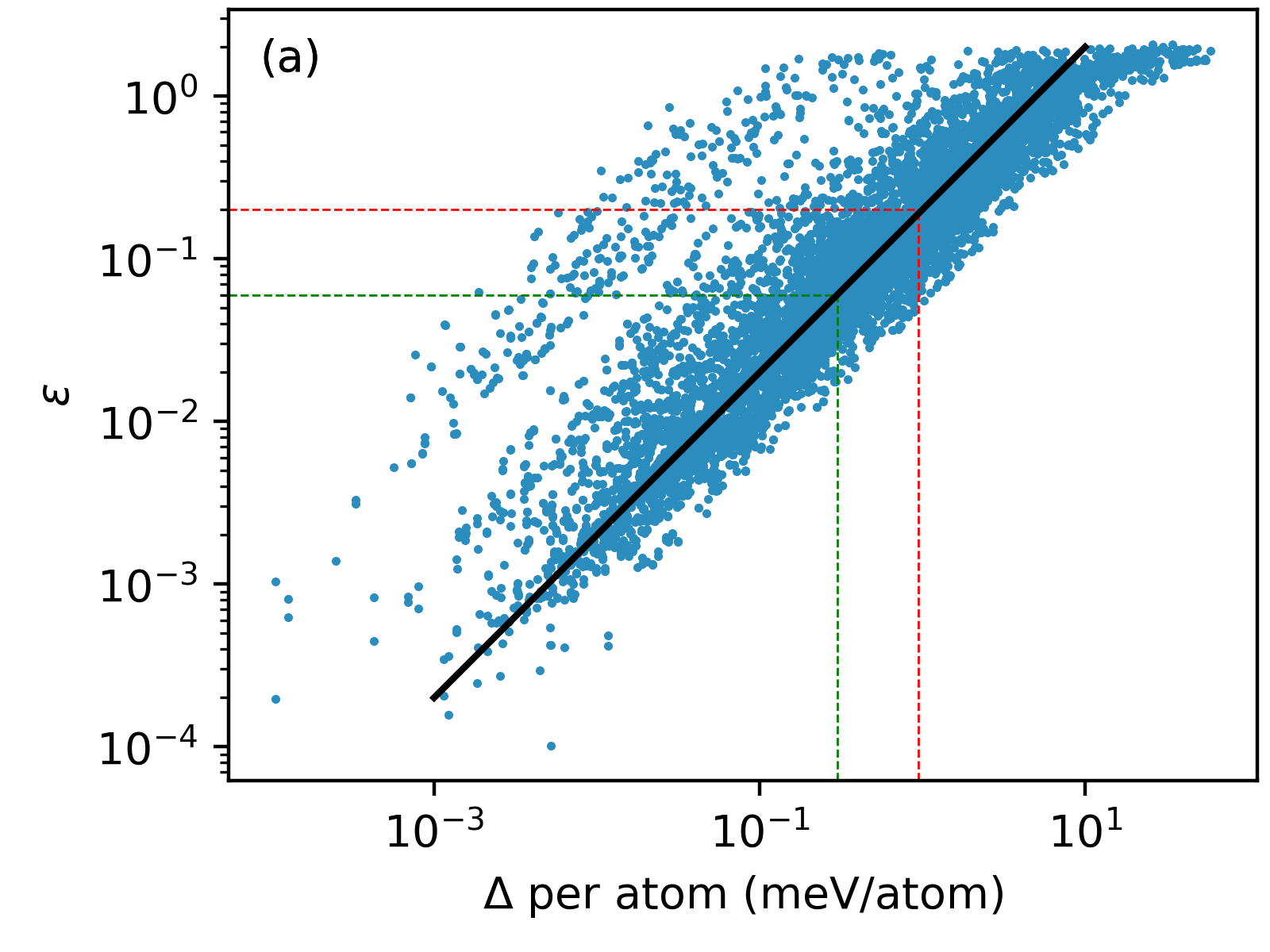}
     \includegraphics[width=0.4\linewidth]{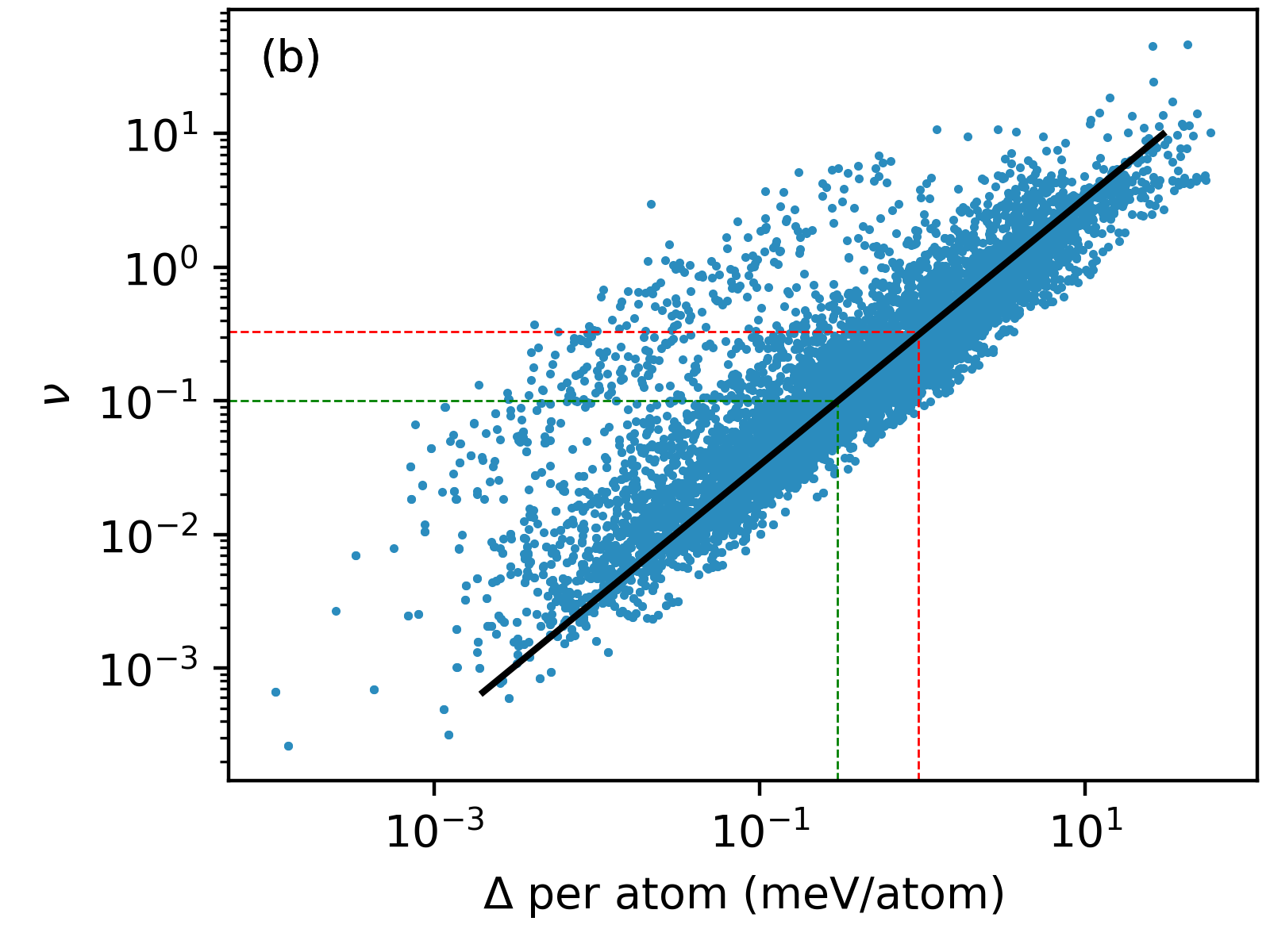}

     \includegraphics[width=0.4\linewidth]{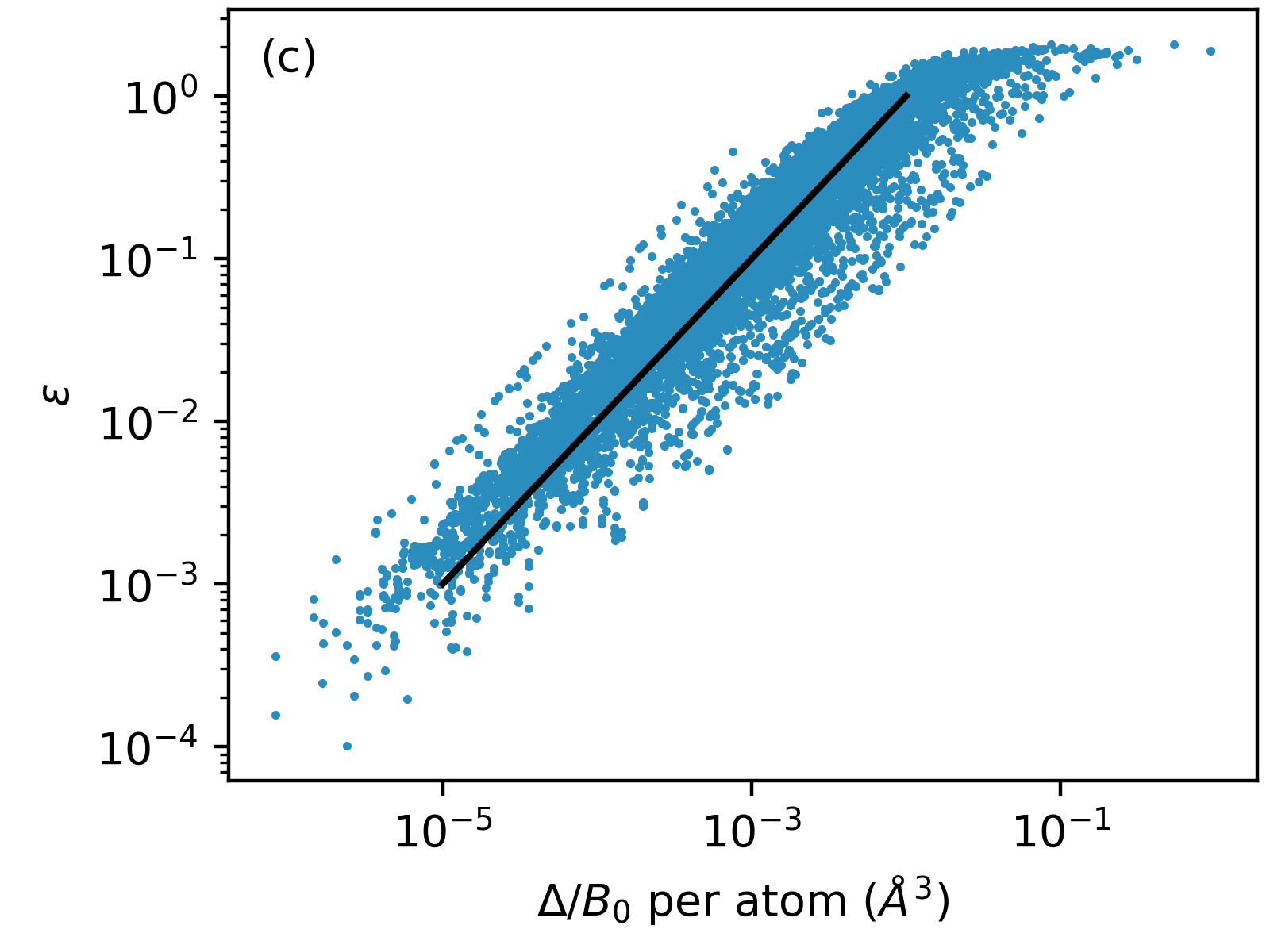}
     \includegraphics[width=0.4\linewidth]{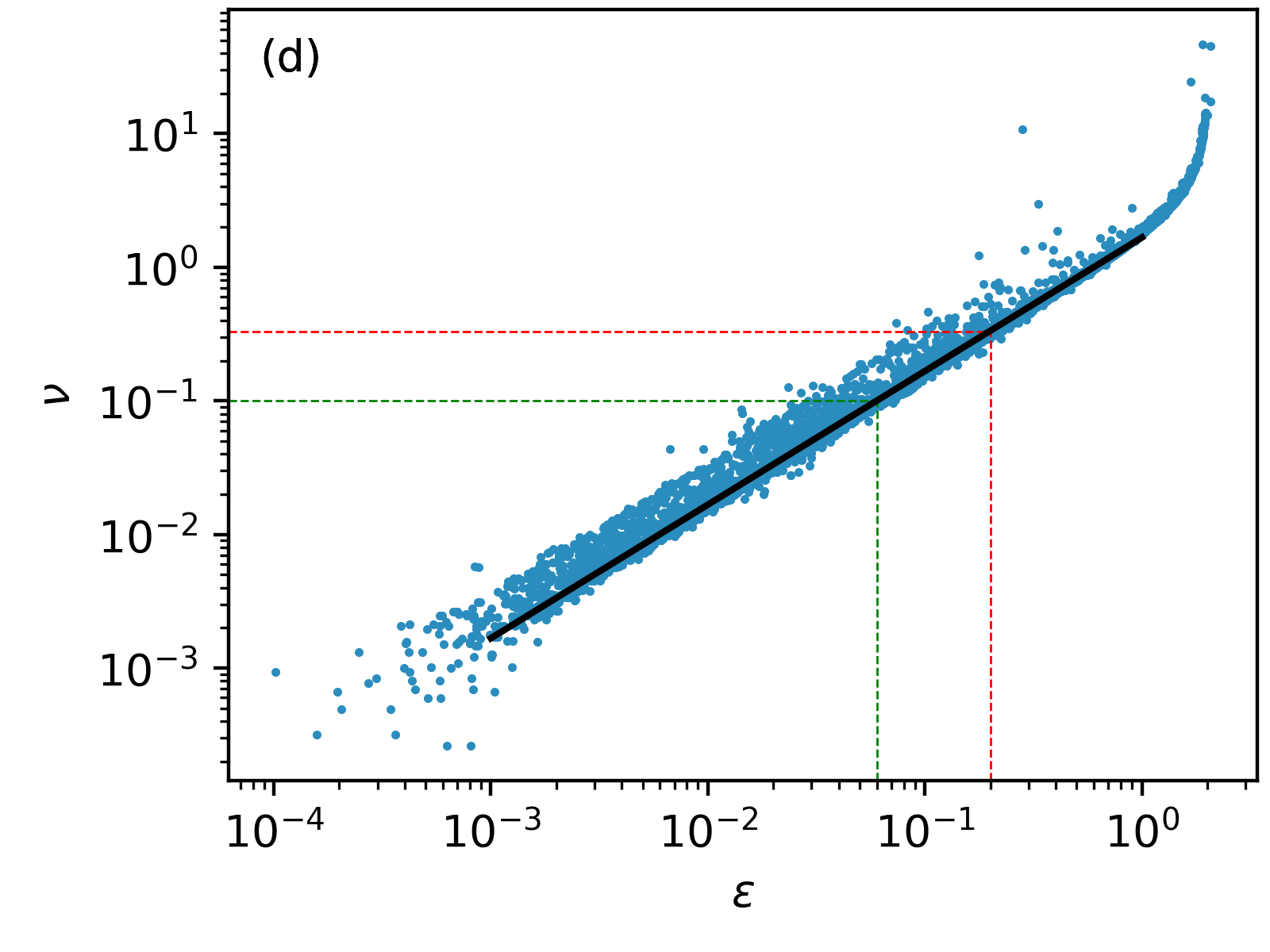}

     \caption{Cross-correlation between the $\Delta$, $\varepsilon$, and $\nu$ metrics for the entire data (unaries and oxides) presented in the main text. Black lines are helpers to indicate a linear relation between the quantities (slope of 1 in a log-log plot). Dashed lines on $\varepsilon$ and $\nu$ axes show the ``excellent'' (green) and ``good'' (red) agreement thresholds recommended in the main text. From the plots, these correspond to values of $\Delta$ of approximately 0.3 and 0.95 meV/atom, respectively. 
    \label{fig:eos-sensitivity-3}}
\end{figure*}

In order to assess the correlation between the various metrics, we present in SI Fig.~\ref{fig:eos-sensitivity-3} for all codes mentioned in the main text their $\Delta$, $\varepsilon$ or $\nu$ with respect to the reference \gls{eos}, for each of the unaries and oxides. These $\Delta$, $\varepsilon$ or $\nu$ are plotted against one of the other metrics.
The results support the presence of an approximately linear correlation between the three metrics ($\Delta$, $\varepsilon$, and $\nu$). However, there is more scattering in the correlation between $\varepsilon$ and $\Delta$ (and, similarly, between $\nu$ and $\Delta$), while  $\varepsilon$ and $\nu$ agree more consistently on a global scale.

We first note that the scattering between $\varepsilon$ and $\Delta$ can be reduced by normalizing $\Delta$ by the bulk modulus, similar to the $\Delta_1$ metric introduced in Ref.~\citenum{Jollet:2014}, as shown in SI Fig.~\ref{fig:eos-sensitivity-3}(c).
We then observe that the new metrics $\varepsilon$ and $\nu$ are instead almost linearly correlated when their values are $\lesssim 1$; for larger discrepancies, the values of $\nu$ tend to grow faster than the values of $\varepsilon$, i.e., $\varepsilon$ becomes relatively less sensitive to further small changes to the \gls{eos} curves if they are already significantly different.

This almost linear correlation can be justified with some approximations.
Let us consider the simple case of two parabolic \gls{eos} curves with the same $B_0$, differing only in the equilibrium volume $V_0$.
This is a valid approximation, since we discussed above that both metrics are mostly sensitive to changes of $V_0$ rather than $B_0$ or $B_1$.
If we call $2V_R = V_M - V_m$ the volume range for the integration in $\varepsilon$ (see also Eq.~\eqref{eq:average-volume-integral} in the main text for the definition of $V_m$ and $V_M$) and $\tilde V$ the average volume of the two curves (with actual minima for $V_0=\tilde V \pm \Delta V$), we obtain $\nu=200 \frac{\Delta V}{\tilde V}$.
We now consider the limit in which $\Delta V\ll V_R$ (i.e., of a small discrepancy of the two curves in the volume range of interest,  corresponding to the regime of small $\varepsilon$ and $\nu$ in which our data show an almost linear relation between the two metrics).

It is then straightforward to show that
$\varepsilon \approx 2\sqrt{15} \frac{\Delta V}{V_R}$. Indeed, writing the two curves as $E_{1,2}(V)=A(V-\tilde V \pm \Delta V)^2$, with $A$ an appropriate coefficient (the same for both curves with our assumptions of same bulk modulus), we get that the integral in the numerator of $\varepsilon$, $\langle [E_1(V) - E_2(V)]^2 \rangle$, is given by:
\begin{align}
\frac 1 {2V_R}\int_{\tilde V - V_R}^{\tilde V + V_R}[A(V - \tilde V - \Delta V)^2 - A(V - \tilde V + \Delta V)^2]^2 dV &= \frac 1 {V_R}\int_{\tilde V}^{\tilde V + V_R}[A(V - \tilde V - \Delta V)^2 - A(V - \tilde V + \Delta V)^2]^2 dV \\
&= \frac 1 {V_R}\int_{\tilde V}^{\tilde V + V_R}[4A(V - \tilde V)\Delta V]^2 dV = \frac{16 A^2 \Delta V^2 V_R^2}{3}.
\end{align}
Similarly, we can obtain (using our assumption $\Delta V\ll V_R$) that 
\begin{align}
\langle E_1(V) \rangle = \frac 1 {2V_R}\int_{\tilde V - V_R}^{\tilde V + V_R} A(V - \tilde V - \Delta V)^2 dV \approx 
\frac 1 {V_R}\int_{\tilde V}^{\tilde V + V_R} A(V - \tilde V)^2 dV = \frac{AV_R^2}3,
\end{align}
and $\langle E_2(V) \rangle = \langle E_1(V) \rangle$.
Using similar steps, one can also obtain $\langle [E_1(V) - \langle E_1 \rangle]^2\rangle = \langle [E_2(V) - \langle E_2 \rangle]^2\rangle \approx \frac {4} {45} A^2 V_R^4$.
Putting all results together, we obtain the final result $\varepsilon \approx 2\sqrt{15} \frac{\Delta V}{V_R}$.

Finally, considering our choice of a $\pm 6\%$ volume range ($V_R\approx 0.06 \tilde V$) gives $\epsilon \approx \frac{100}3\sqrt{15} \frac{\Delta V}{\tilde V}$. 
Therefore, the two metrics are linearly dependent with the ratio of
\begin{equation}
    \left.\frac \nu\varepsilon \right\vert_{\Delta V} = \frac{6}{\sqrt{15}}\approx 1.55,
    \label{eq:nu_eps_ratio_dv}
\end{equation}
which also matches well with the ratio of $\nu$ and $\epsilon$ calculated for a specific value of $V_0$ perturbation shown in SI Fig.~\ref{fig:eos-sensitivity}(b).

A similar analysis can be performed for two parabolic \gls{eos} curves with the same $V_0$, but differing in $B_0$. In this case, $\nu = 10 \frac{\Delta B_0}{\tilde B_0}$, where $\tilde B_0$ is the average $B_0$ of the two curves and $\pm\Delta B_0$ is the difference from the average value for the two curves. It can be shown, similarly to the case of differing $V_0$, that in this case $\varepsilon \approx 3\frac{\Delta B_0}{\tilde B_0}$. The resulting ratio of the two metrics now becomes
\begin{equation}
    \left.\frac \nu\varepsilon \right\vert_{\Delta B_0} \approx \frac{10}{3}\approx 3.33,
    \label{eq:nu_eps_ratio_db0}
\end{equation}
which also matches well with the ratio of $\nu$ and $\epsilon$ in SI Fig.~\ref{fig:eos-sensitivity}(c). The difference by a factor of $2$ between \eqref{eq:nu_eps_ratio_dv} and \eqref{eq:nu_eps_ratio_db0} shows that, compared to $\varepsilon$, $\nu$ is $2$ times more sensitive to $B_0$ variation relative to $V_0$ variation.

\begin{figure*}[h!]
    \centering
     \includegraphics[width=0.9\linewidth]{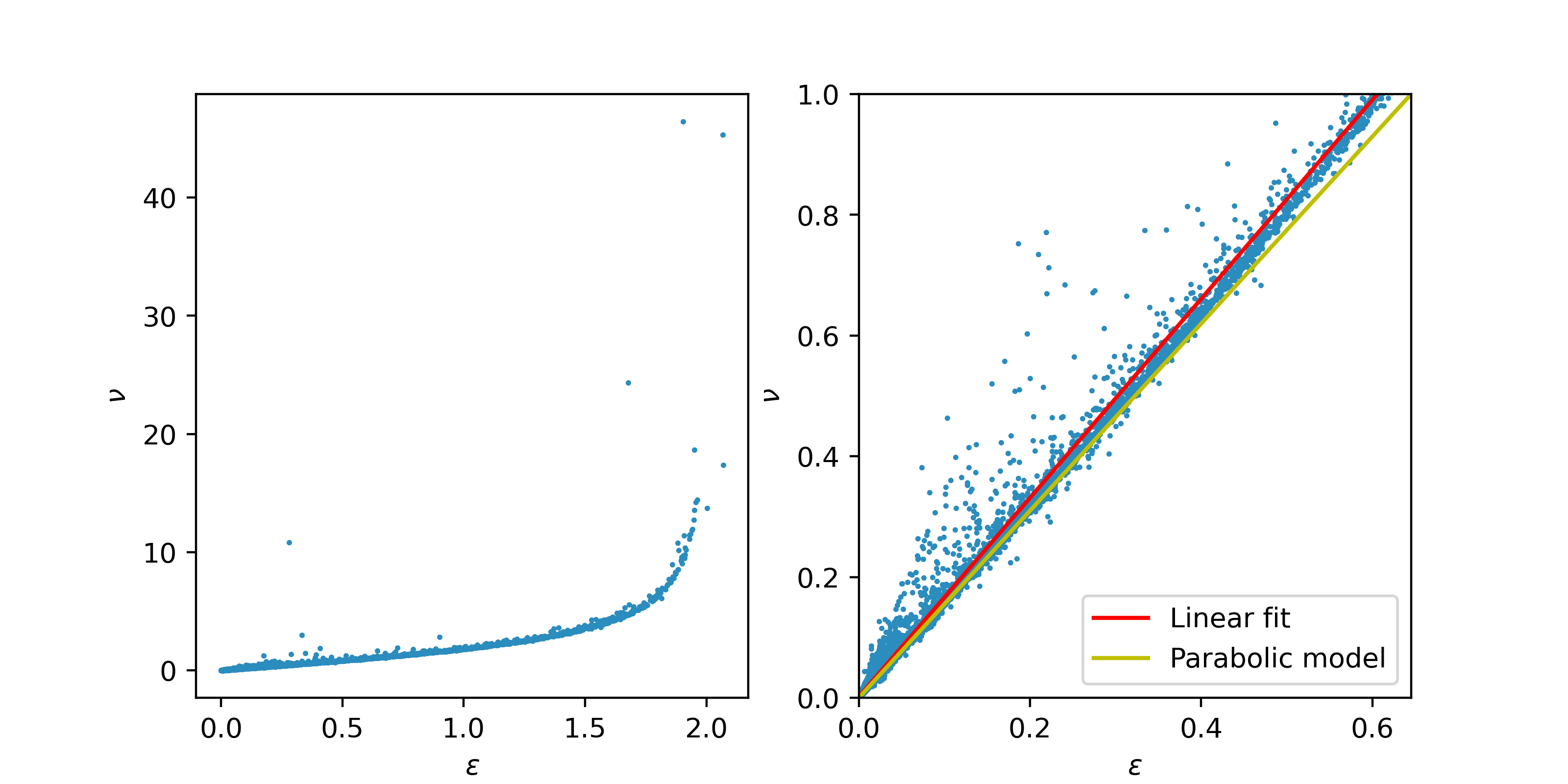}
     \caption{Cross-correlation between the $\varepsilon$ and $\nu$ metrics on a linear-axes plot. Left panel: all data; right panel: zoom on the data with $\nu < 1$ where the relation is approximately linear, together with a linear fit of the data (red) resulting in a slope of $\approx 1.65$, and a line with a slope $\approx 1.55$ (yellow curve) taken from the parabolic model.}
    \label{sifig:epsilon-nu-correlation}
\end{figure*}

To assess the typical ratio of $\nu$ and $\varepsilon$ in our dataset, in SI Fig.~\ref{sifig:epsilon-nu-correlation} we show a cross-correlation plot between $\nu$ and $\epsilon$ for the entire dataset of calculated crystals. One can see that for smaller values of the metrics ($\nu < 1$), the relation is approximately linear, and a numerical fit gives the slope of $\frac{\nu}{\varepsilon}\approx1.65$, which is close to 1.55 found in Eq.~\eqref{eq:nu_eps_ratio_dv}. This is expected, as both metrics give larger weight to $V_0$ errors compared to $B_0$ or $B_1$ errors, and in our dataset errors on $V_0$ (once rescaled with these weights) dominate over $B_0$ and $B_1$ errors.
This is also visible, for instance, from the histograms of Fig.~\ref{fig:ae-histograms} in the main text, where the ratios of the standard deviation of the histograms on $V_0$, $B_0$ and $B_1$ do not follow the 1, $\frac{1}{20}$, $\frac{1}{400}$ ratio of $\nu$.

The cross correlation plots allow to establish a data-driven relation between different metrics. For instance, the two boundaries $\varepsilon=0.06$ and 0.2 selected in this project as ``excellent'' and ``good'' agreement between two \gls{eos}s, and the corresponding thresholds for $\nu$ (0.10 and 0.33) have been chosen according to these cross correlations.
In addition, we can see from SI Fig.~\ref{fig:eos-sensitivity-3}a that these thresholds translate, for $\Delta$, approximately $\Delta \approx 0.3$ and 0.95~meV/atom. This result is comparable with the average $\Delta$ across \gls{ae} codes $\langle\Delta\rangle=0.5-0.9$~meV/atom obtained in the earlier benchmark\cite{Lejaeghere:2016,deltasite} for monoelemental solids, and is consistent with the conclusion obtained there that $\Delta=1$~meV/atom is a threshold under which one can speak about good agreement (for materials with bulk moduli that are not particularly small). This consistency between former and present benchmarks is about the metric; we refer to SI Sec.~\ref{sisec:71-vs-960} for a demonstration of consistency regarding the crystal set, and an illustration of the added value of the present benchmark study.

\clearpage

\section{Detailed results for all computational approaches\label{sisec:periodic-tables-per-code}}

In this section, we report the comparison of each of the computational approaches considered in the main text with the average all-electron reference dataset, using both metrics $\varepsilon$ and $\nu$. For each metric, the same colorbar is used for all approaches, based upon the ranges of agreement identified in SI Sec.~\ref{sisec:perturbations-eos} (in addition, the ratio of the threshold values for $\varepsilon$ and $\nu$ is in agreement with their approximate linear relationship, see SI Sec.~\ref{sisec:correlation-metrics}):

\begin{itemize}
    \item ``excellent agreement'' ($\varepsilon \le 0.06$, $\nu \le 0.10$): a very dark shade of blue (not evolving very much over this narrow interval);

    \item ``good agreement'' ($0.06 < \varepsilon \le 0.20$, $0.10 < \nu \le 0.33$): color evolving from a dark shade of blue to yellow as the values of $\varepsilon$ or $\nu$ increase;

    \item threshold for good agreement ($\varepsilon = 0.20$, $\nu = 0.33$): yellow;

    \item ``noticeably different'' ($0.20 < \varepsilon \le 1.0$, $0.33 < \nu \le 1.65$): color evolving from yellow to red as the values of $\varepsilon$ or $\nu$ increase;

    \item ``clearly different'' ($\varepsilon > 1.0$, $\nu > 1.65$): one uniform darker shade of red, regardless of the value.
\end{itemize}
Crystals that were not computed are left in white. The caption of every plot mentions the number of crystals belonging to each of these categories. The results for all the codes are shown in SI Figs.~\ref{fig:si-ptables-ABINIT} to \ref{fig:si-ptables-WIEN2k}.

\newcommand\singleapproachtemplate[1]{
    \setsepchar{, }
    \readlist\args{#1}
    \begin{center}
    \begin{figure}[h!]\centering
    \resizebox{.9\textwidth}{!}{%
        \includegraphics[height=5cm]{images/periodic-tables/periodic-table-unaries-\args[2]-vs-ae-epsilon}%
        \includegraphics[height=5cm]{images/periodic-tables/periodic-table-oxides-\args[2]-vs-ae-epsilon}%
    }
    \resizebox{.9\textwidth}{!}{%
        \includegraphics[height=5cm]{images/periodic-tables/periodic-table-unaries-\args[2]-vs-ae-nu}%
        \includegraphics[height=5cm]{images/periodic-tables/periodic-table-oxides-\args[2]-vs-ae-nu}%
    }
        \caption[]{Value of the comparison metrics $\varepsilon$ (top) and $\nu$ (bottom) for \args[1] with respect to the average all-electron reference dataset. Left panels: unaries; right panels: oxides. \args[3] out of 960 crystals were calculated. The number of crystals that land in the excellent, good, noticeably different, and clearly different agreement ranges for the $\varepsilon$ metric are \args[4], \args[5], \args[6], \args[7], respectively. For the $\nu$ metric, they are \args[8], \args[9], \args[10], \args[11], respectively.}
        \label{fig:si-ptables-\args[2]}
    \end{figure}
    \end{center}
}

\singleapproachtemplate{\abinitlong, ABINIT, 720, 232, 377, 111, 0, 244, 378, 98, 0}
\singleapproachtemplate{\bigdftlong, BigDFT, 402, 45, 97, 173, 87, 29, 106, 173, 94}
\singleapproachtemplate{\cptwoklong, CP2K-quickstep, 709, 57, 171, 317, 164, 55, 169, 302, 183}
\singleapproachtemplate{\fleurlong, FLEUR, 960, 936, 23, 1, 0, 938, 22, 0, 0}
\singleapproachtemplate{\gpawlong, GPAW, 670, 130, 156, 350, 34, 128, 155, 347, 40}
\singleapproachtemplate{\casteplong, CASTEP, 960, 197, 410, 277, 76, 206, 399, 267, 88}
\singleapproachtemplate{\qelong, QE, 960, 388, 300, 199, 73, 395, 300, 184, 81}
\singleapproachtemplate{\siestalong, SIESTA, 698, 30, 117, 444, 107, 18, 137, 424, 119}
\singleapproachtemplate{\siriuslong, SIRIUS-CP2K, 700, 363, 251, 81, 5, 374, 247, 72, 7}
\singleapproachtemplate{\vasplong, VASP, 960, 403, 348, 200, 9, 419, 341, 189, 11}
\singleapproachtemplate{\wientwoklong, WIEN2k, 960, 936, 23, 1, 0, 938, 22, 0, 0}

\clearpage

\section{Smearing and k-point convergence\label{sisec:kpt-convergence}}
For the study presented in this paper, a fixed choice of k-point integration mesh and smearing has been implemented. In particular, the k-point mesh is a uniform regular grid including the $\Gamma$ point, that guarantees a linear spacing of 0.06 \AA{}$^{-1}$ in each of the three reciprocal-space directions, and the smearing is a Fermi--Dirac type with broadening of 0.0045 Ry.
This choice of parameters is essential in order to compare with the reference dataset presented in this manuscript, as explained in Box 3 of our recommendations. In this section, we present the reasoning for our choice of parameters.

\begin{figure*}[h!]
\centering
 \includegraphics[width=0.8\linewidth] {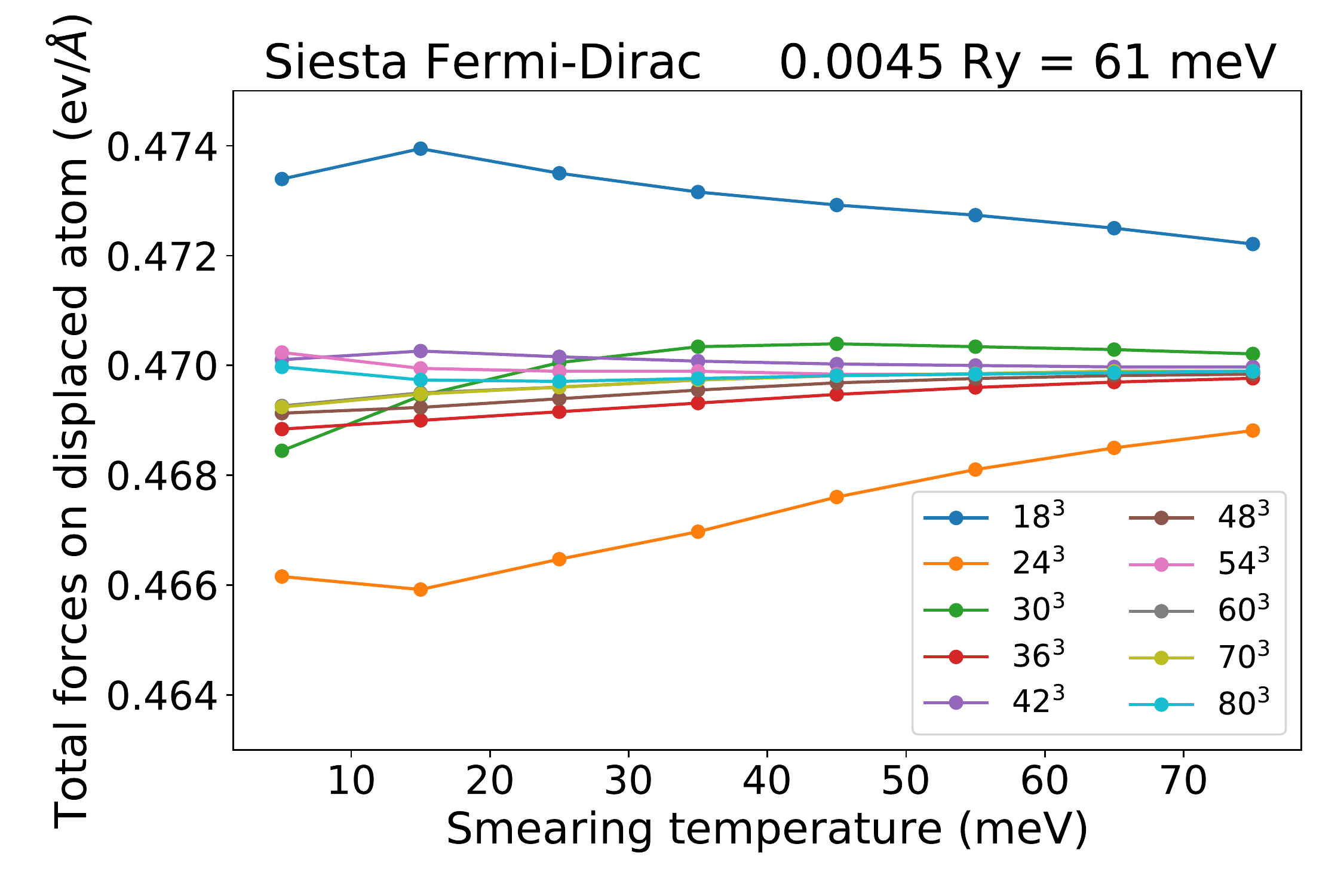}
 \caption{Total force on a displaced atom as a function of the smearing temperature and the k-point integration mesh. The system under investigation is an \gls{fcc} Al with an atom displaced by 0.1 \AA{} in the $x$ direction. The volume per atom is 16.47 \AA{}$^3$, for which a distance of 0.06 \AA{}$^{-1}$ between k-points correspond to a $26 \times 26 \times 26$ mesh. The smearing is a Fermi--Dirac smearing. Calculations of the forces are made with \siesta{} and DZP (double-zeta polarized) basis set.}
 \label{fig:sup3}
\end{figure*}

The Fermi--Dirac smearing is widely used in the community and it is implemented in all the codes that participate in the study. The choice of the broadening has been made according to the recommendation of Ref.~\citenum{dossantos2022}: the smearing parameter should not be too small to avoid sampling errors, nor too large to prevent systematic deviations due to the dependency of the total energy on the smearing broadening. The latter problem is explained in details in Ref.~\citenum{dossantos2022}, that shows the quadratic dependence of the free energy with respect to the Fermi--Dirac smearing temperature. The problem of sampling error is instead demonstrated in SI Fig.~\ref{fig:sup3}, where we analyze a \gls{fcc} aluminum crystal (conventional cell with 4 atoms) with an atom displaced by 0.1 \AA{} with respect to its equilibrium position. The figure reports the magnitude of the force on the displaced atom as a function of the k-point integration mesh and smearing broadening.
The instability of the force for very small broadening is clearly visible. Approaching the zero limit of the smearing, it becomes more and more necessary to have a dense k-point integration mesh in order to maintain 0.001 eV/\AA{} convergence on the forces.
The sampling error is not a peculiar feature of the Fermi--Dirac smearing; any other smearing type suffers from this drawback, as demonstrated in Ref.~\citenum{dossantos2022}.
Our choice of smearing broadening (0.0045 Ry $=\approx$ 61 meV) lays on the extreme right of SI Fig.~\ref{fig:sup3} and in this region a k-point integration mesh of $36\times36\times36$ is sufficient to obtain a converged value of the force within 0.001 eV/\AA{}. 
At the same time, 0.0045 Ry is a small enough value to reduce to a minimum issues due to the dependency of the total energy on the smearing broadening. In any case, for the goal of verification, the exact value is not so important, as long as all codes perform the very same choice of smearing and k-point integration mesh.

\begin{figure*}[h!]
\centering
 \includegraphics[width=0.95\linewidth] {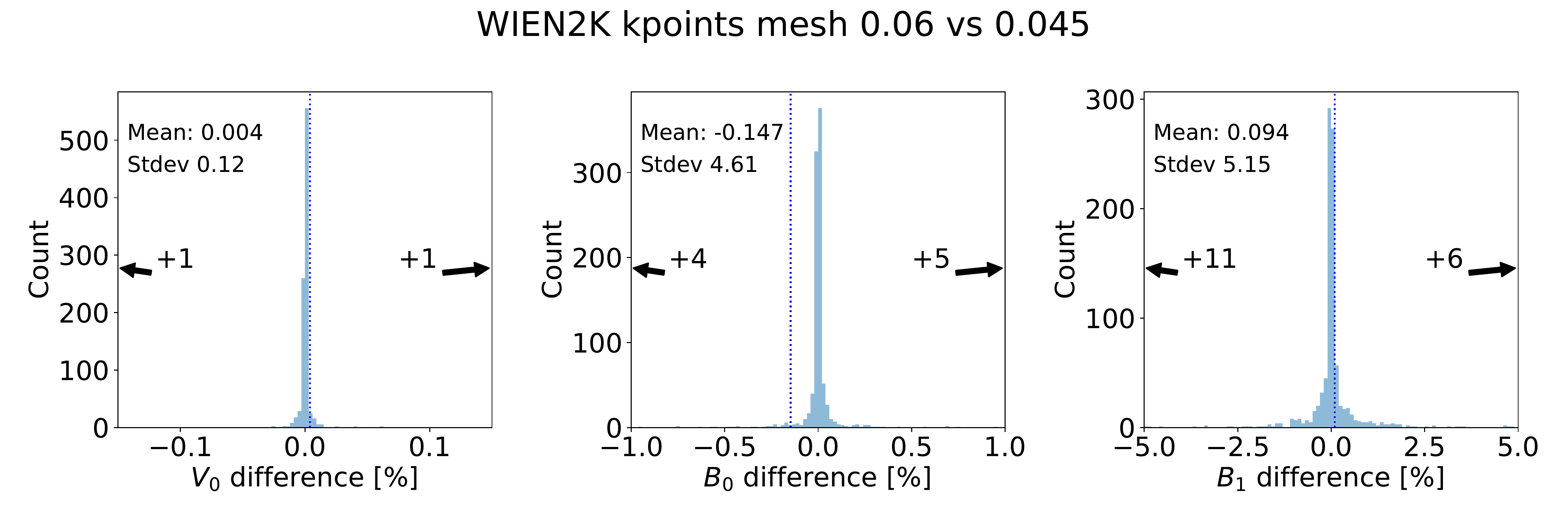}
 \caption{Histograms of the percentage difference between the results obtained with a k-point integration mesh with linear density 0.06 \AA{}$^{-1}$ and 0.045 \AA{}$^{-1}$ for the three \gls{eos} parameters $V_0$, $B_0$, and $B_1$. Results are obtained with \wientwok{} code. Numbers close to the arrows indicate outliers beyond the x-axis range.}
 \label{fig:sup3b}
\end{figure*}

With our choice of the smearing broadening, we expect a rapid convergence of the electronic-structure properties with respect to the the k-point density. We test this assumption looking at the effect of the k-point integration mesh on the estimation of the \gls{eos} parameters. Using \wientwok{} results, we compare calculations with k-point distance of 0.06 \AA{}$^{-1}$ and 0.045 \AA{}$^{-1}$ for all materials in the study. The comparison is reported in SI Fig.~\ref{fig:sup3b}. 
This figure shows an overall discrepancy that is significantly smaller with respect to the \fleur{}-\wientwok{} comparison presented in Fig.~\ref{fig:ae-histograms} in the main text (note that the x-axis range is half of the one in Fig.~\ref{fig:ae-histograms}). Looking at the histograms, we can estimate that the overall agreement is at least a factor of 2 better for $B_1$ with respect to the \fleur{}-\wientwok{} comparison, more than a factor of 4 better for $B_0$ and even one order of magnitude better for $V_0$. The same conclusion cannot be drawn looking at the standard deviations reported in the figure, due to the presence of two important outliers: RbO$_3$ (3.7\% difference in $V_0$, -142.29\% difference in $B_0$ and 149.35\% difference in $B_1$) and HeO (0.16\% difference in $V_0$, 3.33\% difference in $B_0$ and -7.90\% difference in $B_1$). RbO$_3$ and HeO are the only two materials that are not converged with a k-point distance of 0.06 \AA{}$^{-1}$. All other materials are converged within 0.07\% of $V_0$. It is interesting to notice that, even though RbO$_3$ and HeO are not converged, their discrepancy is not so dramatic when comparing \fleur{} and \wientwok{} (Fig.~\ref{fig:ae-histograms} in the main text). This, once more, justifies our recommendation of adopting the same k-point integration mesh for all computational approaches.

For completeness, we mention that the k-point mesh comparison of SI Fig.~\ref{fig:sup3b} has been performed on the crystal-structure set described in the main text for the unaries. For the oxides set, instead, we present results of a previous iteration of the volume refinement. Therefore the structures used for the oxides calculation in SI Fig.~\ref{fig:sup3b} have central volumes that slightly differs with respect to the ones used in the main text.

\clearpage

\section{Band structure of erbium in the diamond crystal structure\label{sisec:Er-dia-bands}}
\begin{figure*}[h!]
\centering
 \includegraphics[width=0.78\linewidth] {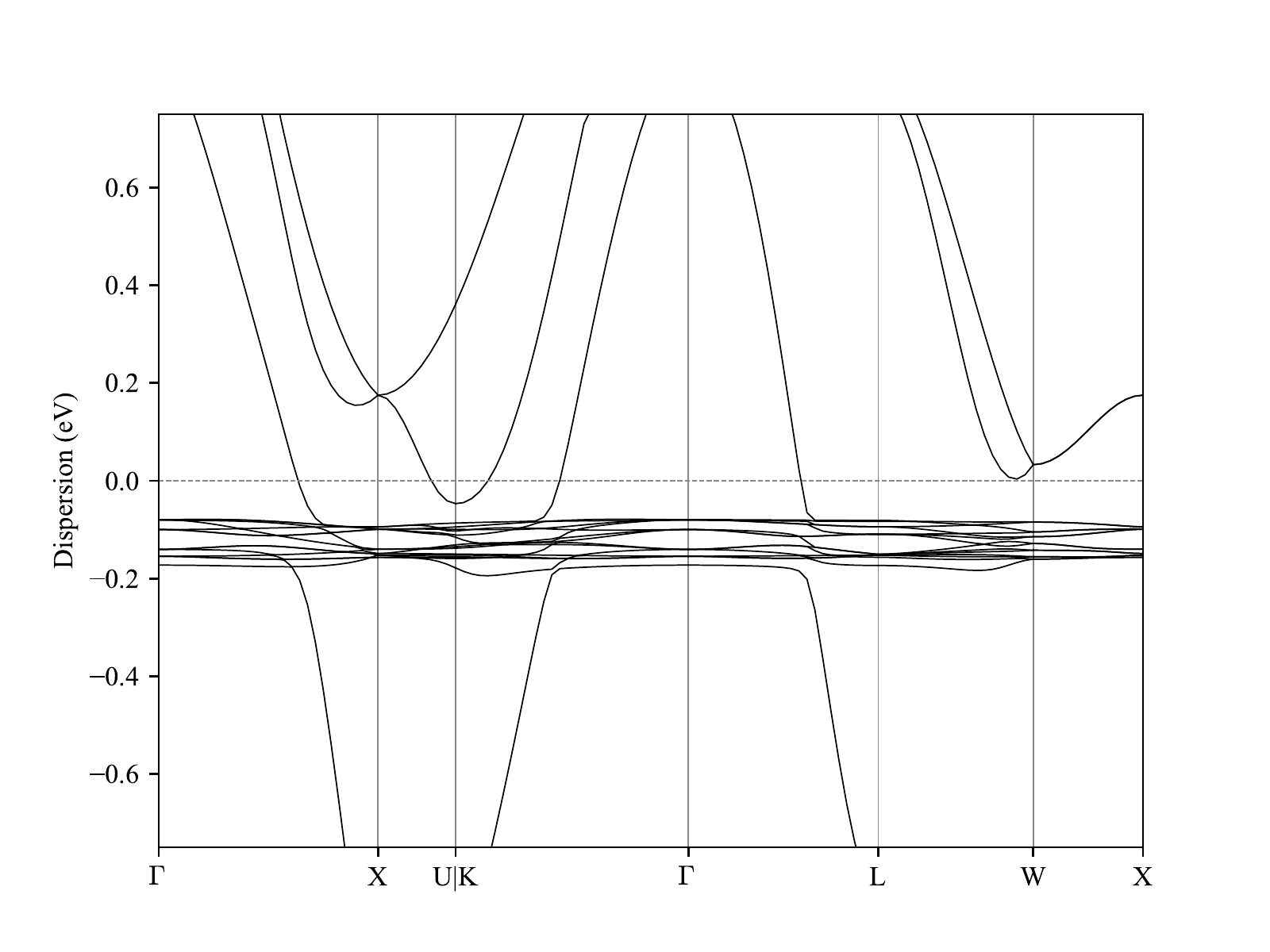}
 \caption{Band structure of erbium in the diamond crystal structure. The zero of energy is set at the Fermi level.}
 \label{fig:er-diamond-bands}
\end{figure*}

In SI Fig.~\ref{fig:er-diamond-bands} we show the band structure of erbium in the diamond crystal structure, obtained using the initial crystal structure (see SI Table~\ref{table-starting-volume}) with conventional cubic lattice parameter of 8.6296~\AA. 
The simulation has been run with the \qe{} code (version 7.0) using the PBE pseudopotential from the SSSP PBE Precision 1.2 library.~\citenum{Topsakal:2014}, and the recommended cutoffs of 40~Ry and 320~Ry for the wavefunctions and charge density, respectively. A $9\times9\times 9$ k-point integration mesh is chosen (note: this is less dense than the mesh recommended in the main text, but sufficient to demonstrate qualitatively the key features of the band structure, that is the goal of this section), and a Fermi--Dirac smearing with 0.0045~Ry of broadening. Input and output files are available in Ref.~\citenum{MCA-ACWF}.

The band structure clearly displays a set of ``almost flat'' bands (with a dispersion of $\sim 0.1$~eV) very close to the Fermi energy in the range between $-0.2$~eV and $0$~eV, originating from the $f$ states of erbium.
For this specific calculation and volume, these $f$ bands are just below the Fermi level, but their position can shift with volume and cross the Fermi level, significantly affecting the nature of the occupied states in the material. This explains the unconventional shape of the \gls{eos} displayed in the main text in Fig.~\ref{fig:various-smearings}.

An interesting note is that the location of the $f$ bands, determining the lowest-energy minimum of the \gls{eos}, also depends on the value $Z$, i.e., on the column of the periodic table.
For our choice of parameters (k-point integration mesh and smearing) erbium is at the boundary between elements favoring the minimum at lower volume (for smaller $Z$) and elements favoring the minimum at higher volume (for higher $Z$). This is clearly visible in SI Fig.~\ref{sifig:first-neigh-dist-unaries} as a jump of the first-neighbor distance for the diamond structures for elements before and after erbium.

\clearpage

\section{Total energy versus free energy\label{sisec:TS-vs-no-TS}}
Here we want to show a couple of examples where the exact choice of the energy to be computed in the \gls{eos}, namely the internal energy $E$, the free energy $E-TS$ (where $-TS$ is the smearing contribution) or the approximation $E-TS/2$, can significantly affect the \gls{eos} curves.

\begin{figure}[ht]
    \centering
    \noindent\includegraphics[width=0.49\linewidth]{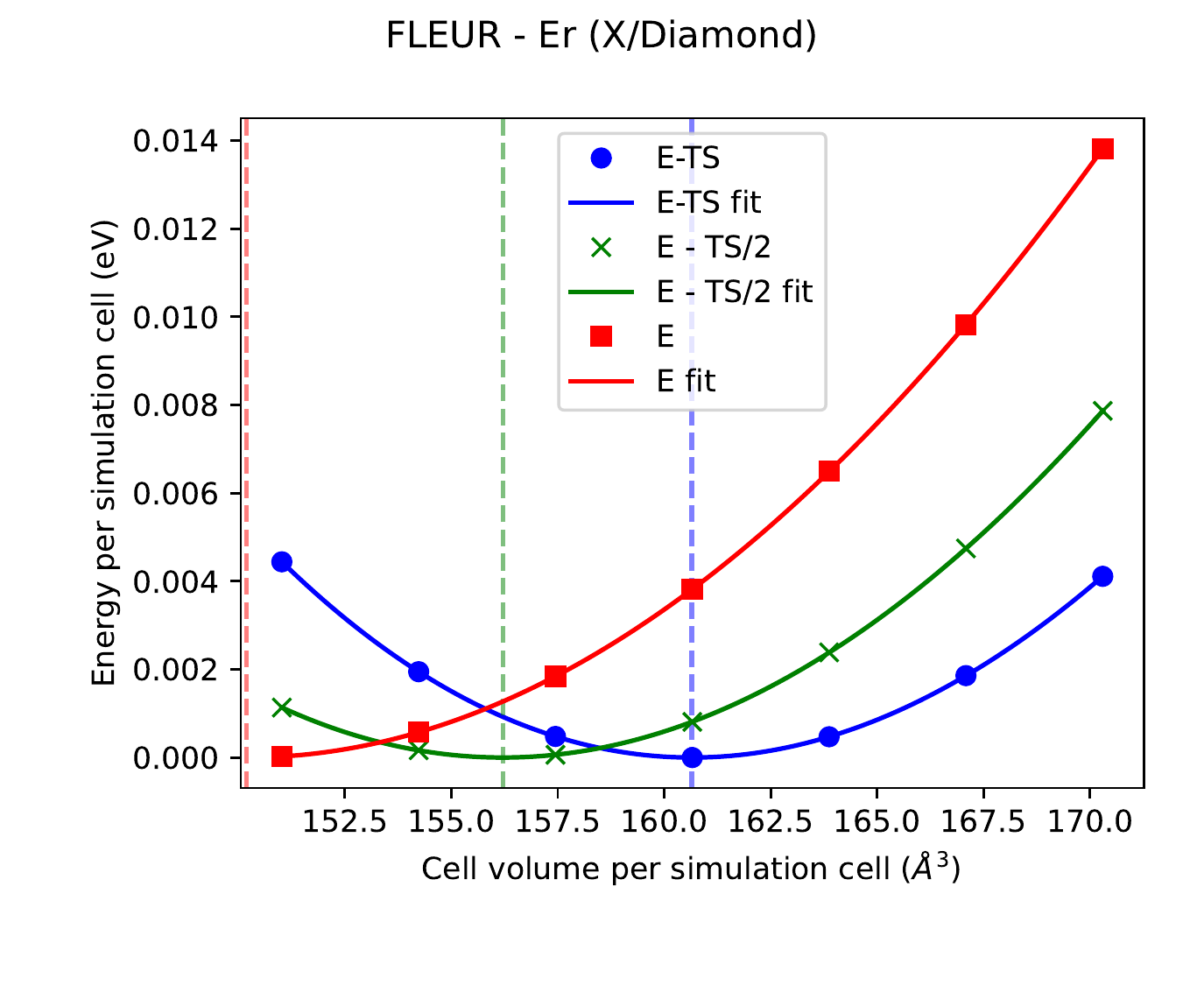}
    \includegraphics[width=0.49\linewidth]{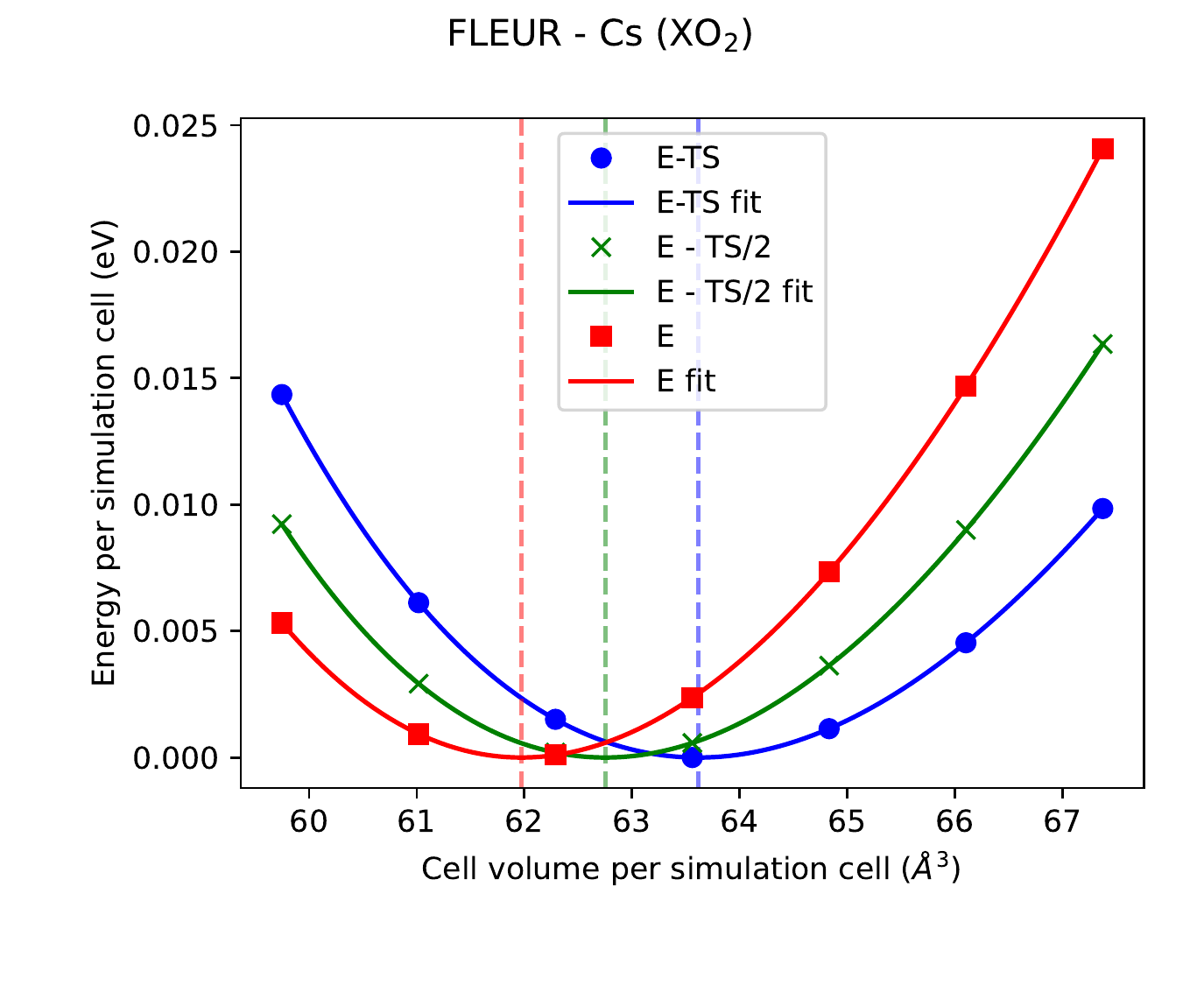}
    \caption{\label{sifig:EminusTSplots}Comparison of \gls{eos} curves for two systems computed using the \fleur{} code, using different quantities on the energy axis. Left panel: Er in the diamond structure; right panel: CsO$_2$.}
\end{figure}

While in many cases the choice of one of these three quantities does not have any effect on the curves (e.g., in the case of large-gap insulators), in SI Fig.~\ref{sifig:EminusTSplots} we show two \gls{eos} curves that we selected as they show a significant deviation (even larger deviations exist for other systems in our dataset).
The first case is erbium in the diamond structure. We already discussed its band structure in SI Sec.~\ref{sisec:Er-dia-bands}: we expect that there are significant flat $f$ bands (thus with high density of states) crossing the Fermi level as a function of volume; therefore, the $-TS$ contribution will also depend on volume, since it originates from the contribution of bands within a small energy range (comparable to the chosen smearing broadening) from the Fermi level.

\begin{figure}[ht]
    \centering
    \includegraphics[width=0.8\linewidth]{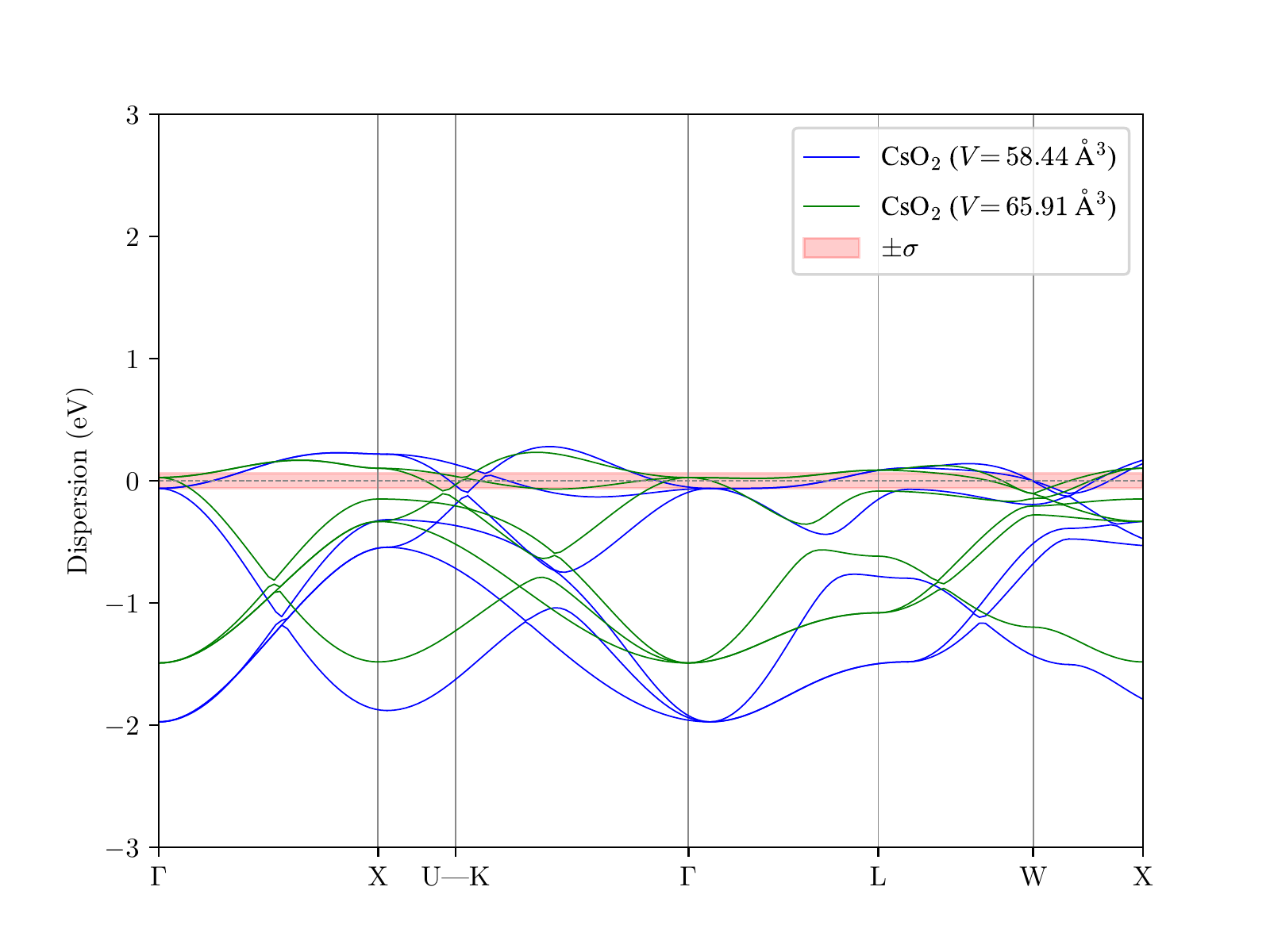}

    \caption{\label{sifig:CsO2bands-volume}Band structure of CsO$_2$ for two different volumes computed with the \qe{} code. The shaded red area indicates an energy range of $\pm 0.0045$~Ry, that is the smearing value recommended in this work.}
\end{figure}

One does not need, however, to consider such pathological cases: even other systems might show important discrepancies between the various curves if the density of states at the Fermi level changes significantly as a function of volume. This is illustrated for instance by CsO$_2$ (right panel in SI Fig.~\ref{sifig:EminusTSplots}), whose band structure for two different volumes is reported in SI Fig.~\ref{sifig:CsO2bands-volume}.
Here, we see that bands with saddle points or almost-flat bands can be found within $\pm \sigma$ from the Fermi level (with $\sigma$ being the value of smearing broadening recommended here). These bands shift with volume and therefore their contribution to $-TS$ will be a function of volume.

\clearpage

\section{Code specific parameters for the pseudopotential codes\label{SI:sec_codes_param}}

In this section, we discuss the technical choices adopted for the verification work and implemented, for each code, in the \texttt{verification-PBE-v1} protocol. 
We remind the reader that a number of parameters (such as the smearing type and broadening or the k-point integration mesh) have been fixed for all codes (see also Box 3 in the main text).
We also remind that the parameters for the two all-electron codes are discussed instead in SI Sec.~\ref{SI:parameters-ae}.

\subsection{\abinitlong{}}

The \abinit{} calculations were performed with the version 9.6.2 of the code\cite{Gonze:2016, Romero:2020, Gonze:2020} and v0.2a2 of the \texttt{aiida-abinit} plugin.
All calculations were run with \texttt{tolvrs = 1e-10}, Fermi--Dirac smearing (\texttt{occopt = 3}) of 0.0045 Rydberg (\texttt{tsmear = 0.00225}), a reasonable number of empty bands (\texttt{fband = 2}), and a minimal k-spacing of 0.06 \AA{}$^{-1}$.
Norm-conserving pseudopotentials from the PseudoDojo-v0.5 scalar-relativistic PBE standard library have been employed; the ``high'' stringency recommended energy cutoffs were used.
The RMM-DIIS diagonalization algorithm\cite{Kresse:1996}({\texttt{rmm\_diis = 1}}) was used for calculations with norm-conserving scalar-relativistic PBE pseudopotentials because of its improved computational efficiency over the default CG method.
Note that {\texttt{rmm\_diis = 1}} means that the first four SCF iterations are performed with the CG method in order to obtain reasonably good trial states before changing to the RMM-DIIS method.
RMM-DIIS is more efficient although less stable than the CG algorithm as there is no explicit orthogonalization while optimizing the trial states.
This led to approximately a 0.7\% error rate (108 out of 15460 total from all calculations performed in the process of this study) in the calculations, however in all cases, running the failed calculations with CG resulted in successful convergence.

\subsection{\bigdftlong}
The version of the BigDFT code employed for these calculation is the 1.9.2. For the great majority of the structures presented here, the pseudopotentials employed in the calculations are norm-conserving Hartwigsen--Goedecker--Hutter\cite{Hartwigsen1998} of the Krack family\cite{krack2005pseudopotentials} (HGH-K). For this verification campaign, rather than choosing the most precise pseudopotentials for a given element, we employed the default pseudopotentials with the least possible number of valence electrons. We have therefore used a set which provides an overestimation of the precision error of this pseudopotential family. 
For comparison we have also included some of the semicore pseudopotentials, see SI Sec.~\ref{BigDFT-SC}. The BigDFT code formalism employs Daubechies wavelets basis sets to express the Kohn-Sham (KS) orbitals, which enable to reach precise converged results for a given pseudopotential with moderate effort with respect to the number of degrees of freedom employed. The wavelets grid spacing was set to a value of $0.3~a_0$, with all the high-resolution degrees of freedom activated, and the k-point integration mesh correspond to a equivalent length of $94~a_0$. Density mixing scheme was employed for electronic convergence, reached for a threshold voxel accuracy of $10^{-12}$ atomic units. Also, 120 empty Kohn--Sham states were included for each k-point. Symmetry operations were also included to limit the calculations to the irreducible k-points.
At the time of developing the workflow, the BigDFT code was migrating its user interface into the PyBigDFT python module, which provides a user interface to the underlying executable. Therefore, to avoid issues in the API modifications, we fixed the PyBigDFT version to a beta release in the \texttt{aiida-bigdft} plugin, which activated limited features with respect to the stable version available nowadays. In particular, not all the structures which were defined with a non-orthogonal unit-cell were transformed in an orthogonal supercell, required by the code. This compatibility problem resulted in less structures treated by this approach. It is planned to release a stable version of the plugin compatible with the AiiDA 2.x API.

\subsection{\casteplong}
\castep{} is a plane-wave pseudopotential code\cite{Clark:2005}, 
the 20.1.1 version is used in this work. 
Calculation parameters closely follow the "precision" setting in the initial common workflow implementation\cite{Huber2021}. 
The cut off energy is fixed at 800 eV for all calculations since energy comparison is needed between different chemical systems. 
The reciprocal space sampling is done through $\Gamma$ centered Monkhorst--Pack grids with a fixed spacing of 0.06 $\text{\AA}^{-1}$ (i.e., 0.00954929 $2\pi\text{\AA}^{-1}$), in line with other codes.
On-the-fly generated (OTFG) core-corrected ultrasoft pseudopotentials from the library \texttt{C19} is used for the study except for the f-block elements (see section \ref{sec:castep-pp} for more details).
The \texttt{C19} library is aimed for general use with a balance between precision and speed,
and it has been the default potential library since \castep{} version 19.1.1.
The modified pseudopotential generation strings for f-block elements are tabulated in Table \ref{tab:castep-otfg}.
The energy convergence threshold for electronic minimization is set to $1\times10^{-8}$ eV per atom.

\subsection{\cptwoklong}
The DFT module \textsc{Quickstep} of the open-source simulation package CP2K is an implementation of the Gaussian and plane wave (GPW) and the all-electron Gaussian augmented plane wave approaches\cite{cp2k,Kuehne:2020}. Therein, the Kohn-Sham orbitals are represented by contracted Gaussian basis functions, whereas the electronic charge density is expanded in plane waves\cite{lippert1997hybrid}. For the former, an accurate molecularly optimized triple-$\zeta$ basis set with two additional sets of polarization functions (TZV2P-MOLOPT) is employed\cite{vandevondele2007gaussian}, whereas for the latter a density cutoff of 2400~Ry is utilized, which differs from a conventional plane wave cutoff by a factor of four. Due to its GPW method, however, CP2K/\textsc{Quickstep} is rather insensitive with respect to high density cutoffs. Furthermore, four multi-grids are used to ensure an efficient mapping of product Gaussians onto the real-space integration grids, so that wide and smooth Gaussian functions are mapped onto a coarser grid than narrow and sharp Gaussians. To control which product Gaussians are mapped onto which level of the multi-grid, a relative cutoff of 80~Ry is applied that defines the plane wave cutoff of a reference grid covered by a Gaussian with unit standard deviation. Separable and norm-conserving Goedecker--Teter--Hutter-type pseudopotentials including scalar relativistic effects are used to describe the interactions between the valence electrons and the ionic cores\cite{goedecker1996separable,krack2005pseudopotentials}. 

\subsection{\gpawlong}
GPAW\cite{GPAW1,GPAW2} is an open-source DFT code developed at the Technical University of Denmark (DTU) and other universities and computer centers, originally created for combining a homogeneous grid basis set with the projector augmented wave (PAW) method. Today, the code also provides a linear combination of atomic orbitals (LCAO) basis\cite{GPAW_lcao} and a plane wave mode. The latter has been applied in this study with a plane-wave cutoff of 800 eV for all calculations, with the exception of systems containing noble gases, where a cutoff of 1200 eV combined with a tighter energy and density convergence was applied. As with other codes, Fermi--Dirac smearing of 0.06122 eV was used. Besides the mentioned parameters, default values (as per GPAW v. 21) were used for all other keywords needed to perform the calculation in order to ensure that the results reflect the most representative user experience.  The applied PAW potentials, included in GPAW's PAW potential suite, were specifically created for the PBE exchange correlation functional, by applying GPAW's setup creator\cite{GPAW-setups}. In particular, we use the pseudopotentials included in the setup release 0.9.20000 available at \url{https://wiki.fysik.dtu.dk/gpaw/setups/setups.html#atomic-paw-setups}.
 GPAW is tightly linked to the atomic simulation environment (ASE)\cite{ASE1,ASE2}, which handles the user interface and is developed independently.

\subsection{\qelong}
All calculations have been run using version 7.0 of the Quantum ESPRESSO code and  version 3.5.1 of the AiiDA Quantum ESPRESSO plugin. For the results presented in the main text, all pseudopotentials were selected from the SSSP PBE Precision 1.3\cite{SSSP_1_3} library, with plane-wave cutoffs corresponding to the largest recommended value from the elements in each structure. In accordance with the \texttt{verification-PBE-v1} protocol, the Brillouin zone sampling was performed using $\Gamma$-centered meshes with a spacing of 0.06 \AA{} and a Fermi--Dirac smearing of 0.0045 Ry. All other inputs parameters were set via the \texttt{precision} protocol as described in the SI of Ref.~\citenum{acwf}, most importantly the energy convergence threshold was set to a very strict 0.1 $\cdot 10^{-9}$ Ry per atom. For the comparison with \abinit{} and \castep{} discussed in~\ref{sisec:plane-wave}, the pseudopotentials were selected from the PseudoDojo SR PBE standard set, version 0.4, in UPF format, after a small modification of the \texttt{.upf} files as described in~\ref{sisec:plane-wave}. Plane-wave cutoffs were obtained from the "high" stringency hints provided by the PseudoDojo table, all other computational parameters were unchanged.

\subsection{\siestalong}
The calculations presented in this work have been carried out with Siesta version Max-1.2.0 (\url{https://gitlab.com/siesta-project/siesta/-/tags/MaX-1.2.0}) powered by the \texttt{aiida-siesta} plugin version 1.2.0. Pseudopotentials from the PseudoDojo FR standard set, version 0.4, in PSML format, have been employed.
The real-space cutoff for the representation of charge densities and potentials is fixed at 900 Ry. The recommended course of action regarding basis sets in Siesta is to perform an optimization considering the key features of the chemical environment of each system. In this project, we have not carried out the optimization for all 960 systems. Instead, we have attempted a partial, per element, optimization, considering only the unary diamond crystals at their central volume. The orbitals thus optimized for each element are then reused for all the other unary and oxide structures involving that element. The optimization starts from a TZDP basis with the addition of an extra $f$ orbital. For alkali metals and alkaline earth metals the addition of a $d$ orbital shell is also necessary. The optimization is performed with the Nelder--Mead algorithm (multidimensional optimization without derivatives), having as variables the first-zeta radius of each orbital and the split norm parameter that controls the ratio between the first and subsequent zetas. We foresee using the information garnered in this verification study to develop further heuristics and guide the development of fully automatic methods to generate basis sets taking into account appropriate chemical environment descriptors.

\subsection{\siriuslong}
\sirius{} is a domain-specific library, which implements pseudopotential plane wave and full potential linearized augmented plane wave methods and is designed for GPU acceleration\cite{sirius}. As such it brings additional functionalities to CP2K such as collinear and non-collinear magnetic systems with or without spin–orbit coupling. It is written in C++14 with the MPI, OpenMP, and CUDA/ROCm programming models. As shown previously, \sirius{}/CP2K allows for energy conserving \textit{ab-initio} molecular dynamics simulations with a constant shift in the order of $\mu$Ha compared to \qe~reference calculations \cite{Kuehne:2020}. All \sirius{}/CP2K simulations were performed using the pseudopotentials of SSSP PBE Precision 1.2\cite{SSSP_1_2} together with a plane-wave cutoff of 55~a.u.$^{-1}$ for the density and potential, as well as 10~a.u.$^{-1}$ for $| \mathbf{k}+\mathbf{G} |$, respectively.

\subsection{\vasplong}
\label{sub:code_params_vasp}
All VASP results in this work have been obtained with VASP version 6.3.0 and AiiDA-VASP 2.1.0, using preferentially the GW VASP PBE potential set version 54.

The input parameters explicitly set for this work were (defaults in brackets): $\mathrm{PREC}=\mathrm{Accurate} (\mathrm{Normal})$, $\mathrm{EDIFF}=\mathrm{1E-7} (\mathrm{1E-4})$, $\mathrm{ALGO}=\mathrm{Normal} (\mathrm{Normal})$, $\mathrm{NELM}=300 (60)$, and $\mathrm{LMAXMIX}=6 (2)$. The last setting ensures that all electronic states up to quantum numbers $l=6$ are included in the density mixer, and it is necessary to change this in order to converge some $d$-electron systems and most lanthanides. The plane wave cutoff was fixed at 1000~eV ($\sim 73.50$ Ry) for all calculations. The PAW method was used\cite{PhysRevB.59.1758}.

We used the recommended GW potential sets, whenever possible, with exception of oxygen, where we chose the hard O\_h\_GW potential instead of the recommended O\_GW since it improved results for the oxides dataset. This potential is required if short bonds to oxygen atoms are encountered. The recommended plane wave cutoff energy for this potential is 765~eV, however, for high precision studies we recommend to increase the cutoff in VASP by 30\% yielding the employed 1000 eV. For the elements where no GW potential is supplied, we use the recommended standard PBE PAW potential. (The recommendations were taken from the \href{https://www.vasp.at/wiki/index.php/Available_PAW_potentials}{VASP-wiki} on April 22nd, 2022).

\clearpage

\section{Precision of plane-wave codes when using the same pseudopotential library\label{sisec:plane-wave}}

In order to assess how much codes implementing the same computational approaches agree among each other, we compare the $\varepsilon$ metric among a subset of the plane-wave codes  considered in this work, when using the same pseudopotential library.
We consider two different cases. In the first case, we compare three different codes (\abinit{}, \qe{}, and \castep{}) using the PseudoDojo v0.4 pseudopotential library. In the second case, we compare  \qe{} and \sirius{}/CP2K using the SSSP PBE precision v1.2 pseudopotential library.

For the first case (PseudoDojo v0.4 library), an initial comparison highlighted some discrepancies among the codes for Cu, Zn, and Ne. %
After investigation, we found that these discrepancies stem from differences in the numerical treatment of the form factors, the local part of the pseudopotential, the model core charge density or the beta projectors with spherical Bessel functions (usage of spline or not, different integration methods, etc.).

While we have not precisely identified the main root of the effect, which would require additional investigation, it is clear that 
the treatment of the long range part of the pseudopotential is rather sensitive to the implementation details.
For this reason, we manually truncated the radial mesh reported in the pseudopotential files (in UPF format, normally truncated at 10 Bobr) after 6 Bohr for all elements, as it is done in the psp8 format used by \abinit{}, in order to have the same radial mesh in all codes.
We verified that, except for the three cases mentioned above, this truncation had no visible effects on the results when compared to all-electron results. 
In the figures below, the tables for which the radial mesh has been truncated will be referred to as ``trim'' as a suffix to the approach label reported in the periodic-table titles.

\begin{figure}[h!]
    \centering
     \resizebox{.9\textwidth}{!}{%
        \includegraphics[height=5cm]{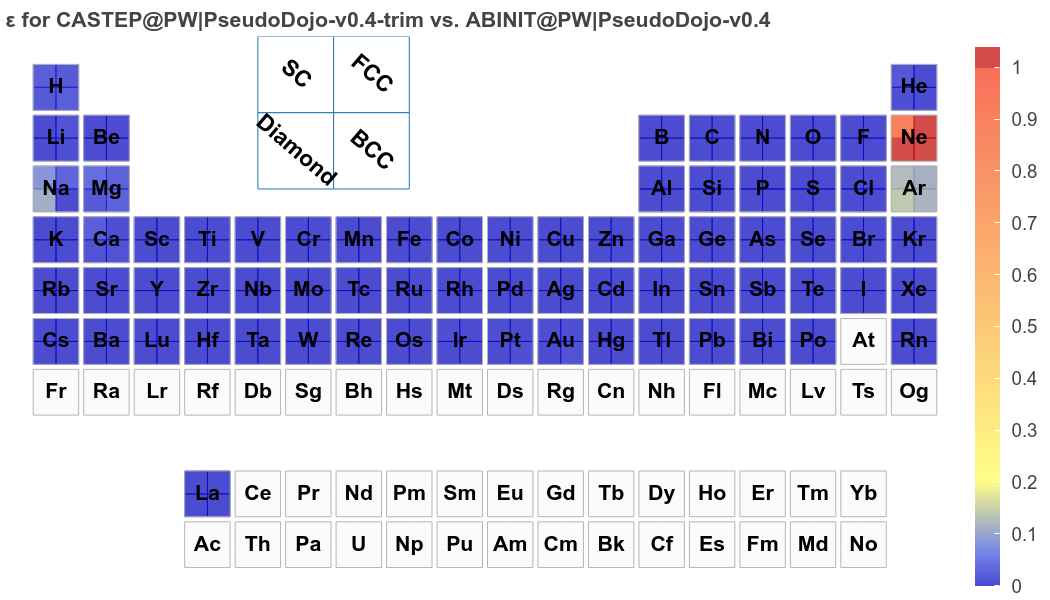}%
        \includegraphics[height=5cm]{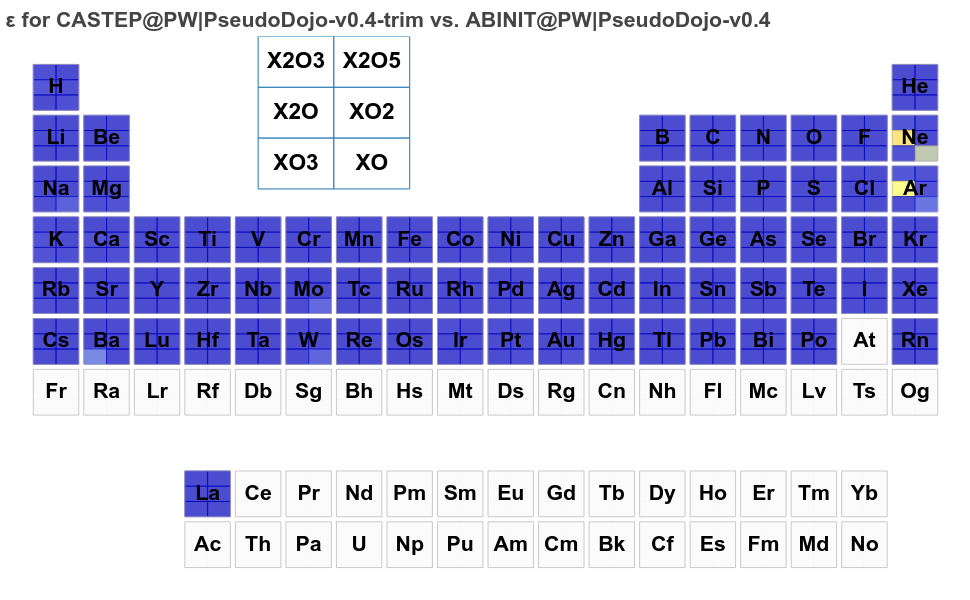}%
    }
    \caption{\label{fig:s161} Value of the comparison metric $\varepsilon$ between the \castep{} code (with truncated radial mesh) and the \abinit{} code for the unaries and oxides set using the same PseudoDojo v0.4 library. 
   }     
\end{figure}

\begin{figure}[h!]
    \centering
     \resizebox{.9\textwidth}{!}{%
        \includegraphics[height=5cm]{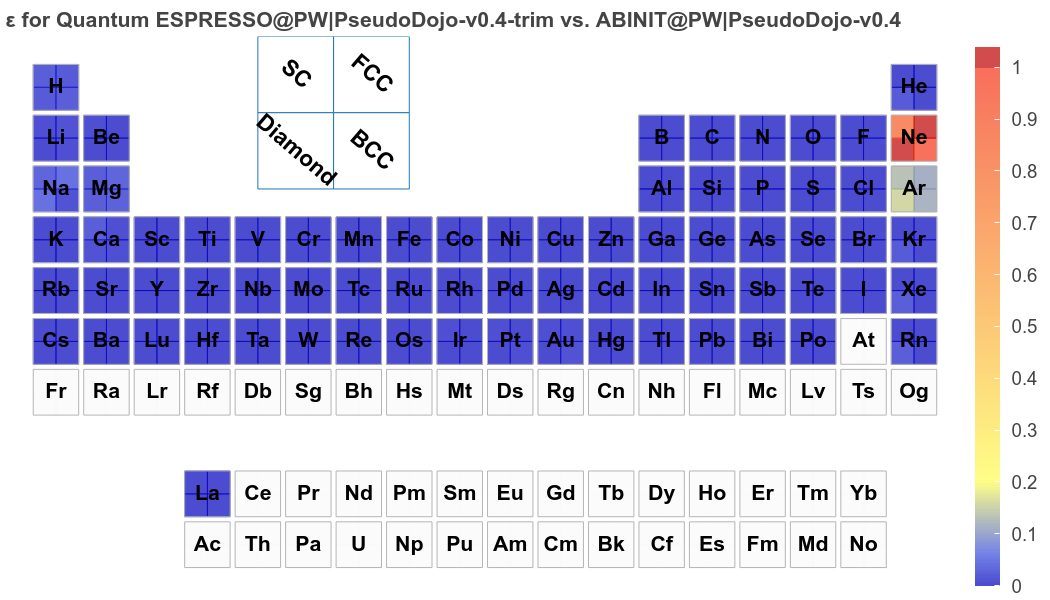}%
        \includegraphics[height=5cm]{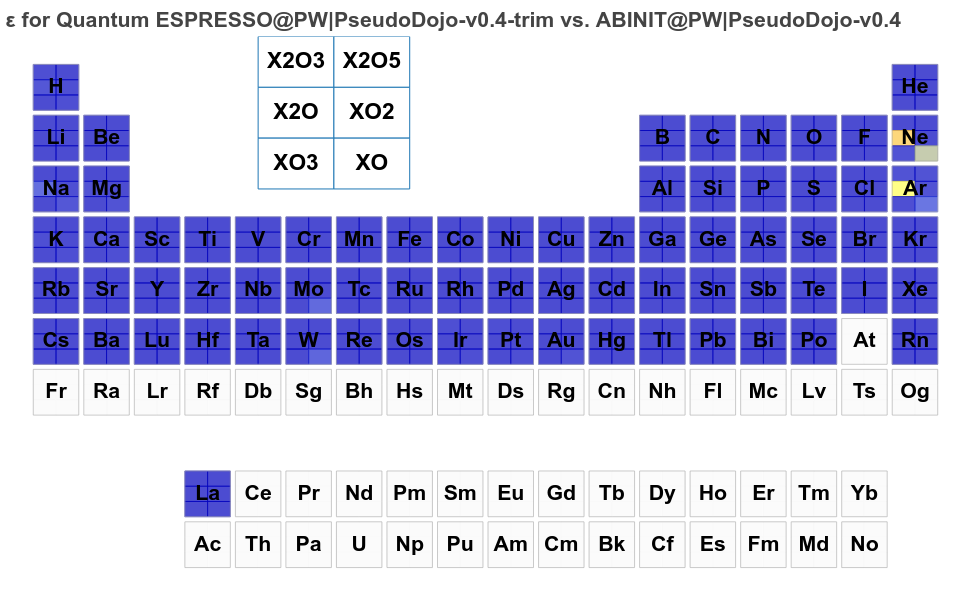}%
    }
    \caption{\label{fig:s162} Value of the comparison metric $\varepsilon$ between the \qe{} code (with truncated radial mesh) and the \abinit{} code for the unaries and oxides set using the same PseudoDojo v0.4 library. 
   }     
\end{figure}

\begin{figure}[h!]
    \centering
     \resizebox{.9\textwidth}{!}{%
        \includegraphics[height=5cm]{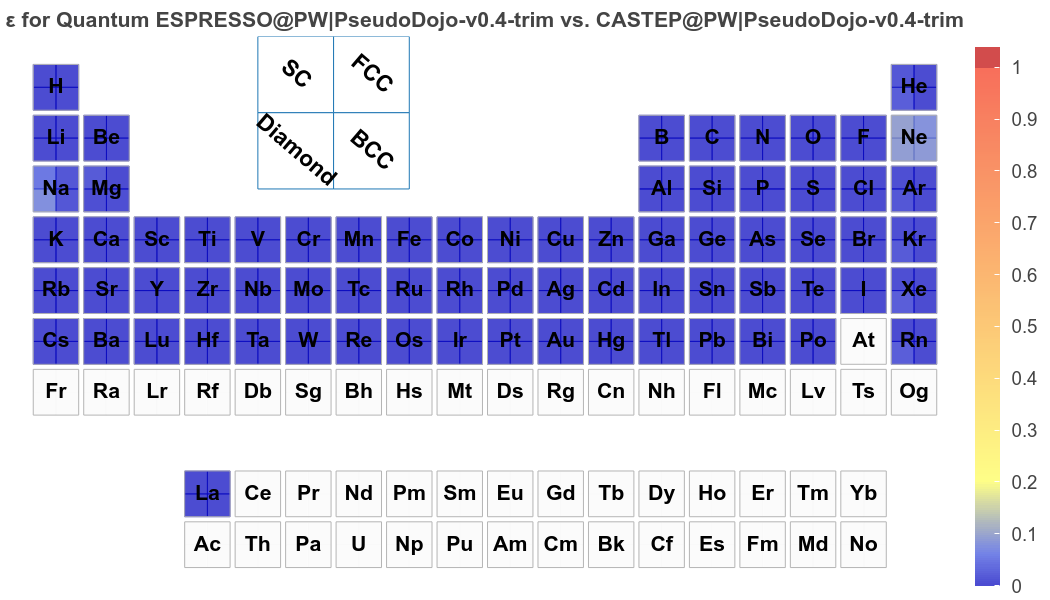}%
        \includegraphics[height=5cm]{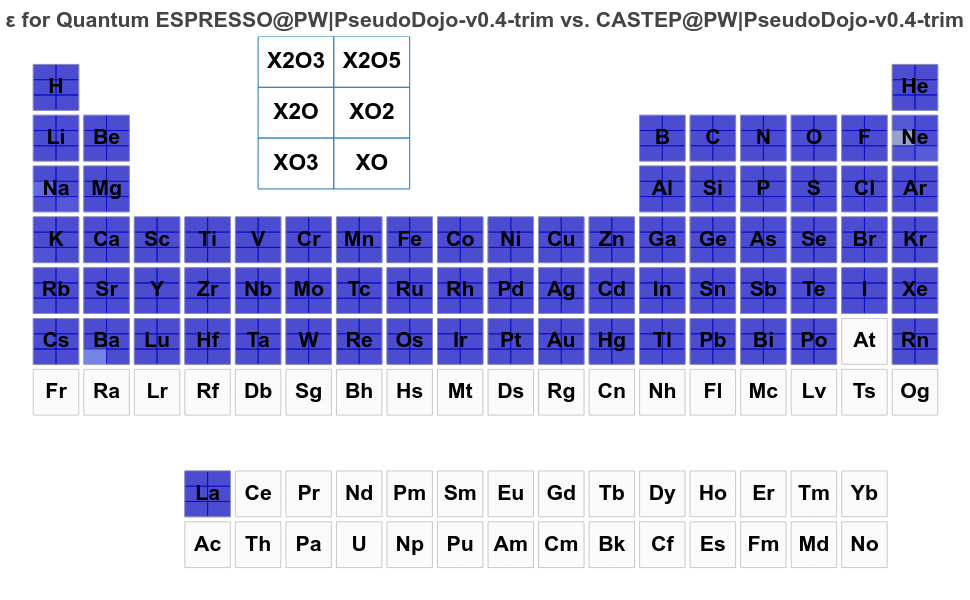}%
    }
    \caption{\label{fig:s163} Value of the comparison metric $\varepsilon$ between the \qe{} and \castep{} codes (both with truncated radial mesh) for the unaries and oxides set using the same PseudoDojo v0.4 library. 
   }     
\end{figure}

Using this simple truncation to avoid numerical instabilities, we show in SI Figs.~\ref{fig:s161}, \ref{fig:s162} and \ref{fig:s163} the six possible pairwise comparisons between the three codes. 
Remarkably, except for two noble gases, the agreement is excellent (we have used the same color scale for $\varepsilon$ as in the main manuscript).
This agreement is on average even better than the agreement between \gls{ae} codes of this work (see Fig.~\ref{fig:ae-histograms} of the main text). We highlight however that this is expected, since the plane-wave codes use the same basis set and, in this SI Section, an almost identical set of numerical parameters, while the two \gls{ae} codes differ in the details of the basis set and numerical parameters.

\begin{figure}[h!]
    \centering
     \resizebox{.9\textwidth}{!}{%
        \includegraphics[height=5cm]{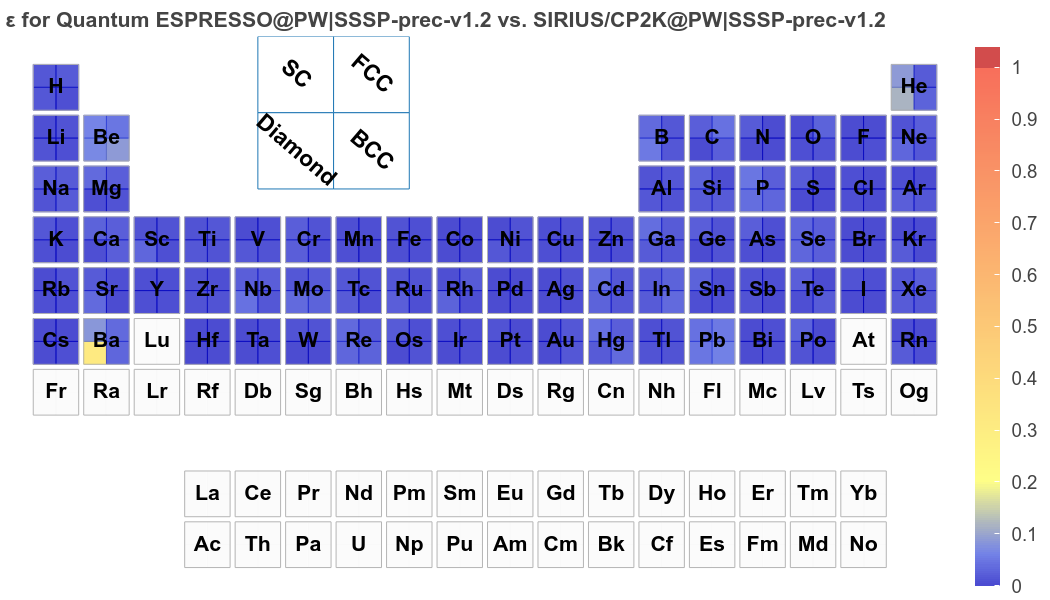}%
        \includegraphics[height=5cm]{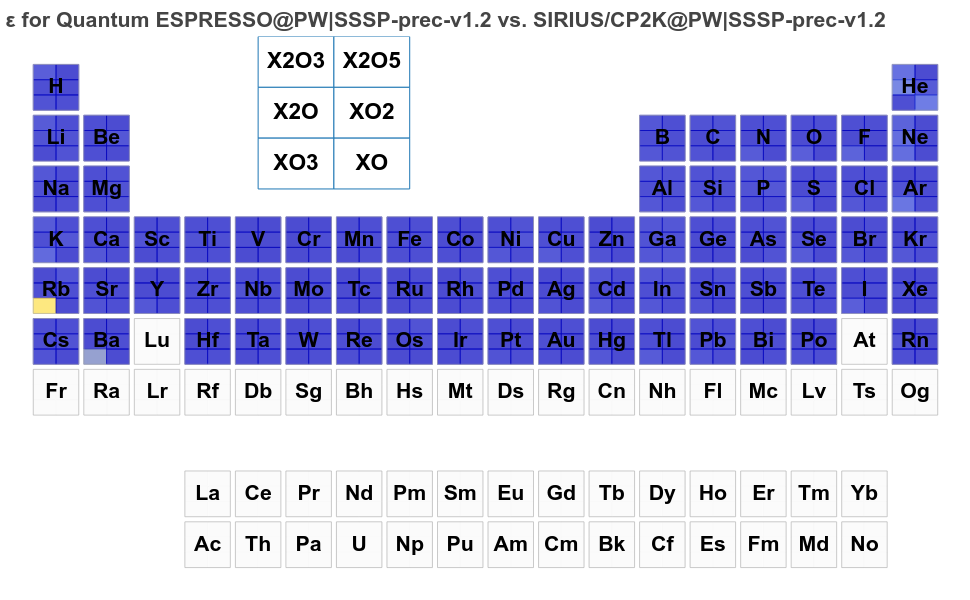}%
    }
    \caption{\label{fig:pt-sssp-qe-vs-cp2ksirus} Value of the comparison metric $\varepsilon$  between the \qe{} and \sirius{}/CP2K software for the unaries and oxides set using the same SSSP PBE precision v1.2 library. 
   }     
\end{figure}

In the second case, we compared the SSSP PBE precision v1.2 pseudopotential library for \qe{} and \sirius{}/CP2K. The comparison of the $\varepsilon$ metric is shown in SI Fig.~\ref{fig:pt-sssp-qe-vs-cp2ksirus}. The match is very good for all systems except for Ba-diamond and RbO$_3$. We note here that in the main text, we used the SSSP PBE precision v1.3 for \qe{}, but we stress that it is equivalent to v1.2 for the subset of chemical elements considered in this section (the difference being that v1.3 also includes actinides).

The overall excellent agreement displayed in this SI Section demonstrates that different codes implementing the same computational approach (including, in addition to the basis set, also the pseudopotential library and other computational parameters) can reproduce the same results.
In addition, our investigation helps identifying the remaining numerical aspects that might produce different outcomes and that merit further investigation, such as the truncation of the radial mesh discussed here.

\clearpage
\section{Transferability of conclusions from the previous smaller crystal set to the current 960 crystal set \label{sisec:71-vs-960}}

In this section we want to show that even if two codes provide nearly identical results for the 71 crystal set of Refs.~\citenum{Lejaeghere:2016,deltasite}, they will not necessarily provide identical results for the larger 960 crystal set of the present paper, demonstrating the value of the larger crystal-structure set.

\begin{figure*}[th!]
    \centering
    \includegraphics[width=0.7\textwidth]{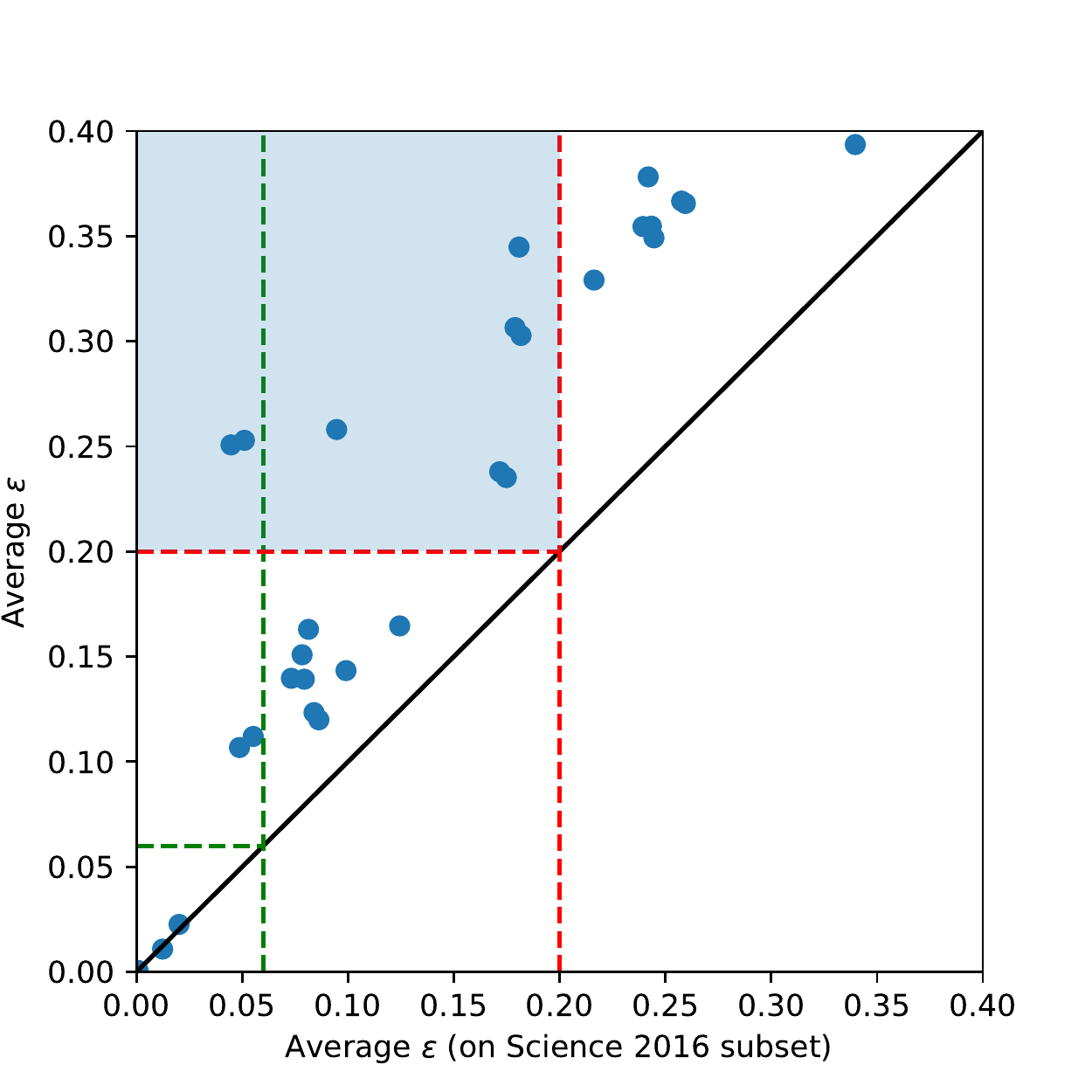}
    \caption{Vertical axis: $\varepsilon$ averaged over all crystals of the unary and oxide set computed in this work, for all pairs of computational approaches considered in the main text. Horizontal axis: $\varepsilon$ averaged over the 29 crystals from these 960 that appear as well in the 71 crystal set from Refs.~\citenum{Lejaeghere:2016,deltasite}. The horizontal and vertical lines indicate the threshold values for excellent (green) and good (red) agreement, as discussed in the main text. The region above the horizontal red line and to the left of the vertical red line, highlighted by the shaded blue area, indicates pairs that are in good agreement according to the small crystal-structure set, but less so when considering the full set. \label{sifig:29-vs-960}}
\end{figure*}

\begin{figure*}[th!]
\begin{center}
    \resizebox{.9\textwidth}{!}{%
		\includegraphics[height=5cm]{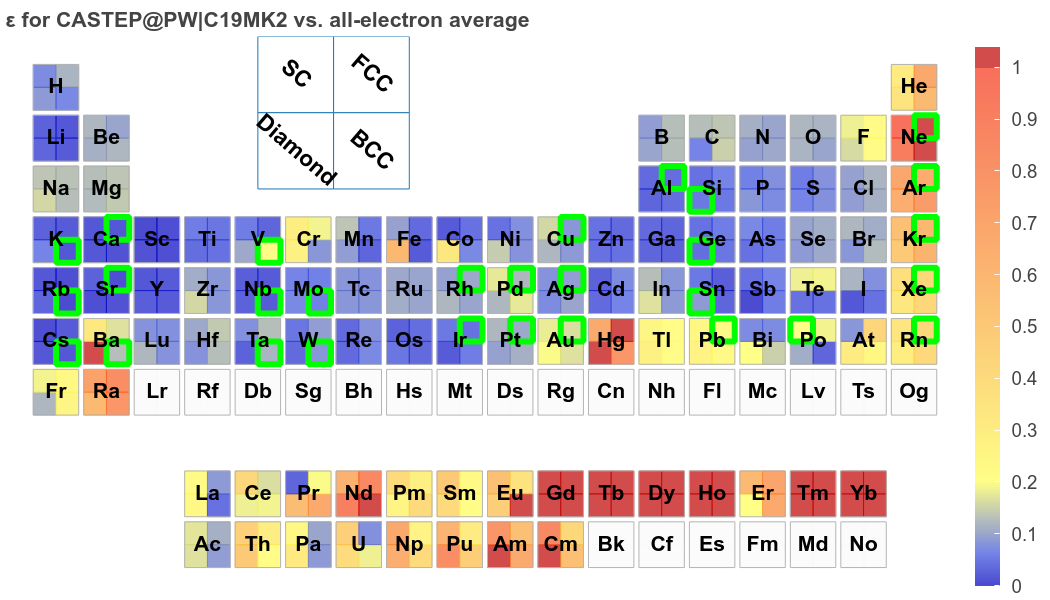}%
		\includegraphics[height=5cm]{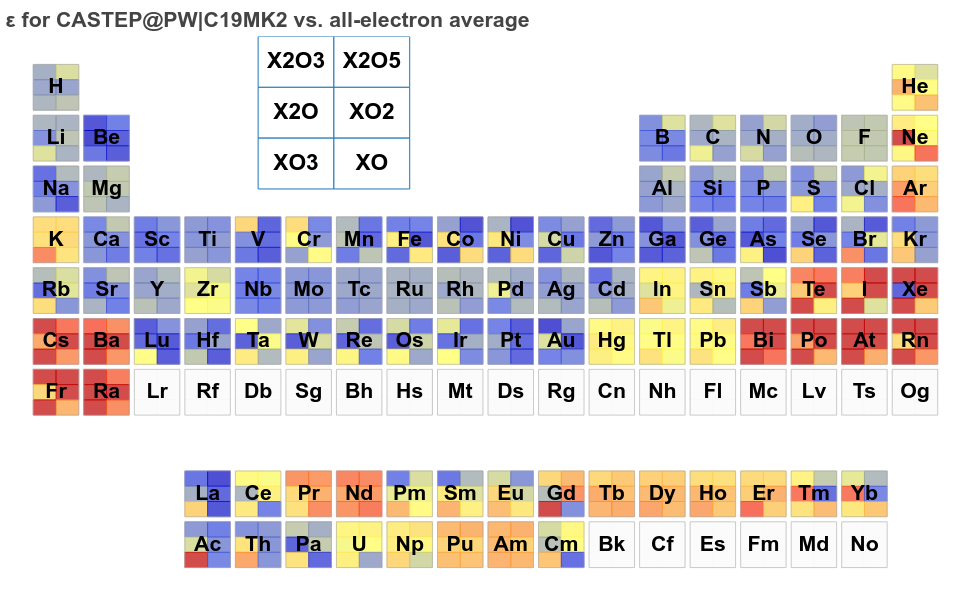}%
	}
	\resizebox{.9\textwidth}{!}{%
        \includegraphics[height=5cm]{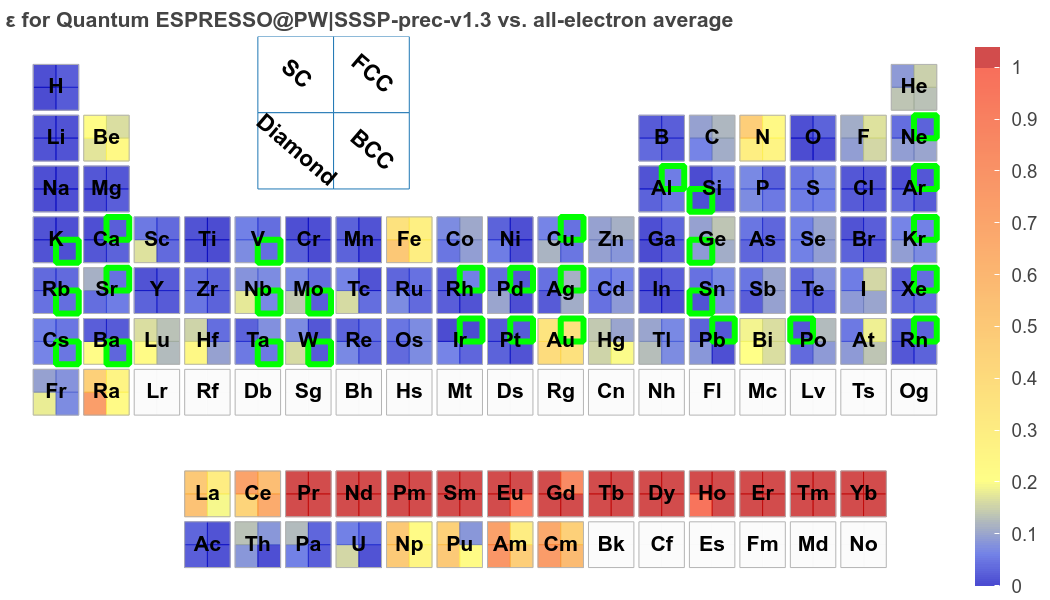}%
        \includegraphics[height=5cm]{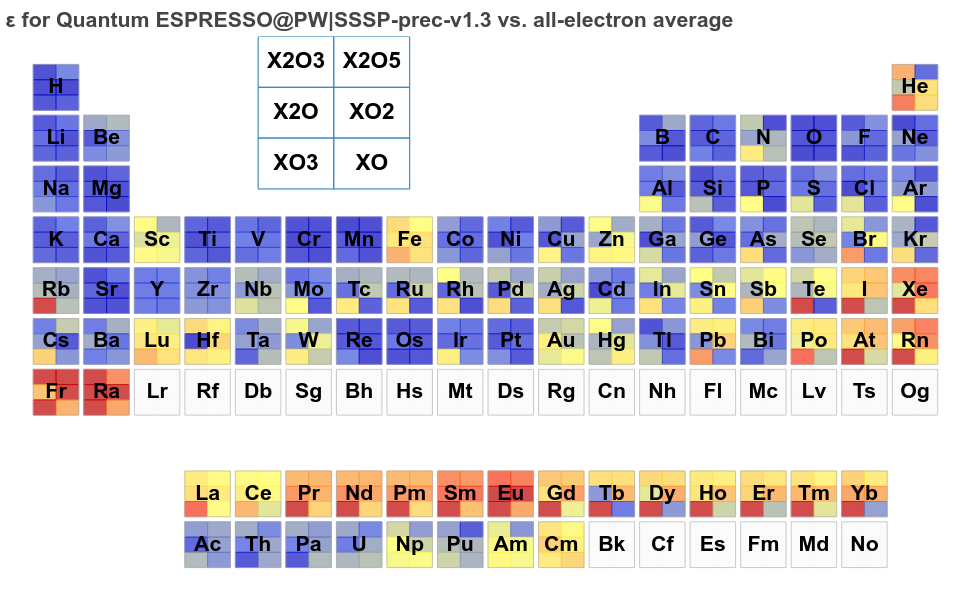}%
    }
\end{center}

 \caption{Same periodic tables for the comparison metric $\varepsilon$ as in SI~\ref{sisec:periodic-tables-per-code} for codes \castep{} and \qe{}, where the 29 crystal structures that appear also in the 71-crystal set of Ref.~\citenum{Lejaeghere:2016} are highlighted. 
    While the 29 structures display values with excellent agreement, other structures result in a less good agreement. These two examples illustrate some of the cases in SI Fig.~\ref{sifig:29-vs-960} for which the agreement is very good on the small crystal set, yet less good on the large crystal set.
 \label{sifig:29-vs-960-periodic}}
\end{figure*}

To perform a quantitative comparison using the same computational approach, we use only data obtained as part of the present study. That means that we do not use the 71 crystal set (that has several crystals that are not part of the present set), but rather the set of 29 crystals common to both sets, as listed in SI Sec.~\ref{sisec:comparing-delta}, as a proxy for the 71 crystal set. SI Fig.~\ref{sifig:29-vs-960} shows on the horizontal axis the $\epsilon$ averaged over these 29 crystals for any pair of approaches or codes used in this work, and on the vertical axis the $\epsilon$ for the same pairs but now averaged over all crystals (up to 960 crystals). The overall correlation in SI Fig.~\ref{sifig:29-vs-960} shows that the 29 crystal set does indeed capture part of the information. However, as expected, the data points fall above the $y=x$ line, demonstrating that the large set probes relevant behavior that is not probed in the small sets previously used. In particular, there are code pairs (see shaded area in SI Fig.~\ref{sifig:29-vs-960}) that have a mutual $\varepsilon \le 0.2$ for the 29 unaries (good agreement), yet have a larger $\varepsilon$ for the large set (and similarly for $\varepsilon \le 0.06$, that signals excellent agreement). SI Fig.~\ref{sifig:29-vs-960-periodic} shows two examples when this happens: panels (a) and (c) demonstrate that while values for $\varepsilon$ are small on the 29 crystals of the small set (boxed cases in SI Fig.~\ref{sifig:29-vs-960-periodic}(a,c)), other crystals outside this set may contribute to a larger average $\varepsilon$. This is typically the case for lanthanides, not included in the earlier 71-crystal benchmark (that stopped at Rn). Moreover, some elements lead to a low $\varepsilon$ for the unaries, but to a larger $\varepsilon$ for the oxides (e.g., Cs-Ba-Fr-Ra and Te-I-Xe-Bi-Po-At-Rn for CASTEP).

The overall conclusion of this analysis is that the 71-crystal set of unaries -- or the 29-crystal set as its proxy -- gives a fair first assessment of the comparative behavior of two DFT codes, while the complete 960-crystal set of unaries and oxides provides a more detailed comparison, both because more elements and more structures per element are included. Conclusions that were based on the benchmarks of Refs.~\citenum{Lejaeghere:2016,deltasite} will therefore still hold, yet can be refined. For testing newly developed pseudopotential libraries, a stepwise approach can be implemented: a first quick test on one unary per element will reveal the largest deviations; once these are fixed, the whole dataset of unaries and oxides can be used to hunt for smaller deviations.

\clearpage

\section{Additional pseudopotential datasets\label{sisec:additional-pseudos}}
This section discusses more in detail some additional datasets obtained using the same codes of the main text, but different basis sets or pseudopotential families, as well as the comparison with earlier versions of pseudopotentials, before those that were optimized here.

\subsection{\abinit{}\label{sisec:abinit-pseudo-improvement}}

During the process of computing the oxide verification equations of state using the initial standard norm-conserving scalar-relativistic PBE PseudoDojo (version 0.4)~\cite{Setten:2018, Hamann:2013}, we observed that the results for around 11 of the pseudopotentials were not in as good agreement with the all-electron equations of state as other elements from the same pseudopotential family.
This led to an investigation into possible improvements to the pseudopotentials for Ba, Bi, I, Pb, Po, Rb, Rn, S, Te, Tl, and Xe.
With the exception of S, we found that the accuracy of the pseudopotentials is significantly improved by including a projector for the $f$ channel.
In the original version, indeed, the local part of the pseudopotential was not able to reproduce the all-electron scattering properties of the $f$ channel in the empty region.

\begin{figure}[h!]
    \centering
    \includegraphics[width=0.95\linewidth]{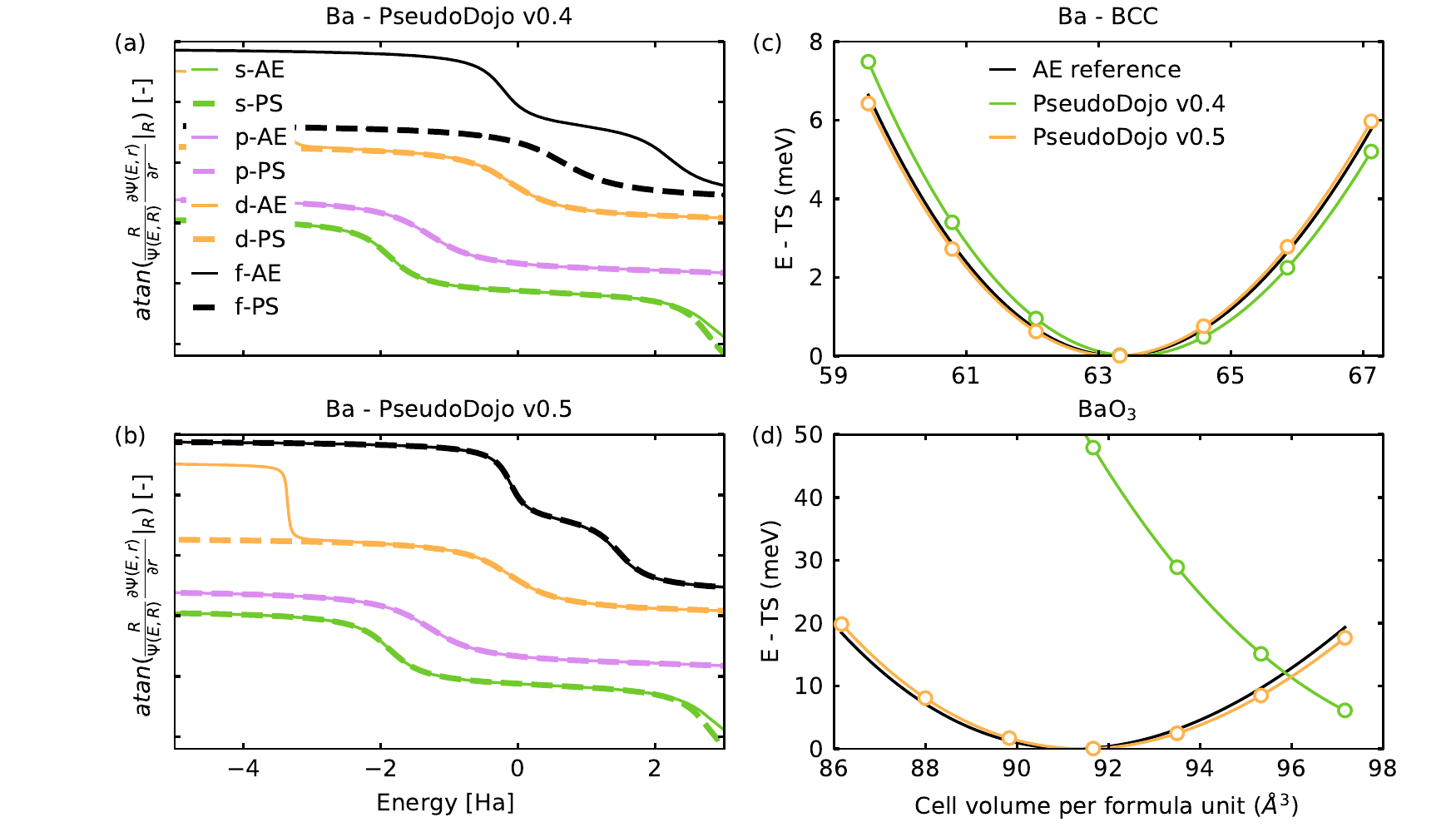}
    \caption{\label{fig:logder-abinit}Comparison of the scattering properties of the all-electron (AE) atom and the pseudized (PS) Hamiltonian for two different Ba pseudopotentials generated without (with) an $f$ projector.
    The subfigures (a) and (b) show the arctangent of the $l$-dependent logarithmic derivative computed for some $R$ greater than the pseudization radius, where $\psi$ is the solution
    of the non-local radial equation regular at the origin (a) without and (b) with $f$ projector.
    Since the transferability of a pseudopotential is directly related to the capability of reproducing the \gls{ae} logarithmic derivative over a wide range of energies,
    the version with $f$ projector is expected to provide more accurate results and we provide two equation of state example in (c) for BCC barium and in (d) for BaO$_3$ where the pseudopotential with the additional projector is more accurate than the one without.
    This is true for all 10 cases tested in this study.
   }    
\end{figure}

A particularly severe case is the one of Ba which is shown in SI Fig.~\ref{fig:logder-abinit}(a) where the pseudized $f$ channel (black dashed line) does not reproduce well the all-electron reference.
As a consequence, the EOS for the stable BCC phase of Ba is slightly off (SI Fig.~\ref{fig:logder-abinit}(c)) but the one for BaO$_3$ is completely wrong, as shown in SI Fig.~\ref{fig:logder-abinit}(d).
This issue is fixed by the additional projector as shown in SI Fig.~\ref{fig:logder-abinit}(b), which then gives excellent agreement with the all-electron EOS.    

It should be noted that the inclusion of the $f$ projector increases the computational cost associated to the application of the non-local part of the Hamiltonian $V_{nl}$.
This is especially true if the computation of $V_{nl}|\psi\rangle$ is obtained by projecting the wavefunction over spherical Harmonics $Y_{lm}$.
We stress, however, that in \abinit{} the projection is implemented by expressing the sum over $m$ in terms of Legendre polynomials (\texttt{useylm = 0} input variable, default option when norm-conserving pseudopotentials are employed).  
In this case, the computational cost of including an additional projector for $l$ is not so high because, contrary to the case when spherical harmonics are used, the number of floating point operations required to apply a projector does not scale with the total number of magnetic quantum numbers $2l+1$.
This study led to the creation and adoption of 11 new pseudopotentials for Ba, Bi, I, Pb, Po, Rb, Rn, S, Te, Tl, and Xe in a new PseudoDojo (version 0.5). Because these pseudopotentials have not been subject to a convergence study with respect to all-electron results, 20 Hartree were added to the ``high'' stringency recommended cutoffs from ONCVPSP for safety.

In addition to norm-conserving pseudopotentials, we also investigated the JTH PAW PBE v1.1~\cite{Jollet:2014} table which has improved versions of the pseudopotentials for  H, Li, Si, Cu, Zn, Ga, Cd, Sb, Lu, Os, Ir and Bi with respect to v1.0 used in the previous AiiDA common workflows study.
Through testing, it was noted that the recommended kinetic energy cutoff values for these PAW pseudopotentials were not sufficient for the desired level of agreement with all-electron codes, so twice the ``high'' recommended cutoffs were used.
The conjugate-gradient (CG) diagonalization algorithm was used for all calculations with PAW pseudopotentials.
Note that there do exist pseudopotentials for lanthanide- and actinide-series elements in this PAW table but they have not been verified against all-electron reference results.
Because of this and because the PAW potentials are included primarily as a point of comparison against the norm-conserving PseudoDojo family (which does not provide lanthanide nor actinide elements), these elements are not included in the reported results and analysis.

\begin{figure}[h!]
    \centering
    \includegraphics[width=0.95\linewidth]{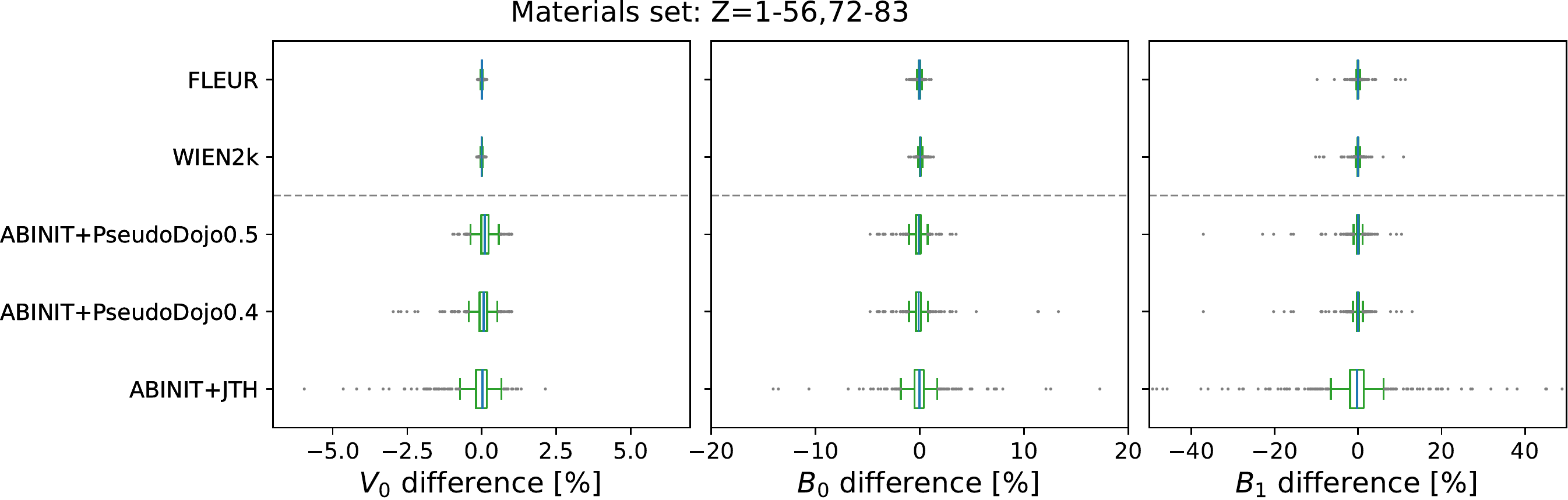}
    \caption{\label{fig:box-abinit}Comparison of the three tested PBE pseudopotential tables using the \abinit{} software.
   }    
\end{figure}

We therefore compare in SI Fig.~\ref{fig:box-abinit} the performance of the three PBE pseudopotential libraries tested here using \abinit{}: the norm-conserving PseudoDojo (version 0.4),  the norm-conserving PseudoDojo (version 0.5) and the PAW JTH v1.1 tables.
We conclude that PseudoDojo (version 0.5) is the most precise one and is therefore used in the main manuscript.

In summary,  unary and oxide verification results were calculated for \abinit{} with three sets of pseudopotentials: JTH PAW PBE v1.1, PseudoDojo norm-conserving standard scalar-relativistic PBE v0.4, and a new PseudoDojo norm-conserving standard scalar-relativistic PBE v0.5 based on v0.4 with improved potentials for the 10 elements listed above. The latter set is reported in the main text of the manuscript.

\subsection{BigDFT} \label{BigDFT-SC}
\begin{figure}[h!]
    \centering
    \includegraphics[width=0.95\linewidth]{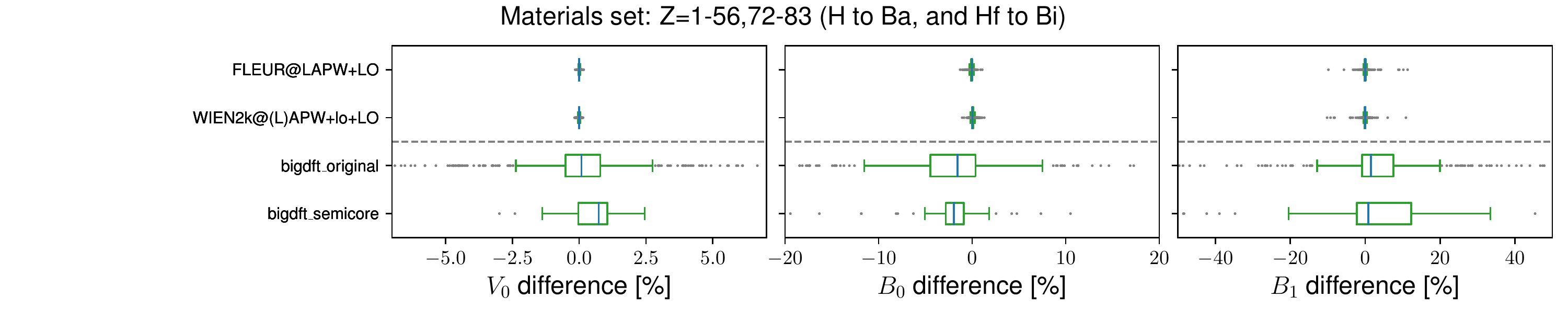}
    \caption{\label{fig:box-bigdft-semicore}Comparative plot displaying the improvement obtained by using HGH-K (Semicore) pseudopotentials over Valence only pseudopotentials.
   }     
\end{figure}

Data production using BigDFT showed clear outliers in comparison to reference codes (\fleur/\wientwok{}) using HGH-K Valence only pseudopotentials. We therefore recalculated the \gls{eos} using semicore pseudopotentials for all the crystals for which this type of pseudopotential was available and that showed a difference in volume $\ge 0.2$ \AA{}$^3$ with respect to the reference \gls{ae} data.
The new calculations resulted, in most cases, in a significant shift towards a closer agreement with the reference \gls{ae}, as shown in SI Fig.~\ref{fig:box-bigdft-semicore}.

\subsection{CASTEP}
\label{sec:castep-pp}

A unique feature of \castep{} is that the pseudopotentials are typically generated on-the-fly  during the calculations, although file-based potentials are still supported.
Each on-the-fly generated (OTFG) potential is defined using a compact configuration string. 
An OTFG library is a collection of predefined configurations.
Many such libraries are built into the \castep{} executable itself, with different focuses.
In this study, the pseudopotentials used are provided by the \texttt{C19} library,
except for the lanthanide and actinide elements.
For these elements, a new set of pseudopotentials are generated to improve agreement with the all-electron reference data, which had not been available during the development of the original \texttt{C19} library. 

The configuration strings for these elements are tabulated in Table \ref{tab:castep-otfg}.
Each field is separated by "|".
The first field is the local angular momentum channel.
The second is the core radius in atomic units and the next three are the recommended cut off energies (Ha) corresponding to the \textit{Coarse}, \textit{Medium}, and \textit{Fine} settings for \castep{}, 
which do not affect a calculation if a cut off energy is specified explicitly (as in this study).
This is followed by the orbitals to be pseudized, which are separated by ":".
Each orbital is defined by a \textit{nl} number and may have suffixes to indicate what kind of projectors should be used.
For example, "60U" indicates that a single ultrasoft projector should be used for the \textit{6s} channel (the default is two ultrasoft projectors).
The "NN" suffix indicates that two norm-conserving projectors should be included.
The "U2U2" suffix indicates two ultrasoft projectors each with a core radius of 2.0.
The "L" suffix pins the local channel, and the "P" suffix indicates that the pseudized channel is not represented by an explicit projector or a local channel.
The parameter "qc" inside the brackets controls the smoothness of the potential.
Occupations of atomic states can be further modified by settings inside a curly bracket.

    \begin{table}[ht]
        \centering
        \scalebox{0.8}{
        \begin{tabular}{c|l|l}
            Element & Old settings & New settings   \\
            La & \texttt{2|2.3|5|6|7|50U:60:51:52(qc=4.5)}  &  \texttt{2|2.3|5|6|7|50U:60:51:52:43\{4f0.1\}(qc=4.5)} \\
            Ce & \texttt{2|2.1|9|10|11|50U:60:51:43:52L(qc=6)}  &  \texttt{2|2.2|8|9|10|50U:60:51:52:43\{5d0.1\}(qc=4.5)} \\
            Pr & \texttt{2|2.1|9|10|11|50U:60:51:43(qc=6)}  & \texttt{2|2.1|10|12|13|50U:60:51:52:43\{5d0.1\}(qc=5)} \\
            Nd & \texttt{2|2.1|9|10|11|50U:60:51:43(qc=6)}  & \texttt{2|2.1|10|12|13|50U:60:51:52:43\{5d0.1\}(qc=5)} \\
            Pm &  \texttt{2|2.1|10|12|13|50U:60:51:43(qc=6)} & \texttt{2|2.1|8|9|11|50U:60:51:52:43\{5d0.1,4f4\}(qc=5.5)} \\
            Sm & \texttt{2|2.1|10|12|13|50U:60:51:43(qc=6)}  & \texttt{2|2.1|9|10|12|50U:60:51:52:43\{5d0.1,4f5\}(qc=5.5)} \\
            Eu & \texttt{2|2.1|10|12|13|50U:60:51:43(qc=6)}  & \texttt{2|2.1|9|10|12|50U:60:51:52:43\{5d0.1,4f6\}(qc=5.5)} \\
            Gd & \texttt{2|2.1|10|12|13|50U:60:51:52L:43(qc=6)}  & \texttt{3|2.1|9|10|12|50U:60:51:52:43(qc=5.5)} \\
            Tb & \texttt{2|2.1|10|12|13|50U:60:51:43(qc=6)}  & \texttt{2|2.2|12|13|15|50U:60:51:52:43\{5d0.1\}(qc=5)} \\
            Dy &  \texttt{2|1.9|12|14|16|50U:60:51:43(qc=6.5)} & \texttt{2|2.0|12|13|15|50U:60:51:52:43\{5d0.1\}(qc=6.5)} \\
            Ho &  \texttt{2|1.9|12|14|16|50U:60:51:43(qc=6.5)} & \texttt{2|2.0|12|13|15|50U:60:51:52:43\{5d0.1\}(qc=6.5)} \\
            Er & \texttt{2|2.1|10|12|13|50U:60:51:43\{6s0.5\}(qc=6)}  & \texttt{2|2.1|10|12|13|50U:60:51:52:43\{6s0.1,5d0.1\}(qc=6)} \\
            Tm & \texttt{2|2.1|10|12|13|50U:60:51:43\{4f12\}(qc=6)}  & \texttt{2|2.1|10|12|13|50U:60:51:52:43\{5d0.1,4f12\}(qc=6)} \\
            Yb & \texttt{2|2.1|10|12|13|50U:60:51:43{4f13}(qc=6)}  & \texttt{2|2.1|10|12|13|50U:60:51:52:43\{5d0.1,4f13\}(qc=6)} \\
            Ac & \texttt{2|2.5|5|6|7|60U:70NN:61:62L} & \texttt{2|2.4|7|7|9|60U:70U2U2:61:62:53\{6d0.1,5f0.1\}(qc=5)} \\
            Th & \texttt{2|2.5|7|7|9|60U:70NN:61:62}  & \texttt{2|2.2|7|7|9|60U:70U2U2:61:62:53\{5f0.1\}(qc=5)} \\
            Pa & \texttt{2|2.1|9|10|11|60U:70:61:53:62P(qc=6)}  & \texttt{2|2.2|8|9|10|60U:70U2U2:61:62:53(qc=5)} \\
            U  & \texttt{2|2.1|10|12|13|60U:70:61:53:62P(qc=6)}  & \texttt{2|2.2|8|9|10|60U:70U2U2:61:62:53(qc=5)} \\
            Np & \texttt{2|2.1|10|12|13|60U:70:61:53:62P(qc=6)}  & \texttt{2|2.2|9|10|12|60U:70U2U2:61:62:53(qc=5)} \\
            Pu & \texttt{2|2.1|10|12|13|60U:70:61:53:62P\{6d1,7s1\}(qc=6)}  & \texttt{2|2.2|9|10|12|60U:70U2U2:61:62:53\{6d0.1\}(qc=5.5)} \\
            Am & \texttt{2|2.1|10|12|13|60U:70:61:53(qc=6)}  & \texttt{2|2.2|9|10|12|60U:70U2U2:61:62:53\{6d0.1\}(qc=5.5)} \\
            Cm & \texttt{2|2.1|10|12|13|60U:70:61:53:62L(qc=6)}  & \texttt{2|2.2|9|10|12|60U:70U2U2:61:62:53(qc=5.5)} \\
        \end{tabular}
        }
        \caption{
Configuration strings of the on-the-fly pseudopotential generation before and after the update for the lanthanide and actinide elements involved in this study.}
        \label{tab:castep-otfg}
    \end{table}

\begin{figure}[ht]
    \centering
    \includegraphics[width=8cm]{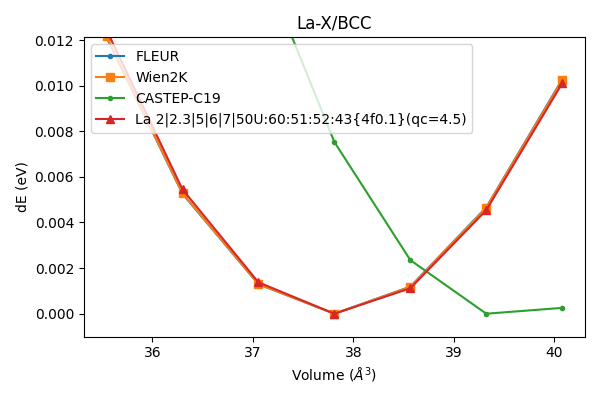}
    \includegraphics[width=8cm]{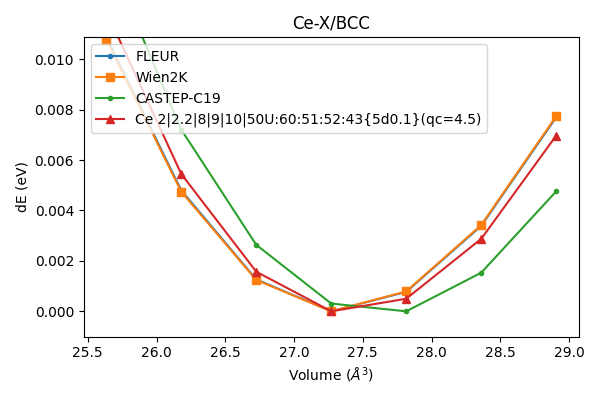}
    \caption{Comparing EOS curves of La and Ce in the BCC configuration. The updated pseudopotentials agree better with the all-electron data compared with those from the \texttt{C19}. }
    \label{fig:castep-c19mk2-ce}
\end{figure}

Examples EOS curves for La and Ce are displayed in SI Fig.~\ref{fig:castep-c19mk2-ce},
showing that the updated pseudopotentials agree better with the all electron data.
The improvements may vary for other elements with $f$ electrons.
One should note that a better fit to the all electron results does not necessarily mean smaller errors compared to the experimental results. 
This is due to the inherent self-interaction errors in semi-local DFT that are present in the description of $f$ electrons. 

For some elements, the improvements are achieved through the inclusion of additional orbital states and l channels/projectors such as La,
where previously the \textit{4f} channel was neglected.
In other cases, the reference atomic calculations include partially occupied atomic states that are otherwise empty in the original configurations.
For example \texttt{\{5d0.1\}} adds 0.1 electrons to the \textit{5d} channel.
In some cases, the updated potentials contain increased core radii and are made softer via a decreased "qc" value as far as possible.
These modifications are applied consistently for elements that neighbors to each other in the periodic table.
We also want to emphasize that the settings used in this study are one-shot updates based on the \texttt{C19} library, rather than the outcomes of iterative optimizations.
With the help of the automated test framework and publicly available all-electron reference data, 
it should be easy to adjust the strings and test them rigorously for further improvements, as required.

\subsection{Quantum ESPRESSO}
SSSP\cite{Prandini:2018} is a library of pseudopotentials that undergoes rigorous verification and ranking procedures. It contains two distinct families of pseudopotentials. The first family (``efficiency'') is composed of relatively soft pseudopotentials that are still sufficiently precise for use in high-throughput calculations. The second family (``precision'') contains pseudopotentials that are extremely precise with respect to all-electron references, even if more computationally expensive.
In our \qe{} calculations, aiming at high precision, we have therefore used the SSSP ``precision'' library.
Version 1.1.2 of SSSP, available before this work, was verified only on the unary configurations presented in Ref.~\citenum{Lejaeghere:2016,deltasite}. With the additional results of this work, we have identified that certain pseudopotentials were not the best selection, and that more precise pseudopotentials are available in libraries that were not included in the previous generation of the SSSP library.

    \begin{table}[ht]
        \centering
        \scalebox{0.9}{
        \begin{tabular}{cll}
            \textbf{Element} & \textbf{SSSP precision v1.1.2} & \textbf{SSSP precision v1.3}   \\ \hline
            Te & \texttt{~6|US~|GBRV|v1|uspp}  &  \texttt{~6|US~|~PSL|v1.0.0-low|ld1} \\ 
            Na & \texttt{~9|NC~|~~PD|v4-std|oncvpsp3}  &  \texttt{~9|PAW|~PSL|v1.0.0-low|ld1} \\
            Cu & \texttt{19|NC~|~~PD|v4-std|oncvpsp3}  &  \texttt{11|PAW|~PSL|v1.0.0-low|ld1} \\
            Cs & \texttt{~9|US~|GBRV|v1|uspp}  &  \texttt{~9|NC~|~~PD|v4-str|oncvpsp3} \\
            Cd & \texttt{12|US~|~PSL|v0.3.1|ld1}  &  \texttt{20|PAW|~PSL|v1.0.0-high|ld1} \\
            Ba & \texttt{10|PAW|~PSL|v1.0.0-high|ld1}  &  \texttt{10|NC~|~~PD|v5|oncvpsp4} \\
            As & \texttt{~5|US~|~PSL|v0.2|ld1}  &  \texttt{15|NC~|~~PD|v4-std|oncvpsp3} \\
            I & \texttt{17|PAW|~PSL|v1.0.0-high|ld1}  &  \texttt{17|NC~|~~PD|v4-std|oncvpsp3} \\
            Hg & \texttt{20|NC~|SG15|v0|oncvpsp3}  &  \texttt{12|US~|GBRV|v1|uspp} \\
            Ne & \texttt{~8|NC~|SG15|v0|oncvpsp3}  &  \texttt{~8|PAW|~PSL|v1.0.0-high|ld1} \\
            Ar & \texttt{~8|NC~|SG15|v0|oncvpsp3}  &  \texttt{~8|PAW|~PSL|v1.0.0-high|ld1} \\
            Kr & \texttt{~8|NC~|SG15|v0|oncvpsp3}  &  \texttt{18|PAW|~PSL|v1.0.0-high|ld1} \\
            Xe & \texttt{18|NC~|SG15|v0|oncvpsp3}  &  \texttt{18|PAW|~PSL|v1.0.0-high|ld1} \\
            Rn & \texttt{18|NC~|SG15|v0|oncvpsp3}  &  \texttt{18|PAW|~PSL|v1.0.0-high|ld1} \\
            Ir & \texttt{15|NC~|GBRV|v1|uspp}  &  \texttt{31|US~|~PSL|v1.0.0-high|ld1} \\
        \end{tabular}
        }
        \caption{
List of pseudopotentials modified between SSSP v1.1.2 and SSSP v1.3. Each element in the table is composed of 5 strings separated by a | symbol, respectively: the number of electrons in the valence ($Z$), the type of pseudopotential (NC: norm-conserving, PAW: projector-augmented wave, US: ultrasoft), the source library (SG15: from Ref.~\citenum{schlipf2015optimization}, PSL: PSlibrary\cite{dal2014pseudopotentials}, GBRV: from Ref.~\citenum{garrity2014pseudopotentials}, PD: PseudoDojo\cite{Setten:2018}), an internal version number that identifies the pseudopotential inside the given source library and the code used to generate them (oncvpsp3, oncvpsp4: version 3 and version 4 of the ONCVPSP code\cite{Hamann:2013}, ld1: the ld1.x code of Quantum ESPRESSO\cite{Giannozzi:2017}, uspp: the UltraSoft PseudoPotential (USPP) generation code\cite{vanderbilt1990soft}).\label{tab:qe-sssp-vs}}
    \end{table}

\begin{figure}[ht]
    \centering
    \includegraphics[width=8cm]{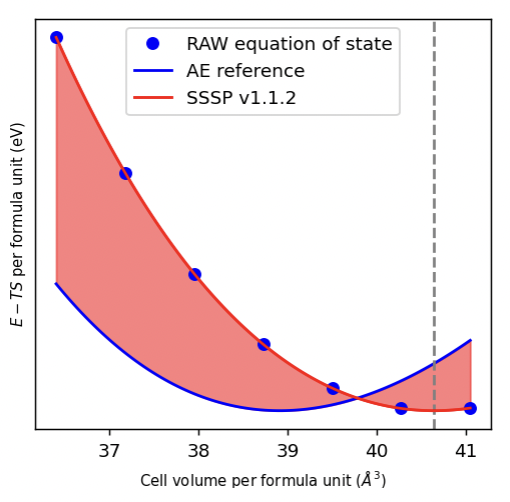}
    \includegraphics[width=8cm]{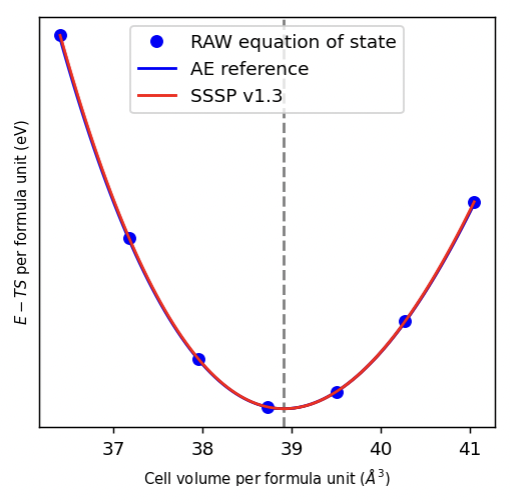}
    \caption{Left: EOS of NeO$_2$ using the Ne pseudopotential from \texttt{SSSP v1.1.2}. Right: EOS of NeO$_2$ using the Ne pseudopotential from \texttt{SSSP v1.3}.}
    \label{fig:neo2-qe-sssp}
\end{figure}

The pseudopotentials that were updated are summarized in Table~\ref{tab:qe-sssp-vs}. In particular, 
the pseudopotentials of the noble gases (Ne, Ar, Kr, Xe, Rn) were previously obtained using the SG15\cite{schlipf2015optimization} library; in SSSP v1.3, we have replaced them with ``PAW-high’’ pseudopotentials from the PSLibrary\cite{dal2014pseudopotentials} based on our more recent verification results as they provide significantly better agreement with the all-electron reference \gls{eos} curves. An example of the improvement in the case of NeO$_2$ is shown in SI Fig.~\ref{fig:neo2-qe-sssp}.

The pseudopotentials of I, Hg were obtained from SG15\cite{schlipf2015optimization} as well. In SSSP v1.3, the pseudopotential of I is replaced with the one from PseudoDojo library and the pseudopotential of Hg is replaced with the on from GBRV\cite{garrity2014pseudopotentials} library. It is not only because these two pseudopotentials from SG15 library are less precise, but also because these two pseudopotentials lead to the electronic step convergence issue during the calculation. Using new pseudopotentials makes all equation of state calculation finished without issues. 

The SSSP v1.1.2 libraries for Ir have been updated with the latest pseudopotential from Pslibrary US v1.0.0 in the ``high’’ family\cite{dal2014pseudopotentials}. It should be noted that the original Ir pseudopotential from the GBRV library contains a ghost state at 10eV.

For Ba, in SSSP v1.3 we select the new pseudopotential from the PseudoDojo v0.5 library\cite{PseudoDojoSite} (generated in the context of this work, see SI Sec.~\ref{sisec:abinit-pseudo-improvement}) that includes an $f$ projector.

\begin{figure}[h!]
    \centering
    \includegraphics[width=8cm]{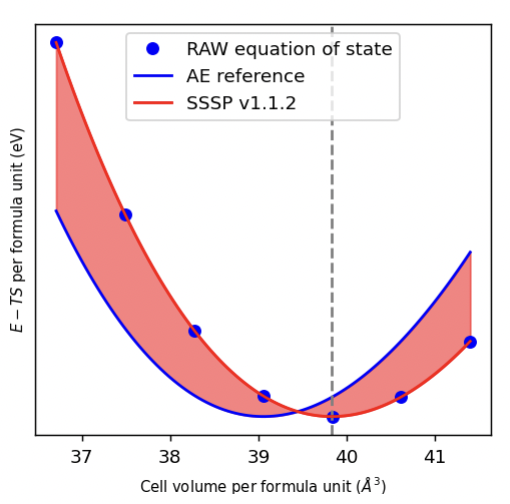}
    \includegraphics[width=8cm]{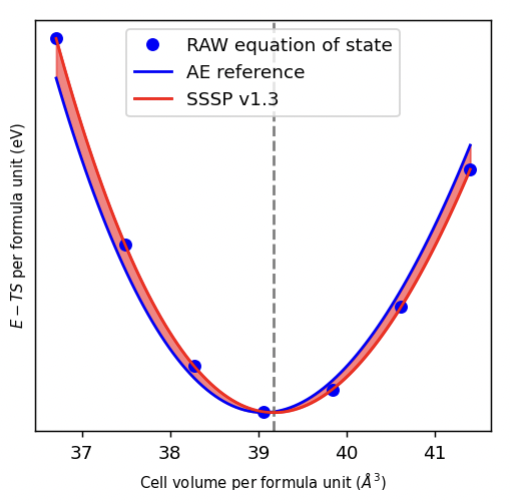}
    \caption{Left: EOS of TeO$_2$ using the Te pseudopotential from \texttt{SSSP v1.1.2}. Right: EOS of TeO$_2$ using the Te pseudopotential from \texttt{SSSP v1.3}.}
    \label{fig:teo2-qe-sssp}
\end{figure}

For Te, Na, and Cu we consider in SSSP v1.3 the PAW or ultrasoft pseudopotentials from the PSlibrary\cite{dal2014pseudopotentials} in the ``low’’ family. These pseudopotentials have fewer semicore states and larger cut-off radii, making them possibly less accurate, but optimized for lower kinetic energy cut-offs. Tests on ten configurations, including oxides, revealed that some pseudopotentials from the ``low’’ family are actually even more accurate than those from the ``high’’ family (the latter have more semicore states and smaller cut-off radii). We note that PSlibrary suggests to use ``high’’ pseudopotentials only for special applications, while ``low’’ ones can yield sufficient precision for regular calculations. As an example of the results with the new pseudos, we show the results for TeO$_2$ for the two versions of SSSP in SI Fig.~\ref{fig:teo2-qe-sssp}.

Starting from v1.3, the SSSP library also includes pseudopotentials for actinides (Th-Lr) developed in Ref.~\citenum{sachs2021dft} as well as the pseudopotentials for Ac, At, Ra, and Fr from the ``high’’ family of PSlibrary\cite{dal2014pseudopotentials}.

\begin{figure}[h!]
    \centering
    \includegraphics[width=16cm]{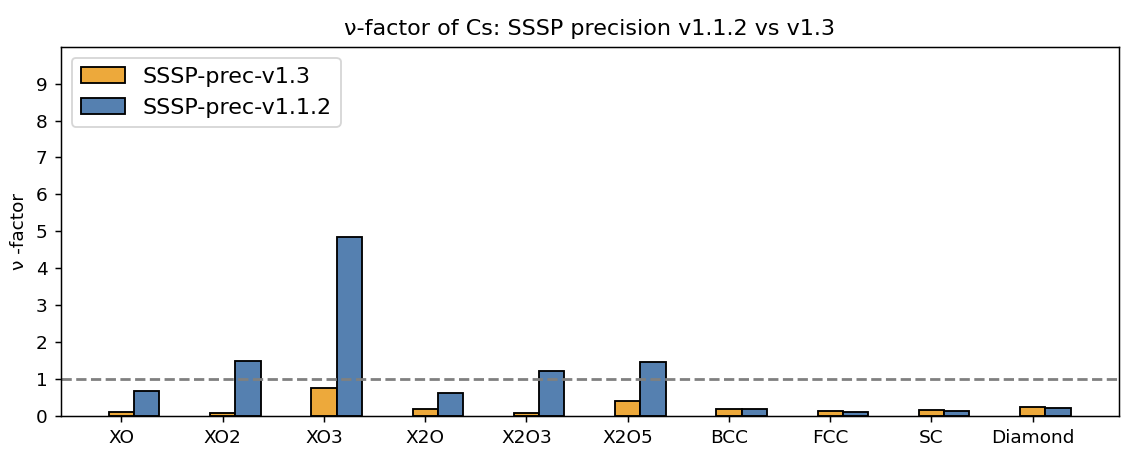}
    \caption{Comparison (via the $\nu$ metric) between \qe{} results and the all-electron reference from \wientwok{} for the different crystal structures of Cs, revealing that while the old pseudopotential was already quite precise for unaries, only the new pseudopotential in SSSP v1.3 (from the PseudoDojo library) provides high precision results for oxides.}
    \label{fig:cs-sssp-update-compare}
\end{figure}

Finally, in the case of Cs, As, while the test of the SSSP v1.1.2 pseudopotentials on unaries was resulting in good-quality \gls{eos} curves, we obtained significant disagreements for oxides. We therefore replaced the corresponding pseudopotentials with others that provided more precise agreements. We illustrate the improvements in the case of Cs in SI Fig.~\ref{fig:cs-sssp-update-compare}.

\subsection{VASP}\label{VASP_SI}
The lanthanides Pr, Nd, Pm, Sm, Eu, Gd, Tb, Dy, Ho, Er, Tm, and Yb were updated during the preparation of the data, since the initial ones showed quite significant deviations from the all-electron results.
For elements containing $4f$ electrons, VASP recommends to use potentials with the $f$ electrons placed in the frozen core, to avoid the well known self-interaction errors resulting from DFT.
VASP therefore provides well-tested potentials with frozen $f$ electrons for the lanthanides, with a valency of 2 or 3 (for Er, Eu, and Yb, both valences are available as separate potentials). Since this study has settled on treating $f$ electrons explicitly as valence, we used the potentials that place all $f$ electrons in the valence. Using semi-local functionals, these potentials lead to significant over-binding and too small unit cells (compared to experiment as well as compared to PAW potentials that place the $f$ electrons in the core). However, the aim of this work is code comparison primarily. The lanthanide potentials used in this work have been generated by G. Kresse. They are using a much smaller core radius of 2.2~a.u.~than the previous versions. Additionally, the reference electronic configuration was altered, by placing 0.5 electrons (instead of 1 electron) from the 4$f$ shell into the 5$d$ shell. Generally, two projectors were used for the $f$ shells. Due to contracting $f$ shells towards the right of the series (Tm, Er, and Yb) a third projector was found to be necessary to obtain an accurate description of the $f$ scattering properties. The new lanthanide potentials will be released on the VASP portal in the 6.4 PBE PAW potential set with an \_h suffix. 
For future reference the specific potential mapping that was used is $\mathrm{RECOMMENDED\_ACWF\_LANTH}$, where $\mathrm{LANTH}$ indicates that the new lanthanide potentials have been used.

\begin{figure}[ht]
  \centering
  \begin{subfigure}{0.49\linewidth}
  \caption{}
  \includegraphics[width=1.0\linewidth]{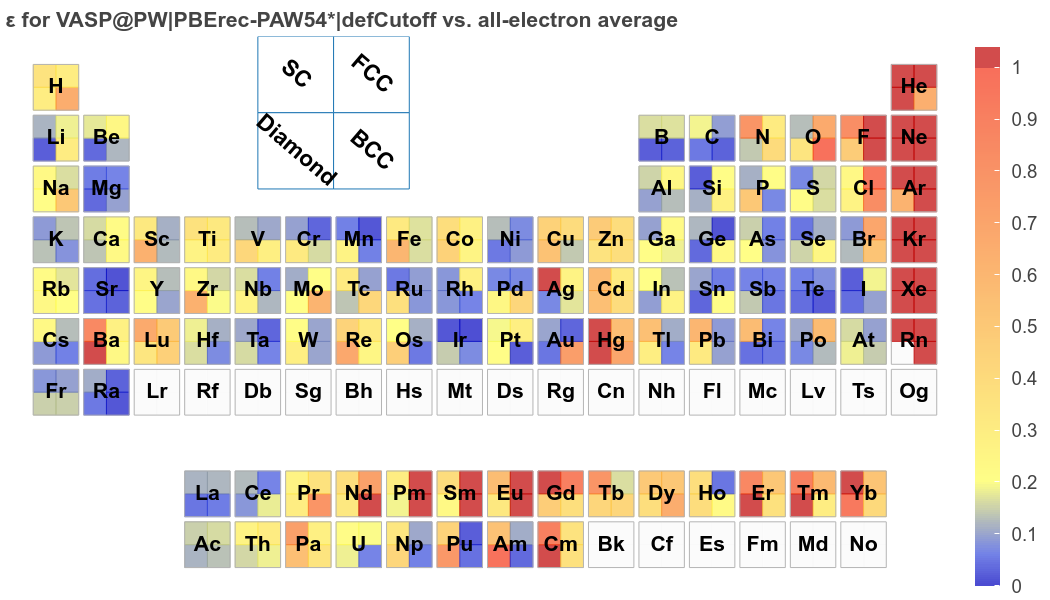}
  \end{subfigure}
  \begin{subfigure}{0.49\linewidth}
    \caption{}
    \includegraphics[width=1.0\linewidth]{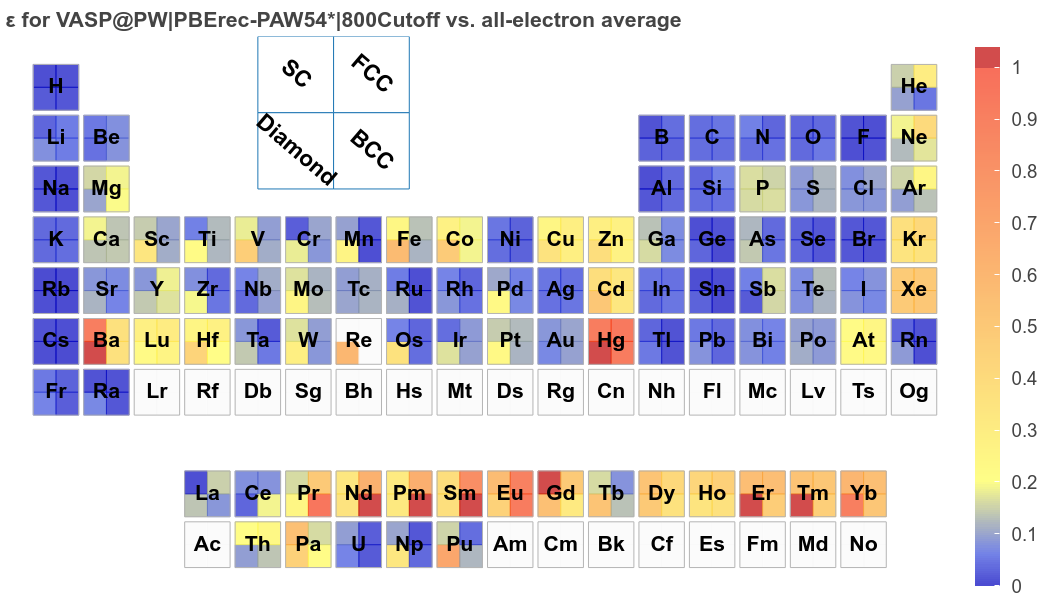}
  \end{subfigure}
    \begin{subfigure}{0.49\linewidth}
    \caption{}
    \includegraphics[width=1.0\linewidth]{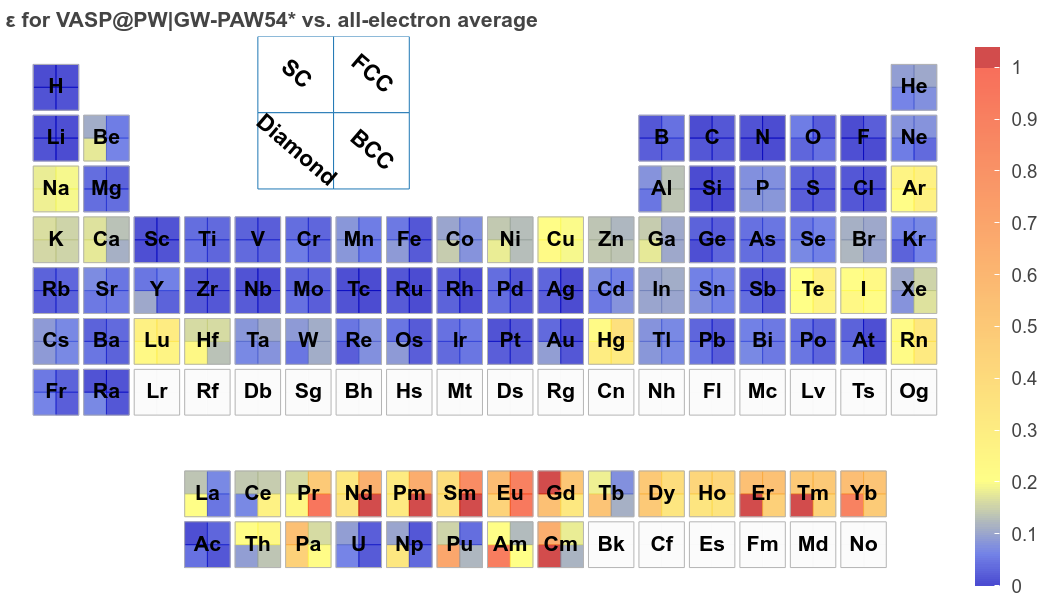}
  \end{subfigure}
  \caption{Value of the comparison metric $\varepsilon$ for the unaries using three different settings for VASP with the averaged \gls{ae} results as reference. The datasets in (a) and (b) use the recommended PBE potentials. Computational parameters are mostly left at default values in (a), while (b) increases the energy cutoff to 800~eV and adopts the other parameters described in section~\ref{sub:code_params_vasp}. In (c) we show the final dataset with recommended GW potentials, if available, and 1000~eV energy cutoff.}
    \label{fig:unaries_parameters_comp_vasp}
\end{figure}

For the unaries we have compared the elevated settings optimized for this study with the default settings of VASP (keeping only $\mathrm{LMAXMIX}=6$) and the chosen recommended GW potential set with the recommended PBE potential set. The resulting $\varepsilon$ metrics are plotted in SI Fig.~\ref{fig:unaries_parameters_comp_vasp}.

In the first panel of SI Fig.~\ref{fig:unaries_parameters_comp_vasp} we used the recommended PAW potentials for PBE (as given on the VASP website, for the lanthanides the new PAW potentials were used as detailed above), with the default cutoffs (dependent on the system) and precision setting ($\mathrm{PREC}=\mathrm{Normal}$). In panel (b) we used the same potentials, but $\mathrm{PREC}=\mathrm{Accurate}$ and a plane wave cutoff of 800~eV alongside the other settings detailed in section~\ref{sub:code_params_vasp}, while panel (c) corresponds to the final results presented in the paper using the final settings (1000 eV) and the recommended GW potentials.

 Clearly, changing the default parameters substantially improves the agreement with the \gls{ae} reference calculations. The noble gases in particular are dramatically improved by using the more accurate settings. But many other weakly bonded structures are improved as well. If then also the GW potentials are used, some elements do improve again by quite a lot (Ba, Hg, Xe, Cd,\ldots), while others improve only slightly (e.g., P, S, As,\dots) and some get slightly worse (e.g., Na, K, Rn,\ldots). For the oxides, the improvement is significant for materials with short oxygen bonds, when using the O\_h\_GW potential instead of the standard PBE one (not shown).

\clearpage

\section{Consistency check with previous benchmarks for the all-electron data\label{sisec:comparing-delta}}
All calculations in this work have been run independently and from scratch by expert users of the respective codes, following a strict protocol. As the all-electron calculations \wientwok{} and \fleur{} serve as references here, it is useful to compare them to results obtained for the same crystals in an earlier benchmark\cite{Lejaeghere:2016,deltasite}. Table~\ref{sitab:comparison-science-paper} shows this comparison for the parameters of the Birch-Murnaghan EOS: both for \fleur{} and \wientwok{}, there are only small relative differences between the parameters obtained in Refs.~\citenum{Lejaeghere:2016,deltasite} and the ones obtained in the present work. The only crystal with significant relative deviations is Ne (FCC), which is not surprising given its very shallow equation of state (small value of the bulk modulus, which is known\cite{volume-error} to lead to a large uncertainty in the volume). This demonstrates good agreement between the previous and current data, obtained from independent calculations. 

\begin{figure}
    \centering
    \includegraphics[width=0.7\linewidth]{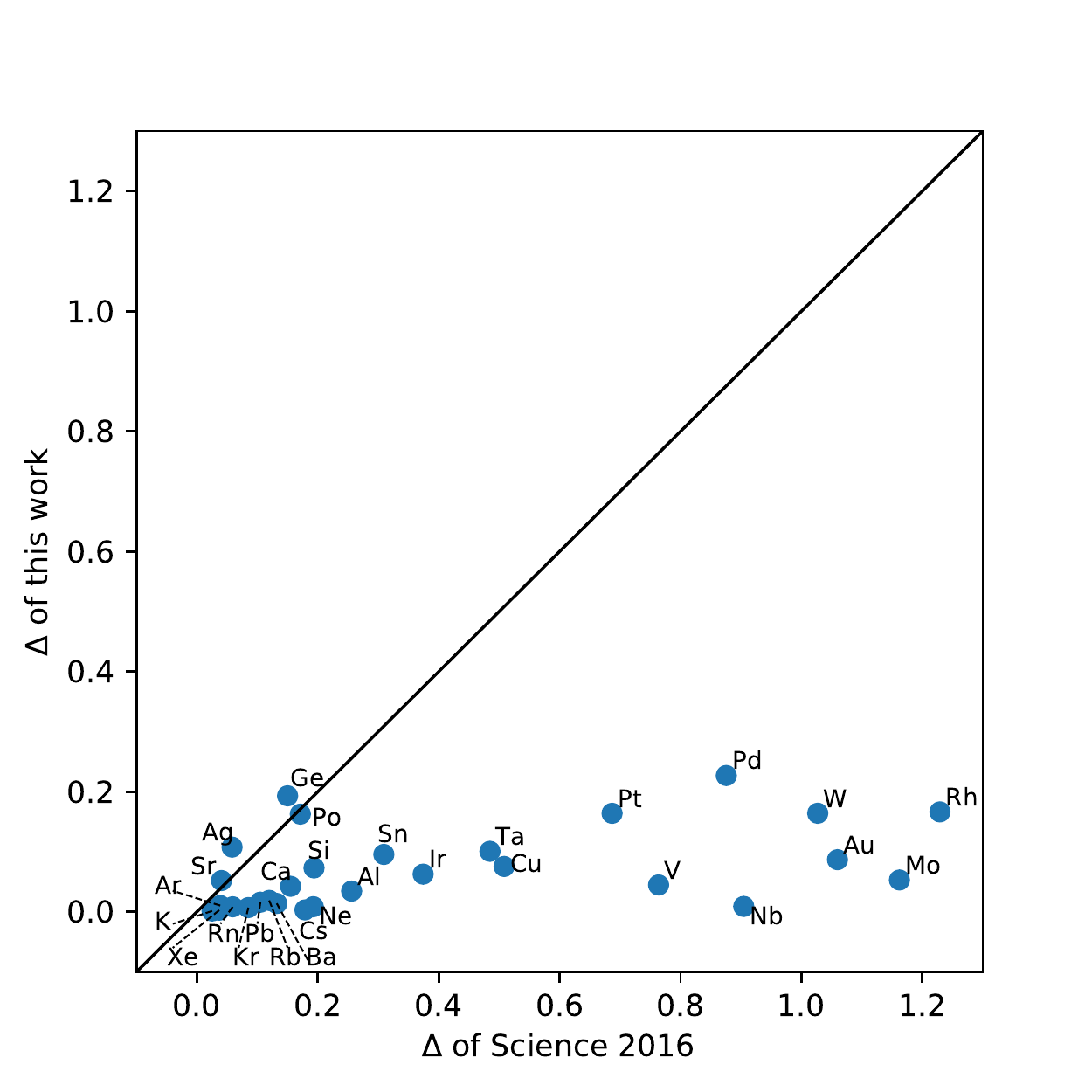}
    \caption{\label{fig:old-new-Delta} Correlation of the $\Delta$ metric on the 29 crystals listed in Tab.~\ref{sitab:comparison-science-paper}. $x$ axis: $\Delta$ metric for these crystals between \fleur{} and \wientwok{}, using the results from Refs.~\citenum{Lejaeghere:2016,deltasite}. $y$ axis: same metric $\Delta$ for the same crystals and the same two codes, but using data from the present work. The black solid line indicates $y=x$. We note that $\Delta \le 1$~meV/atom was considered in Ref.~\citenum{Lejaeghere:2016} to indicate a good agreement.}    
\end{figure}

A different view on the same data is represented in SI Fig.~\ref{fig:old-new-Delta}. It shows on the horizontal axis the difference between the \fleur{} and \wientwok{} results for these 29 crystals, expressed by the $\Delta$ metric, as obtained in Refs.~\citenum{Lejaeghere:2016,deltasite}. On the vertical axis, the $\Delta$ metric for the same crystal and the same two codes is shown, now using the data obtained in the present work. Although the range on the horizontal axis does not extend much beyond the threshold of good agreement of $\Delta= 1$~meV/at that was used in Refs.~\citenum{Lejaeghere:2016,deltasite}, the range on the vertical axis is significantly smaller. 

The conclusion of this analysis is that although the agreement between the all-electron codes \fleur{} and \wientwok{} was already very good in Refs.~\citenum{Lejaeghere:2016,deltasite} ($\Delta \le 1$~meV/at), and although the relative differences in the observable properties that can be derived from the EOS are very small (Tab.~\ref{sitab:comparison-science-paper}), the agreement between the two all-electron codes in the present work is definitely even better than it was in previous works (SI Fig.~\ref{fig:old-new-Delta}).

\begin{table}[h]
    \caption{\label{sitab:comparison-science-paper}Table comparing the $V_0$, $B_0$ and $B_1$ parameters for the subset of 29 structures of those suggested in Ref.~\citenum{Lejaeghere:2016,deltasite} that are also present in our set of unaries, namely those that have a cubic FCC, BCC, SC or Diamond structure and that were treated without spin polarization. In particular, Fe(BCC), Mn(FCC), Ni(FCC) and Cr(BCC) have not been included because they are treated including spin polarization in Ref.~\citenum{Lejaeghere:2016,deltasite}; all other structures not shown here were not considered in a cubic structure in Ref.~\citenum{Lejaeghere:2016,deltasite}.
    The table reports the absolute percentage error on each parameter with respect to the data for the same code and the same crystal reported in Ref.~\citenum{Lejaeghere:2016,deltasite}.}
    \centering\footnotesize
    \begin{tabular}{ll|rrr|rrr}
        && \multicolumn{3}{c|}{FLEUR} & \multicolumn{3}{c}{WIEN2k} \\
        Element & Structure & $V_0$ error [\%] & $B_0$ error [\%] & $B_1$ error [\%] & $V_0$ error [\%] & $B_0$ error [\%] & $B_1$ error [\%] \\ \hline
        Ag & FCC        &   0.01 &  1.59 &   1.98 &   0.06 &  0.86 &   6.97 \\ 
        Al & FCC        &   0.01 &  1.22 &   5.69 &   0.10 &  0.72 &   1.15 \\ 
        Ar & FCC        &   0.55 &  5.61 &  29.04 &   0.28 &  1.28 &   2.71 \\ 
        Au & FCC        &   0.20 &  1.62 &  10.40 &   0.02 &  0.27 &   2.95 \\ 
        Ba & BCC        &   0.16 &  2.18 &   9.86 &   0.26 &  0.05 &  30.67 \\ 
        Ca & FCC        &   0.11 &  0.24 &   5.91 &   0.00 &  1.50 &   0.59 \\ 
        Cs & BCC        &   0.19 &  0.39 &   4.82 &   0.20 &  1.45 &  38.97 \\ 
        Cu & FCC        &   0.16 &  0.06 &   0.71 &   0.05 &  0.10 &   4.14 \\ 
        Ge & Diamond    &   0.09 &  0.34 &   2.68 &   0.02 &  0.31 &   2.60 \\ 
        Ir & FCC        &   0.00 &  0.82 &   0.07 &   0.03 &  0.12 &   1.38 \\ 
        K  & BCC        &   0.15 &  0.47 &   0.47 &   0.14 &  0.14 &  21.68 \\ 
        Kr & FCC        &   0.37 &  4.43 &  38.54 &   0.62 &  5.21 &  35.59 \\ 
        Mo & BCC        &   0.09 &  0.03 &   5.32 &   0.04 &  0.12 &   2.87 \\ 
        Nb & BCC        &   0.12 &  0.59 &  12.32 &   0.03 &  0.97 &   3.71 \\ 
        Ne & FCC        &   2.69 &  0.41 &  41.55 &   0.29 & 13.74 &  99.20 \\ 
        Pb & FCC        &   0.15 &  0.45 &   1.86 &   0.09 &  0.16 &   4.52 \\ 
        Pd & FCC        &   0.03 &  0.77 &   3.34 &   0.08 &  0.35 &   0.65 \\ 
        Po & SC         &   0.03 &  0.60 &   9.91 &   0.04 &  0.13 &   0.39 \\ 
        Pt & FCC        &   0.19 &  1.19 &   6.52 &   0.08 &  0.52 &   0.02 \\ 
        Rb & BCC        &   0.07 &  0.86 &   0.83 &   0.39 &  0.55 &  53.53 \\ 
        Rh & FCC        &   0.07 &  0.63 &   0.91 &   0.07 &  0.34 &   2.65 \\ 
        Rn & FCC        &   0.04 &  7.79 &  14.45 &   0.48 &  4.23 &  34.33 \\ 
        Si & Diamond    &   0.04 &  0.03 &   0.68 &   0.03 &  0.02 &   0.07 \\ 
        Sn & Diamond    &   0.01 &  0.47 &   3.56 &   0.06 &  0.71 &   4.94 \\ 
        Sr & FCC        &   0.71 &  3.27 &  37.98 &   0.68 &  3.43 &   5.89 \\ 
        Ta & BCC        &   0.02 &  1.62 &   8.91 &   0.03 &  1.11 &   1.17 \\ 
        V  & BCC        &   0.07 &  1.20 &   1.87 &   0.07 &  0.16 &   2.31 \\ 
        W  & BCC        &   0.04 &  0.72 &   6.96 &   0.03 &  0.04 &   2.63 \\ 
        Xe & FCC        &   0.11 &  4.06 &   6.18 &   0.36 &  2.00 &  12.28 \\ 
        \end{tabular}
\end{table}

\clearpage 

\section{Discrepancies of formation energies computed from the current dataset\label{sisec:formation-energies}}
\begin{figure}[h!]
    \centering
    \includegraphics[width=0.6\linewidth]{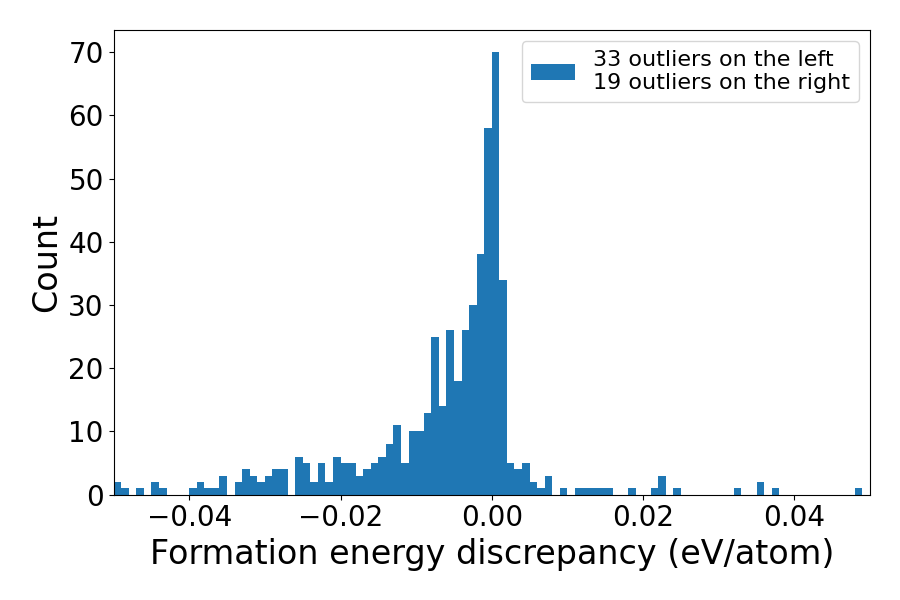}
    \caption{\label{SIfig:formation-energies}Histogram of the discrepancy between the formation energy obtained from our reference dataset for the two codes  \fleur{} and \wientwok{}.
    We note that no correction is applied to the formation energies (as it is typically done, for instance, for oxygen\cite{Wang2021}). However, since we are only considering differences in formation energies between two codes, these corrections cancel out so they do not need to be considered explicitly. The number of outliers outside of the visible $x$ axis range is reported in the top right corner of the figure.
   }     
\end{figure}

In SI Fig.~\ref{SIfig:formation-energies} we report a histogram of the difference of the formation energy obtained using the data for the two all-electron codes \fleur{} and \wientwok{}, computed from the minimum-energy value of the \gls{eos} data curves.
The histogram is obtained considering the formation energy of all X$_{n}$O$_m$ oxides, using the lowest-energy unary of element X and of oxygen as the two endpoints (in the case of oxygen, the lowest-energy non-magnetic unary in our dataset is the simple cubic structure).
The majority of the datapoints are in the visible $x$ axis range, i.e., with an (absolute) discrepancy smaller than 50 meV/atom.
Nevertheless, several outliers are present: 52 out of the 576 materials considered have a discrepancy larger than 50 meV/atom. The most outstanding outliers are AtO$_3$, PoO$_3$, BiO$_3$, Am$_2$O, Pu$_2$O, AmO and PuO. They have a discrepancy larger than 1 eV/atom. For compounds containing Am and Pu, the discrepancy is also due by the fact that the unary identified as having lowest energy is different between the two codes.

As we discuss in the main text, the reason for this discrepancy is that our workflows have been designed to guarantee consistent simulation parameters among calculations for a given material at different volumes. However, when considering different structures, we did not enforce any consistency between simulation parameters, e.g., the choices of atomic radii for the \gls{ae} codes might be different in different systems. Especially, changes to the core/valence separation from structure to structure may lead to larger discrepancies in this comparison, because the differing relativistic descriptions of core and valence electrons lead to different energy contributions.
Therefore, we recommend not to use our dataset to generate plots like the one of SI Fig.~\ref{SIfig:formation-energies} or, more generally, to avoid performing data analysis that considers energy differences between different structures. In this case, instead, one should design new appropriate workflows that can ensure the consistency of simulation parameters among all relevant calculations.

\ifsupplementaryonly

\bibliography{../biblio}

\clearpage

\begin{thebibliography}{100}
    \urlstyle{rm}
    \expandafter\ifx\csname url\endcsname\relax
      \def\url#1{\texttt{#1}}\fi
    \expandafter\ifx\csname urlprefix\endcsname\relax\def\urlprefix{URL }\fi
    \expandafter\ifx\csname doiprefix\endcsname\relax\def\doiprefix{DOI: }\fi
    \providecommand{\bibinfo}[2]{#2}
    \providecommand{\eprint}[2][]{\url{#2}}
    
    \bibitem{Alberi_2019}
    \bibinfo{author}{Alberi, K.} \emph{et~al.}
    \newblock \bibinfo{journal}{\bibinfo{title}{The 2019 materials by design
      roadmap}}.
    \newblock {\emph{\JournalTitle{Journal of Physics D: Applied Physics}}}
      \textbf{\bibinfo{volume}{52}}, \bibinfo{pages}{013001},
      \doiprefix\url{10.1088/1361-6463/aad926} (\bibinfo{year}{2018}).
    
    \bibitem{Marzari2021}
    \bibinfo{author}{Marzari, N.}, \bibinfo{author}{Ferretti, A.} \&
      \bibinfo{author}{Wolverton, C.}
    \newblock \bibinfo{journal}{\bibinfo{title}{Electronic-structure methods for
      materials design}}.
    \newblock {\emph{\JournalTitle{Nature Materials}}}
      \textbf{\bibinfo{volume}{20}}, \bibinfo{pages}{736--749},
      \doiprefix\url{10.1038/s41563-021-01013-3} (\bibinfo{year}{2021}).
    
    \bibitem{Pizzi:2016}
    \bibinfo{author}{Pizzi, G.}, \bibinfo{author}{Cepellotti, A.},
      \bibinfo{author}{Sabatini, R.}, \bibinfo{author}{Marzari, N.} \&
      \bibinfo{author}{Kozinsky, B.}
    \newblock \bibinfo{journal}{\bibinfo{title}{{AiiDA}: automated interactive
      infrastructure and database for computational science}}.
    \newblock {\emph{\JournalTitle{Computational Materials Science}}}
      \textbf{\bibinfo{volume}{111}}, \bibinfo{pages}{218--230},
      \doiprefix\url{10.1016/j.commatsci.2015.09.013} (\bibinfo{year}{2016}).
    
    \bibitem{Huber:2020}
    \bibinfo{author}{Huber, S.~P.} \emph{et~al.}
    \newblock \bibinfo{journal}{\bibinfo{title}{{AiiDA} 1.0, a scalable
      computational infrastructure for automated reproducible workflows and data
      provenance}}.
    \newblock {\emph{\JournalTitle{Scientific Data}}} \textbf{\bibinfo{volume}{7}},
      \doiprefix\url{10.1038/s41597-020-00638-4} (\bibinfo{year}{2020}).
    
    \bibitem{ONG2013314}
    \bibinfo{author}{Ong, S.~P.} \emph{et~al.}
    \newblock \bibinfo{journal}{\bibinfo{title}{Python materials genomics
      (pymatgen): A robust, open-source python library for materials analysis}}.
    \newblock {\emph{\JournalTitle{Computational Materials Science}}}
      \textbf{\bibinfo{volume}{68}}, \bibinfo{pages}{314--319},
      \doiprefix\url{https://doi.org/10.1016/j.commatsci.2012.10.028}
      (\bibinfo{year}{2013}).
    
    \bibitem{CPE:CPE3505}
    \bibinfo{author}{Jain, A.} \emph{et~al.}
    \newblock \bibinfo{journal}{\bibinfo{title}{Fireworks: a dynamic workflow
      system designed for high-throughput applications}}.
    \newblock {\emph{\JournalTitle{Concurrency and Computation: Practice and
      Experience}}} \textbf{\bibinfo{volume}{27}}, \bibinfo{pages}{5037--5059},
      \doiprefix\url{10.1002/cpe.3505} (\bibinfo{year}{2015}).
    
    \bibitem{atomate}
    \bibinfo{author}{Mathew, K.} \emph{et~al.}
    \newblock \bibinfo{journal}{\bibinfo{title}{Atomate: A high-level interface to
      generate, execute, and analyze computational materials science workflows}}.
    \newblock {\emph{\JournalTitle{Computational Materials Science}}}
      \textbf{\bibinfo{volume}{139}}, \bibinfo{pages}{140--152},
      \doiprefix\url{10.1016/j.commatsci.2017.07.030} (\bibinfo{year}{2017}).
    
    \bibitem{ASE1}
    \bibinfo{author}{Bahn, S.~R.} \& \bibinfo{author}{Jacobsen, K.~W.}
    \newblock \bibinfo{journal}{\bibinfo{title}{{An object-oriented scripting
      interface to a legacy electronic structure code}}}.
    \newblock {\emph{\JournalTitle{Computing in Science and Engineering}}}
      \textbf{\bibinfo{volume}{4}}, \bibinfo{pages}{56--66},
      \doiprefix\url{10.1109/5992.998641} (\bibinfo{year}{2002}).
    
    \bibitem{ASE2}
    \bibinfo{author}{{Hjorth Larsen}, A.} \emph{et~al.}
    \newblock \bibinfo{journal}{\bibinfo{title}{{The atomic simulation
      environment—a Python library for working with atoms}}}.
    \newblock {\emph{\JournalTitle{Journal of Physics: Condensed Matter}}}
      \textbf{\bibinfo{volume}{29}}, \bibinfo{pages}{273002},
      \doiprefix\url{10.1088/1361-648X/AA680E} (\bibinfo{year}{2017}).
    
    \bibitem{CURTAROLO2012218}
    \bibinfo{author}{Curtarolo, S.} \emph{et~al.}
    \newblock \bibinfo{journal}{\bibinfo{title}{Aflow: An automatic framework for
      high-throughput materials discovery}}.
    \newblock {\emph{\JournalTitle{Computational Materials Science}}}
      \textbf{\bibinfo{volume}{58}}, \bibinfo{pages}{218--226},
      \doiprefix\url{https://doi.org/10.1016/j.commatsci.2012.02.005}
      (\bibinfo{year}{2012}).
    
    \bibitem{pyiron-paper}
    \bibinfo{author}{Janssen, J.} \emph{et~al.}
    \newblock \bibinfo{journal}{\bibinfo{title}{pyiron: An integrated development
      environment for computational materials science}}.
    \newblock {\emph{\JournalTitle{Computational Materials Science}}}
      \textbf{\bibinfo{volume}{163}}, \bibinfo{pages}{24 -- 36},
      \doiprefix\url{https://doi.org/10.1016/j.commatsci.2018.07.043}
      (\bibinfo{year}{2019}).
    
    \bibitem{HTTK-Armiento2020}
    \bibinfo{author}{Armiento, R.}
    \newblock \emph{\bibinfo{title}{Database-Driven High-Throughput Calculations
      and Machine Learning Models for Materials Design}}, \bibinfo{pages}{377--395}
      (\bibinfo{publisher}{Springer International Publishing},
      \bibinfo{address}{Cham}, \bibinfo{year}{2020}).
    
    \bibitem{Gonze:2020}
    \bibinfo{author}{Gonze, X.} \emph{et~al.}
    \newblock \bibinfo{journal}{\bibinfo{title}{{The Abinit project: Impact,
      environment and recent developments}}}.
    \newblock {\emph{\JournalTitle{Computer Physics Communications}}}
      \textbf{\bibinfo{volume}{248}}, \bibinfo{pages}{107042},
      \doiprefix\url{10.1016/j.cpc.2019.107042} (\bibinfo{year}{2020}).
    
    \bibitem{mat-project}
    \bibinfo{author}{Jain, A.} \emph{et~al.}
    \newblock \bibinfo{journal}{\bibinfo{title}{{Commentary: The Materials Project:
      A materials genome approach to accelerating materials innovation}}}.
    \newblock {\emph{\JournalTitle{APL Materials}}} \textbf{\bibinfo{volume}{1}},
      \bibinfo{pages}{011002}, \doiprefix\url{10.1063/1.4812323}
      (\bibinfo{year}{2013}).
    
    \bibitem{CMR}
    \bibinfo{author}{Landis, D.~D.} \emph{et~al.}
    \newblock \bibinfo{journal}{\bibinfo{title}{{The Computational Materials
      Repository}}}.
    \newblock {\emph{\JournalTitle{Computing in Science \& Engineering}}}
      \textbf{\bibinfo{volume}{14}}, \bibinfo{pages}{51--57},
      \doiprefix\url{10.1109/MCSE.2012.16} (\bibinfo{year}{2012}).
    
    \bibitem{OQMD-Kirklin2015}
    \bibinfo{author}{Kirklin, S.} \emph{et~al.}
    \newblock \bibinfo{journal}{\bibinfo{title}{The open quantum materials database
      (oqmd): assessing the accuracy of dft formation energies}}.
    \newblock {\emph{\JournalTitle{npj Computational Materials}}}
      \textbf{\bibinfo{volume}{1}}, \bibinfo{pages}{15010},
      \doiprefix\url{10.1038/npjcompumats.2015.10} (\bibinfo{year}{2015}).
    
    \bibitem{TCOD-Merkys2017}
    \bibinfo{author}{Merkys, A.} \emph{et~al.}
    \newblock \bibinfo{journal}{\bibinfo{title}{A posteriori metadata from
      automated provenance tracking: Integration of {AiiDA} and {TCOD}}}.
    \newblock {\emph{\JournalTitle{Journal of Cheminformatics}}}
      \textbf{\bibinfo{volume}{9}}, \bibinfo{pages}{56},
      \doiprefix\url{10.1186/s13321-017-0242-y} (\bibinfo{year}{2017}).
    \newblock \eprint{1706.08704v3}.
    
    \bibitem{NREL-MatDB}
    \bibinfo{author}{Stevanovi\ifmmode~\acute{c}\else \'{c}\fi{}, V.},
      \bibinfo{author}{Lany, S.}, \bibinfo{author}{Zhang, X.} \&
      \bibinfo{author}{Zunger, A.}
    \newblock \bibinfo{journal}{\bibinfo{title}{Correcting density functional
      theory for accurate predictions of compound enthalpies of formation: Fitted
      elemental-phase reference energies}}.
    \newblock {\emph{\JournalTitle{Phys. Rev. B}}} \textbf{\bibinfo{volume}{85}},
      \bibinfo{pages}{115104}, \doiprefix\url{10.1103/PhysRevB.85.115104}
      (\bibinfo{year}{2012}).
    
    \bibitem{Talirz:2020}
    \bibinfo{author}{Talirz, L.} \emph{et~al.}
    \newblock \bibinfo{journal}{\bibinfo{title}{Materials cloud, a platform for
      open computational science}}.
    \newblock {\emph{\JournalTitle{Scientific Data}}} \textbf{\bibinfo{volume}{7}},
      \doiprefix\url{10.1038/s41597-020-00637-5} (\bibinfo{year}{2020}).
    
    \bibitem{NOMAD-Draxl2018}
    \bibinfo{author}{Draxl, C.} \& \bibinfo{author}{Scheffler, M.}
    \newblock \bibinfo{journal}{\bibinfo{title}{Nomad: The fair concept for big
      data-driven materials science}}.
    \newblock {\emph{\JournalTitle{MRS Bulletin}}} \textbf{\bibinfo{volume}{43}},
      \bibinfo{pages}{676--682}, \doiprefix\url{10.1557/mrs.2018.208}
      (\bibinfo{year}{2018}).
    
    \bibitem{CURTAROLO2012227}
    \bibinfo{author}{Curtarolo, S.} \emph{et~al.}
    \newblock \bibinfo{journal}{\bibinfo{title}{Aflowlib.org: A distributed
      materials properties repository from high-throughput ab initio
      calculations}}.
    \newblock {\emph{\JournalTitle{Computational Materials Science}}}
      \textbf{\bibinfo{volume}{58}}, \bibinfo{pages}{227--235},
      \doiprefix\url{https://doi.org/10.1016/j.commatsci.2012.02.002}
      (\bibinfo{year}{2012}).
    
    \bibitem{Wilkinson:2016}
    \bibinfo{author}{Wilkinson, M.~D.} \emph{et~al.}
    \newblock \bibinfo{journal}{\bibinfo{title}{The {FAIR} guiding principles for
      scientific data management and stewardship}}.
    \newblock {\emph{\JournalTitle{Scientific Data}}} \textbf{\bibinfo{volume}{3}},
      \doiprefix\url{10.1038/sdata.2016.18} (\bibinfo{year}{2016}).
    
    \bibitem{Andersen_2021}
    \bibinfo{author}{Andersen, C.~W.} \emph{et~al.}
    \newblock \bibinfo{journal}{\bibinfo{title}{{OPTIMADE}, an {API} for exchanging
      materials data}}.
    \newblock {\emph{\JournalTitle{Scientific Data}}} \textbf{\bibinfo{volume}{8}},
      \doiprefix\url{10.1038/s41597-021-00974-z} (\bibinfo{year}{2021}).
    
    \bibitem{ieee-8055462}
    \bibinfo{journal}{\bibinfo{title}{{IEEE Standard for System, Software, and
      Hardware Verification and Validation}}}.
    \newblock {\emph{\JournalTitle{IEEE Std 1012-2016 (Revision of IEEE Std
      1012-2012/ Incorporates IEEE Std 1012-2016/Cor1-2017)}}}
      \bibinfo{pages}{1--260}, \doiprefix\url{10.1109/IEEESTD.2017.8055462}
      (\bibinfo{year}{2017}).
    
    \bibitem{Wang2021}
    \bibinfo{author}{Wang, A.} \emph{et~al.}
    \newblock \bibinfo{journal}{\bibinfo{title}{A framework for quantifying
      uncertainty in dft energy corrections.}}
    \newblock {\emph{\JournalTitle{Sci. Reports}}} \textbf{\bibinfo{volume}{11}},
      \bibinfo{pages}{15496}, \doiprefix\url{10.1038/s41598-021-94550-5}
      (\bibinfo{year}{2021}).
    
    \bibitem{Carbogno2022}
    \bibinfo{author}{Carbogno, C.} \emph{et~al.}
    \newblock \bibinfo{journal}{\bibinfo{title}{Numerical quality control for
      dft-based materials databases}}.
    \newblock {\emph{\JournalTitle{npj Computational Materials}}}
      \textbf{\bibinfo{volume}{8}}, \bibinfo{pages}{69},
      \doiprefix\url{10.1038/s41524-022-00744-4} (\bibinfo{year}{2022}).
    
    \bibitem{PONCE2014341}
    \bibinfo{author}{Poncé, S.} \emph{et~al.}
    \newblock \bibinfo{journal}{\bibinfo{title}{Verification of first-principles
      codes: Comparison of total energies, phonon frequencies, electron–phonon
      coupling and zero-point motion correction to the gap between abinit and
      qe/yambo}}.
    \newblock {\emph{\JournalTitle{Computational Materials Science}}}
      \textbf{\bibinfo{volume}{83}}, \bibinfo{pages}{341--348},
      \doiprefix\url{https://doi.org/10.1016/j.commatsci.2013.11.031}
      (\bibinfo{year}{2014}).
    
    \bibitem{popleNobel}
    \bibinfo{author}{Pople, J.}
    \newblock \bibinfo{title}{{Nobel Lecture}}.
    \newblock \bibinfo{howpublished}{NobelPrize.org,
      \url{https://www.nobelprize.org/prizes/chemistry/1998/pople/lecture/}.}
    
    \bibitem{Lejaeghere:2016}
    \bibinfo{author}{Lejaeghere, K.} \emph{et~al.}
    \newblock \bibinfo{journal}{\bibinfo{title}{Reproducibility in density
      functional theory calculations of solids}}.
    \newblock {\emph{\JournalTitle{Science}}} \textbf{\bibinfo{volume}{351}},
      \bibinfo{pages}{aad3000}, \doiprefix\url{10.1126/science.aad3000}
      (\bibinfo{year}{2016}).
    
    \bibitem{deltasite}
    \bibinfo{title}{71-crystal benchmark results from
      ref.~\citenum{Lejaeghere:2016} plus data collected later on,
      https://molmod.ugent.be/deltacodesdft}.
    
    \bibitem{fleurCode}
    \bibinfo{title}{{The FLEUR project}}.
    \newblock \bibinfo{howpublished}{\url{https://www.flapw.de/}}.
    
    \bibitem{fleurSource}
    \bibinfo{author}{Wortmann, D.} \emph{et~al.}
    \newblock \bibinfo{title}{Fleur}.
    \newblock \bibinfo{howpublished}{Zenodo},
      \doiprefix\url{10.5281/zenodo.7576163} (\bibinfo{year}{2023}).
    
    \bibitem{WIEN2k}
    \bibinfo{author}{Blaha, P.} \emph{et~al.}
    \newblock \bibinfo{title}{Wien2k: An augmented plane wave plus local orbitals
      program for calculating crystal properties}.
    \newblock \bibinfo{howpublished}{\url{http://www.wien2k.at/}}.
    
    \bibitem{WIEN20}
    \bibinfo{author}{Blaha, P.} \emph{et~al.}
    \newblock \bibinfo{journal}{\bibinfo{title}{Wien2k: An apw+lo program for
      calculating the properties of solids}}.
    \newblock {\emph{\JournalTitle{J. Chem. Phys.}}}
      \textbf{\bibinfo{volume}{152}}, \bibinfo{pages}{074101},
      \doiprefix\url{10.1063/1.5143061} (\bibinfo{year}{2020}).
    
    \bibitem{Uhrin:2021}
    \bibinfo{author}{Uhrin, M.}, \bibinfo{author}{Huber, S.~P.},
      \bibinfo{author}{Yu, J.}, \bibinfo{author}{Marzari, N.} \&
      \bibinfo{author}{Pizzi, G.}
    \newblock \bibinfo{journal}{\bibinfo{title}{Workflows in {AiiDA}: Engineering a
      high-throughput, event-based engine for robust and modular computational
      workflows}}.
    \newblock {\emph{\JournalTitle{Computational Materials Science}}}
      \textbf{\bibinfo{volume}{187}}, \bibinfo{pages}{110086},
      \doiprefix\url{10.1016/j.commatsci.2020.110086} (\bibinfo{year}{2021}).
    
    \bibitem{Huber2021}
    \bibinfo{author}{Huber, S.~P.} \emph{et~al.}
    \newblock \bibinfo{journal}{\bibinfo{title}{Common workflows for computing
      material properties using different quantum engines}}.
    \newblock {\emph{\JournalTitle{npj Computational Materials}}}
      \textbf{\bibinfo{volume}{7}}, \bibinfo{pages}{136},
      \doiprefix\url{10.1038/s41524-021-00594-6} (\bibinfo{year}{2021}).
    
    \bibitem{acwf}
    \bibinfo{author}{Huber, S.~P.} \emph{et~al.}
    \newblock \bibinfo{title}{{AiiDA common workflows (ACWF) package}}
      (\bibinfo{year}{2021}).
    \newblock
      \bibinfo{note}{\url{https://github.com/aiidateam/aiida-common-workflows}}.
    
    \bibitem{Lejaeghere:2014}
    \bibinfo{author}{Lejaeghere, K.}, \bibinfo{author}{Speybroeck, V.~V.},
      \bibinfo{author}{Oost, G.~V.} \& \bibinfo{author}{Cottenier, S.}
    \newblock \bibinfo{journal}{\bibinfo{title}{{Error Estimates for Solid-State
      Density-Functional Theory Predictions: An Overview by Means of the
      Ground-State Elemental Crystals}}}.
    \newblock {\emph{\JournalTitle{Critical Reviews in Solid State and Materials
      Sciences}}} \textbf{\bibinfo{volume}{39}}, \bibinfo{pages}{1--24},
      \doiprefix\url{10.1080/10408436.2013.772503} (\bibinfo{year}{2013}).
    
    \bibitem{GPP}
    \bibinfo{author}{Grosso, G.} \& \bibinfo{author}{Pastori~Parravicini, G.}
    \newblock \emph{\bibinfo{title}{Solid State Physics}}
      (\bibinfo{publisher}{Academic Press}, \bibinfo{address}{Amsterdam},
      \bibinfo{year}{2013}), \bibinfo{edition}{second} edn.
    
    \bibitem{RMARTIN1}
    \bibinfo{author}{Martin, R.~M.}
    \newblock \emph{\bibinfo{title}{Electronic Structure: Basic Theory and
      Practical Methods}} (\bibinfo{publisher}{Cambridge University Press},
      \bibinfo{address}{Cambridge}, \bibinfo{year}{2013}),
      \bibinfo{edition}{second} edn.
    
    \bibitem{COHENLOUIE}
    \bibinfo{author}{Cohen, M.~L.} \& \bibinfo{author}{Louie, S.~G.}
    \newblock \emph{\bibinfo{title}{Fundamentals of Condensed Matter Physics}}
      (\bibinfo{publisher}{Cambridge University Press},
      \bibinfo{address}{Cambridge}, \bibinfo{year}{2016}).
    
    \bibitem{Gonze:2016}
    \bibinfo{author}{Gonze, X.} \emph{et~al.}
    \newblock \bibinfo{journal}{\bibinfo{title}{Recent developments in the {ABINIT}
      software package}}.
    \newblock {\emph{\JournalTitle{Computer Physics Communications}}}
      \textbf{\bibinfo{volume}{205}}, \bibinfo{pages}{106--131},
      \doiprefix\url{10.1016/j.cpc.2016.04.003} (\bibinfo{year}{2016}).
    
    \bibitem{Romero:2020}
    \bibinfo{author}{Romero, A.~H.} \emph{et~al.}
    \newblock \bibinfo{journal}{\bibinfo{title}{{ABINIT}: Overview and focus on
      selected capabilities}}.
    \newblock {\emph{\JournalTitle{The Journal of Chemical Physics}}}
      \textbf{\bibinfo{volume}{152}}, \bibinfo{pages}{124102},
      \doiprefix\url{10.1063/1.5144261} (\bibinfo{year}{2020}).
    
    \bibitem{Ratcliff2020}
    \bibinfo{author}{Ratcliff, L.~E.} \emph{et~al.}
    \newblock \bibinfo{journal}{\bibinfo{title}{Flexibilities of wavelets as a
      computational basis set for large-scale electronic structure calculations}}.
    \newblock {\emph{\JournalTitle{The Journal of Chemical Physics}}}
      \textbf{\bibinfo{volume}{152}}, \bibinfo{pages}{194110},
      \doiprefix\url{10.1063/5.0004792} (\bibinfo{year}{2020}).
    
    \bibitem{Clark:2005}
    \bibinfo{author}{Clark, S.~J.} \emph{et~al.}
    \newblock \bibinfo{journal}{\bibinfo{title}{First principles methods using
      {CASTEP}}}.
    \newblock {\emph{\JournalTitle{Zeitschrift f\"ur Kristallographie - Crystalline
      Materials}}} \textbf{\bibinfo{volume}{220}},
      \doiprefix\url{10.1524/zkri.220.5.567.65075} (\bibinfo{year}{2005}).
    
    \bibitem{cp2k}
    \bibinfo{title}{{The CP2K simulation package}}.
    \newblock \bibinfo{howpublished}{\url{https://www.cp2k.org}}.
    
    \bibitem{Kuehne:2020}
    \bibinfo{author}{Kühne, T.~D.} \emph{et~al.}
    \newblock \bibinfo{journal}{\bibinfo{title}{{CP}2k: An electronic structure and
      molecular dynamics software package - quickstep: Efficient and accurate
      electronic structure calculations}}.
    \newblock {\emph{\JournalTitle{The Journal of Chemical Physics}}}
      \textbf{\bibinfo{volume}{152}}, \bibinfo{pages}{194103},
      \doiprefix\url{10.1063/5.0007045} (\bibinfo{year}{2020}).
    
    \bibitem{GPAW1}
    \bibinfo{author}{Mortensen, J.~J.}, \bibinfo{author}{Hansen, L.~B.} \&
      \bibinfo{author}{Jacobsen, K.~W.}
    \newblock \bibinfo{journal}{\bibinfo{title}{{Real-space grid implementation of
      the projector augmented wave method}}}.
    \newblock {\emph{\JournalTitle{Physical Review B}}}
      \textbf{\bibinfo{volume}{71}}, \bibinfo{pages}{035109},
      \doiprefix\url{10.1103/PhysRevB.71.035109} (\bibinfo{year}{2005}).
    
    \bibitem{GPAW2}
    \bibinfo{author}{Enkovaara, J.} \emph{et~al.}
    \newblock \bibinfo{journal}{\bibinfo{title}{{Electronic structure calculations
      with GPAW: a real-space implementation of the projector augmented-wave
      method}}}.
    \newblock {\emph{\JournalTitle{Journal of Physics: Condensed Matter}}}
      \textbf{\bibinfo{volume}{22}}, \bibinfo{pages}{253202}
      (\bibinfo{year}{2010}).
    
    \bibitem{Giannozzi:2009}
    \bibinfo{author}{Giannozzi, P.} \emph{et~al.}
    \newblock \bibinfo{journal}{\bibinfo{title}{{QUANTUM} {ESPRESSO}: a modular and
      open-source software project for quantum simulations of materials}}.
    \newblock {\emph{\JournalTitle{Journal of Physics: Condensed Matter}}}
      \textbf{\bibinfo{volume}{21}}, \bibinfo{pages}{395502},
      \doiprefix\url{10.1088/0953-8984/21/39/395502} (\bibinfo{year}{2009}).
    
    \bibitem{Giannozzi:2017}
    \bibinfo{author}{Giannozzi, P.} \emph{et~al.}
    \newblock \bibinfo{journal}{\bibinfo{title}{Advanced capabilities for materials
      modelling with quantum {ESPRESSO}}}.
    \newblock {\emph{\JournalTitle{Journal of Physics: Condensed Matter}}}
      \textbf{\bibinfo{volume}{29}}, \bibinfo{pages}{465901},
      \doiprefix\url{10.1088/1361-648x/aa8f79} (\bibinfo{year}{2017}).
    
    \bibitem{Soler:2002}
    \bibinfo{author}{Soler, J.~M.} \emph{et~al.}
    \newblock \bibinfo{journal}{\bibinfo{title}{{The {SIESTA} method for ab initio
      order-N materials simulation}}}.
    \newblock {\emph{\JournalTitle{Journal of Physics: Condensed Matter}}}
      \textbf{\bibinfo{volume}{14}}, \bibinfo{pages}{2745--2779},
      \doiprefix\url{10.1088/0953-8984/14/11/302} (\bibinfo{year}{2002}).
    
    \bibitem{Garcia:2020}
    \bibinfo{author}{Garc{\'{\i}}a, A.} \emph{et~al.}
    \newblock \bibinfo{journal}{\bibinfo{title}{Siesta: Recent developments and
      applications}}.
    \newblock {\emph{\JournalTitle{The Journal of Chemical Physics}}}
      \textbf{\bibinfo{volume}{152}}, \bibinfo{pages}{204108},
      \doiprefix\url{10.1063/5.0005077} (\bibinfo{year}{2020}).
    
    \bibitem{sirius}
    \bibinfo{title}{{The SIRIUS domain-specific library for electronic-structure
      calculations}}.
    \newblock
      \bibinfo{howpublished}{\url{https://github.com/electronic-structure/SIRIUS}}.
    
    \bibitem{Kresse:1996}
    \bibinfo{author}{Kresse, G.} \& \bibinfo{author}{Furthmüller, J.}
    \newblock \bibinfo{journal}{\bibinfo{title}{Efficient iterative schemes for ab
      initio total-energy calculations using a plane-wave basis set}}.
    \newblock {\emph{\JournalTitle{Physical Review B}}}
      \textbf{\bibinfo{volume}{54}}, \bibinfo{pages}{11169--11186},
      \doiprefix\url{10.1103/physrevb.54.11169} (\bibinfo{year}{1996}).
    
    \bibitem{Kresse:1999}
    \bibinfo{author}{Kresse, G.} \& \bibinfo{author}{Joubert, D.}
    \newblock \bibinfo{journal}{\bibinfo{title}{From ultrasoft pseudopotentials to
      the projector augmented-wave method}}.
    \newblock {\emph{\JournalTitle{Physical Review B}}}
      \textbf{\bibinfo{volume}{59}}, \bibinfo{pages}{1758--1775},
      \doiprefix\url{10.1103/physrevb.59.1758} (\bibinfo{year}{1999}).
    
    \bibitem{Perdew:1996}
    \bibinfo{author}{Perdew, J.~P.}, \bibinfo{author}{Burke, K.} \&
      \bibinfo{author}{Ernzerhof, M.}
    \newblock \bibinfo{journal}{\bibinfo{title}{Generalized gradient approximation
      made simple}}.
    \newblock {\emph{\JournalTitle{Physical Review Letters}}}
      \textbf{\bibinfo{volume}{77}}, \bibinfo{pages}{3865--3868},
      \doiprefix\url{10.1103/physrevlett.77.3865} (\bibinfo{year}{1996}).
    
    \bibitem{VanpouckeDannyEP:2013aJComputChem}
    \bibinfo{author}{Vanpoucke, D. E.~P.}, \bibinfo{author}{Bultinck, P.} \&
      \bibinfo{author}{Van~Driessche, I.}
    \newblock \bibinfo{journal}{\bibinfo{title}{{Extending Hirshfeld-I to bulk and
      periodic materials}}}.
    \newblock {\emph{\JournalTitle{J. Comput. Chem.}}}
      \textbf{\bibinfo{volume}{34}}, \bibinfo{pages}{405--417},
      \doiprefix\url{10.1002/jcc.23088} (\bibinfo{year}{2013}).
    
    \bibitem{VanpouckeDannyEP:2013bJComputChem}
    \bibinfo{author}{Vanpoucke, D. E.~P.}, \bibinfo{author}{Van~Driessche, I.} \&
      \bibinfo{author}{Bultinck, P.}
    \newblock \bibinfo{journal}{\bibinfo{title}{Reply to `{C}omment on
      ``{E}xtending {H}irshfeld-{I} to bulk and periodic materials'' '}}.
    \newblock {\emph{\JournalTitle{J. Comput. Chem.}}}
      \textbf{\bibinfo{volume}{34}}, \bibinfo{pages}{422--427},
      \doiprefix\url{10.1002/jcc.23193} (\bibinfo{year}{2013}).
    
    \bibitem{BultinckHI2007}
    \bibinfo{author}{Bultinck, P.}, \bibinfo{author}{Van~Alsenoy, C.},
      \bibinfo{author}{Ayers, P.~W.} \& \bibinfo{author}{Carbó-Dorca, R.}
    \newblock \bibinfo{journal}{\bibinfo{title}{Critical analysis and extension of
      the hirshfeld atoms in molecules}}.
    \newblock {\emph{\JournalTitle{J. Chem. Phys.}}}
      \textbf{\bibinfo{volume}{126}}, \bibinfo{pages}{144111},
      \doiprefix\url{10.1063/1.2715563} (\bibinfo{year}{2007}).
    
    \bibitem{Birch1947}
    \bibinfo{author}{Birch, F.}
    \newblock \bibinfo{journal}{\bibinfo{title}{Finite elastic strain of cubic
      crystals}}.
    \newblock {\emph{\JournalTitle{Phys. Rev.}}} \textbf{\bibinfo{volume}{71}},
      \bibinfo{pages}{809--824}, \doiprefix\url{10.1103/PhysRev.71.809}
      (\bibinfo{year}{1947}).
    
    \bibitem{MCA-ACWF}
    \bibinfo{author}{Bosoni, E.} \emph{et~al.}
    \newblock \bibinfo{journal}{\bibinfo{title}{How to verify the precision of
      density-functional-theory implementations via reproducible and universal
      workflows}}.
    \newblock {\emph{\JournalTitle{Materials Cloud Archive}}}
      \doiprefix\url{https://doi.org/10.24435/materialscloud:s4-3h}
      (\bibinfo{year}{2023}).
    \newblock \bibinfo{note}{\url{https://doi.org/10.24435/materialscloud:s4-3h}}.
    
    \bibitem{Jollet:2014}
    \bibinfo{author}{Jollet, F.}, \bibinfo{author}{Torrent, M.} \&
      \bibinfo{author}{Holzwarth, N.}
    \newblock \bibinfo{journal}{\bibinfo{title}{Generation of projector
      augmented-wave atomic data: {A} 71 element validated table in the {XML}
      format}}.
    \newblock {\emph{\JournalTitle{Comput. Phys. Commun.}}}
      \textbf{\bibinfo{volume}{185}}, \bibinfo{pages}{1246--1254},
      \doiprefix\url{10.1016/j.cpc.2013.12.023} (\bibinfo{year}{2014}).
    
    \bibitem{R2test}
    \bibinfo{author}{Glantz, S.~A.}, \bibinfo{author}{Slinker, B.~K.} \&
      \bibinfo{author}{Neilands, T.~B.}
    \newblock \emph{\bibinfo{title}{Selecting the ``Best'' Regression Model, in
      Primer of Applied Regression and Analysis of Variance, third edition}}
      (\bibinfo{publisher}{McGraw-Hill Education}, \bibinfo{address}{New York},
      \bibinfo{year}{2017}).
    
    \bibitem{MV-cold-smearing}
    \bibinfo{author}{Marzari, N.}, \bibinfo{author}{Vanderbilt, D.},
      \bibinfo{author}{De~Vita, A.} \& \bibinfo{author}{Payne, M.~C.}
    \newblock \bibinfo{journal}{\bibinfo{title}{{Thermal Contraction and
      Disordering of the Al(110) Surface}}}.
    \newblock {\emph{\JournalTitle{Phys. Rev. Lett.}}}
      \textbf{\bibinfo{volume}{82}}, \bibinfo{pages}{3296--3299},
      \doiprefix\url{10.1103/PhysRevLett.82.3296} (\bibinfo{year}{1999}).
    
    \bibitem{dossantos2022}
    \bibinfo{author}{dos Santos, F.~J.} \& \bibinfo{author}{Marzari, N.}
    \newblock \bibinfo{journal}{\bibinfo{title}{Fermi energy determination for
      advanced smearing techniques}}.
    \newblock {\emph{\JournalTitle{arXiv}}} \textbf{\bibinfo{volume}{2212.07988}},
      \doiprefix\url{10.48550/ARXIV.2212.07988} (\bibinfo{year}{2022}).
    
    \bibitem{Gillan1989}
    \bibinfo{author}{Gillan, M.~J.}
    \newblock \bibinfo{journal}{\bibinfo{title}{Calculation of the vacancy
      formation energy in aluminium}}.
    \newblock {\emph{\JournalTitle{J. Phys.. Cond. Matter}}}
      \textbf{\bibinfo{volume}{1}}, \bibinfo{pages}{689},
      \doiprefix\url{10.1088/0953-8984/1/4/005} (\bibinfo{year}{1989}).
    
    \bibitem{Setten:2018}
    \bibinfo{author}{van Setten, M.} \emph{et~al.}
    \newblock \bibinfo{journal}{\bibinfo{title}{The {PseudoDojo}: Training and
      grading a 85 element optimized norm-conserving pseudopotential table}}.
    \newblock {\emph{\JournalTitle{Computer Physics Communications}}}
      \textbf{\bibinfo{volume}{226}}, \bibinfo{pages}{39--54},
      \doiprefix\url{10.1016/j.cpc.2018.01.012} (\bibinfo{year}{2018}).
    
    \bibitem{PseudoDojoSite}
    \bibinfo{title}{{The PseudoDojo website}}.
    \newblock \bibinfo{howpublished}{\url{http://www.pseudo-dojo.org/}}.
    
    \bibitem{Hartwigsen1998}
    \bibinfo{author}{Hartwigsen, C.}, \bibinfo{author}{Goedecker, S.} \&
      \bibinfo{author}{Hutter, J.}
    \newblock \bibinfo{journal}{\bibinfo{title}{{Relativistic separable dual-space
      Gaussian pseudopotentials from H to Rn}}}.
    \newblock {\emph{\JournalTitle{Phys. Rev. B}}} \textbf{\bibinfo{volume}{58}},
      \bibinfo{pages}{3641--3662}, \doiprefix\url{10.1103/PhysRevB.58.3641}
      (\bibinfo{year}{1998}).
    
    \bibitem{goedecker1996separable}
    \bibinfo{author}{Goedecker, S.}, \bibinfo{author}{Teter, M.} \&
      \bibinfo{author}{Hutter, J.}
    \newblock \bibinfo{journal}{\bibinfo{title}{{Separable dual-space Gaussian
      pseudopotentials}}}.
    \newblock {\emph{\JournalTitle{Physical Review B}}}
      \textbf{\bibinfo{volume}{54}}, \bibinfo{pages}{1703} (\bibinfo{year}{1996}).
    
    \bibitem{krack2005pseudopotentials}
    \bibinfo{author}{Krack, M.}
    \newblock \bibinfo{journal}{\bibinfo{title}{{Pseudopotentials for H to Kr
      optimized for gradient-corrected exchange-correlation functionals}}}.
    \newblock {\emph{\JournalTitle{Theoretical Chemistry Accounts}}}
      \textbf{\bibinfo{volume}{114}}, \bibinfo{pages}{145--152}
      (\bibinfo{year}{2005}).
    
    \bibitem{vandevondele2007gaussian}
    \bibinfo{author}{VandeVondele, J.} \& \bibinfo{author}{Hutter, J.}
    \newblock \bibinfo{journal}{\bibinfo{title}{Gaussian basis sets for accurate
      calculations on molecular systems in gas and condensed phases}}.
    \newblock {\emph{\JournalTitle{The Journal of chemical physics}}}
      \textbf{\bibinfo{volume}{127}}, \bibinfo{pages}{114105}
      (\bibinfo{year}{2007}).
    
    \bibitem{gpaw-setups-link}
    \bibinfo{title}{{GPAW atomic PAW setups}}.
    \newblock
      \bibinfo{howpublished}{\url{https://wiki.fysik.dtu.dk/gpaw/setups/setups.html\#atomic-paw-setups}}.
    
    \bibitem{Prandini:2018}
    \bibinfo{author}{Prandini, G.}, \bibinfo{author}{Marrazzo, A.},
      \bibinfo{author}{Castelli, I.~E.}, \bibinfo{author}{Mounet, N.} \&
      \bibinfo{author}{Marzari, N.}
    \newblock \bibinfo{journal}{\bibinfo{title}{Precision and efficiency in
      solid-state pseudopotential calculations}}.
    \newblock {\emph{\JournalTitle{npj Computational Materials}}}
      \textbf{\bibinfo{volume}{4}}, \doiprefix\url{10.1038/s41524-018-0127-2}
      (\bibinfo{year}{2018}).
    
    \bibitem{SSSP_1_3}
    \bibinfo{author}{Prandini, G.} \emph{et~al.}
    \newblock \bibinfo{journal}{\bibinfo{title}{A standard solid state
      pseudopotentials (sssp) library optimized for precision and efficiency}}.
    \newblock {\emph{\JournalTitle{Materials Cloud Archive}}}
      \textbf{\bibinfo{volume}{2023.65}} (\bibinfo{year}{2022}).
    \newblock \bibinfo{note}{Version v11,
      \url{https://doi.org/10.24435/materialscloud:f3-ym}}.
    
    \bibitem{Garcia:2018}
    \bibinfo{author}{García, A.}, \bibinfo{author}{Verstraete, M.~J.},
      \bibinfo{author}{Pouillon, Y.~n.} \& \bibinfo{author}{Junquera, J.}
    \newblock \bibinfo{journal}{\bibinfo{title}{The psml format and library for
      norm-conserving pseudopotential data curation and interoperability}}.
    \newblock {\emph{\JournalTitle{Comput. Phys. Commun.}}}
      \textbf{\bibinfo{volume}{227}}, \bibinfo{pages}{51 -- 71},
      \doiprefix\url{https://doi.org/10.1016/j.cpc.2018.02.011}
      (\bibinfo{year}{2018}).
    
    \bibitem{Topsakal:2014}
    \bibinfo{author}{Topsakal, M.} \& \bibinfo{author}{Wentzcovitch, R.}
    \newblock \bibinfo{journal}{\bibinfo{title}{Accurate projected augmented wave
      (paw) datasets for rare-earth elements (re=la--lu)}}.
    \newblock {\emph{\JournalTitle{Computational Materials Science}}}
      \textbf{\bibinfo{volume}{95}}, \bibinfo{pages}{263--270},
      \doiprefix\url{https://doi.org/10.1016/j.commatsci.2014.07.030}
      (\bibinfo{year}{2014}).
    
    \bibitem{SSSP_1_1_2}
    \bibinfo{author}{Prandini, G.} \emph{et~al.}
    \newblock \bibinfo{journal}{\bibinfo{title}{A standard solid state
      pseudopotentials (sssp) library optimized for precision and efficiency}}.
    \newblock {\emph{\JournalTitle{Materials Cloud Archive}}}
      \textbf{\bibinfo{volume}{2021.76}} (\bibinfo{year}{2021}).
    \newblock \bibinfo{note}{Version v7,
      \url{https://doi.org/10.24435/materialscloud:rz-77}}.
    
    \bibitem{SSSP_1_2}
    \bibinfo{author}{Prandini, G.} \emph{et~al.}
    \newblock \bibinfo{journal}{\bibinfo{title}{A standard solid state
      pseudopotentials (sssp) library optimized for precision and efficiency}}.
    \newblock {\emph{\JournalTitle{Materials Cloud Archive}}}
      \textbf{\bibinfo{volume}{2022.159}} (\bibinfo{year}{2022}).
    \newblock \bibinfo{note}{Version v8,
      \url{https://doi.org/10.24435/materialscloud:3v-xt}}.
    
    \bibitem{sachs2021dft}
    \bibinfo{author}{Sachs, M.} \emph{et~al.}
    \newblock \bibinfo{journal}{\bibinfo{title}{Dft-guided crystal structure
      redetermination and lattice dynamics of the intermetallic actinoid compound
      uir}}.
    \newblock {\emph{\JournalTitle{Inorganic Chemistry}}}
      \textbf{\bibinfo{volume}{60}}, \bibinfo{pages}{16686--16699}
      (\bibinfo{year}{2021}).
    
    \bibitem{dal2014pseudopotentials}
    \bibinfo{author}{Dal~Corso, A.}
    \newblock \bibinfo{journal}{\bibinfo{title}{{Pseudopotentials periodic table:
      From H to Pu}}}.
    \newblock {\emph{\JournalTitle{Computational Materials Science}}}
      \textbf{\bibinfo{volume}{95}}, \bibinfo{pages}{337--350}
      (\bibinfo{year}{2014}).
    
    \bibitem{thornig2021jureca}
    \bibinfo{author}{Th{\"o}rnig, P.}
    \newblock \bibinfo{journal}{\bibinfo{title}{{JURECA: Data Centric and Booster
      Modules implementing the Modular Supercomputing Architecture at J{\"u}lich
      Supercomputing Centre}}}.
    \newblock {\emph{\JournalTitle{Journal of large-scale research facilities
      JLSRF}}} \textbf{\bibinfo{volume}{7}}, \bibinfo{pages}{182},
      \doiprefix\url{10.17815/jlsrf-7-182} (\bibinfo{year}{2021}).
    
    \bibitem{ICSD}
    \bibinfo{author}{{Igor Levin}}.
    \newblock \bibinfo{title}{{NIST Inorganic Crystal Structure Database (ICSD),
      National Institute of Standards and Technology}} (\bibinfo{year}{2018}).
    
    \bibitem{Kokalj1999}
    \bibinfo{author}{Kokalj, A.}
    \newblock \bibinfo{journal}{\bibinfo{title}{Xcrysden—a new program for
      displaying crystalline structures and electron densities}}.
    \newblock {\emph{\JournalTitle{Journal of Molecular Graphics and Modelling}}}
      \textbf{\bibinfo{volume}{17}}, \bibinfo{pages}{176--179},
      \doiprefix\url{https://doi.org/10.1016/S1093-3263(99)00028-5}
      (\bibinfo{year}{1999}).
    
    \bibitem{Mulliken_a}
    \bibinfo{author}{Mulliken, R.~S.}
    \newblock \bibinfo{journal}{\bibinfo{title}{Electronic structures of molecules
      xi. electroaffinity, molecular orbitals and dipole moments}}.
    \newblock {\emph{\JournalTitle{J. Chem. Phys.}}} \textbf{\bibinfo{volume}{3}},
      \bibinfo{pages}{573--585}, \doiprefix\url{10.1063/1.1749731}
      (\bibinfo{year}{1935}).
    
    \bibitem{Mulliken_b}
    \bibinfo{author}{Mulliken, R.~S.}
    \newblock \bibinfo{journal}{\bibinfo{title}{Electronic population analysis on
      lcao–mo molecular wave functions. i}}.
    \newblock {\emph{\JournalTitle{J. Chem. Phys.}}} \textbf{\bibinfo{volume}{23}},
      \bibinfo{pages}{1833--1840}, \doiprefix\url{10.1063/1.1740588}
      (\bibinfo{year}{1955}).
    
    \bibitem{Hirshfeld1977}
    \bibinfo{author}{Hirshfeld, F.~L.}
    \newblock \bibinfo{journal}{\bibinfo{title}{Bonded-atom fragments for
      describing molecular charge densities}}.
    \newblock {\emph{\JournalTitle{Theor. Chim. Acta}}}
      \textbf{\bibinfo{volume}{44}}, \bibinfo{pages}{129--138},
      \doiprefix\url{10.1007/BF00549096} (\bibinfo{year}{1977}).
    
    \bibitem{Bader1991}
    \bibinfo{author}{Bader, R. F.~W.}
    \newblock \bibinfo{journal}{\bibinfo{title}{A quantum theory of molecular
      structure and its applications}}.
    \newblock {\emph{\JournalTitle{Chem. Rev.}}} \textbf{\bibinfo{volume}{91}},
      \bibinfo{pages}{893--928}, \doiprefix\url{10.1021/cr00005a013}
      (\bibinfo{year}{1991}).
    
    \bibitem{Becke1988}
    \bibinfo{author}{Becke, A.~D.}
    \newblock \bibinfo{journal}{\bibinfo{title}{A multicenter numerical integration
      scheme for polyatomic molecules}}.
    \newblock {\emph{\JournalTitle{J. Chem. Phys.}}} \textbf{\bibinfo{volume}{88}},
      \bibinfo{pages}{2547--2553}, \doiprefix\url{10.1063/1.454033}
      (\bibinfo{year}{1988}).
    
    \bibitem{HIVE}
    \bibinfo{author}{{Vanpoucke, D.E.P.}}
    \newblock \bibinfo{title}{{HIVE-tools}}.
    \newblock
      \bibinfo{howpublished}{\url{https://github.com/DannyVanpoucke/HIVE4-tools}}.
    
    \bibitem{LebedevLaikov}
    \bibinfo{author}{Lebedev, V.~I.} \& \bibinfo{author}{Laikov, D.}
    \newblock \bibinfo{journal}{\bibinfo{title}{A quadrature formula for the sphere
      of the 131st algebraic order of accuracy}}.
    \newblock {\emph{\JournalTitle{Doklady Math.}}} \textbf{\bibinfo{volume}{59}},
      \bibinfo{pages}{477--481} (\bibinfo{year}{1988}).
    
    \bibitem{BultinckHIbasisset2007}
    \bibinfo{author}{Bultinck, P.}, \bibinfo{author}{Ayers, P.~W.},
      \bibinfo{author}{Fias, S.}, \bibinfo{author}{Tiels, K.} \&
      \bibinfo{author}{{Van Alsenoy}, C.}
    \newblock \bibinfo{journal}{\bibinfo{title}{Uniqueness and basis set dependence
      of iterative hirshfeld charges}}.
    \newblock {\emph{\JournalTitle{Chem. Phys. Let.}}}
      \textbf{\bibinfo{volume}{444}}, \bibinfo{pages}{205--208},
      \doiprefix\url{https://doi.org/10.1016/j.cplett.2007.07.014}
      (\bibinfo{year}{2007}).
    
    \bibitem{WolffisJJVanpouckeDEP:MicroMesoMater2019}
    \bibinfo{author}{Wolffis, J.~J.}, \bibinfo{author}{Vanpoucke, D. E.~P.},
      \bibinfo{author}{Sharma, A.}, \bibinfo{author}{Lawler, K.~V.} \&
      \bibinfo{author}{Forster, P.~M.}
    \newblock \bibinfo{journal}{\bibinfo{title}{Predicting partial atomic charges
      in siliceous zeolites}}.
    \newblock {\emph{\JournalTitle{Microporous and Mesoporous Materials}}}
      \textbf{\bibinfo{volume}{277}}, \bibinfo{pages}{184--196},
      \doiprefix\url{10.1016/j.micromeso.2018.10.028} (\bibinfo{year}{2019}).
    
    \bibitem{BeukenVanpoucke:AngewChemIntEd2015}
    \bibinfo{author}{Bueken, B.} \emph{et~al.}
    \newblock \bibinfo{journal}{\bibinfo{title}{A {F}lexible {P}hotoactive
      {T}itanium {M}etal--{O}rganic {F}ramework {B}ased on a
      {[Ti}$^{IV}_3$($\mu_3$-{O)(O)}$_2${(COO)}$_6$] {C}luster}}.
    \newblock {\emph{\JournalTitle{Angew. Chem. Int. Ed.}}}
      \textbf{\bibinfo{volume}{54}}, \bibinfo{pages}{13912--13917},
      \doiprefix\url{10.1002/anie.201505512} (\bibinfo{year}{2015}).
    
    \bibitem{Verstraelen2012}
    \bibinfo{author}{Verstraelen, T.}, \bibinfo{author}{Sukhomlinov, S.~V.},
      \bibinfo{author}{Van~Speybroeck, V.}, \bibinfo{author}{Waroquier, M.} \&
      \bibinfo{author}{Smirnov, K.~S.}
    \newblock \bibinfo{journal}{\bibinfo{title}{Computation of charge distribution
      and electrostatic potential in silicates with the use of chemical potential
      equalization models}}.
    \newblock {\emph{\JournalTitle{J. Phys. Chem. C}}}
      \textbf{\bibinfo{volume}{116}}, \bibinfo{pages}{490--504},
      \doiprefix\url{10.1021/jp210129r} (\bibinfo{year}{2012}).
    
    \bibitem{HirBook}
    \bibinfo{title}{{2 - Chemical Periodicity and the Periodic Table}}.
    \newblock In \bibinfo{editor}{Greenwood, N.} \& \bibinfo{editor}{Earnshaw, A.}
      (eds.) \emph{\bibinfo{booktitle}{Chemistry of the Elements}},
      \bibinfo{pages}{27--28},
      \doiprefix\url{https://doi.org/10.1016/B978-0-7506-3365-9.50008-0}
      (\bibinfo{publisher}{Butterworth-Heinemann}, \bibinfo{address}{Oxford},
      \bibinfo{year}{1997}), \bibinfo{edition}{second} edn.
    
    \bibitem{volume-error}
    \bibinfo{author}{Lejaeghere, K.}, \bibinfo{author}{Vanduyfhuys, L.},
      \bibinfo{author}{Verstraelen, T.}, \bibinfo{author}{{Van Speybroeck}, V.} \&
      \bibinfo{author}{Cottenier, S.}
    \newblock \bibinfo{journal}{\bibinfo{title}{Is the error on first-principles
      volume predictions absolute or relative?}}
    \newblock {\emph{\JournalTitle{Computational Materials Science}}}
      \textbf{\bibinfo{volume}{117}}, \bibinfo{pages}{390--396},
      \doiprefix\url{https://doi.org/10.1016/j.commatsci.2016.01.039}
      (\bibinfo{year}{2016}).
    
    \bibitem{PhysRevB.12.3060}
    \bibinfo{author}{Andersen, O.~K.}
    \newblock \bibinfo{journal}{\bibinfo{title}{Linear methods in band theory}}.
    \newblock {\emph{\JournalTitle{Phys. Rev. B}}} \textbf{\bibinfo{volume}{12}},
      \bibinfo{pages}{3060--3083}, \doiprefix\url{10.1103/PhysRevB.12.3060}
      (\bibinfo{year}{1975}).
    
    \bibitem{PhysRevB.24.864}
    \bibinfo{author}{Wimmer, E.}, \bibinfo{author}{Krakauer, H.},
      \bibinfo{author}{Weinert, M.} \& \bibinfo{author}{Freeman, A.~J.}
    \newblock \bibinfo{journal}{\bibinfo{title}{{Full-potential self-consistent
      linearized-augmented-plane-wave method for calculating the electronic
      structure of molecules and surfaces: ${\mathrm{O}}_{2}$ molecule}}}.
    \newblock {\emph{\JournalTitle{Phys. Rev. B}}} \textbf{\bibinfo{volume}{24}},
      \bibinfo{pages}{864--875}, \doiprefix\url{10.1103/PhysRevB.24.864}
      (\bibinfo{year}{1981}).
    
    \bibitem{PhysRevB.43.6388}
    \bibinfo{author}{Singh, D.}
    \newblock \bibinfo{journal}{\bibinfo{title}{Ground-state properties of
      lanthanum: Treatment of extended-core states}}.
    \newblock {\emph{\JournalTitle{Phys. Rev. B}}} \textbf{\bibinfo{volume}{43}},
      \bibinfo{pages}{6388--6392}, \doiprefix\url{10.1103/PhysRevB.43.6388}
      (\bibinfo{year}{1991}).
    
    \bibitem{Michalicek20132670}
    \bibinfo{author}{Michalicek, G.}, \bibinfo{author}{Betzinger, M.},
      \bibinfo{author}{Friedrich, C.} \& \bibinfo{author}{Bl\"ugel, S.}
    \newblock \bibinfo{journal}{\bibinfo{title}{Elimination of the linearization
      error and improved basis-set convergence within the {FLAPW} method}}.
    \newblock {\emph{\JournalTitle{Computer Physics Communications}}}
      \textbf{\bibinfo{volume}{184}}, \bibinfo{pages}{2670 -- 2679},
      \doiprefix\url{http://dx.doi.org/10.1016/j.cpc.2013.07.002}
      (\bibinfo{year}{2013}).
    
    \bibitem{PhysRevB.74.045104}
    \bibinfo{author}{Friedrich, C.}, \bibinfo{author}{Schindlmayr, A.},
      \bibinfo{author}{Bl\"ugel, S.} \& \bibinfo{author}{Kotani, T.}
    \newblock \bibinfo{journal}{\bibinfo{title}{Elimination of the linearization
      error in \textit{GW} calculations based on the linearized
      augmented-plane-wave method}}.
    \newblock {\emph{\JournalTitle{Phys. Rev. B}}} \textbf{\bibinfo{volume}{74}},
      \bibinfo{pages}{045104}, \doiprefix\url{10.1103/PhysRevB.74.045104}
      (\bibinfo{year}{2006}).
    
    \bibitem{PhysRevB.83.045105}
    \bibinfo{author}{Betzinger, M.}, \bibinfo{author}{Friedrich, C.},
      \bibinfo{author}{Bl\"ugel, S.} \& \bibinfo{author}{G\"orling, A.}
    \newblock \bibinfo{journal}{\bibinfo{title}{{Local exact exchange potentials
      within the all-electron FLAPW method and a comparison with pseudopotential
      results}}}.
    \newblock {\emph{\JournalTitle{Phys. Rev. B}}} \textbf{\bibinfo{volume}{83}},
      \bibinfo{pages}{045105}, \doiprefix\url{10.1103/PhysRevB.83.045105}
      (\bibinfo{year}{2011}).
    
    \bibitem{Broeder:2019}
    \bibinfo{author}{Br\"oder, J.}, \bibinfo{author}{Wortmann, D.} \&
      \bibinfo{author}{Bl\"ugel, S.}
    \newblock \bibinfo{title}{{U}sing the {A}ii{DA}-{FLEUR} package for
      all-electron abinitio electronic structure data generation and processing in
      materials science}.
    \newblock In \emph{\bibinfo{booktitle}{In Extreme Data Workshop 2018
      Proceedings}}, vol.~\bibinfo{volume}{40} of \emph{\bibinfo{series}{IAS
      Series}}, \bibinfo{pages}{p 43--48} (\bibinfo{publisher}{Forschungszentrum
      J\"ulich}, \bibinfo{address}{J\"ulich}, \bibinfo{year}{2019}).
    
    \bibitem{AiiDA.FLEUR.1.3}
    \bibinfo{author}{Bröder, J.} \emph{et~al.}
    \newblock \bibinfo{title}{{JuDFTteam/aiida-fleur: AiiDA-FLEUR}}.
    \newblock \bibinfo{howpublished}{Zenodo},
      \doiprefix\url{10.5281/zenodo.6420726} (\bibinfo{year}{2022}).
    
    \bibitem{Madsen01}
    \bibinfo{author}{Madsen, G. K.~H.}, \bibinfo{author}{Blaha, P.},
      \bibinfo{author}{Schwarz, K.}, \bibinfo{author}{Sj\"ostedt, E.} \&
      \bibinfo{author}{Nordstr\"om, L.}
    \newblock \bibinfo{journal}{\bibinfo{title}{Efficient linearization of the
      augmented plane-wave method}}.
    \newblock {\emph{\JournalTitle{Phys.~Rev.~B}}} \textbf{\bibinfo{volume}{64}},
      \bibinfo{pages}{195134}, \doiprefix\url{10.1103/PhysRevB.64.195134}
      (\bibinfo{year}{2001}).
    
    \bibitem{Karsai17}
    \bibinfo{author}{Karsai, F.}, \bibinfo{author}{Tran, F.} \&
      \bibinfo{author}{Blaha, P.}
    \newblock \bibinfo{journal}{\bibinfo{title}{On the importance of local orbitals
      using second energy derivatives for $d$ and $f$ electrons}}.
    \newblock {\emph{\JournalTitle{Comput. Phys. Commun.}}}
      \textbf{\bibinfo{volume}{220}}, \bibinfo{pages}{230},
      \doiprefix\url{10.1016/j.cpc.2017.07.008} (\bibinfo{year}{2017}).
    
    \bibitem{lippert1997hybrid}
    \bibinfo{author}{Lippert, G.}, \bibinfo{author}{Parrinello, M.} \&
      \bibinfo{author}{Hutter, J.}
    \newblock \bibinfo{journal}{\bibinfo{title}{{A hybrid Gaussian and plane wave
      density functional scheme}}}.
    \newblock {\emph{\JournalTitle{Molecular Physics}}}
      \textbf{\bibinfo{volume}{92}}, \bibinfo{pages}{477--488}
      (\bibinfo{year}{1997}).
    
    \bibitem{GPAW_lcao}
    \bibinfo{author}{Larsen, A.~H.}, \bibinfo{author}{Vanin, M.},
      \bibinfo{author}{Mortensen, J.~J.}, \bibinfo{author}{Thygesen, K.~S.} \&
      \bibinfo{author}{Jacobsen, K.~W.}
    \newblock \bibinfo{journal}{\bibinfo{title}{{Localized atomic basis set in the
      projector augmented wave method}}}.
    \newblock {\emph{\JournalTitle{Physical Review B}}}
      \textbf{\bibinfo{volume}{80}}, \bibinfo{pages}{195112},
      \doiprefix\url{10.1103/PhysRevB.80.195112} (\bibinfo{year}{2009}).
    
    \bibitem{GPAW-setups}
    \bibinfo{title}{{Setup generation — GPAW}} (\bibinfo{year}{2022}).
    \newblock
      \bibinfo{note}{\url{https://wiki.fysik.dtu.dk/gpaw/setups/generation_of_setups.html}}.
    
    \bibitem{PhysRevB.59.1758}
    \bibinfo{author}{Kresse, G.} \& \bibinfo{author}{Joubert, D.}
    \newblock \bibinfo{journal}{\bibinfo{title}{From ultrasoft pseudopotentials to
      the projector augmented-wave method}}.
    \newblock {\emph{\JournalTitle{Phys. Rev. B}}} \textbf{\bibinfo{volume}{59}},
      \bibinfo{pages}{1758--1775}, \doiprefix\url{10.1103/PhysRevB.59.1758}
      (\bibinfo{year}{1999}).
    
    \bibitem{Hamann:2013}
    \bibinfo{author}{Hamann, D.~R.}
    \newblock \bibinfo{journal}{\bibinfo{title}{Optimized norm-conserving
      vanderbilt pseudopotentials}}.
    \newblock {\emph{\JournalTitle{Phys. Rev. B}}} \textbf{\bibinfo{volume}{88}},
      \bibinfo{pages}{085117}, \doiprefix\url{10.1103/physrevb.88.085117}
      (\bibinfo{year}{2013}).
    
    \bibitem{schlipf2015optimization}
    \bibinfo{author}{Schlipf, M.} \& \bibinfo{author}{Gygi, F.}
    \newblock \bibinfo{journal}{\bibinfo{title}{Optimization algorithm for the
      generation of oncv pseudopotentials}}.
    \newblock {\emph{\JournalTitle{Computer Physics Communications}}}
      \textbf{\bibinfo{volume}{196}}, \bibinfo{pages}{36--44}
      (\bibinfo{year}{2015}).
    
    \bibitem{garrity2014pseudopotentials}
    \bibinfo{author}{Garrity, K.~F.}, \bibinfo{author}{Bennett, J.~W.},
      \bibinfo{author}{Rabe, K.~M.} \& \bibinfo{author}{Vanderbilt, D.}
    \newblock \bibinfo{journal}{\bibinfo{title}{Pseudopotentials for
      high-throughput dft calculations}}.
    \newblock {\emph{\JournalTitle{Computational Materials Science}}}
      \textbf{\bibinfo{volume}{81}}, \bibinfo{pages}{446--452}
      (\bibinfo{year}{2014}).
    
    \bibitem{vanderbilt1990soft}
    \bibinfo{author}{Vanderbilt, D.}
    \newblock \bibinfo{journal}{\bibinfo{title}{Soft self-consistent
      pseudopotentials in a generalized eigenvalue formalism}}.
    \newblock {\emph{\JournalTitle{Physical Review B}}}
      \textbf{\bibinfo{volume}{41}}, \bibinfo{pages}{7892} (\bibinfo{year}{1990}).
    
    \end{thebibliography}

\end{document}
\fi
\fi

\end{document}